\documentclass{article}
\usepackage{epubtk}    
\usepackage{booktabs}  
\usepackage{graphicx,color}  
\usepackage{tensor,amssymb,amsmath}
\usepackage{tabularx}
\usepackage{makeidx} 

\makeindex

\newcommand{\e}{\mathrm{e}} 
\newcommand{\drm}{\mathrm{d}}
\newcommand{\irm}{\mathrm{i}} 
\newcommand{\beq}{\begin{equation}}
\newcommand{\eeq}{\end{equation}}
\newcommand{\bdm}{\begin{displaymath}}
\newcommand{\edm}{\end{displaymath}}

\DeclareFontFamily{OT1}{pzc}{}
\DeclareFontShape{OT1}{pzc}{m}{it}{<-> s * [1.10] pzcmi7t}{}
\DeclareMathAlphabet{\mathpzc}{OT1}{pzc}{m}{it}

\begin{document}

\title{Terrestrial Gravity Fluctuations}

\author{\epubtkAuthorData{Jan Harms}{%
Universit\`a degli Studi di Urbino 'Carlo Bo'\\
61029 Urbino, Italy}{%
jan.harms@fi.infn.it}{%
http://www.fi.infn.it}%
}

\date{\today}
\maketitle

\begin{abstract}
Different forms of fluctuations of the terrestrial gravity field are observed by gravity experiments. For example, atmospheric pressure fluctuations generate a gravity-noise foreground in measurements with super-conducting gravimeters. Gravity changes caused by high-magnitude earthquakes have been detected with the satellite gravity experiment GRACE, and we expect high-frequency terrestrial gravity fluctuations produced by ambient seismic fields to limit the sensitivity of ground-based gravitational-wave (GW) detectors. Accordingly, terrestrial gravity fluctuations are considered noise and signal depending on the experiment. Here, we will focus on ground-based gravimetry. This field is rapidly progressing through the development of GW detectors. The technology is pushed to its current limits in the advanced generation of the LIGO and Virgo detectors, targeting gravity strain sensitivities better than $10^{-23}\,\rm Hz^{-1/2}$ above a few tens of a Hz. Alternative designs for GW detectors evolving from traditional gravity gradiometers such as torsion bars, atom interferometers, and superconducting gradiometers are currently being developed to extend the detection band to frequencies below 1\,Hz. The goal of this article is to provide the analytical framework to describe terrestrial gravity perturbations in these experiments. Models of terrestrial gravity perturbations related to seismic fields, atmospheric disturbances, and vibrating, rotating or moving objects, are derived and analyzed. The models are then used to evaluate passive and active gravity noise mitigation strategies in GW detectors, or alternatively, to describe their potential use in geophysics. The article reviews the current state of the field, and also presents new analyses especially with respect to the impact of seismic scattering on gravity perturbations, active gravity noise cancellation, and time-domain models of gravity perturbations from atmospheric and seismic point sources. Our understanding of terrestrial gravity fluctuations will have great impact on the future development of GW detectors and high-precision gravimetry in general, and many open questions need to be answered still as emphasized in this article.
\end{abstract}

\epubtkKeywords{Terrestrial gravity, Newtonian noise, Wiener filter, Mitigation}

\tableofcontents

\newpage
\begin{center}
{\bf\Large Notation} \\[0.5cm]
\renewcommand{\arraystretch}{1.5}
\begin{tabular*}{\textwidth}{p{4.5cm}|l}
\hline\hline
$c=299792458\,$m/s & speed of light \\
$G=6.674\times 10^{-11}\,$N\,${\rm m}^2$/${\rm kg}^2$ & gravitational constant \\
\hline\hline
\end{tabular*}\\[0.5cm]

\renewcommand{\arraystretch}{1.5}
\begin{tabular*}{\textwidth}{p{4.5cm}|l}
\hline\hline
$\vec r,\, \vec e_r$ & position vector, and corresponding unit vector \\
$x,\,y,\,z$ & Cartesian coordinates \\
$r,\,\theta,\,\phi$ & spherical coordinates \\
$\varrho,\,\phi,\,z$ & cylindrical coordinates \\
$\drm\Omega\equiv\drm\phi\,\drm\theta\sin(\theta)$ & solid angle \\
$\delta_{ij}$ & Kronecker delta \\
$\delta(\cdot)$ & Dirac $\delta$ distribution \\
$\Re$ & real part of a complex number \\
$\partial_x^n$ & $n$-th partial derivative with respect to $x$ \\
$\nabla$ & nabla operator, e.~g.~$(\partial_x,\partial_y,\partial_z)$ \\
$\vec{\xi}(\vec{r},t)$ & displacement field \\
$\phi_{\rm s}(\vec{r},t)$ & potential of seismic compressional waves \\
$\vec\psi_{\rm s}(\vec{r},t)$ & potential of seismic shear waves \\
$\rho_0$ & time-averaged mass density \\
$\alpha,\,\beta$ & compressional-wave and shear-wave speed \\
$\mu$ & shear modulus \\
\hline\hline
\end{tabular*}\\[0.5cm]

\renewcommand{\arraystretch}{1.5}
\begin{tabular*}{\textwidth}{p{4.5cm}|l}
\hline\hline
$\otimes$ & dyadic product \\
$\mathbf{M},\,\vec v,\,s$ & matrix/tensor, vector, scalar \\
$P_l(x)$ & Legendre polynomial \\
$P_l^m(x)$ & associated Legendre polynomial \\
$Y_l^m(x)$ & scalar surface spherical harmonics \\
$J_n(x)$ & Bessel function of the first kind \\
$K_n(x)$ & modified Bessel function of the second kind \\
$j_n(x)$ & spherical Bessel function of the first kind \\
$Y_n(x)$ & Bessel function of the second kind \\
$y_n(x)$ & spherical Bessel function of the second kind \\
$H_n(x)$ & Hankel function or Bessel function of the third kind \\
$h^{(2)}_n(x)$ & spherical Hankel function of the second kind\\
$X_l^m$ & exterior spherical multipole moment \\
$N_l^m$ & interior spherical multipole moment \\
\hline\hline
\end{tabular*}
\end{center}
\clearpage


\section{Introduction}
\label{sec:intro} 
In the coming years, we will see a transition in the field of high-precision gravimetry from observations of slow lasting changes of the gravity field to the experimental study of fast gravity fluctuations. The latter will be realized by the advanced generation of the US-based LIGO \cite{LSC2015} and Europe-based Virgo \cite{AcEA2015} gravitational-wave (GW) detectors. Their goal is to directly observe for the first time GWs that are produced by astrophysical sources such as inspiraling and merging neutron-star or black-hole binaries. Feasibility of the laser-interferometric detector concept has been demonstrated successfully with the first generation of detectors, which, in addition to the initial LIGO and Virgo detectors, also includes the GEO600 \cite{LuEA2010} and TAMA300 \cite{Tat2008} detectors, and several prototypes around the world. The impact of these projects onto the field is two-fold. First of all, the direct detection of GWs will be a milestone in science opening a new window to our universe, and marking the beginning of a new era in observational astronomy. Second, several groups around the world have already started to adapt the technology to novel interferometer concepts \cite{DiEA2013,ShEA2014}, with potential applications not only in GW science, but also geophysics. The basic measurement scheme is always the same: the relative displacement of test masses is monitored by using ultra-stable lasers. Progress in this field is strongly dependent on how well the motion of the test masses can be shielded from the environment. Test masses are placed in vacuum and are either freely falling (e.~g.~atom clouds \cite{PCC2001}), or suspended and seismically isolated (e.~g.~high-quality glass or crystal mirrors as used in all of the detectors listed above). The best seismic isolations realized so far are effective above a few Hz, which limits the frequency range of detectable gravity fluctuations. Nonetheless, low-frequency concepts are continuously improving, and it is conceivable that future detectors will be sufficiently sensitive to detect GWs well below a Hz \cite{HaEA2013}. 

Terrestrial gravity perturbations were identified as a potential noise source already in the first concept laid out for a laser-interferometric GW detector \cite{Wei1972}. Today, this form of noise is known as ``terrestrial gravitational noise'', ``Newtonian noise'', or ``gravity-gradient noise''. It has never been observed in GW detectors, but it is predicted to limit the sensitivity of the advanced GW detectors at low frequencies. The most important source of gravity noise comes from fluctuating seismic fields \cite{Sau1984}. Gravity perturbations from atmospheric disturbances such as pressure and temperature fluctuations can become significant at lower frequencies \cite{Cre2008}. Anthropogenic sources of gravity perturbations are easier to avoid, but could also be relevant at lower frequencies \cite{ThWi1999}. Today, we only have one example of a direct observation of gravity fluctuations, i.~e.~from pressure fluctuations of the atmosphere in high-precision gravimeters \cite{Neu2010}. Therefore, almost our entire understanding of gravity fluctuations is based on models. Nonetheless, potential sensitivity limits of future large-scale GW detectors need to be identified and characterized well in advance, and so there is a need to continuously improve our understanding of terrestrial gravity noise. Based on our current understanding, the preferred option is to construct future GW detectors underground to avoid the most dominant Newtonian-noise contributions. This choice was made for the next-generation Japanese GW detector KAGRA, which is currently being constructed underground at the Kamioka site \cite{AsEA2013}, and also as part of a design study for the Einstein Telescope in Europe \cite{PuEA2010}. While the benefit from underground construction with respect to gravity noise is expected to be substantial in GW detectors sensitive above a few Hz \cite{BeEA2012}, it can be argued that it is less effective at lower frequencies \cite{HaEA2013}.   

Alternative mitigation strategies includes coherent noise cancellation \cite{Cel2000}. The idea is to monitor the sources of gravity perturbations using auxiliary sensors such as microphones and seismometers, and to use their data to generate a coherent prediction of gravity noise. This technique is successfully applied in gravimeters to reduce the foreground of atmospheric gravity noise using collocated pressure sensors \cite{Neu2010}. It is also noteworthy that the models of the atmospheric gravity noise are consistent with observations. This should give us some confidence at least that coherent Newtonian-noise cancellation can also be achieved in GW detectors. It is evident though that a model-based prediction of the performance of coherent noise cancellation schemes is prone to systematic errors as long as the properties of the sources are not fully understood. Ongoing experiments at the Sanford Underground Research Facility with the goal to characterize seismic fields in three dimensions are expected to deliver first data from an underground seismometer array in 2015 (see \cite{HaEA2010} for results from an initial stage of the experiment). While most people would argue that constructing GW detectors underground is always advantageous, it is still necessary to estimate how much is gained and whether the science case strongly profits from it. This is a complicated problem that needs to be answered as part of a site selection process. 

More recently, high-precision gravity strainmeters have been considered as monitors of geophysical signals \cite{HaEA2015}. Analytical models have been calculated, which allow us to predict gravity transients from seismic sources such as earthquakes. It was suggested to implement gravity strainmeters in existing earthquake-early warning systems to increase warning times. It is also conceivable that an alternative method to estimate source parameters using gravity signals will improve our understanding of seismic sources. Potential applications must still be investigated in greater detail, but the study already demonstrates that the idea to use GW technology to realize new geophysical sensors seems feasible. As explained in \cite{CoHa2014}, gravitational forces start to dominate the dynamics of seismic phenomena below about 1\,mHz (which coincides approximately with a similar transition in atmospheric dynamics where gravity waves start to dominate over other forms of oscillations \cite{Tos2014}). Seismic isolation would be ineffective below 1\,mHz since the gravitational acceleration of a test mass produced by seismic displacement becomes comparable to the seismic acceleration itself. Therefore, we claim that 10\,mHz is about the lowest frequency at which ground-based gravity strainmeters will ever be able to detect GWs, and consequently, modelling terrestrial gravity perturbations in these detectors can focus on frequencies above 10\,mHz.

This article is divided into six main sections. Section \ref{sec:gravmeasure} serves as an introduction to gravity measurements focussing on the response mechanisms and basic properties of gravity sensors. Section \ref{sec:ambient} describes models of gravity perturbations from ambient seismic fields. The results can be used to estimate noise spectra at the surface and underground. A subsection is devoted to the problem of noise estimation in low-frequency GW detectors, which differs from high-frequency estimates mostly in that gravity perturbations are strongly correlated between different test masses. In the low-frequency regime, the gravity noise is best described as gravity-gradient noise. Section \ref{sec:pointsources} is devoted to time domain models of transient gravity perturbations from seismic point sources. The formalism is applied to point forces and shear dislocations. The latter allows us to estimate gravity perturbations from earthquakes. Atmospheric models of gravity perturbations are presented in Section \ref{sec:atmos}. This includes gravity perturbations from atmospheric temperature fields, infrasound fields, shock waves, and acoustic noise from turbulence. The solution for shock waves is calculated in time domain using the methods of Section \ref{sec:pointsources}. A theoretical framework to calculate gravity perturbations from objects is given in Section \ref{sec:objects}. Since many different types of objects can be potential sources of gravity perturbations, the discussion focusses on the development of a general method instead of summarizing all of the calculations that have been done in the past. Finally, Section \ref{sec:mitigate} discusses possible passive and active noise mitigation strategies. Due to the complexity of the problem, most of the section is devoted to active noise cancellation providing the required analysis tools and showing limitations of this technique. Site selection is the main topic under passive mitigation, and is discussed in the context of reducing environmental noise and criteria relevant to active noise cancellation. Each of these sections ends with a summary and a discussion of open problems. While this article is meant to be a review of the current state of the field, it also presents new analyses especially with respect to the impact of seismic scattering on gravity perturbations (Sections \ref{sec:scattercomp} and \ref{sec:scattershear}), active gravity noise cancellation (Section \ref{sec:arrayNNP}), and time-domain models of gravity perturbations from atmospheric and seismic point sources (Sections \ref{sec:forcegrav}, \ref{sec:sourcehalf}, and \ref{sec:shockNN}). 

Even though evident to experts, it is worth emphasizing that all calculations carried out in this article have a common starting point, namely Newton's universal law of gravitation. It states that the attractive gravitational force $\vec F$ between two point masses $m_1,\,m_2$ is given by
\beq
\vec F=-G\dfrac{m_1m_2}{r^2}\vec e_r,
\label{eq:newtonlaw}
\eeq
where $G=6.672\times 10^{-11}\,$N\,${\rm m}^2$/${\rm kg}^2$ is the gravitational constant. Equation (\ref{eq:newtonlaw}) gives rise to many complex phenomena on Earth such as inner-core oscillations \cite{Sli1961}, atmospheric gravity waves \cite{SFV1987}, ocean waves \cite{HEG1995,You1999}, and coseismic gravity changes \cite{MaHe2011}. Due to its importance, we will honor the eponym by referring to gravity noise as Newtonian noise in the following. It is thereby clarified that the gravity noise models considered in this article are non-relativistic, and propagation effects of gravity changes are neglected. While there could be interesting scenarios where this approximation is not fully justified (e.~g.~whenever a gravity perturbation can be sensed by several sensors and differences in arrival times can be resolved), it certainly holds in any of the problems discussed in this article. We now invite the reader to enjoy the rest of the article, and hope that it proves to be useful.

\section{Gravity Measurements}
\label{sec:gravmeasure}
In this section, we describe the relevant mechanisms by which a gravity sensor can couple to gravity perturbations, and give an overview of the most widely used measurement schemes: the (relative) gravimeter \cite{CHR2013,ZhEA2011}, the gravity gradiometer \cite{MPC2002}, and the gravity strainmeter. The last category includes the large-scale GW detectors Virgo \cite{Vir2011}, LIGO \cite{LSC2010}, GEO600 \cite{LuEA2010}, KAGRA \cite{AsEA2013}, and a new generation of torsion-bar antennas currently under development \cite{AnEA2010b}. Also atom interferometers can potentially be used as gravity strainmeters in the future \cite{DiEA2008b}. Strictly speaking, none of the sensors only responds to a single field quantity (such as changes in gravity acceleration or gravity strain), but there is always a dominant response mechanism in each case, which justifies to give the sensor a specific name. A clear distinction between gravity gradiometers and gravity strainmeters has never been made to our knowledge. Therefore the sections on these two measurement principles will introduce a definition, and it is by no means the only possible one. Later on in this article, we almost exclusively discuss gravity models relevant to gravity strainmeters since the focus lies on gravity fluctuations above 10\,mHz. Today, the sensitivity near 10\,mHz of gravimeters towards gravity fluctuations is still competitive to or exceeds the sensitivity of gravity strainmeters, but this is likely going to change in the future so that we can expect strainmeters to become the technology of choice for gravity observations above 10\,mHz \cite{HaEA2013}. The following sections provide further details on this statement. Space-borne gravity experiments such as GRACE \cite{WaEA2004} will not be included in this overview. The measurement principle of GRACE is similar to that of gravity strainmeters, but only very slow changes of Earth gravity field can be observed, and for this reason it is beyond the scope of this article. 

The different response mechanisms to terrestrial gravity perturbations are summarized in Section \ref{sec:gravresp}. While we will identify the tidal forces acting on the test masses as dominant coupling mechanism, other couplings may well be relevant depending on the experiment. The Shapiro time delay will be discussed as the only relativistic effect. Higher-order relativistic effects are neglected. All other coupling mechanisms can be calculated using Newtonian theory including tidal forces, coupling in static non-uniform gravity fields, and coupling through ground displacement induced by gravity fluctuations. In Sections \ref{sec:superg} to \ref{sec:gravstrain}, the different measurement schemes are explained including a brief summary of the sensitivity limitations (choosing one of a few possible experimental realizations in each case). As mentioned before, we will mostly develop gravity models relevant to gravity strainmeters in the remainder of the article. Therefore, the detailed discussion of alternative gravimetry concepts mostly serves to highlight important differences between these concepts, and to develop a deeper understanding of the instruments and their role in gravity measurements. 

\subsection{Gravity response mechanisms}
\label{sec:gravresp}

\subsubsection{Gravity acceleration and tidal forces}
\label{sec:respgh}
We will start with the simplest mechanism of all, the acceleration of a test mass in the gravity field. Instruments that measure the acceleration are called gravimeters. A test mass inside a gravimeter can be freely falling such as atom clouds \cite{ZhEA2011} or, as suggested as possible future development, even macroscopic objects \cite{FrEA2014}. Typically though, test masses are supported mechanically or magnetically constraining motion in some of its degrees of freedom. A test mass suspended from strings responds to changes in the horizontal gravity acceleration. A test mass attached at the end of a cantilever with horizontal equilibrium position responds to changes in vertical gravity acceleration. The support fulfills two purposes. First, it counteracts the static gravitational force in a way that the test mass can respond to changes in the gravity field along a chosen degree of freedom. Second, it isolates the test mass from vibrations. Response to signals and isolation performance depend on frequency. If the support is modelled as a linear, harmonic oscillator, then the test mass response to gravity changes extends over all frequencies, but the response is strongly suppressed below the oscillators resonance frequency. The response function between the gravity perturbation $\delta g(\omega)$ and induced test mass acceleration $\delta a(\omega)$ assumes the form\index{response! gravity acceleration}
\beq
\delta a(\omega)=\frac{\omega^2}{\omega^2-\omega_0^2+\irm\gamma\omega}\delta g(\omega)\equiv R(\omega;\omega_0,\gamma)\delta g(\omega),
\label{eq:accresp}
\eeq
where we have introduced a viscous damping parameter $\gamma$, and $\omega_0$ is the resonance frequency. Well below resonance, the response is proportional to $\omega^2$, while it is constant well above resonance. Above resonance, the supported test mass responds like a freely falling 
mass, at least with respect to ``soft'' directions of the support. The test-mass response to vibrations $\delta \alpha(\omega)$ of the support is given by\index{vibration isolation}
\beq
\delta a(\omega)=\frac{\omega_0^2-\irm\gamma\omega}{\omega_0^2-\omega^2-\irm\gamma\omega}\delta \alpha(\omega)\equiv S(\omega;\omega_0,\gamma)\delta \alpha(\omega),
\label{eq:isolvibr}
\eeq
This applies for example to horizontal vibrations of the suspension points of strings that hold a test mass, or to vertical vibrations of the clamps of a horizontal cantilever with attached test mass. Well above resonance, vibrations are suppressed by $\omega^{-2}$, while no vibration isolation is provided below resonance. The situation is somewhat more complicated in realistic models of the support especially due to internal modes of the mechanical system (see for example \cite{GoSa1994}), or due to coupling of degrees of freedom \cite{MaEA2014}. Large mechanical support structures can feature internal resonances at relatively low frequencies, which can interfere to some extent with the desired performance of the mechanical support \cite{Win2002}. While Equations (\ref{eq:accresp}) and (\ref{eq:isolvibr}) summarize the properties of isolation and response relevant for this paper, details of the readout method can fundamentally impact an instrument's response to gravity fluctuations and its susceptibility to seismic noise, as explained in Sections \ref{sec:superg} to \ref{sec:gravstrain}.

Next, we discuss the response to tidal forces. In Newtonian theory, tidal forces cause a relative acceleration $\delta g_{12}(\omega)$ between two freely falling test masses according to 
\beq
\begin{split}
\delta \vec g_{12}(\omega)&= -\nabla \psi(\vec r_2,\omega)+\nabla \psi(\vec r_1,\omega)\\
&\approx-(\nabla\otimes\nabla \psi(\vec r_1,\omega))\cdot \vec r_{12},
\end{split}
\label{eq:resptide}
\eeq
where $\psi(\vec r,\omega)$ is the Fourier amplitude of the gravity potential. The last equation holds if the distance $r_{12}$ between the test masses is sufficiently small, which also depends on the frequency. The term $-\nabla\otimes\nabla \psi(\vec r,t)$ is called gravity-gradient tensor\index{gravity gradient}. In Newtonian approximation, the second time integral of this tensor corresponds to gravity strain $\mathbf h(\vec r,t)$, which is discussed in more detail in Section \ref{sec:gravstrain}. Its trace needs to vanish in empty space since the gravity potential fulfills the Poisson equation. Tidal forces produce the dominant signals in gravity gradiometers and gravity strainmeters, which measure the differential acceleration or associated relative displacement between two test masses (see Sections \ref{sec:gradio} and \ref{sec:gravstrain}). If the test masses used for a tidal measurement are supported, then typically the supports are designed to be as similar as possible, so that the response in Equation (\ref{eq:accresp}) holds for both test masses approximately with the same parameter values for the resonance frequencies (and to a lesser extent also for the damping). For the purpose of response calibration, it is less important to know the parameter values exactly if the signal is meant to be observed well above the resonance frequency where the response is approximately equal to 1 independent of the resonance frequency and damping (here, ``well above'' resonance also depends on the damping parameter, and in realistic models, the signal frequency also needs to be ``well below'' internal resonances of the mechanical support). 

\subsubsection{Shapiro time delay}\index{Shapiro time delay}
\label{sec:Shapiro}
Another possible gravity response is through the Shapiro time delay \cite{BMS2010}. This effect is not universally present in all gravity sensors, and depends on the readout mechanism. Today, the best sensitivities are achieved by reflecting laser beams from test masses in interferometric configurations. If the test mass is displaced by gravity fluctuations, then it imprints a phase shift onto the reflected laser, which can be observed in laser interferometers, or using phasemeters. We will give further details on this in Section \ref{sec:gravstrain}. In Newtonian gravity, the acceleration of test masses is the only predicted response to gravity fluctuations. However, from general relativity we know that gravity also affects the propagation of light. The leading-order term is the Shapiro time delay, which produces a phase shift of the laser beam with respect to a laser propagating in flat space. It can be calculated from the weak-field spacetime metric (see chapter 18 in \cite{MTW1973}):
\beq
\drm s^2=-(1+2\psi(\vec r,t)/c^2)(c\drm t)^2+(1-2\psi(\vec r,t)/c^2)|\drm \vec r\,|^2
\label{eq:weakfield}
\eeq
Here, $c$ is the speed of light, $\drm s$ is the so-called line element of a path in spacetime, and $\psi(\vec r,t)/c^2\ll 1$. Additionally, for this metric to hold, motion of particles in the source of the gravity potential responsible for changes of the gravity potential need to be much slower than the speed of light, and also stresses inside the source must be much smaller than its mass energy density. All conditions are fulfilled in the case of Earth gravity field. Light follows \emph{null geodesics} with $\drm s^2=0$. For the spacetime metric in Equation (\ref{eq:weakfield}), we can immediately write
\beq
\begin{split}
\left|\frac{\drm \vec r}{\drm t}\right| &= c\sqrt{\frac{1+2\psi(\vec r,t)/c^2}{1-2\psi(\vec r,t)/c^2}}\\
&\approx c(1+2\psi(\vec r,t)/c^2)
\end{split}
\label{eq:null}
\eeq
As we will find out, this equation can directly be used to calculate the time delay as an integral along a straight line in terms of the coordinates $\vec r$, but this is not immediately clear since light bends in a gravity field. So one may wonder if integration along the proper light path instead of a straight line yields additional significant corrections. The so-called geodesic equation must be used to calculate the path. It is a set of four differential equations, one for each coordinate $t,\,\vec r$ in terms of a parameter $\lambda$. The weak-field geodesic equation is obtained from the metric in Equation (\ref{eq:weakfield}):\index{geodesic equation}
\beq
\begin{split}
\frac{\drm^2 t}{\drm\lambda^2} &= -\frac{2}{c^2}\frac{\drm t}{\drm\lambda}\frac{\drm \vec r}{\drm\lambda}\cdot\nabla\psi(\vec r,t),\\
\frac{\drm^2 \vec r}{\drm\lambda^2} &= \frac{2}{c^2}\frac{\drm \vec r}{\drm\lambda}\times\left(\frac{\drm \vec r}{\drm\lambda}\times\nabla\psi(\vec r,t)\right),
\end{split}
\label{eq:geodesic}
\eeq
where we have made use of Equation (\ref{eq:null}) and the slow-motion condition $|\dot\psi(\vec r,t)|/c\ll |\nabla\psi(\vec r,t)|$. The coordinates $t,\,\vec r$ are to be understood as functions of $\lambda$. Since the deviation of a straight path is due to a weak gravity potential, we can solve these equations by perturbation theory introducing expansions $\vec r=\vec r^{\,(0)}+\vec r^{\,(1)}+\ldots$ and $t=t^{(0)}+t^{(1)}+\ldots$. The superscript indicates the order in $\psi/c^2$. The unperturbed path has the simple parametrization
\beq
\vec r^{\,(0)}(\lambda)=c\vec e_0\,\lambda+\vec r_0,\quad t^{(0)}(\lambda)=\lambda+t_0
\label{eq:zeroorder}
\eeq
We have chosen integration constants such that unperturbed time $t^{(0)}$ and parameter $\lambda$ can be used interchangeably (apart from a shift by $t_0$). Inserting these expressions into the right-hand side of Equation (\ref{eq:geodesic}), we obtain
\beq
\begin{split}
\frac{\drm^2 t^{(1)}}{\drm\lambda^2} &= -\frac{2}{c}\vec e_0\cdot\nabla\psi(\vec r^{\,(0)},t^{(0)}),\\
\frac{\drm^2 \vec r^{\,(1)}}{\drm\lambda^2} &= 2\vec e_0\times\left(\vec e_0\times\nabla\psi(\vec r^{\,(0)},t^{(0)})\right)=2\vec e_0\cdot\left(\vec e_0\cdot\nabla\psi(\vec r^{\,(0)},t^{(0)})\right)-2\nabla\psi(\vec r^{\,(0)},t^{(0)}),
\end{split}
\label{eq:pertgeo}
\eeq
As we can see, up to linear order in $\psi(\vec r,t)$, the deviation $\vec r^{\,(1)}(\lambda)$ is in orthogonal direction to the unperturbed path $\vec r^{\,(0)}(\lambda)$, which means that the deviation can be neglected in the calculation of the time delay. After some transformations, it is possible to derive Equation (\ref{eq:null}) from Equation (\ref{eq:pertgeo}), and this time we find explicitly that the right-hand-side of the equation only depends on the unperturbed coordinates \footnote{It should be emphasized that in general, the null constraint given by Equation (\ref{eq:null}) cannot be obtained from the geodesic equation since the geodesic equation is valid for all freely falling objects (massive and massless). The reason that the null constraint can be derived from Equation (\ref{eq:pertgeo}) is that we used the null constraint together with the geodesic equation to obtain Equation (\ref{eq:pertgeo}), which is therefore valid only for massless particles.}. In other words, we can integrate the time delay along a straight line as defined in Equation (\ref{eq:zeroorder}), and so the total phase integrated over a travel distance $L$ is given by
\beq
\begin{split}
\Delta\phi(\vec r_0,t_0) &= \frac{\omega_0}{c}\int\limits_0^{L/c}\drm \lambda\frac{\drm t}{\drm \lambda}\\
&= \frac{\omega_0 L}{c}-\frac{2\omega_0}{c^2}\int\limits_0^{L/c}\drm \lambda\, \psi(\vec r^{\,(0)}(\lambda),t^{(0)}(\lambda))
\end{split}
\eeq
In static gravity fields, the phase shift doubles if the light is sent back since not only the direction of integration changes, but also the sign of the expression substituted for $\drm t/\drm\lambda$. 

\subsubsection{Gravity induced ground motion}
\label{sec:gravground}
As we will learn in Section \ref{sec:ambient}, seismic fields produce gravity perturbations either through density fluctuations of the ground, or by displacing interfaces between two materials of different density. It is also well-known in seismology that seismic fields can be affected significantly by self-gravity. Self-gravity means that the gravity perturbation produced by a seismic field acts back on the seismic field. The effect is most significant at low frequency where gravity induced acceleration competes against acceleration from elastic forces. In seismology, low-frequency seismic fields are best described in terms of Earth's normal modes \cite{DaTr1998}.\index{normal modes!Earth} Normal modes exist as toroidal modes and spheroidal modes. Spheroidal modes are influenced by self-gravity, toroidal modes are not. For example, predictions of frequencies and shapes of spheroidal modes based on Earth models such as PREM (Preliminary Reference Earth Model) \cite{DzAn1981} are inaccurate if self-gravity effects are excluded. What this practically means is that in addition to displacement amplitudes, gravity becomes a dynamical variable in the elastodynamic equations that determine the normal-mode properties. Therefore, seismic displacement and gravity perturbation cannot be separated in normal-mode formalism (although self-gravity can be neglected in calculations of spheroidal modes at sufficiently high frequency). 

In certain situations, it is necessary or at least more intuitive to separate gravity from seismic fields. An exotic example is Earth's response to GWs \cite{Dys1969,CoHa2014,CoHa2014c,Ben1983,CoHa2014b}. Another example is the seismic response to gravity perturbations produced by strong seismic events at large distance to the source as described in Section \ref{sec:pointsources}. It is more challenging to analyze this scenario using normal-mode formalism. The sum over all normal modes excited by the seismic event (each of which describing a global displacement field) must lead to destructive interference of seismic displacement at large distances (where seismic waves have not yet arrived), but not of the gravity amplitudes since gravity is immediately perturbed everywhere. It can be easier to first calculate the gravity perturbation from the seismic perturbation, and then to calculate the response of the seismic field to the gravity perturbation at larger distance. This method will be adopted in this section. Gravity fields will be represented as arbitrary force or tidal fields (detailed models are presented in later sections), and we simply calculate the response of the seismic field. Normal-mode formalism can be avoided only at sufficiently high frequencies where the curvature of Earth does not significantly influence the response (i.~e.~well above 10\,mHz). In this section, we will model the ground as homogeneous half space, but also more complex geologies can in principle be assumed. 

Gravity can be introduced in two ways into the elastodynamic equations, as a conservative force $-\nabla\psi$ \cite{Run1980,Wan2005}, or as tidal strain $h$. The latter method was described first by Dyson to calculate Earth's response to GWs \cite{Dys1969}. The approach also works for Newtonian gravity, with the difference that the tidal field produced by a GW is necessarily a quadrupole field with only two degrees of freedom (polarizations), while tidal fields produced by terrestrial sources are less constrained. Certainly, GWs can only be fully described in the framework of general relativity, which means that their representation as a Newtonian tidal field cannot be used to explain all possible observations \cite{MTW1973}. Nonetheless, important here is that Dyson's method can be extended to Newtonian tidal fields. Without gravity, the elastodynamic equations for small seismic displacement can be written as
\beq
\rho\partial_t^2\vec\xi(\vec r,t)=\nabla\cdot\boldsymbol{\sigma}(\vec r,t),
\label{eq:elastic}
\eeq
where $\vec\xi(\vec r,t)$ is the seismic displacement field, and $\boldsymbol{\sigma}(\vec r,t)$ is the stress tensor \cite{AkRi2009}. In the absence of other forces, the stress is determined by the seismic field. In the case of a homogeneous and isotropic medium, the stress tensor for small seismic displacement can be written as
\beq
\begin{split}
\boldsymbol{\sigma}_\epsilon(\vec r,t) &= \lambda{\rm Tr}(\boldsymbol{\epsilon}(\vec r,t))\mathbf{1}+2\mu \boldsymbol{\epsilon}(\vec r,t)\\
\epsilon_{ij}(\vec r,t) &= \frac{1}{2}\left(\partial_i\xi_j(\vec r,t)+\partial_j\xi_i(\vec r,t)\right)
\end{split}
\label{eq:homoelast}
\eeq
The quantity $\boldsymbol{\epsilon}(\vec r,t)$ is known as seismic strain tensor\index{seismic strain}, and $\lambda,\,\mu$ are the Lam\'e constants (see Section \ref{sec:seismic})\index{Lam\'e constants}. Its trace is equal to the divergence of the displacement field. Dyson introduced the tidal field from first principles using Lagrangian mechanics, but we can follow a simpler approach. Equation (\ref{eq:homoelast}) means that a stress field builds up in response to a seismic strain field, and the divergence of the stress field acts as a force producing seismic displacement. The same happens in response to a tidal field, which we represent as gravity strain $\mathbf{h}(\vec r,t)$. A strain field changes the distance between two freely falling test masses separated by $\vec L$ by $\delta\vec L(\vec r,t)=\mathbf{h}(\vec r,t)\cdot\vec L$. For sufficiently small distances $L$, the strain field can be substituted by the second time integral of the gravity-gradient tensor $-\nabla\otimes\nabla\psi(\vec r,t)$. If the masses are not freely falling, then the strain field acts as an additional force. The corresponding contribution to the material's stress tensor can be written
\beq
\begin{split}
\boldsymbol{\sigma}_h(\vec r,t) &= -\lambda{\rm Tr}(\boldsymbol{h}(\vec r,t))\mathbf{1}-2\mu \boldsymbol{h}(\vec r,t)\\
\partial_t^2\boldsymbol{\sigma}_h(\vec r,t) &= \lambda(\Delta\psi(\vec r,t))\mathbf{1}+2\mu\nabla\otimes\nabla\psi(\vec r,t)
\end{split}
\eeq
Since we assume that the gravity field is produced by a distant source, the local contribution to gravity perturbations is neglected, which means that the gravity potential obeys the Laplace equation, $\Delta\psi(\vec r,t)=0$. Calculating the divergence of the stress tensor according to Equation (\ref{eq:elastic}), we find that the gravity term vanishes! This means that a homogeneous and isotropic medium does not respond to gravity strain fields. However, we have to be more careful here. Our goal is to calculate the response of a half-space to gravity strain. Even if the half-space is homogeneous, the Lam\'e constants change discontinuously across the surface. Hence, at the surface, the divergence of the stress tensor reads
\beq
\partial_t^2(\nabla\cdot\boldsymbol{\sigma}_h(\vec r,t))=2(\nabla\mu)\cdot(\nabla\otimes\nabla\psi(\vec r,t))=-2(\nabla\mu)\cdot\partial_t^2\boldsymbol{h}(\vec r,t)
\eeq
In other words, tidal fields produce a force onto an elastic medium via gradients in the shear modulus (second Lam\'e constant). The gradient of the shear modulus can be written in terms of a Dirac delta function, $\nabla\mu=-\mu\delta(z)\vec e_n$, for a flat surface at $z=0$ with unit normal vector $\vec e_n$. The response to gravity strain fields is obtained applying the boundary condition of vanishing surface traction\index{traction}, $\vec e_n\cdot\boldsymbol{\sigma}(\vec r,t)=0$: 
\beq
\lambda {\rm Tr}(\boldsymbol{\epsilon}(\vec r,t))\vec e_n+2\mu\,\vec e_n\cdot(\boldsymbol{\epsilon}(\vec r,t)-\boldsymbol{h}(\vec r,t))=0
\eeq
Once the seismic strain field is calculated, it can be used to obtain the seismic stress, which determines the displacement field $\vec \xi(\vec r,t)$ according to Equation (\ref{eq:elastic}). In this way, one can for example calculate that a seismometer or gravimeter can observe GWs by monitoring surface displacement as was first calculated by Dyson \cite{Dys1969}.

\subsubsection{Coupling in non-uniform, static gravity fields}\index{gravity gradient!coupling}
\label{sec:gravgrad}
If the gravity field is static, but non-uniform, then displacement $\vec\xi(t)$ of the test mass in this field due to a non-gravitational fluctuating force is associated with a changing gravity acceleration according to
\beq
\delta \vec a(\vec r,t)=(\nabla\otimes\vec g(\vec r\,))\cdot\vec\xi(t)
\label{eq:gravgrad}
\eeq 
We introduce a characteristic length $\lambda$, over which gravity acceleration varies significantly. Hence, we can rewrite the last equation in terms of the associated test-mass displacement $\zeta$
\beq
\zeta(\omega)\sim \frac{g}{\omega^2}\frac{\xi(\omega)}{\lambda},
\eeq
where we have neglected directional dependence and numerical factors. The acceleration change from motion in static, inhomogeneous fields is generally more significant at low frequencies. Let us consider the specific case of a suspended test mass. It responds to fluctuations in horizontal gravity acceleration. The test mass follows the motion of the suspension point in vertical direction (i.~e.~no seismic isolation), while seismic noise in horizontal direction is suppressed according to Equation (\ref{eq:isolvibr}). Accordingly, it is possible that the unsuppressed vertical ($z$-axis) seismic noise $\xi_z(t)$ coupling into the horizontal ($x$-axis) motion of the test mass through the term $\partial_x g_z=\partial_z g_x$ dominates over the gravity response term in Equation (\ref{eq:accresp}). Due to additional coupling mechanisms between vertical and horizontal motion in real seismic-isolation systems, test masses especially in GW detectors are also isolated in vertical direction, but without achieving the same noise suppression as in horizontal direction. For example, the requirements on vertical test-mass displacement for Advanced LIGO are a factor 1000 less stringent than on the horizontal displacement \cite{BaEA2013}. Requirements can be set on the vertical isolation by estimating the coupling of vertical motion into horizontal motion, which needs to take the gravity-gradient coupling of Equation (\ref{eq:gravgrad}) into account. Although, because of the frequency dependence, gravity-gradient effects are more significant in low-frequency detectors, such as the space-borne GW detector LISA \cite{Sch2003}.

Next, we calculate an estimate of gravity gradients in the vicinity of test masses in large-scale GW detectors, and see if the gravity-gradient coupling matters compared to mechanical vertical-to-horizontal coupling. 
\epubtkImage{}{
    \begin{figure}[htbp]
    \centerline{\includegraphics[width=0.4\textwidth]{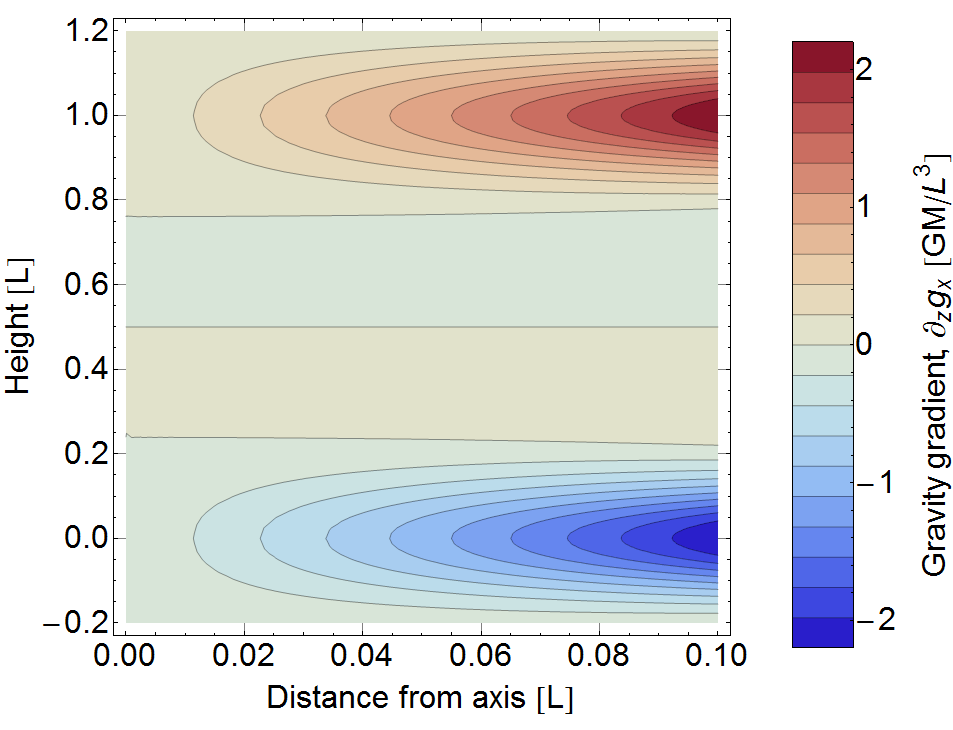}\hspace*{0.3cm}
                \includegraphics[width=0.4\textwidth]{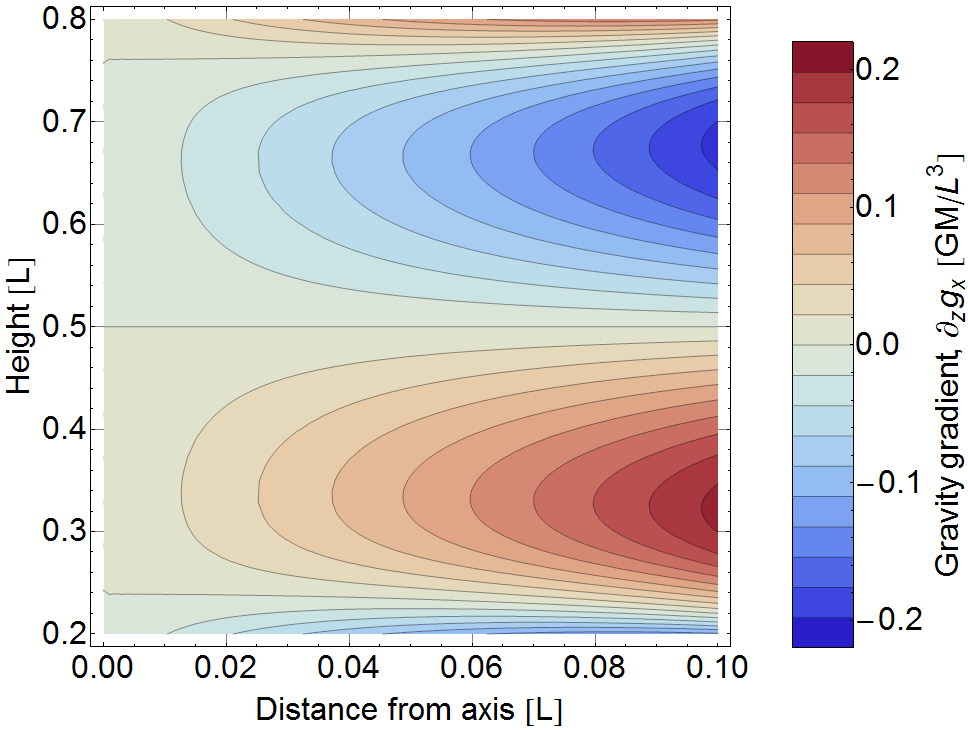}}
    \caption[Gravity gradients inside hollow cylinder]{Gravity gradients inside hollow cylinder. The total height of the cylinder is $L$, and $M$ is its total mass. The radius of the cylinder is $0.3 L$. The axes correspond to the distance of the test mass from the symmetry axis of the cylinder, and its height above one of the cylinders ends. The plot on the right is simply a zoom of the left plot into the intermediate heights.}
    \label{fig:cylingrad}
    \end{figure}}
One contribution to gravity gradients will come from the vacuum chamber surrounding the test mass. We approximate the shape of the chamber as a hollow cylinder with open ends (open ends just to simplify the calculation). In our calculation, the test mass can be offset from the cylinder axis and be located at any distance to the cylinder ends (we refer to this coordinate as height). The gravity field can be expressed in terms of elliptic integrals, but the explicit solution is not of concern here. Instead, let us take a look at the results in Figure \ref{fig:cylingrad}. Gravity gradients $\partial_z g_x$ vanish if the test mass is located on the symmetry axis or at height $L/2$. There are also two additional $\partial_z g_x=0$ contour lines starting at the symmetry axis at heights $\sim 0.24$ and $\sim 0.76$. Let us assume that the test mass is at height $0.3 L$, a distance $0.05L$ from the cylinder axis, the total mass of the cylinder is $M=5000\,$kg, and the cylinder height is $L=4\,$m. In this case, the gravity-gradient induced vertical-to-horizontal coupling factor at 20\,Hz is 
\beq
\zeta/\xi\sim 0.1 \frac{GM}{L^3\omega^2}\sim 3\times 10^{-14}
\eeq
This means that gravity-gradient induced coupling is extremely weak, and lies well below estimates of mechanical coupling (of order 0.001 in Advanced LIGO \footnote{According to pages 2 and 25 of second attachment to \url{https://alog.ligo-wa.caltech.edu/aLOG/index.php?callRep=6760}}). Even though the vacuum chamber was modelled with a very simple shape, and additional asymmetries in the mass distribution around the test mass may increase gravity gradients, it still seems very unlikely that the coupling would be significant. As mentioned before, one certainly needs to pay more attention when calculating the coupling at lower frequencies. The best procedure is of course to have a 3D model of the near test-mass infrastructure available and to use it for a precise calculation of the gravity-gradient field.

\subsection{Gravimeters}
\label{sec:superg}\index{gravimeter}
Gravimeters are instruments that measure the displacement of a test mass with respect to a non-inertial reference rigidly connected to the ground. The test mass is typically supported mechanically or magnetically (atom-interferometric gravimeters are an exception), which means that the test-mass response to gravity is altered with respect to a freely falling test mass. We will use Equation (\ref{eq:accresp}) as a simplified response model. There are various possibilities to measure the displacement of a test mass. The most widespread displacement sensors are based on capacitive readout, as for example used in superconducting gravimeters (see Figure \ref{fig:superg} and \cite{HCW2007}). Sensitive displacement measurements are in principle also possible with optical readout systems; a method that is (necessarily) implemented in atom-interferometric gravimeters \cite{PCC2001}, and prototype seismometers \cite{BeEA2014} (we will explain the distinction between seismometers and gravimeters below). As will become clear in Section \ref{sec:gravstrain}, optical readout is better suited for displacement measurements over long baselines, as required for the most sensitive gravity strain measurements, while the capacitive readout should be designed with the smallest possible distance between the test mass and the non-inertial reference \cite{JoRi1973}. 

Let us take a closer look at the basic measurement scheme of a superconducting gravimeter shown in Figure \ref{fig:superg}. The central part is formed by a spherical superconducting shell that is levitated by superconducting coils. Superconductivity provides stability of the measurement, and also avoids some forms of noise (see \cite{HCW2007} for details). In this gravimeter design, the lower coil is responsible mostly to balance the mean gravitational force acting on the sphere, while the upper coil modifies the magnetic gradient such that a certain ``spring constant" of the magnetic levitation is realized. In other words, the current in the upper coil determines the resonance frequency in Equation (\ref{eq:accresp}).
\epubtkImage{}{%
    \begin{figure}[htbp]
    \centerline{\includegraphics[width=0.6\textwidth]{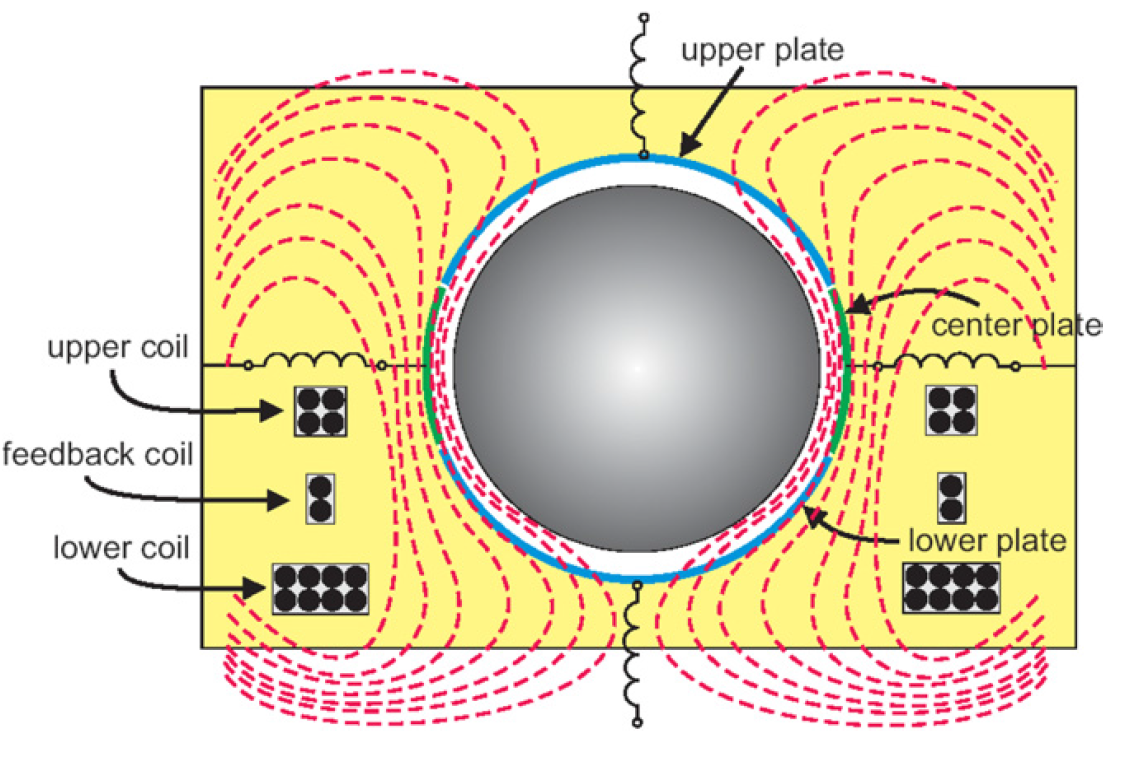}}
    \caption[Superconducting gravimeter]{Sketch of a levitated sphere serving as test mass in a superconducting gravimeter. Dashed lines indicate magnetic field lines. Coils are used for levitation and precise positioning of the sphere. From Hinderer et al \cite{HCW2007}.}
    \label{fig:superg}
    \end{figure}}
Capacitor plates are distributed around the sphere. Whenever a force acts on the sphere, the small signal produced in the capacitive readout is used to immediately cancel this force by a feedback coil. In this way, the sphere is kept at a constant location with respect to the external frame. This illustrates a common concept in all gravimeters. The displacement sensors can only respond to relative displacement between a test mass and a surrounding structure. If small gravity fluctuations are to be measured, then it is not sufficient to realize low-noise readout systems, but also vibrations of the surrounding structure forming the reference frame must be as small as possible. In general, as we will further explore in the coming sections, gravity fluctuations are increasingly dominant with decreasing frequency. At about 1\,mHz, gravity acceleration associated with fluctuating seismic fields become comparable to seismic acceleration, and also atmospheric gravity noise starts to be significant \cite{CHR2013}. At higher frequencies, seismic acceleration is much stronger than typical gravity fluctuations, which means that the gravimeter effectively operates as a seismometer. In summary, at sufficiently low frequencies, the gravimeter senses gravity accelerations of the test mass with respect to a relatively quiet reference, while at higher frequencies, the gravimeter senses seismic accelerations of the reference with respect to a test mass subject to relatively small gravity fluctuations. In superconducting gravimeters, the third important contribution to the response is caused by vertical motion $\xi(t)$ of a levitated sphere against a static gravity gradient (see Section \ref{sec:gravgrad}). As explained above, feedback control suppresses relative motion between sphere and gravimeter frame, which causes the sphere to move as if attached to the frame or ground. In the presence of a static gravity gradient $\partial_z g_z$, the motion of the sphere against this gradient leads to a change in gravity, which alters the feedback force (and therefore the recorded signal). The full contribution from gravitational, $\delta a(t)$, and seismic, $\ddot \xi(t)=\delta\alpha(t)$, accelerations can therefore be written
\beq
s(t)=\delta a(t)-\delta\alpha(t)+(\partial_z g_z)\xi(t)
\label{eq:gravdata}
\eeq
It is easy to verify, using Equations (\ref{eq:accresp}) and (\ref{eq:isolvibr}), that the relative amplitude of gravity and seismic fluctuations from the first two terms is independent of the test-mass support. Therefore, vertical seismic displacement of the reference frame must be considered fundamental noise of gravimeters and can only be avoided by choosing a quiet measurement site. Obviously, Equation (\ref{eq:gravdata}) is based on a simplified support model. One of the important design goals of the mechanical support is to minimize \emph{additional} noise due to non-linearities and cross-coupling. As is explained further in Section \ref{sec:gradio}, it is also not possible to suppress seismic noise in \emph{gravimeters} by subtracting the disturbance using data from a collocated seismometer. Doing so inevitably turns the gravimeter into a gravity gradiometer. 

Gravimeters target signals that typically lie well below 1\,mHz. Mechanical or magnetic supports of test masses have resonance frequencies at best slightly below 10\,mHz along horizontal directions, and typically above 0.1\,Hz in the vertical direction \cite{BeEA1997,WLB1999} \footnote{Winterflood explains in his thesis why vertical resonance frequencies are higher than horizontal, and why this does not necessarily have to be so \cite{Win2002}.}. Well below resonance frequency, the response function can be approximated as $\omega^2/\omega_0^2$. At first, it may look as if the gravimeter should not be sensitive to very low-frequency fluctuations since the response becomes very weak. However, the strength of gravity fluctuations also strongly increases with decreasing frequency, which compensates the small response. It is clear though that if the resonance frequency was sufficiently high, then the response would become so weak that the gravity signal would not stand out above other instrumental noise anymore. The test-mass support would be too stiff. The sensitivity of the gravimeter depends on the resonance frequency of the support and the intrinsic instrumental noise. With respect to seismic noise, the stiffness of the support has no influence as explained before (the test mass can also fall freely as in atom interferometers).
\epubtkImage{}{%
    \begin{figure}[htbp]
    \centerline{\includegraphics[width=0.6\textwidth]{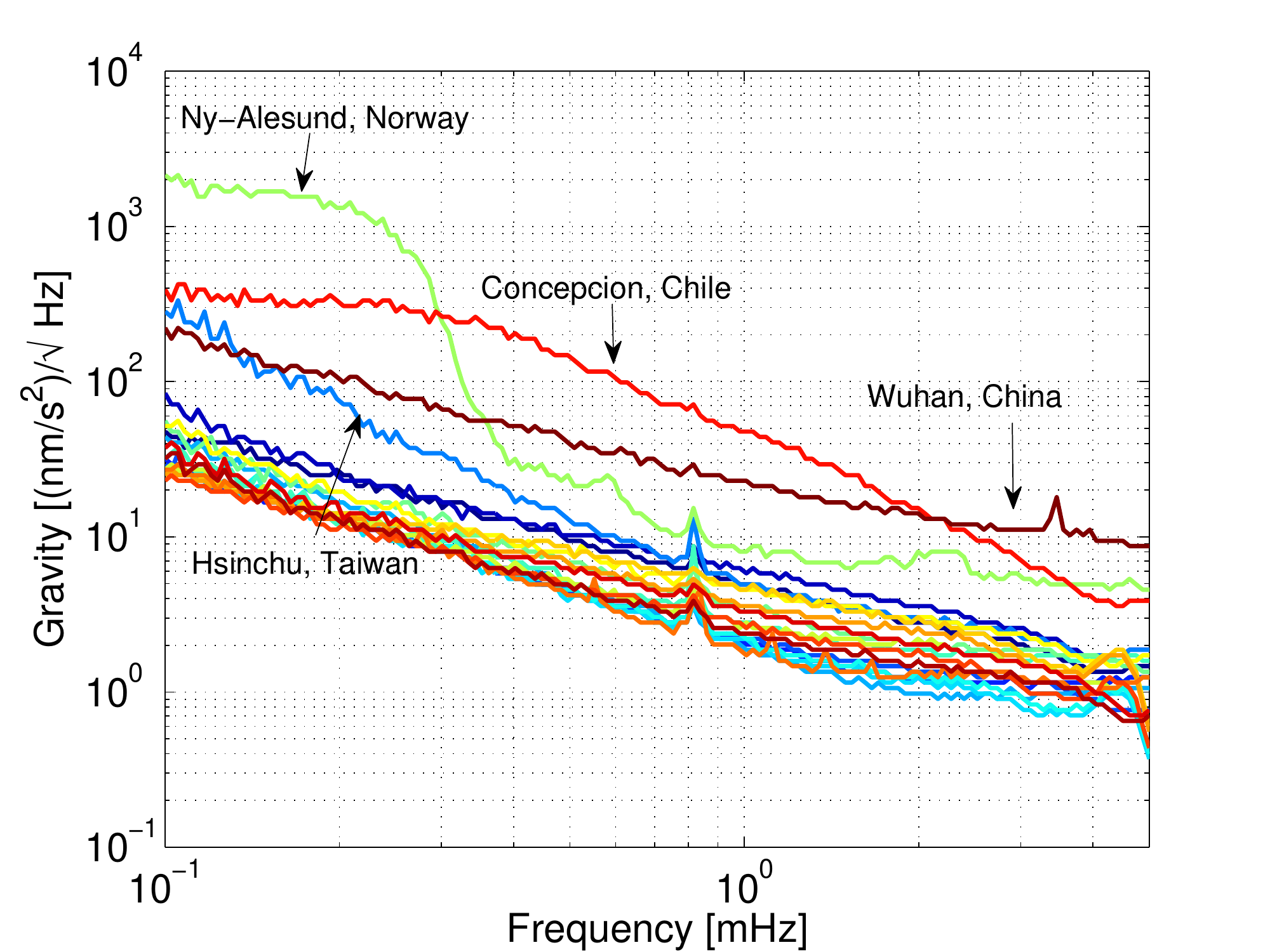}}
    \caption[Superconducting gravimeter]{Median spectra of superconducting gravimeters of the GGP \cite{CoHa2014b}.}
    \label{fig:supernoise}
    \end{figure}}
For superconducting gravimeters of the Global Geodynamics Project (GGP) \cite{CrHi2010}, the median spectra are shown in Figure \ref{fig:supernoise}. Between 0.1\,mHz and 1\,mHz, atmospheric gravity perturbations typically dominate, while instrumental noise is the largest contribution between 1\,mHz and 5\,mHz \cite{HCW2007}. The smallest signal amplitudes that have been measured by integrating long-duration signals is about $10^{-12}\,\rm m/s^2$. A detailed study of noise in superconducting gravimeters over a larger frequency range can be found in \cite{RoEA2003}. Note that in some cases, it is not fit to categorize seismic and gravity fluctuations as noise and signal. For example, Earth's spherical normal modes coherently excite seismic and gravity fluctuations, and the individual contributions in Equation (\ref{eq:gravdata}) have to be understood only to accurately translate data into normal-mode amplitudes \cite{DaTr1998}.

\subsection{Gravity gradiometers}
\label{sec:gradio}\index{gravity gradiometer}
It is not the purpose of this section to give a complete overview of the different gradiometer designs. Gradiometers find many practical applications, for example in navigation and resource exploration, often with the goal to measure static or slowly changing gravity gradients, which do not concern us here. For example, we will not discuss rotating gradiometers, and instead focus on gradiometers consisting of stationary test masses. While the former are ideally suited to measure static or slowly changing gravity gradients with high precision especially under noisy conditions, the latter design has advantages when measuring weak tidal fluctuations. In the following, we only refer to the stationary design. A gravity gradiometer measures the relative acceleration between two test masses each responding to fluctuations of the gravity field \cite{Jek2014,MPC2002}. The test masses have to be located close to each other so that the approximation in Equation (\ref{eq:resptide}) holds. The proximity of the test masses is used here as the defining property of gradiometers. They are therefore a special type of gravity strainmeter (see Section \ref{sec:gravstrain}), which denotes any type of instrument that measures relative gravitational acceleration (including the even more general concept of measuring space-time strain). 

Gravity gradiometers can be realized in two versions. First, one can read out the position of two test masses with respect to the same rigid, non-inertial reference. The two channels, each of which can be considered a gravimeter, are subsequently subtracted. This scheme is for example realized in dual-sphere designs of superconducting gravity gradiometers \cite{HaEA2000} or in atom-interferometric gravity gradiometers \cite{SoEA2014}\index{gravity gradiometer! superconducting, dual-sphere}. 
\epubtkImage{}{%
    \begin{figure}[htbp]
    \centerline{\includegraphics[width=0.6\textwidth]{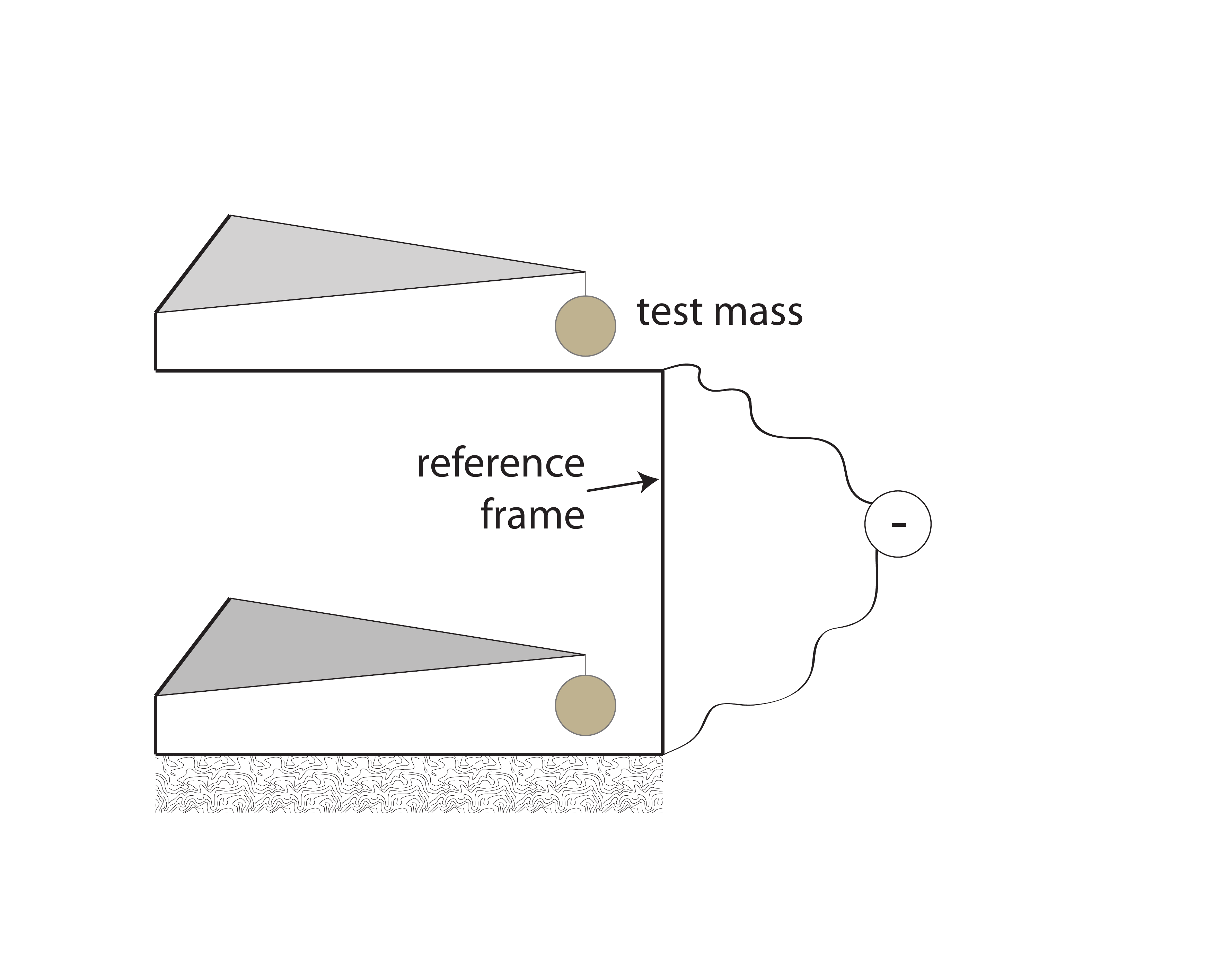}}
    \caption[Gravity gradiometer]{Basic scheme of a gravity gradiometer for measurements along the vertical direction. Two test masses are supported by horizontal cantilevers (superconducting magnets, $\ldots$). Acceleration of both test masses is measured against the same non-inertial reference frame, which is connected to the ground. Each measurement constitutes one gravimeter. Subtraction of the two channels yields a gravity gradiometer.}
    \label{fig:gradiometer}
    \end{figure}}
It is schematically shown in Figure \ref{fig:gradiometer}. Let us first consider the dual-sphere design of a superconducting gradiometer. If the reference is perfectly stiff, and if we assume as before that there are no cross-couplings between degrees of freedom and the response is linear, then the subtraction of the two gravity channels cancels all of the seismic noise, leaving only the instrumental noise and the differential gravity signal given by the second line of Equation (\ref{eq:resptide}). Even in real setups, the reduction of seismic noise can be many orders of magnitude since the two spheres are close to each other, and the two readouts pick up (almost) the same seismic noise \cite{MPC2002}. This does not mean though that gradiometers are necessarily more sensitive instruments to monitor gravity fields. A large part of the gravity signal (the common-mode part) is subtracted together with the seismic noise, and the challenge is now passed from finding a seismically quiet site to developing an instrument with lowest possible intrinsic noise. 

The atom-interferometric gradiometer differs in some important details from the superconducting gradiometer\index{gravity gradiometer!atom interferometric}. The test masses are realized by ultracold atom clouds, which are (nearly) freely falling provided that magnetic shielding of the atoms is sufficient, and interaction between atoms can be neglected. Interactions of a pair of atom clouds with a laser beam constitute the basic gravity gradiometer scheme. Even though the test masses are freely falling, the readout is not generally immune to seismic noise \cite{Har2011,BaTh2012}. The laser beam interacting with the atom clouds originates from a source subject to seismic disturbances, and interacts with optics that require seismic isolation. Schemes have been proposed that could lead to a large reduction of seismic noise \cite{NaTi2011,GrEA2013}, but their effectiveness has not been tested in experiments yet. Since the differential position (or tidal) measurement is performed using a laser beam, the natural application of atom-interferometer technology is as gravity strainmeter (as explained before, laser beams are favorable for differential position measurements over long baselines). Nonetheless, the technology is currently insufficiently developed to realize large-baseline experiments, and we can therefore focus on its application in gradiometry. Let us take a closer look at the response of atom-interferometric gradiometers to seismic noise. In atom-interferometric detectors (excluding the new schemes proposed in \cite{NaTi2011,GrEA2013}), one can show that seismic acceleration $\delta\alpha(\omega)$ of the optics or laser source limits the sensitivity of a tidal measurement according to
\beq
\delta a_{12}(\omega)\sim \frac{\omega L}{c}\delta\alpha(\omega),
\label{eq:seismicatom}
\eeq
where $L$ is the separation of the two atom clouds, and $c$ is the speed of light. It should be emphasized that the seismic noise remains, even if all optics and the laser source are all linked to the same infinitely stiff frame. In addition to this noise term, other coupling mechanisms may play a role, which can however be suppressed by engineering efforts. The noise-reduction factor $\omega L/c$ needs to be compared with the common-mode suppression of seismic noise in superconducting gravity gradiometers, which depends on the stiffness of the instrument frame, and on contamination from cross coupling of degrees-of-freedom. While the seismic noise in Equation (\ref{eq:seismicatom}) is a fundamental noise contribution in (conventional) atom-interferometric gradiometers, the noise suppression in superconducting gradiometers depends more strongly on the engineering effort (at least, we venture to claim that common-mode suppression achieved in current instrument designs is well below what is fundamentally possible). 

To conclude this section, we discuss in more detail the connection between gravity gradiometers and seismically (actively or passively) isolated gravimeters. As we have explained in Section \ref{sec:superg}, the sensitivity limitation of gravimeters by seismic noise is independent of the mechanical support of the test mass (assuming an ideal, linear support). The main purpose of the mechanical support is to maximize the response of the test mass to gravity fluctuations, and thereby increase the signal with respect to instrumental noise other than seismic noise. Here we will explain that even a seismic isolation of the gravimeter cannot overcome this noise limitation, at least not without fundamentally changing its response to gravity fluctuations. Let us first consider the case of a passively seismically isolated gravimeter. For example, we can imagine that the gravimeter is suspended from the tip of a strong horizontal cantilever. The system can be modelled as two oscillators in a chain, with a light test mass $m$ supported by a heavy mass $M$ representing the gravimeter (reference) frame, which is itself supported from a point rigidly connected to Earth. The two supports are modelled as harmonic oscillators. As before, we neglect cross coupling between degrees of freedom. Linearizing the response of the gravimeter frame and test mass for small accelerations, and further neglecting terms proportional to $m/M$, one finds the gravimeter response to gravity fluctuations:
\beq
\begin{split}
\delta a(\omega) &= R(\omega;\omega_2,\gamma_2)\left(\delta g_2(\omega)-R(\omega;\omega_1,\gamma_1)\delta g_1(\omega)\right)\\
&= R(\omega;\omega_2,\gamma_2)\left(\delta g_2(\omega)-\delta g_1(\omega)+S(\omega;\omega_1,\gamma_1)\delta g_1(\omega)\right)
\end{split}
\label{eq:gravisolpass}
\eeq
Here, $\omega_1,\,\gamma_1$ are the resonance frequency and damping of the gravimeter support, while $\omega_2,\,\gamma_2$ are the resonance frequency and damping of the test-mass support. The response and isolation functions $R(\cdot),\,S(\cdot)$ are defined in Equations (\ref{eq:accresp}) and (\ref{eq:isolvibr}). Remember that Equation (\ref{eq:gravisolpass}) is obtained as a differential measurement of test-mass acceleration versus acceleration of the reference frame. Therefore, $\delta g_1(\omega)$ denotes the gravity fluctuation at the center-of-mass of the gravimeter frame, and $\delta g_2(\omega)$ at the test mass. An infinitely stiff gravimeter suspension, $\omega_1\rightarrow\infty$, yields $R(\omega;\omega_1,\gamma_1)=0$, and the response turns into the form of the non-isolated gravimeter. The seismic isolation is determined by
\beq
\delta a(\omega)=-R(\omega;\omega_2,\gamma_2)S(\omega;\omega_1,\gamma_1)\delta \alpha(\omega)
\eeq
We can summarize the last two equations as follows. At frequencies well above $\omega_1$, the seismically isolated gravimeter responds like a gravity gradiometer, and seismic noise is strongly suppressed. The deviation from the pure gradiometer response $\sim \delta g_2(\omega)-\delta g_1(\omega)$ is determined by the same function $S(\omega;\omega_1,\gamma_1)$ that describes the seismic isolation. In other words, if the gravity gradient was negligible, then we ended up with the conventional gravimeter response, with signals suppressed by the seismic isolation function. Well below $\omega_1$, the seismically isolated gravimeter responds like a conventional gravimeter without seismic-noise reduction. If the centers of the masses $m$ (test mass) and $M$ (reference frame) coincide, and therefore $\delta g_1(\omega)=\delta g_2(\omega)$, then the response is again like a conventional gravimeter, but this time suppressed by the isolation function $S(\omega;\omega_1,\gamma_1)$. 

Let us compare the passively isolated gravimeter with an actively isolated gravimeter. In active isolation, the idea is to place the gravimeter on a stiff platform whose orientation can be controlled by actuators. Without actuation, the platform simply follows local surface motion. There are two ways to realize an active isolation. One way is to place a seismometer next to the platform onto the ground, and use its data to subtract ground motion from the platform. The actuators cancel the seismic forces. This scheme is called feed-forward noise cancellation\index{active noise cancellation!seismic}. Feed-forward cancellation of gravity noise is discussed at length in Section \ref{sec:cohcancel}, which provides details on its implementation and limitations. The second possibility is to place the seismometer together with the gravimeter onto the platform, and to suppress seismic noise in a feedback configuration \cite{AbCh2000,AbEA2004}. In the following, we discuss the feed-forward technique as an example since it is easier to analyze (for example, feedback control can be unstable \cite{AbCh2000}). As before, we focus on gravity and seismic fluctuations. The seismometer's intrinsic noise plays an important role in active isolation limiting its performance, but we are only interested in the modification of the gravimeter's response. Since there is no fundamental difference in how a seismometer and a gravimeter respond to seismic and gravity fluctuations, we know from Section \ref{sec:superg} that the seismometer output is proportional to $\delta g_1(\omega)-\delta\alpha(\omega)$, i.~e.~using a single test mass for acceleration measurements, seismic and gravity perturbations contribute in the same way. A transfer function needs to be multiplied to the acceleration signals, which accounts for the mechanical support and possibly also electronic circuits involved in the seismometer readout. To cancel the seismic noise of the platform that carries the gravimeter, the effect of all transfer functions needs to be reversed by a matched feed-forward filter. The output of the filter is then equal to $\delta g_1(\omega)-\delta\alpha(\omega)$ and is added to the motion of the platform using actuators cancelling the seismic noise and adding the seismometer's gravity signal. In this case, the seismometer's gravity signal takes the place of the seismic noise in Equation (\ref{eq:isolvibr}). The complete gravity response of the actively isolated gravimeter then reads
\beq
\delta a(\omega) = R(\omega;\omega_2,\gamma_2)(\delta g_2(\omega)-\delta g_1(\omega))
\label{eq:gravisolact}
\eeq
The response is identical to a gravity gradiometer, where $\omega_2,\gamma_2$ are the resonance frequency and damping of the gravimeter's test-mass support. In reality, instrumental noise of the seismometer will limit the isolation performance and introduce additional noise into Equation (\ref{eq:gravisolact}). Nonetheless, Equations (\ref{eq:gravisolpass}) and (\ref{eq:gravisolact}) show that any form of seismic isolation turns a gravimeter into a gravity gradiometer at frequencies where seismic isolation is effective. For the passive seismic isolation, this means that the gravimeter responds like a gradiometer at frequencies well above the resonance frequency $\omega_1$ of the gravimeter support, while it behaves like a conventional gravimeter below $\omega_1$. From these results it is clear that the design of seismic isolations and the gravity response can in general not be treated independently. As we will see in Section \ref{sec:gravstrain} though, tidal measurements can profit strongly from seismic isolation especially when common-mode suppression of seismic noise like in gradiometers is insufficient or completely absent.

\subsection{Gravity strainmeters}
\label{sec:gravstrain}\index{strainmeter!gravity}
Gravity strain is an unusual concept in gravimetry that stems from our modern understanding of gravity in the framework of general relativity. From an observational point of view, it is not much different from elastic strain. Fluctuating gravity strain causes a change in distance between two freely falling test masses, while seismic or elastic strain causes a change in distance between two test masses bolted to an elastic medium. Fundamentally, gravity strain corresponds to a perturbation of the metric that determines the geometrical properties of spacetime \cite{MTW1973}. It should be emphasized though that there are important differences between seismic and gravity strain, which can play a role in certain experiments \cite{KaCh2004}. To understand this better, we need to talk briefly about GWs, before we can return to a Newtonian description of gravity strain.

Gravitational waves are weak perturbations of spacetime propagating at the speed of light\index{gravitational wave}. Freely falling test masses change their distance in the field of a GW. When the length of the GW is much larger than the separation between the test masses, it is possible to interpret this change as if caused by a Newtonian force. We call this the long-wavelength regime. Since we are interested in the low-frequency response of gravity strainmeters throughout this article (i.~e.~frequencies well below 100\,Hz), this condition is always fulfilled for Earth-bound experiments. The effect of a gravity-strain field $\mathbf h(\vec r,t)$ on a pair of test masses can then be represented as an equivalent Newtonian tidal field
\beq
\delta a_{12}(\vec r,t)=\frac{1}{2}L \vec e_{12}^\top\cdot\ddot{\mathbf h}(\vec r,t)\cdot \vec e_{12}
\label{eq:tidalstrain}
\eeq
Here, $\delta a_{12}(\vec r,t)$ is the relative acceleration between two freely falling test masses, $L$ is the distance between them, and $\vec e_{12}$ is the unit vector pointing from one to the other test mass. As can be seen, the gravity-strain field is represented by a $3\times 3$ tensor. It contains the space-components of a 4-dimensional metric perturbation of spacetime, and determines all properties of GWs \footnote{In order to identify components of the metric perturbation with tidal forces acting on test masses, one needs to choose specific spacetime coordinates, the so-called transverse-traceless gauge \cite{MTW1973}.}. Note the factor $1/2$ in Equation (\ref{eq:tidalstrain}), which is a consequence of $\mathbf h(\vec r,t)$ being defined as the space components of a metric perturbation. This factor does not appear in similar equations for example of seismic strain. 

The strain field of a GW takes the form of a quadrupole oscillation with two possible polarizations commonly denoted $\times$(cross)-polarization and $+$(plus)-polarization. The arrows in Figure \ref{fig:polarizeGW} indicate the lines of the equivalent tidal field of Equation (\ref{eq:tidalstrain}).
\epubtkImage{}{%
    \begin{figure}[htbp]
    \centerline{\includegraphics[width=0.6\textwidth]{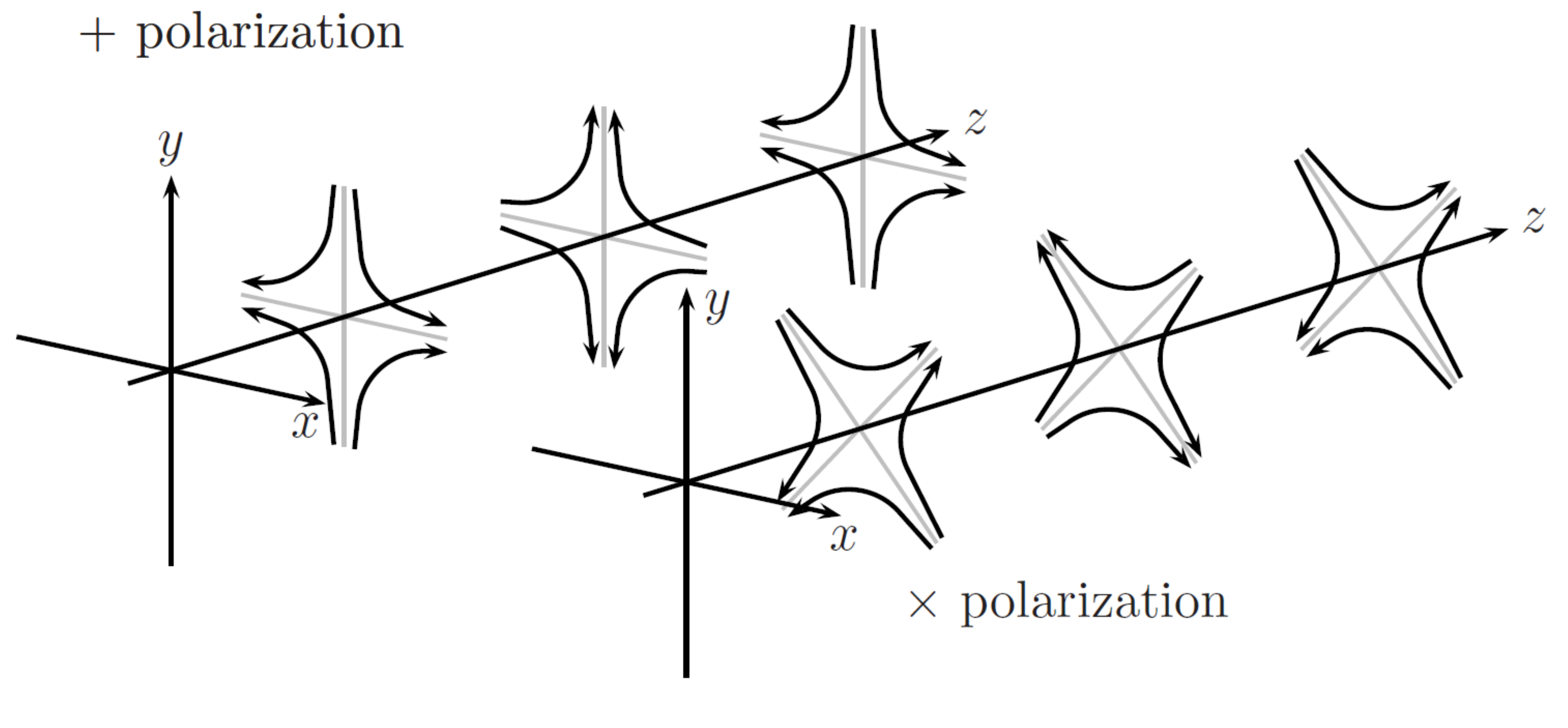}}
    \caption[Polarizations of a gravitational wave]{Polarizations of a gravitational wave.}
    \label{fig:polarizeGW}
    \end{figure}}
Consequently, to (directly) observe GWs, one can follow two possible schemes: (1) the conventional method, which is a measurement of the \emph{relative displacement} of suspended test masses typically carried out along two perpendicular baselines (arms); and (2) measurement of the \emph{relative rotation} between two suspended bars. Figure \ref{fig:measureGW} illustrates the two cases. In either case, the response of a gravity strainmeter is obtained by projecting the gravity strain tensor onto a combination of two unit vectors, $\vec e_1$ and $\vec e_2$, that characterize the orientation of the detector, such as the directions of two bars in a rotational gravity strain meter, or of two arms of a conventional gravity strain meter. This requires us to define two different gravity strain projections. The projection for the rotational strain measurement is given by
\begin{equation}
h_\times(\vec r\,,t)=(\vec e_1^{\top}\cdot {\mathbf h}(\vec r\,,t)\cdot\vec e_1^{\,\rm r}-\vec e_2^{\top}\cdot {\mathbf h}(\vec r\,,t)\cdot\vec e_2^{\,\rm r})/2,
\label{eq:hx}
\end{equation}
where the subscript $\times$ indicates that the detector responds to the $\times$-polarization assuming that the $x,y$-axes (see Figure \ref{fig:polarizeGW}) are oriented along two perpendicular bars. The vectors $\vec e_1^{\,\rm r}$ and $\vec e_2^{\,\rm r}$ are rotated counter-clockwise by 90$^\circ$ with respect to $\vec e_1$ and $\vec e_2$. In the case of perpendicular bars $\vec e_1^{\,\rm r}=\vec e_2$ and $\vec e_2^{\,\rm r}=-\vec e_1$. The corresponding projection for the conventional gravity strain meter reads
\begin{equation}
h_+(\vec r\,,t)=(\vec e_1^{\top}\cdot {\mathbf h}(\vec r\,,t)\cdot\vec e_1-\vec e_2^{\top}\cdot {\mathbf h}(\vec r\,,t)\cdot\vec e_2)/2
\label{eq:hp}
\end{equation}
The subscript $+$ indicates that the detector responds to the $+$-polarization provided that the $x,\,y$-axes are oriented along two perpendicular baselines (arms) of the detector. The two schemes are shown in Figure \ref{fig:measureGW}.
\epubtkImage{}{%
    \begin{figure}[htbp]
    \centerline{\includegraphics[width=0.4\textwidth]{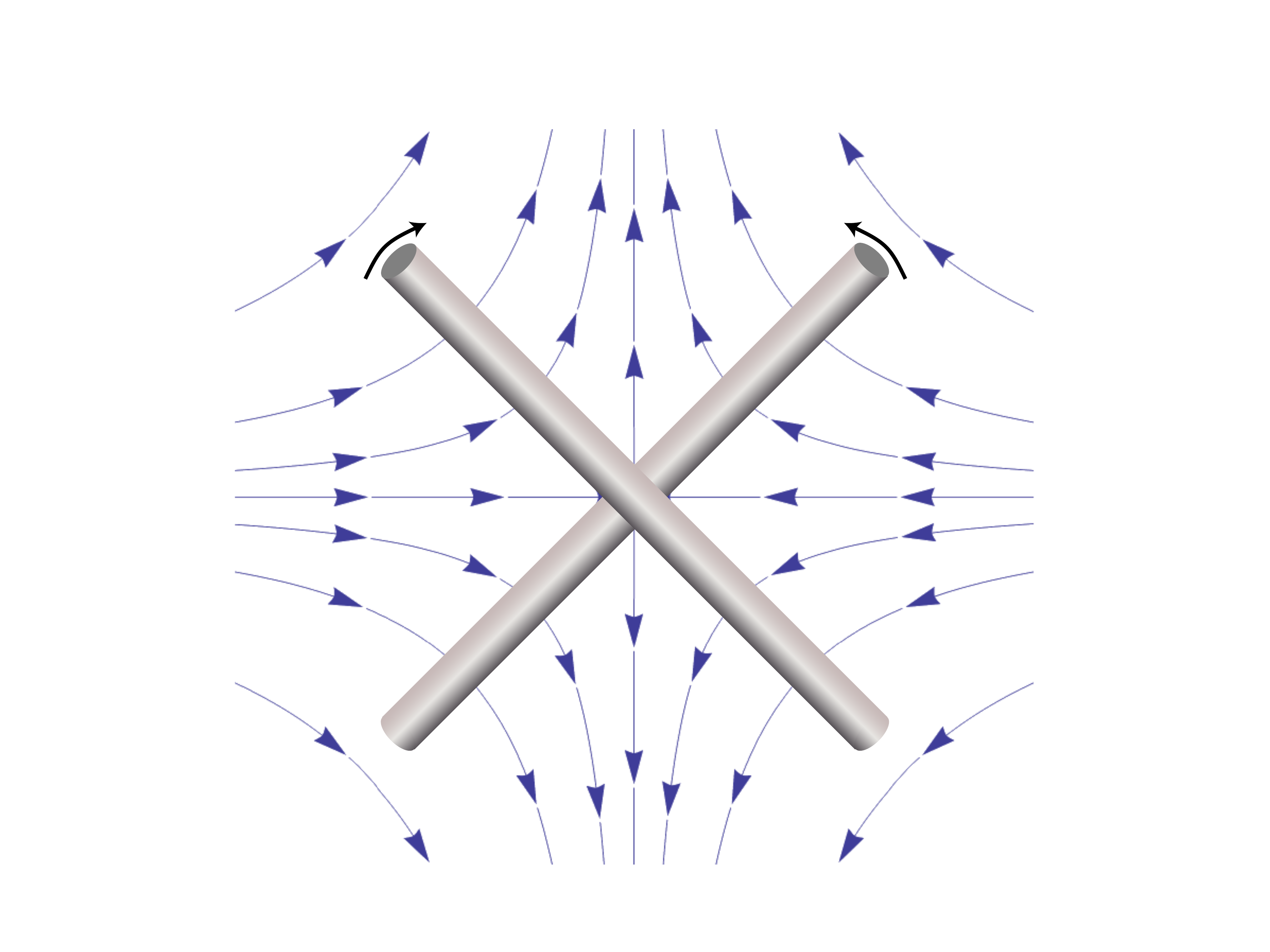}
                \includegraphics[width=0.4\textwidth]{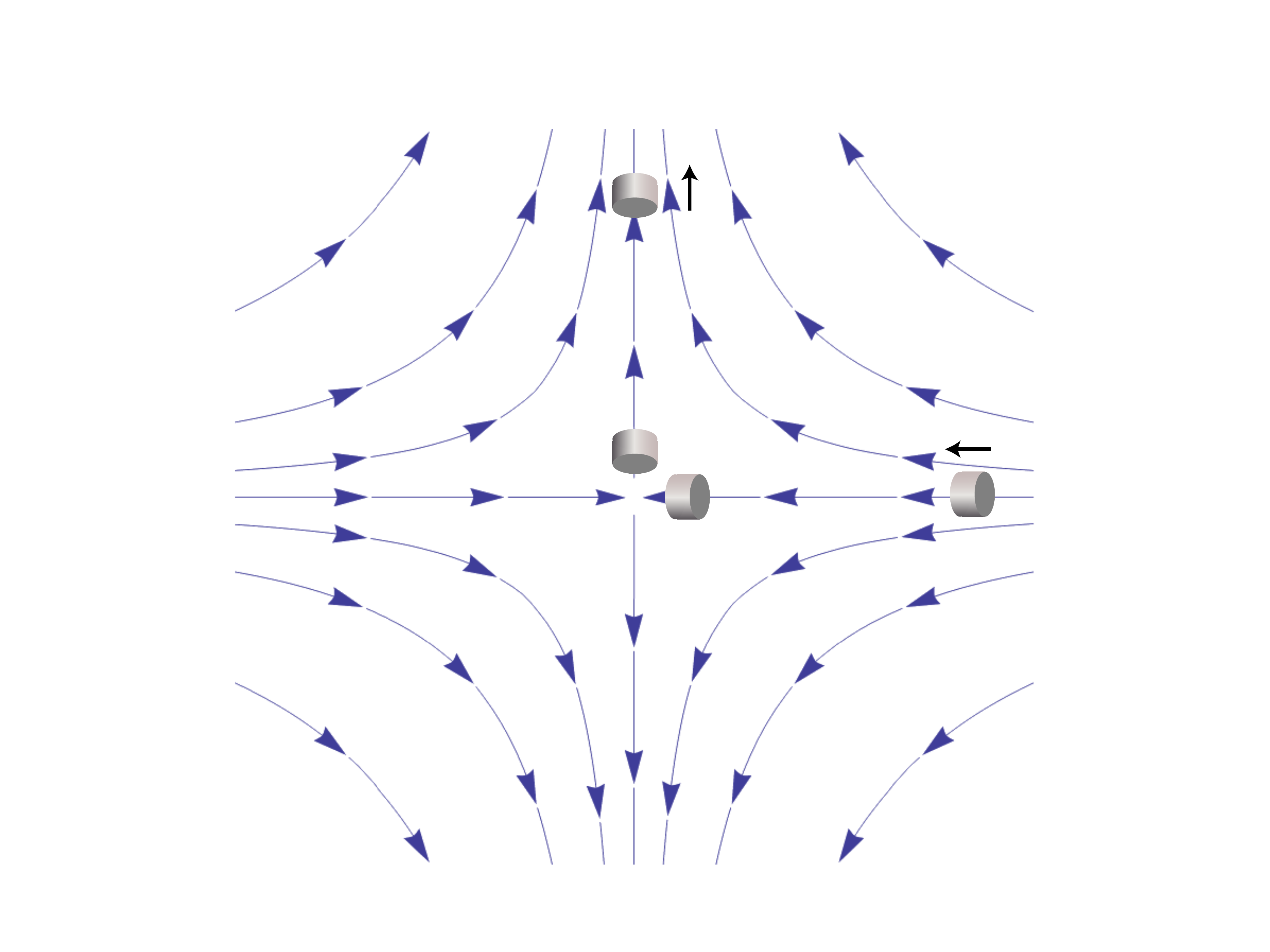}}
    \caption[Gravity strainmeter]{Sketches of the relative rotational and displacement measurement schemes.}
    \label{fig:measureGW}
    \end{figure}}
The most sensitive GW detectors are based on the conventional method, and distance between test masses is measured by means of laser interferometry. The LIGO and Virgo detectors have achieved strain sensitivities of better than $10^{-22}$\,Hz$^{-1/2}$ between about 50\,Hz and 1000\,Hz in past science runs and are currently being commissioned in their advanced configurations \cite{LSC2010,AcEA2015}. The rotational scheme is realized in torsion-bar antennas, which are considered as possible technology for sub-Hz GW detection \cite{ShEA2014,EdEA2014}\index{torsion-bar antenna}. However, with achieved strain sensitivity of about $10^{-8}$\,Hz$^{-1/2}$ near 0.1\,Hz, the torsion-bar detectors are far from the sensitivity we expect to be necessary for GW detection \cite{HaEA2013}. 

Let us now return to the discussion of the previous sections on the role of seismic isolation and its impact on gravity response. Gravity strainmeters profit from seismic isolation more than gravimeters or gravity gradiometers. We have shown in Section \ref{sec:superg} that seismically isolated gravimeters are effectively gravity gradiometers. So in this case, seismic isolation changes the response of the instrument in a fundamental way, and it does not make sense to talk of seismically isolated gravimeters. Seismic isolation could in principle be beneficial for gravity gradiometers (i.~e.~the acceleration of two test masses is measured with respect to a common rigid, seismically isolated reference frame), but the common-mode rejection of seismic noise (and gravity signals) due to the differential readout is typically so high that other instrumental noise becomes dominant. So it is possible that some gradiometers would profit from seismic isolation, but it is not generally true. Let us now consider the case of a gravity strainmeter. As explained in Section \ref{sec:gradio}, we distinguish gradiometers and strainmeters by the distance of their test masses. For example, the distance of the LIGO or Virgo test masses is 4\,km and 3\,km respectively. Seismic noise and terrestrial gravity fluctuations are insignificantly correlated between the two test masses within the detectors' most sensitive frequency band (above 10\,Hz). Therefore, the approximation in Equation (\ref{eq:resptide}) does not apply. Certainly, the distinction between gravity gradiometers and strainmeters remains somewhat arbitrary since at any frequency the approximation in Equation (\ref{eq:resptide}) can hold for one type of gravity fluctuation, while it does not hold for another. Let us adopt a more practical definition at this point. \emph{Whenever the design of the instrument places the test masses as distant as possible from each other given current technology, then we call such an instrument strainmeter}\index{strainmeter!gravity (practical definition)}. In the following, we will discuss seismic isolation and gravity response for three strainmeter designs, the laser-interferometric, atom-interferometric, and superconducting strainmeters. It should be emphasized that the atom-interferometric and superconducting concepts are still in the beginning of their development and have not been realized yet with scientifically interesting sensitivities. 

\paragraph{Laser-interferometric strainmeters} The most sensitive gravity strainmeters, namely the large-scale GW detectors, use laser interferometry to read out the relative displacement between mirror pairs forming the test masses. Each test mass in these detectors is suspended from a seismically isolated platform, with the suspension itself providing additional seismic isolation. Section \ref{sec:respgh} introduced a simplified response and isolation model based on a harmonic oscillator characterized by a resonance frequency $\omega_0$ and viscous damping $\gamma$ \footnote{In reality, the dominant damping mechanism in suspension systems is not viscous damping, but structural damping characterized by the so-called loss angle $\phi$, which quantifies the imaginary part of the elastic modulus \cite{Sau1992}.}. In a multi-stage isolation and suspension system as realized in GW detectors (see for example \cite{BrEA2005,MaEA2014}), coupling between multiple oscillators cannot be neglected, and is fundamental to the seismic isolation performance, but the basic features can still be explained with the simplified isolation and response model of Equations (\ref{eq:accresp}) and (\ref{eq:isolvibr}). The signal output of the interferometer is proportional to the relative displacement between test masses. Since seismic noise is approximately uncorrelated between two distant test masses, the differential measurement itself cannot reject seismic noise as in gravity gradiometers. Without seismic isolation, the dominant signal would be seismic strain, i.~e.~the distance change between test masses due to elastic deformation of the ground, with a value of about $10^{-15}$\,Hz$^{-1/2}$ at 50\,Hz (assuming kilometer-scale arm lengths)\index{seismic strain}. At the same time, without seismically isolated test masses, the gravity signal can only come from the ground response to gravity fluctuations as described in Section \ref{sec:gravground}, and from the Shapiro time delay as described in Section \ref{sec:Shapiro}. These signals would lie well below the seismic noise.  Consequently, to achieve the sensitivities of past science runs, the seismic isolation of the large-scale GW detectors had to suppress seismic noise by at least 7 orders of magnitude, and test masses had to be supported so that they can (quasi-)freely respond to gravity-strain fluctuations in the targeted frequency band (which, according to Equations (\ref{eq:accresp}) and (\ref{eq:isolvibr}), is achieved automatically with the seismic isolation). Stacking multiple stages of seismic isolation enhances the gravity response negligibly, while it is essential to achieve the required seismic-noise suppression. Using laser beams, long-baseline strainmeters can be realized, which increases the gravity response according to Equation (\ref{eq:resptide}). The price to be paid is that seismic noise needs to be suppressed by a sophisticated isolation and suspension system since it is uncorrelated between test masses and therefore not rejected in the differential measurement. As a final note, the most sensitive torsion-bar antennas also implement a laser-interferometric readout of the relative rotation of the suspended bars \cite{ShEA2014}, and concerning the gravity response and seismic isolation, they can be modelled very similarly to conventional strainmeters. However, the suppression of seismic noise is impeded by mechanical cross-coupling, since a torsion bar has many soft degrees of freedom that can interact resonantly within the detection band. This problem spoils to some extent the big advantage of torsion bars to realize a very low-frequency torsion resonance, which determines the fundamental response and seismic isolation performance\index{torsion-bar antenna}. Nonetheless, cross-coupling can in principle be reduced by precise engineering, and additional seismic pre-isolation of the suspension point of the torsion bar can lead to significant noise reduction.

\paragraph{Atom-interferometric strainmeters} \index{strainmeter!gravity (atom interferometric)}In this design, the test masses consist of freely-falling ultracold atom clouds. A laser beam interacting with the atoms serves as a common phase reference, which the test-mass displacement can be measured against. The laser phase is measured locally via atom interferometry by the same freely-falling atom clouds \cite{ChEA2008}. Subtraction of two of these measurements forms the strainmeter output. The gravity response is fundamentally the same as for the laser-interferometric design since it is based on the relative displacement of atom clouds. Seismic noise couples into the strain measurement through the laser. If displacement noise of the laser or laser optics has amplitude $\xi(\omega)$, then the corresponding strain noise in atom-interferometric strainmeters is of order $\omega \xi(\omega)/c$, where $c$ is the speed of light, and $\omega$ the signal frequency \cite{BaTh2012}. While this noise is lower than the corresponding term $\xi(\omega)/L$ in laser-interferometric detectors ($L$ being the distance between test masses), seismic isolation is still required. As we know from previous discussions, seismic isolation causes the optics to respond to gravity fluctuations. However, the signal contribution from the optics is weaker by a factor $\omega L/c$ compared to the contribution from distance changes between atom clouds. Here, $L$ is the distance between two freely-falling atom clouds, which also corresponds approximately to the extent of the optical system. This signal suppression is very strong for any Earth-bound atom-interferometric detector (targeting sub-Hz gravity fluctuations), and we can neglect signal contributions from the optics. Here we also assumed that there are no control forces acting on the optics, which could further suppress their signal response, if for example the distance between optics is one of the controlled parameters. Nonetheless, seismic isolation is required, not only to suppress seismic noise from distance changes between laser optics, which amounts to $\omega \xi(\omega)/c\sim 10^{-17}\,$Hz$^{-1/2}$ at 0.1\,Hz without seismic isolation (too high at least for GW detection \cite{HaEA2013}), but also to suppress seismic-noise contributions through additional channels (e.~g.~tilting optics in combination with laser-wavefront aberrations \cite{HoEA2011b}). The additional channels dominate in current experiments, which are already seismic-noise limited with strain noise many orders of magnitude higher than $10^{-17}\,$Hz$^{-1/2}$ \cite{DiEA2013}. It is to be expected though that improvements of the atom-interferometer technology will suppress the additional channels relaxing the requirement on seismic isolation. 

\paragraph{Superconducting strainmeters} \index{strainmeter!gravity (superconducting)} The response of superconducting strainmeters to gravity-strain fluctuations is based on the differential displacement of magnetically levitated spheres. The displacement of individual spheres is monitored locally via a capacitive readout (see Section \ref{sec:superg}). Subtracting local readouts of test-mass displacement from each other constitutes the basic strainmeter scheme \cite{Pai1976}. The common reference for the local readouts is a rigid, material frame. The stiffness of the frame is a crucial parameter facilitating the common-mode rejection of seismic noise. Even in the absence of seismic noise, the quality of the reference frame is ultimately limited by thermally excited vibrations\index{thermal noise} of the frame \footnote{It should not be forgotten that thermal noise also plays a role in the other two detector designs, but it is a more severe problem for superconducting gravimeters since the mechanical structure supporting the thermal vibrations is much larger. Any method to lower thermal noise, such as cooling of the structure, or lowering its mechanical loss is a greater effort.} (similar to the situation with torsion-bar antennas \cite{HaEA2013}). However, since strainmeters are very large (by definition), vibrational eigenmodes of the frame can have low resonance frequencies impeding the common-mode rejection of seismic noise. In fact, it is unclear if a significant seismic-noise reduction can be achieved by means of mechanical rigidity. Therefore, seismic isolation of the strainmeter frame is necessary. In this case, each local readout is effectively a gravity-strain measurement, since the gravity response of the test mass is measured against a reference frame that also responds to gravity fluctuations (see discussion of seismically isolated gravimeters in Section \ref{sec:gradio}). Another solution could be to substitute the mechanical structure by an optically rigid body as suggested in \cite{HaEA2013} for a low-frequency laser-interferometric detector. The idea is to connect different parts of a structure via laser links in all degrees of freedom. The stiffness of the link is defined by the control system that forces the different parts to keep their relative positions and orientations. Optical rigidity in all degrees of freedom has not been realized experimentally yet, but first experiments known as suspension point or platform interferometers have been conducted to control some degrees of freedom in the relative orientation of two mechanical structures \cite{AsEA2004,DaEA2012}. This approach would certainly add complexity to the experiment, especially in full-tensor configurations of superconducting gravity strainmeters, where six different mechanical structures have to be optically linked \cite{MPC2002}.

\section{Gravity Perturbations from Seismic Fields}
\label{sec:ambient} 
\index{Newtonian noise!seismic}

Already in the first design draft of a laser-interferometric GW detector laid out by Rainer Weiss, gravity perturbations from seismic fields were recognized as a potential noise contribution \cite{Wei1972}. He expressed the transfer function between ground motion and gravitational displacement noise of a test mass as effective isolation factor, highlighting the fact that gravitational coupling can be understood as additional link that circumvents seismic isolation. The equations that he used already had the correct dependence on ground displacement, density and seismic wavelength, but it took another decade, before Peter Saulson presented a more detailed calculation of numerical factors \cite{Sau1984}. He divided the half space below a test mass into volumes of correlated density fluctuations, and assigned a mean displacement to each of these volumes. Fluctuations were assumed to be uncorrelated between different volumes. The total gravity perturbation was then obtained as an incoherent sum over these volumes. The same scheme was carried out for gravity perturbations associated with vertical surface displacement. The sizes of volumes and surface areas of correlated density perturbations were determined by the length of seismic waves, but Saulson did not make explicit use of the wave nature of the seismic field that produces the density perturbations. As a result, also Saulson had to concede that certain steps in his calculation ``cannot be regarded as exact''. The next step forward was marked by two papers that were published almost simultaneously by groups from the LIGO and Virgo communities \cite{HuTh1998,BeEA1998}. In these papers, the wave nature of the seismic field was taken into account, producing for the first time accurate predictions of Newtonian noise. They understood that the dominant contribution to Newtonian noise would come from seismic surface waves, more specifically Rayleigh waves. The Rayleigh field produces density perturbations beneath the surface, and correlated surface displacement at the same time. The coherent summation of these effects was directly obtained, and since then, models of Newtonian noise from Rayleigh waves have not improved apart from a simplification of the formalism. 

Nonetheless, Newtonian-noise models are not only important to estimate a noise spectrum with sufficient accuracy. More detailed models are required to analyze Newtonian-noise mitigation, which is discussed in Section \ref{sec:mitigate}. Especially the effect of seismic scattering on gravity perturbations needs to be quantified. A first analytical calculation of gravity perturbations from seismic waves scattered from a spherical cavity is presented in Sections \ref{sec:scattercomp} and \ref{sec:scattershear}. In general, much of the recent research on Newtonian-noise modelling was carried out to identify possible limitations in Newtonian-noise mitigation. Among others, this has led to two major new developments in the field. First, finite-element simulations were added to the set of tools \cite{HaEA2009a,BeEA2010c}. We will give a brief summary in Section \ref{sec:numsim}. The advantage is that several steps of a complex analysis can be combined such as simulations of a seismic field, simulations of seismic measurements, and simulations of noise mitigation. Second, since seismic sources can be close to the test masses, it is clear that the seismic field cannot always be described as a superposition of propagating plane seismic waves. For this reason, analytical work has begun to base calculations of gravity perturbations on simple models of seismic sources, which can give rise to complex seismic fields \cite{HaEA2015}. Since this work also inspired potential applications in geophysics and seismology, we devote Section \ref{sec:pointsources} entirely to this new theory. Last but not least, ideas for new detector concepts have evolved over the last decade, which will make it possible to monitor gravity strain perturbations at frequencies below 1\,Hz. This means that our models of seismic Newtonian noise (as for all other types of Newtonian noise) need to be extended to lower frequencies, which is not always a trivial task. We will discuss aspects of this problem in Section \ref{sec:lowfNNRay}.

\subsection{Seismic waves}
\label{sec:seismic}
In this section, we describe the properties of seismic waves relevant for calculations of gravity perturbations. The reader interested in further details is advised to study one of the classic books on seismology, for example Aki \& Richards \cite{AkRi2009}. The formalism that will be introduced is most suited to describe physics in infinite or half-spaces with simple modifications such as spherical cavities, or small perturbations of a flat surface topography. At frequencies well below 10\,mHz where the finite size of Earth starts to affect significantly the properties of the seismic field, seismic motion is best described by Earth's normal modes \cite{DaTr1998}. It should also be noted that in the approximation used in the following, the gravity field does not act back on the seismic field. This is in contrast to the theory of Earth's normal modes, which includes the gravity potential and its derivative in the elastodynamic equations.

Seismic waves can generally be divided into shear waves, compressional waves, and surface waves. Compressional waves\index{seismic waves!compressional or P} produce displacement along the direction of propagation. They are sometimes given the alternative name ``P-waves'', which arises from the field of seismology. The P stands for \emph{primary} and means that these waves are the first to arrive after an earthquake (i.~e.~they are the fastest waves). These waves are characterized by a frequency $\omega$ and a wave vector $\vec k^{\,\rm P}$. While one typically assumes $\omega=k^{\rm P}\alpha$ with compressional wave speed $\alpha$, this does not have to hold in general, and many results presented in the following sections do not require a fixed relation between frequency and wavenumber. The displacement field of a plane compressional wave can be written
\beq
\vec\xi^{\,\rm P}(\vec r,t)=\vec e_k\xi_0^{\rm P}(\vec k^{\,\rm P},\omega)\exp(\irm(\vec k^{\,\rm P}\cdot\vec r\,-\omega t))
\label{eq:comprPW}
\eeq
The index 'P' is introduced to distinguish between displacements of shear and compressional waves, and $\vec e_k\equiv\vec k^{\,\rm P}/k^{\rm P}$. In media with vanishing shear modulus such as liquids and gases, compressional waves are also called sound waves. There are many ways to express the P-wave speed in terms of other material constants, but a widely used definition is in terms of the Lam\'e constants $\lambda,\,\mu$:\index{Lam\'e constants}
\beq
\alpha = \sqrt{\frac{\lambda+2\mu}{\rho}}
\eeq
The Lam\'e constant $\mu$ is also known as shear modulus, and $\rho$ is the density of the medium. Shear waves\index{seismic waves!shear or S} produce transversal displacement and do not exist in media with vanishing shear modulus. They are also known as ``S-waves'', where S stands for \emph{secondary} since it is the seismic phase to follow the P-wave arrival after earthquakes. The shear-wave displacement $\vec\xi^{\;\rm S}(\vec r,t)$ of a single plane wave can be expressed in terms of a polarization vector $\vec e_p$:
\beq
\vec\xi^{\,\rm S}(\vec r,t)=\vec e_p\xi_0^{\rm S}(\vec k^{\,\rm S},\omega)\exp(\irm(\vec k^{\,\rm S}\cdot\vec r\,-\omega t))
\label{eq:shearPW}
\eeq
with $\vec e_p\cdot\vec k^{\,\rm S}=0$. The S-wave speed in terms of the Lam\'e constants reads
\beq
\beta = \sqrt{\frac{\mu}{\rho}}
\eeq
Both wave types, compressional and shear, will be referred to as \emph{body waves} since they can propagate through media in all directions. Clearly, inside inhomogeneous media, all material constants are functions of the position vector $\vec r$. Another useful relation between the two seismic speeds is given by
\beq
\beta = \alpha\cdot\sqrt{\frac{1-2\nu}{2-2\nu}},
\label{eq:speedPS}
\eeq
where $\nu$ is the Poisson's ratio \index{Poisson's ratio}of the medium. It should be mentioned that there are situations when a wave field cannot be described as a superposition of compressional and shear waves. This is for example the case in the near field of a seismic source. In the remainder of this section, we will calculate gravity perturbations for cases where the distinction between compressional and shear waves is meaningful. The more complicated case of gravity perturbations from seismic fields near their sources is considered in Section \ref{sec:pointsources}.

An elegant way to represent a seismic displacement field $\vec \xi(\vec{r},t)$ is in terms of its seismic or Lam\'e potentials $\phi_{\rm s}(\vec{r},t)$, $\vec\psi_{\rm s}(\vec{r},t)$ \cite{AkRi2009}: \index{potential!seismic, or Lam\'e}
\beq
\vec{\xi}(\vec{r},t)=\nabla\phi_{\rm s}(\vec{r},t)+\nabla\times\vec\psi_{\rm s}(\vec{r},t)
\label{eq:seispot}
\eeq
with $\nabla\cdot\vec\psi_{\rm s}(\vec{r},t)=0$. The rotation of the first term vanishes, which is characteristic for compressional waves. The divergence of the second term vanishes, which is characteristic for shear waves. Therefore, the scalar potential $\phi_{\rm s}(\vec{r},t)$ will be called P-wave potential, and $\vec \psi_{\rm s}(\vec{r},t)$ S-wave potential. As will become clear in the following, many integrals involving the seismic field $\vec{\xi}(\vec{r},t)$ simplify greatly when using the seismic potentials to represent the field. It is possible to rewrite the shear-wave potential in terms of two scalar quantities in Cartesian coordinates \cite{Sas1985}:
\beq
\vec\psi_{\rm s}(\vec{r},t)=\nabla\times(0,0,\psi_{\rm s}(\vec{r},t))+(0,0,\chi_{\rm s}(x,y,t))
\label{eq:shearpot}
\eeq
This form can lead to useful simplifications. For example, if seismic displacement is relevant only in $z$-direction, then it suffices to calculate the contribution from the scalar potential $\psi_{\rm s}(\vec{r},t)$.

Next we will introduce the Rayleigh waves\index{seismic waves!Rayleigh or Rf}. These are surface waves and in fact the only seismic waves that can propagate on surfaces of homogeneous media. In the presence of an interface between two types of media, the set of possible solutions of interface waves is much richer as described in detail in \cite{Pil1972}. In this paper, we will not deal specifically with the general solutions of interface waves, but it should be noted that gravity perturbations from at least one of the types, the Stoneley waves\index{seismic waves!Stoneley}, can be calculated using the same equations derived later for the Rayleigh waves. The definition of Rayleigh waves does not require a plane surface, but let us consider the case of a homogeneous half space for simplicity. The direction normal to the surface corresponds to the $z$-axis of the coordinate system, and will also be called vertical direction. The normal vector is denoted as $\vec e_z$. Rayleigh waves propagate along a horizontal direction $\vec e_k$. A wave vector $\vec k$ can be split into its vertical $\vec k_z$ and horizontal components $\vec k_\varrho$. The vertical wavenumbers are defined as
\beq
k_z^{\rm P}(k_\varrho)=\sqrt{(k^{\rm P})^2-k_\varrho^2},\qquad k_z^{\rm S}(k_\varrho)=\sqrt{(k^{\rm S})^2-k_\varrho^2}
\label{eq:wavek}
\eeq
Even though Rayleigh waves are surface waves, their displacement field extends evanescently (i.~e.~with exponential amplitude fall-off from the surface) throughout the entire medium. They can be considered as analytical extension of a situation where body waves are reflected from the surface in the sense that we can allow the horizontal wavenumber $k_\varrho$ to be larger than $k^{\rm P}$ and $k^{\rm S}$. In this case, the vertical wavenumbers have imaginary values. Hence, in the case of Rayleigh waves, it is convenient to define new wave parameters as:
\beq
q_z^{\rm S}=\sqrt{k_\varrho^2-(k^{\rm S})^2},\qquad q_z^{\rm P}=\sqrt{k_\varrho^2-(k^{\rm P})^2}
\label{eq:verticalk}
\eeq
Here, $k_\varrho$ is the horizontal wavenumber of the Rayleigh wave. Note that the order of terms in the square-roots are reversed with respect to the case of body waves as in Equation (\ref{eq:wavek}). Rewriting the equations in \cite{HaNa1998} in terms of the horizontal and vertical wavenumbers, the horizontal and vertical amplitudes of the three-dimensional displacement field of a Rayleigh wave reads
\beq
\begin{split}
\xi_k(\vec{r},t) &= A\cdot\left(k_\varrho\e^{q_z^{\rm P}z}-\zeta q_z^{\rm S}\e^{q_z^{\rm S}z}\right)\cdot\sin(\vec k_\varrho\cdot\vec \varrho-\omega t)\\
\xi_z(\vec{r},t) &= A\cdot\left(q_z^{\rm P}\e^{q_z^{\rm P}z}-\zeta k_\varrho\e^{q_z^{\rm S}z}\right)\cdot\cos(\vec k_\varrho\cdot\vec \varrho-\omega t)
\end{split}
\label{eq:Rayfield}
\eeq
with $\zeta(k_\varrho)\equiv\sqrt{q_z^{\rm P}/q_z^{\rm S}}$. The speed $c_{\rm R}=k_\varrho/\omega$ of the fundamental Rayleigh wave obeys the equation \index{Rayleigh pole}
\beq
\begin{split}
&R\left((c_{\rm R}/\beta)^2\right)=0,\\ 
&R(x)=x^3-8x^2+8x\frac{2-\nu}{1-\nu}-\frac{8}{1-\nu}
\end{split}
\label{eq:speedR}
\eeq
The real-valued solution to this equation is known as Rayleigh pole since the same function appears in the denominator of surface reflection coefficients. Note that the horizontal and vertical displacements are phase shifted by 90$^\circ$, which gives rise to elliptical particle motions. Therefore, arbitrary time series of vertical displacement are related to horizontal displacement via the Hilbert transform\index{Hilbert transform}. The displacement vector is constructed according to
\beq
\vec\xi(\vec r,t) = \xi_k(\vec r,t)\vec e_k+\xi_z(\vec r,t)\vec e_z
\eeq
In the case of a stratified medium, this wave type is also known as fundamental Rayleigh wave to distinguish them from \emph{higher-order Rayleigh waves} that can exist in these media \cite{HuTh1998}. For this reason, we will occasionally refer to Rayleigh waves as Rf-waves. According to Equations (\ref{eq:speedPS}) \& (\ref{eq:speedR}), given a shear-wave speed $\beta$, the compressional-wave speed $\alpha$ and Rayleigh-wave speed $c_{\rm R}$ are functions of the Poisson's ratio only. Figure \ref{fig:wavespeeds} shows the values of the wave speeds in units of $\beta$. 
\epubtkImage{}{%
    \begin{figure}[htbp]
    \centerline{\includegraphics[width=0.6\textwidth]{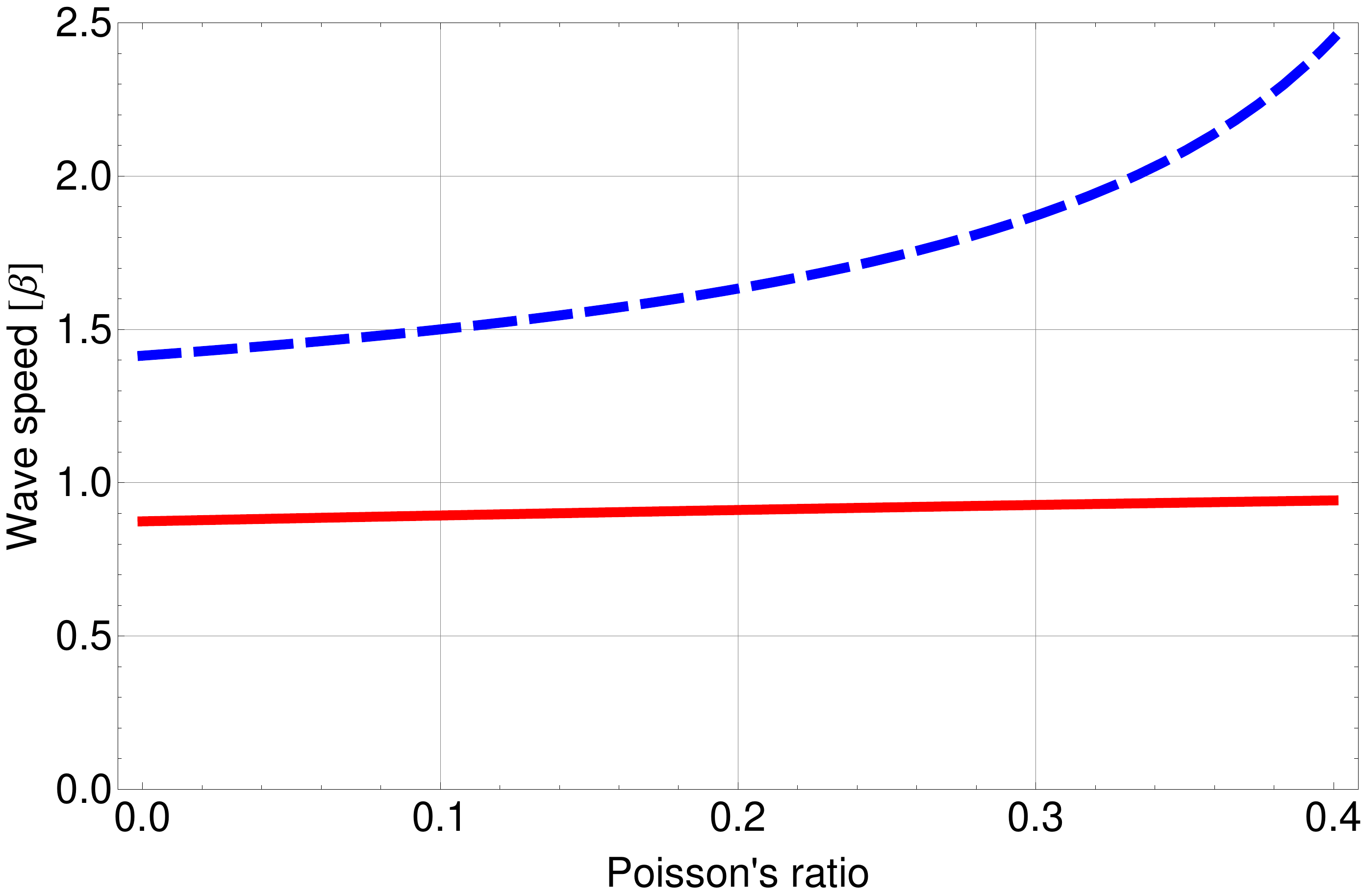}}
    \caption[Seismic wave speeds]{Rayleigh speed (solid line) and P-wave speed (dashed line) in units of S-wave speed $\beta$ as a function of Poisson's ratio.}
\label{fig:wavespeeds}
    \end{figure}}
As can be seen, for a given shear-wave speed the Rayleigh-wave speed (shown as solid line), depends only weakly on the Poisson's ratio. The P-wave speed however varies more strongly, and in fact grows indefinitely with Poisson's ratio approaching the value $\nu=0.5$. 

\subsection{Basics of seismic gravity perturbations}
\label{eq;basicsgrav}
In this section, we derive the basic equations that describe the connection between seismic fields and associated gravity perturbations. Expressions will first be derived in terms of the seismic displacement field $\vec\xi(\vec r,t)$, then in terms of seismic potentials $\phi_{\rm s}(\vec r,t),\,\vec\psi_{\rm s}(\vec r,t)$, and this section concludes with a discussion of gravity perturbations in transform domain.

\subsubsection{Gravity perturbations from seismic displacement}
The starting point is the continuity equation\index{continuity equation}, which gives an expression for the density perturbation caused by seismic displacement:
\beq
\delta\rho(\vec{r},t) = -\nabla\cdot(\rho(\vec r\,)\vec\xi(\vec r,t))
\label{eq:densinh}
\eeq
Here it is assumed that the seismic density perturbations are much smaller than the unperturbed density $\delta\rho(\vec{r},t)\ll\rho(\vec r\,)$ so that self-induced seismic scattering is insignificant. The perturbation of the gravity potential can now be written
\beq
\begin{split}
\delta\phi (\vec{r}_0,t) &= -G\int\drm V \frac{\delta\rho(\vec r,t)}{|\vec r-\vec r_0|}\\
&= G\int\drm V \frac{\nabla\cdot(\rho(\vec r\,)\vec\xi(\vec r,t))}{|\vec r-\vec r_0|}\\
&= -G\int\drm V\,\rho(\vec r\,)\vec\xi(\vec r,t)\cdot\nabla\frac{1}{|\vec r-\vec r_0|}
\end{split}
\label{eq:totalNNinh}
\eeq
Note that in the last step integration by parts did not lead to surface terms since any type of geology can be described as having infinite size. For example, a half space would correspond to an infinite space with vanishing density above surface. Carrying out the gradient operation, we obtain the gravity perturbation in dipole form\index{gravity perturbation!dipole form}
\beq
\delta\phi(\vec r_0,t) = G\int\drm V\rho(\vec r\,)\vec \xi(\vec r,t)\cdot\frac{\vec r-\vec r_0}{|\vec r-\vec r_0|^3}
\label{eq:dipolephi}
\eeq
and the corresponding perturbation of gravity acceleration reads
\beq
\begin{split}
\delta\vec a(\vec r_0,t) &= -G\int\drm V\rho(\vec r\,)(\vec \xi(\vec r,t)\cdot\nabla_0)\cdot\frac{\vec r-\vec r_0}{|\vec r-\vec r_0|^3}\\
&= G\int\drm V\rho(\vec r\,)\frac{1}{|\vec r-\vec r_0|^3}\left(\vec\xi(\vec r,t)-3(\vec e_{rr_0}\cdot\vec\xi(\vec r,t))\vec e_{rr_0}\right)
\end{split}
\label{eq:dipoleacc}
\eeq
with $\vec e_{rr_0}\equiv (\vec r-\vec r_0)/|\vec r-\vec r_0|$, and $\nabla_0$ denotes the gradient operation with respect to $\vec r_0$. In this form, it is straight-forward to implement gravity perturbations in finite-element simulations (see Section \ref{sec:numsim}), where each finite element is given a mass $\rho(\vec r\,)\delta V$. This equation is valid whenever the continuity Equation (\ref{eq:densinh}) holds, and describes gravity perturbations inside infinite media as well as media with surfaces. 

Especially in the case of a homogeneous medium with surface, treating bulk and surface contributions to gravity perturbations separately can often simplify complex calculations. The continuity equation with constant (unperturbed) density $\rho_0=\rho(\vec r\,)$ describes density perturbations inside the medium contained in the volume $\mathcal V$, which directly yields the bulk term:
\beq
\delta\phi_{\rm bulk}(\vec{r}_0,t)=G\rho_0\int\limits_{\mathcal V}\drm V \frac{\nabla\cdot\vec\xi(\vec r,t)}{|\vec r-\vec r_0|}
\label{eq:bulkNN}
\eeq 
The surface term can be constructed by noting that it is the displacement normal to the surface that generates gravity perturbations:
\beq
\delta\phi_{\rm surf}(\vec{r}_0,t) = -G\rho_0\int\drm S \frac{\vec n(\vec{r})\cdot\vec{\xi}(\vec{r},t)}{|\vec r-\vec r_0|}
\label{eq:surfNN}
\eeq
Note that also this equation is true only for small displacements since the surface normal is assumed to change negligibly due to seismic waves. The sum of bulk and surface terms is equal to the expression in Equation (\ref{eq:dipolephi}) with constant mass density. 

The same results can also be obtained using an explicit expression of the density $\rho(\vec r\,)$ that includes the density change at the surface in the form of a Heaviside function $\Theta(\cdot)$. The surface is solution to an equation $\sigma(\vec r\,)=0$, with $\sigma(\vec r\,)$ being normalized such that $\nabla\sigma(\vec r\,)$ is the unit normal vector $\vec n(\vec r\,)$ of the surface pointing from the medium outwards into the empty space. For a homogeneous medium with density $\rho_0$, the density of the entire space can be written as
\beq
\rho(\vec r\,)=\rho_0\Theta(-\sigma(\vec r\,))
\eeq
Inserting this expression into Equation (\ref{eq:totalNNinh}), one obtains
\beq
\begin{split}
\delta\phi (\vec{r}_0,t) &= G\rho_0\int\drm V \frac{\nabla\cdot(\Theta(-\sigma(\vec r\,))\vec\xi(\vec r,t))}{|\vec r-\vec r_0|}\\
&= G\rho_0\int\drm V \frac{\Theta(-\sigma(\vec r\,))\nabla\cdot\vec\xi(\vec r,t)-\delta(-\sigma(\vec r\,))\vec n(\vec r\,)\cdot\vec\xi(\vec r,t)}{|\vec r-\vec r_0|}
\end{split}
\eeq
The first part of the infinite-space integral can be rewritten as the integral in Equation (\ref{eq:bulkNN}) over the volume $\mathcal V$ of the medium, while the second part translates into the surface integral in Equation (\ref{eq:surfNN}).

\subsubsection{Gravity perturbations in terms of seismic potentials}
In the last part of this section, results will be expressed in terms of the seismic potentials. This is helpful to connect this work to geophysical publications where solutions to seismic fields are often derived for these potentials. In many cases, it also greatly simplifies the calculation of gravity perturbations. In order to simplify the notation, the equations are derived for a homogeneous medium. Expressing the displacement field in terms of its potentials according to Equation (\ref{eq:seispot}), the bulk contribution reads
\beq
\delta\phi_{\rm bulk}(\vec{r}_0,t)=G\rho_0\int\limits_{\mathcal V}\drm V\,\frac{\Delta\phi_{\rm s}(\vec r,t)}{|\vec r-\vec r_0|},
\eeq
This expression can be transformed via integration by parts into
\beq
\delta\phi_{\rm bulk}(\vec{r}_0,t)=G\rho_0\int\drm S\,\vec n(\vec r\,)\cdot\left[\frac{\nabla\phi_{\rm s}(\vec r,t)}{|\vec r-\vec r_0|}-\phi_{\rm s}(\vec r,t)\nabla\frac{1}{|\vec r-\vec r_0|}\right]-4\pi G\rho_0\phi_{\rm s}(\vec r_0,t).
\label{eq:bodyinparts}
\eeq
One integral was solved explicitly by using
\beq
\Delta\frac{1}{|\vec r-\vec r_0|}=-4\pi\delta(\vec r-\vec r_0)
\eeq
The contribution $\delta\phi_{\rm surf}(\vec{r}_0,t)$ from the surface can also be rewritten in terms of seismic potentials
\beq
\delta\phi_{\rm surf}(\vec{r}_0,t)=-G\rho_0\int\drm S\,\vec n(\vec r\,)\cdot\frac{\nabla\phi_{\rm s}(\vec r,t)+\nabla\times\vec\psi_{\rm s}(\vec r,t)}{|\vec r-\vec r_0|}.
\eeq
As can be seen, terms in the bulk and surface contributions cancel, and so we get for the gravity potential
\beq
\begin{split}
\delta\phi(\vec{r}_0,t) &=\delta\phi_{\rm bulk}(\vec{r}_0,t)+\delta\phi_{\rm surf}(\vec{r}_0,t) \\
&= -G\rho_0\int\drm S\,\vec n(\vec r\,)\cdot\left[\frac{\nabla\times\vec\psi_{\rm s}(\vec r,t)}{|\vec r-\vec r_0|}+\phi_{\rm s}(\vec r,t)\nabla\frac{1}{|\vec r-\vec r_0|}\right]-4\pi G\rho_0\phi_{\rm s}(\vec r_0,t)\\
&= -G\rho_0\int\drm S\,\vec n(\vec r\,)\cdot\left[\vec\psi_{\rm s}(\vec r,t)\times\nabla\frac{1}{|\vec r-\vec r_0|}+\phi_{\rm s}(\vec r,t)\nabla\frac{1}{|\vec r-\vec r_0|}\right]-4\pi G\rho_0\phi_{\rm s}(\vec r_0,t)
\end{split}
\label{eq:gravHelm}
\eeq
The last equation follows from the fact that the boundary of a boundary is zero after application of Stokes' theorem (the surface $S$ being the boundary of a body with volume $V$). The seismic potentials vanish above surface, and therefore the gravity perturbation in empty space is the result of a surface integral. This is a very important conclusion and useful to theoretical investigations, but of limited practical relevance since the integral depends on the seismic potential $\phi_{\rm s}(\vec r,t)$, which cannot be measured or inferred in general from measurements. The shear-wave potential enters as $\nabla\times\vec\psi_{\rm s}(\vec r,t)$, which is equal to the (observable) shear-wave displacement. In the absence of a surface, the solution simplifies to
\beq
\delta\phi(\vec{r}_0,t) =-4\pi G\rho_0\phi_{\rm s}(\vec r_0,t).
\label{eq:gravP}
\eeq
The latter result is remarkable as it states the proportionality of gravity and seismic potentials in infinite media. If a solution of a seismic field is given for its seismic potentials, then one can immediately write down the gravity perturbation without further calculations. We will make use of it in Section \ref{sec:pointsources} to calculate gravity perturbations from seismic point sources.

\subsubsection{Gravity perturbations in transform domain}
In certain situations, it is favorable to consider gravity perturbations in transform domain. For example, in calculations of gravity perturbations in a half space, it can be convenient to express solutions in terms of the displacement amplitudes $\vec\xi(\vec k_\varrho,z,t)$, and in infinite space in terms of $\vec\xi(\vec k,t)$. As shown in Section \ref{sec:sourcehalf}, it is also possible to obtain concise solutions for the half-space problem using cylindrical harmonics, but in the following, we only consider plane-wave harmonics.

The transform-domain equations for gravity perturbations from seismic fields in a half space, with the surface at $z=0$, are obtained by calculating the Fourier transforms of Equations (\ref{eq:bulkNN}) and (\ref{eq:surfNN}) with respect to $x_0,\,y_0$. This yields the bulk term
\beq
\begin{split}
\delta\phi_{\rm bulk}(\vec k_\varrho,z_0,t)&=2\pi G\rho_0\frac{1}{k_\varrho}\int\limits_{-\infty}^0\drm z \, \e^{-k_\varrho|z-z_0|}\left[\partial_z\xi_z(\vec k_\varrho,z,t)-\irm\vec k_\varrho\cdot\vec \xi_\varrho(\vec k_\varrho,z,t)\right]\\
&=2\pi G\rho_0\frac{1}{k_\varrho}\Bigg[\xi_z(\vec k_\varrho,0,t)\e^{-k_\varrho|z_0|}\\
&\qquad\qquad-\int\limits_{-\infty}^0\drm z \, \e^{-k_\varrho|z-z_0|}\left(-k_\varrho{\rm sgn}(z-z_0)\xi_z(\vec k_\varrho,z,t)-\irm\vec k_\varrho\cdot\vec \xi_\varrho(\vec k_\varrho,z,t)\right)\Bigg],
\end{split}
\label{eq:bulkNNtr}
\eeq 
where sgn denotes the signum function. The surface term reads
\beq
\delta\phi_{\rm surf}(\vec k_\varrho,z_0,t) = -2\pi G\rho_0\frac{1}{k_\varrho}\xi_z(\vec k_\varrho,0,t)\e^{-k_\varrho|z_0|}
\label{eq:surfNNtr}
\eeq
Hence, the total perturbation of the gravity potential is given by
\beq
\delta\phi(\vec k_\varrho,z_0,t)=
2\pi G\rho_0\int\limits_{-\infty}^0\drm z \, \e^{-k_\varrho|z-z_0|}\left({\rm sgn}(z-z_0)\xi_z(\vec k_\varrho,z,t)+\frac{\irm}{k_\varrho}\vec k_\varrho\cdot\vec \xi_\varrho(\vec k_\varrho,z,t)\right).
\eeq
This equation is valid above surface as well as underground. Expanding the seismic field into plane waves, the integral over the coordinate $z$ is straight-forward to calculate.

Using seismic potentials as defined in Equations (\ref{eq:seispot}) and (\ref{eq:shearpot}) instead of the displacement field, one obtains
\beq
\begin{split}
\delta\phi(\vec k_\varrho,z_0,t)&=
2\pi G\rho_0\int\limits_{-\infty}^0\drm z \, \e^{-k_\varrho|z-z_0|}\left({\rm sgn}(z-z_0)(\partial_z\phi_{\rm s}(z)+k_\varrho^2\psi_{\rm s}(z))-k_\varrho(\phi_{\rm s}(z)+\partial_z\psi_{\rm s}(z))\right)\\
&=-2\pi G\rho_0\Bigg[\e^{-k_\varrho|z_0|}\left({\rm sgn}(z_0)\phi_{\rm s}(0)+k_\varrho\psi_{\rm s}(0)\right)+2\phi_{\rm s}(z_0)\Bigg],
\end{split}
\eeq
with $\phi_{\rm s}(z>0)=0$, and dependence of the potentials on $\vec k_\varrho$ and $t$ is omitted. This equation is particularly useful since seismologists often define their fields in terms of seismic potentials, and it is then possible to directly write down the perturbation of the gravity potential in transform domain without solving any integrals.

The corresponding expressions in infinite space are obtained by calculating the Fourier transforms of Equations (\ref{eq:bulkNN}) and (\ref{eq:surfNN}) with respect to $x_0,\,y_0$ and $z_0$. Since there are no surface terms, the result is simply
\beq
\delta\phi(\vec k,t)=-4\pi\irm G\rho_0\frac{1}{k^2}\vec k\cdot\vec\xi(\vec k,t)
\eeq
Substituting the displacement field by its seismic potentials, we find immediately the transform-domain version of Equation (\ref{eq:gravP}).

\subsection{Seismic gravity perturbations inside infinite, homogeneous media with spherical cavity}
\label{sec:scatterNN}
Test masses of underground detectors, as for example KAGRA \cite{AsEA2013}, will be located inside large chambers hosting corner and end stations of the interferometer. Calculation of gravity perturbations based on a spherical chamber model can be carried out explicitly and provides at least some understanding of the problem. This case was first investigated by Harms et al.~\cite{HaEA2009b}. In their work, contributions from normal displacement of cavity walls were taken into account, but scattering of incoming seismic waves from the cavity was neglected. In this section, we will outline the main results of their paper in Section \ref{sec:bodynoscatt}, and present for the first time a calculation of gravity perturbations from seismic waves scattered from a spherical cavity in Sections \ref{sec:scattercomp} and \ref{sec:scattershear}.

\subsubsection{Gravity perturbations without scattering}
\label{sec:bodynoscatt}
The first step is to calculate an explicit solution of the integral in Equation (\ref{eq:dipoleacc}) for plane seismic waves. The plane-wave solution will be incomplete, since scattering of the incident wave from the cavity is neglected. However, as will be shown later, scattering can be neglected assuming realistic dimensions of a cavity. We will start with the gravity perturbation from a plane compressional wave as defined in Equation (\ref{eq:comprPW}). Inserting this expression into Equation (\ref{eq:dipoleacc}), which includes bulk as well as surface gravity perturbations, the integral over the infinite medium excluding a cavity of radius $a$ can be solved. The gravity acceleration at the center $\vec r=\vec 0$ of the cavity is given by
\beq
\delta\vec a^{\,\rm P}(\vec 0,t) = 8\pi G\rho_0\vec\xi^{\;\rm P}(\vec 0,t)\frac{j_1(k^{\rm P}a)}{(k^{\rm P}a)},
\label{eq:planePcav}
\eeq
where $j_n(\cdot)$ is the spherical Bessel function. In the case that the length of the seismic wave is much larger than the cavity radius, the ratio can be approximated according to
\beq
\frac{j_1(ka)}{(ka)}\approx \frac{1}{3}\left[1-\frac{1}{10}(ka)^2\right],
\label{eq:approxGG}
\eeq
which neglects terms of order $\mathcal{O}((ka)^4)$, and the result in the limit of vanishing cavity radius simplifies to
\beq
\delta\vec a^{\,\rm P}(\vec 0,t) = \frac{8\pi}{3} G\rho_0\vec\xi^{\,\rm P}(\vec 0,t)
\label{eq:longewaveP}
\eeq
Since the gravity perturbation and therefore the seismic displacement is evaluated at the center of the cavity, the seismic displacement cannot be observed strictly speaking. Placing a seismometer at the cavity walls, an error of order $(ka)^2$ is made in the modelling of the gravity perturbation. The numerical factor in this equation is smaller by $-4\pi/3$ compared to the factor in Equation (\ref{eq:gravP}). This means that the bulk gravity perturbation is partially cancelled by cavity-surface contributions, which can be verified by directly evaluating the surface term:
\beq
\delta\vec a_{\rm surf}^{\,\rm P}(\vec 0,t) = -4\pi G\rho_0\vec\xi^{\;\rm P}(\vec 0,t)
\cdot\left(j_0(k^{\rm P}a)-2\dfrac{j_1(k^{\rm P}a)}{(k^{\rm P}a)}\right)
\label{eq:surfP}
\eeq
The long-wavelength limit $ka\rightarrow 0$ of the expression in brackets is 1/3, which is consistent with Equations (\ref{eq:longewaveP}) and (\ref{eq:gravP}). If the seismic field consisted only of pressure waves propagating in a homogeneous medium, then Equation (\ref{eq:longewaveP}) would mean that a seismometer placed at the test mass monitors all information required to estimate the corresponding gravity perturbations. 

A concise form of Equation (\ref{eq:planePcav}) can still be maintained if shear waves, which produce NN exclusively through surface displacement, are added to the total displacement $\vec\xi(\vec r,t)=\vec\xi^{\;\rm P}(\vec r,t)+\vec\xi^{\;\rm S}(\vec r,t)$. Inserting the plane-wave expression of Equation (\ref{eq:shearPW}) into Equation (\ref{eq:dipoleacc}), and adding the solution to the compressional-wave contribution, one obtains
\beq
\delta\vec a(\vec 0,t) = 4\pi G\rho_0\left(2\vec\xi^{\;\rm P}(\vec 0,t)\frac{j_1(k^{\rm P}a)}{(k^{\rm P}a)}-\vec\xi^{\;\rm S}(\vec 0,t)\frac{j_1(k^{\rm S}a)}{(k^{\rm S}a)}\right)
\label{eq:planeNNcav}
\eeq
The shear-wave contribution has the same dependence on cavity radius as the compressional-wave contribution, even though the shear term is purely due to cavity-surface displacement.

We can take a look at the gravity perturbation as a function of cavity radius. Figure \ref{fig:cavNNr} shows the perturbation from P-waves and S-waves using Equation (\ref{eq:planeNNcav}). It is assumed that P-waves have a factor 1.8 higher speed than S-waves.
\epubtkImage{}{ 
    \begin{figure}[htbp]
    \centerline{\includegraphics[width=0.6\textwidth]{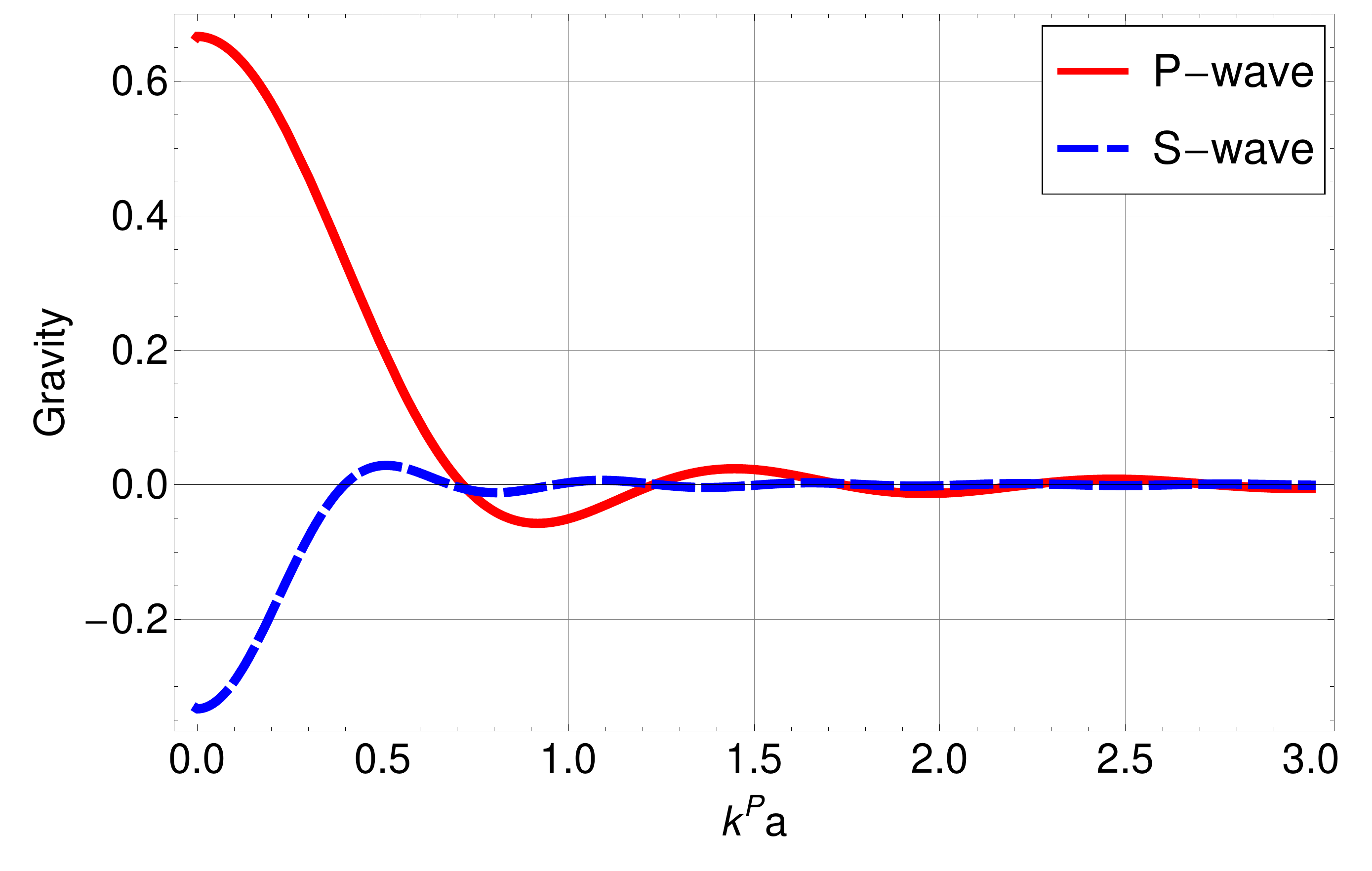}}
    \caption[Gravity perturbation inside cavity]{The plot shows the gravity perturbation at the center of a cavity as a function of cavity radius in units of seismic wavelength.}
\label{fig:cavNNr}
    \end{figure}}
If the cavity has a radius of about $0.4\lambda$, then gravity perturbation is reduced by about a factor 2. Keeping in mind that the highest interesting frequency of Newtonian noise is about 30\,Hz, and that compressional waves have a speed of about 4\,km/s, the minimal cavity radius should be about 50\,m to show a significant effect on gravity noise. Building such cavities would be a major and very expensive effort, and therefore, increasing cavity size does not seem to be a good option to mitigate underground Newtonian noise. 

\subsubsection{Incident compressional wave}
\label{sec:scattercomp}
The fact that the shear term in Equation (\ref{eq:planeNNcav}) has the opposite sign of the compressional term does not mean that gravity perturbations are reduced since noise in both components is typically uncorrelated. However, as explained in more detail in Section \ref{sec:bodyhalf}, compressional and shear waves are partially converted into each other at reflection from interfaces, which leads to correlated shear and compressional displacement. So one may wonder if a detailed calculation of the problem including scattering effects yields different numerical factors due to partial cancellation or coherent enhancement of gravity perturbations. This problem will be solved now and outlined in greater detail since it is algebraically more complex. The calculation is based on an explicit solution of the seismic field for a compressional wave incident on a spherical obstacle \cite{YiTr1956}. The part of the seismic field that is produced by the spherical cavity has spherical symmetry. Therefore, it can be written in the form:
\beq
\vec\xi_{\rm cav}(\vec r,t)=[\nabla\phi_{\rm s}(\vec r\,)+\nabla\times(\nabla\times(\psi_{\rm s}(\vec r\,)\,\vec r\,))]\e^{-\irm\omega t}
\label{eq:seismicSphSym}
\eeq
Since the seismic field can be expressed in terms of scalar potentials, it is possible to expand the incident plane wave according to Equation (\ref{eq:pwscalar}). The outgoing scattered field is then obtained by fulfilling the boundary conditions at the cavity walls. For hollow cavities, the boundary condition states that the stress tensor produced by the seismic field projected onto the cavity normal, which yields a vector known as traction\index{traction}, must vanish \cite{AkRi2009}. In spherical coordinates, the potentials of the scattered waves can be expanded according to\index{scattering!compressional waves}
\beq
\begin{split}
\phi_{\rm s}(r,\cos(\theta)) &= \xi_0\sum\limits_{l=0}^\infty A_l(a)h_l^{(2)}(k_{\rm P}r)P_l(\cos(\theta))\\
\psi_{\rm s}(r,\cos(\theta)) &= \xi_0\sum\limits_{l=0}^\infty B_l(a)h_l^{(2)}(k_{\rm S}r)P_l(\cos(\theta))
\end{split}
\eeq
where $k_{\rm P},\,k_{\rm S}$ are the wave numbers of compressional and shear waves respectively, $\theta$ is the angle between the direction of propagation of the scattered wave with respect to the direction of the incident compressional wave, $\xi_0$ is the displacement amplitude of the incoming compressional wave, and the origin of the coordinate system lies at the center of the cavity. The spherical Hankel functions of the second kind $h_n^{(2)}(\cdot)$ are defined in terms of the spherical Bessel functions of the first and second kind as:\index{Hankel function!spherical, second kind}
\beq
h_n^{(2)}(x) \equiv j_n(x)-\irm y_n(x)
\eeq
The expansion or scattering coefficients $A_l,\,B_l$ need to be determined from boundary conditions at the cavity surface, which was presented in detail in \cite{YiTr1956}. Here we just mention that for small cavities, i.~e.~in the Rayleigh-scattering regime with $\{k_{\rm P},k_{\rm S}\}\cdot a\ll 1$, the dependence of the scattering coefficients on the cavity radius $a$ is $(k_{\rm P}a)^3$ or higher order.

In order to understand the gravity perturbations from shear and compressional components, we consider bulk and surface contributions separately. The bulk contribution of Equation (\ref{eq:bulkNN}) assumes the form 
\beq
\delta\vec a_{\rm bulk}(\vec{r}_0,t) = -G\rho_0\e^{-\irm\omega t}\int\limits_{\mathcal V}\drm V \frac{-k_{\rm P}^2\phi_{\rm s}(r,\cos(\theta))}{|\vec r-\vec r_0|^2}\vec e_{rr_0},
\eeq
where we have used the fact that the P-wave potential obeys the Helmholtz \index{Helmholtz equation} equation:
\beq
(\Delta+k_{\rm P}^2)\phi_{\rm s}(\vec r\,) = 0
\eeq
According to Equation (\ref{eq:surfNN}), the surface contribution reads
\beq
\delta\vec a_{\rm surf}(\vec r_0,t) = G\rho_0\int\drm S \frac{\vec\xi_{\rm cav}(\vec{r},t)\cdot\vec e_r}{|\vec r-\vec r_0|^2}\vec e_{rr_0}
\eeq
The last expression can be further simplified by making use of the identity
\beq
\vec\xi_{\rm cav}(\vec r,t)\cdot\vec e_r=\left(\partial_r\phi_{\rm s}(\vec r\,)-\frac{1}{r}\partial_u\left[(1-u^2)\partial_u\psi_{\rm s}(\vec r\,)\right]\right)\e^{-\irm\omega t}
\eeq
with $u\equiv\cos(\theta)$. If the gravity perturbations are to be calculated at the center $\vec r_0=\vec 0$ of the spherical cavity, then the integrals are easily evaluated by substituting powers of $\cos(\theta)$ according to the right-hand-side of Table \ref{tab:legendre}, and making use of the orthogonality relation in Equation (\ref{eq:legorth}). We first outline the calculation for the bulk term. Identifying the $z$-axis with the direction of propagation of the incoming wave, one obtains:
\beq
\begin{split}
\delta a_{\rm bulk}^z(\vec 0,t) &= 2\pi G\rho_0k_{\rm P}^2\e^{-\irm\omega t}\int\limits_a^\infty\drm r\int\limits_{-1}^1\drm u\, u\,\phi_{\rm s}(r,u)\\
& = 2\pi G\rho_0k_{\rm P}^2\e^{-\irm\omega t}\xi_0\sum\limits_{l=0}^\infty A_l(a)\int\limits_a^\infty\drm r\, h_l^{(2)}(k_{\rm P}r)\int\limits_{-1}^1\drm uP_1(u)P_l(u)\\
& = \frac{4\pi}{3} G\rho_0k_{\rm P}^2\xi_0A_1(a)\e^{-\irm\omega t}\int\limits_a^\infty\drm r\,h_1^{(2)}(k_{\rm P}r)\\
& = \frac{4\pi}{3a} G\rho_0\xi_0A_1(a)\e^{-\irm \omega t}(k_{\rm P}a)h_0^{(2)}(k_{\rm P}a)
\end{split}
\label{eq:scattPbulk}
\eeq
The perturbations along $x,\,y$ vanish. Also the surface contribution is readily obtained with integration by parts:
\beq
\begin{split}
\delta a_{\rm surf}^z(\vec 0,t) &= -2\pi G\rho_0\xi_0\e^{-\irm\omega t}\int\limits_{-1}^1\drm u\,u \left(\partial_r\phi_{\rm s}-\frac{1}{a}\partial_u\left[(1-u^2)\partial_u\psi_{\rm s}\right]\right)_{r=a}\\
&= -2\pi G\rho_0\xi_0\e^{-\irm\omega t}\int\limits_{-1}^1\drm u\,\left(P_1(u)(\partial_r\phi_{\rm s})+\frac{2}{a}P_1(u)\psi_{\rm s}\right)_{r=a}\\
& = -2\pi G\rho_0\xi_0\e^{-\irm\omega t}\sum\limits_{l=0}^\infty\int\limits_{-1}^1\drm u\,\bigg(A_l(a)(\partial_rh_l^{(2)}(k_{\rm P}r))_{r=a}+\frac{2B_l(a)h_l^{(2)}(k_{\rm S}a)}{a}\bigg)P_1(u)
P_l(u)\\
& = \frac{4\pi}{3a} G\rho_0\xi_0\e^{-\irm\omega t}\bigg(A_1(a)(2h_1^{(2)}(k_{\rm P}a)-(k_{\rm P}a)h_0^{(2)}(k_{\rm P}a))-2B_1(a)h_1^{(2)}(k_{\rm S}a)\bigg)
\end{split}
\label{eq:scattPsurf}
\eeq
Again, perturbations along $x,\,y$ vanish. Adding the bulk and surface contributions, we finally obtain
\beq
\delta a^z(\vec 0,t) = \frac{8\pi}{3a} G\rho_0\xi_0\e^{-\irm\omega t}\bigg(A_1(a)h_1^{(2)}(k_{\rm P}a)-B_1(a)h_1^{(2)}(k_{\rm S}a)\bigg)
\eeq
This expression can be evaluated in the Rayleigh regime $\{k_{\rm P},\,k_{\rm S}\}\cdot a\ll 1$ using the following approximations of the scattering coefficients $A_1(a),B_1(a)$ given in \cite{YiTr1956}:
\beq
\begin{split}
A_1(a) &= \frac{\irm}{3k_{\rm P}}(k_{\rm P}a)^3\left[1-\frac{1}{45}\left(11(k_{\rm P}a)^2+15(k_{\rm S}a)^2\right)\right]\\
B_1(a) &= \frac{\irm}{3k_{\rm S}}(k_{\rm S}a)^3\left[1-\frac{1}{18}\left(5(k_{\rm P}a)^2+6(k_{\rm S}a)^2\right)\right]
\end{split}
\eeq

\beq
\delta a^z(\vec 0,t) = \frac{4\pi}{9} G\rho_0\xi_0\e^{-\irm\omega t}\left((k_{\rm S}a)^2-\frac{16(k_{\rm P}a)^2}{15}\right)
\label{eq:scatterPNN}
\eeq
This solution needs to be added to the contribution in Equation (\ref{eq:planePcav}) from the unperturbed incident wave. The gravity perturbation associated with the scattered waves is in phase with the perturbation from the incoming compressional wave. The perturbation in Equation (\ref{eq:scatterPNN}) vanishes in the limit $a\rightarrow 0$, which may seem intuitive, but notice that the surface contribution of the incoming wave does not vanish in the same limit. Instead, it is a consequence of perfect cancellation of leading order terms from scattered shear and compressional waves. Therefore, this result shows explicitly that neglecting contributions from scattered waves has no influence on leading order terms of the full gravity perturbation, at least if the incident wave is of compressional type. 

\subsubsection{Incident shear wave}
\label{sec:scattershear}
The calculation of the seismic field scattered from a spherical cavity with incident shear-wave can be found in \cite{KoJo1996}. Although it is in principle possible to solve this problem in terms of scalar seismic potentials, we choose to represent the fields directly in vector form using vector spherical harmonics. We assume that the polarization vector of the incident shear wave is $\vec e_x$, while the propagation direction is along $\vec e_z$. The explicit expression of the incident field is given in Equation (\ref{eq:expandPWtrans}). The scattered field can be expanded according to
\beq
\vec\xi_{\rm s}(\vec r,t) = \xi_0\e^{-\irm\omega t}\sum\limits_{l,m}\left(y_{lm}(r)\vec Y_l^m(\theta,\phi)+s_{lm}(r)\vec \Psi_l^m(\theta,\phi)+p_{lm}(r)\vec \Phi_l^m(\theta,\phi)\right)
\eeq
We will not further specify the radial functions. The expressions can be found in \cite{KoJo1996} (after converting their vector spherical harmonics into the ones used here). As for the incident P-wave, we will carry out the calculation of the gravity perturbations at the center of the cavity. Let us first calculate the bulk integral of Equation (\ref{eq:bulkNN}), using the divergence relations in Equation (\ref{eq:divvecharm}) and the integral Equation (\ref{eq:intvecharm}):
\beq
\begin{split}
\delta\vec a_{\rm bulk}(\vec 0,t) &= -G\rho_0\int\limits_{\mathcal V}\drm V \frac{\nabla\cdot\vec\xi(\vec r,t)}{r^2}\vec e_r\\
&= -G\rho_0\xi_0\e^{-\irm\omega t}\int\limits_a^\infty\drm r \int\drm\Omega\sum\limits_{l,m}\left(\nabla\cdot(y_{lm}(r)\vec Y_l^m(\theta,\phi))+\nabla\cdot(s_{lm}(r)\vec \Psi_l^m(\theta,\phi))\right)\vec e_r\\
&= -G\rho_0\xi_0\e^{-\irm\omega t}\int\limits_a^\infty\drm r \int\drm\Omega\sum\limits_{l,m}\left(\partial_ry_{lm}(r)+\frac{2}{r}y_{lm}(r)-\frac{\sqrt{l(l+1)}}{r}s_{lm}(r)\right) \vec Y_l^m(\theta,\phi)\\
&= -G\rho_0\xi_0\e^{-\irm\omega t}\int\limits_a^\infty\drm r \sqrt{\frac{2\pi}{3}}\left((\mathcal Y_1^{-1}(r)-\mathcal Y_1^1(r))\vec e_x-\irm(\mathcal Y_1^{-1}(r)+\mathcal Y_1^1(r))\vec e_y+\sqrt{2}\mathcal Y_1^0(r)\vec e_z\right)\\
&= -G\rho_0\xi_0\e^{-\irm\omega t}\sqrt{\frac{2\pi}{3}}\vec e_x\int\limits_a^\infty\drm r (\mathcal Y_1^{-1}(r)-\mathcal Y_1^1(r))
\end{split}
\eeq 
where the term in brackets in the third line was defined as $\mathcal Y_l^m(r)$. The last equation follows from the definition of the coefficients $y_{lm},\,s_{lm}$ in Equation (C.2) of \cite{KoJo1996}, but it should also be clear from symmetry considerations that gravity perturbation can be non-zero only along the displacement direction of the incident wave. The term under the last integral takes the form
\beq
\mathcal Y_1^{-1}(r)-\mathcal Y_1^1(r)=-\frac{1}{r}\sqrt{6\pi}(k^{\rm P}r)a^{\rm SP}_1h^{(2)}_1(k^{\rm P}r)
\eeq
The scattering coefficient $a^{\rm SP}_1$ corresponds to the relative amplitude of the $l=1$ scattered $P$-wave to the $l=1$ amplitude of the incident S-wave. It can be calculated using equations from \cite{KoJo1996} (note that the explicit solutions given in the appendix are wrong). Inserting this expression into the last equation, we finally obtain
\beq
\delta\vec a_{\rm bulk}(\vec 0,t) =2\pi G\rho_0\xi_0\e^{-\irm\omega t}a^{\rm SP}_1h_0^{(2)}(k^{\rm P}a)\vec e_x 
\eeq
This result is very similar to Equation (\ref{eq:scattPbulk}), just that the scattering coefficients are defined slightly differently. We can now repeat the exercise for the surface term:
\beq
\begin{split}
\delta\vec a_{\rm surf}(\vec 0,t) &= G\rho_0\int\drm S \frac{\vec n(\vec{r})\cdot\vec{\xi}(\vec{r},t)}{r^2}\vec e_r\\
&= -G\rho_0\int\drm\Omega (\vec e_r\cdot\vec{\xi}(r=a,\theta,\phi,t))\vec e_r\\
&= -G\rho_0\xi_0\e^{-\irm\omega t}\sqrt{\frac{2\pi}{3}}\left((y_{1,-1}(a)-y_{1,1}(a))\vec e_x-\irm(y_{1,-1}(a)+y_{1,1}(a))\vec e_y+\sqrt{2}y_{1,0}(a)\vec e_z\right)\\
&= -G\rho_0\xi_0\e^{-\irm\omega t}\sqrt{\frac{2\pi}{3}}(y_{1,-1}(a)-y_{1,1}(a))\vec e_x\\
&= -G\rho_0\xi_0\e^{-\irm\omega t}2\pi\left(\frac{a_1^{\rm SP}}{k^{\rm P}a}(-2h_1^{(2)}(k^{\rm P}a)+(k^{\rm P}a)h_0^{(2)}(k^{\rm P}a))-\frac{2b_1^{\rm SS}}{k^{\rm S}a}h_1^{(2)}(k^{\rm S}a)\right)\vec e_x
\end{split}
\eeq 
As in Section \ref{sec:bodynoscatt}, the surface term contains P-wave contributions quantified by the scattering coefficient $a_1^{\rm SP}$, and S-wave contributions quantified by $b_1^{\rm SS}$. Again, the result is formally very similar to Equation (\ref{eq:scattPsurf}) with incident P-wave. Adding the surface and bulk term, we finally obtain
\beq
\delta\vec a(\vec 0,t) = 4\pi G\rho_0\xi_0\e^{-\irm\omega t}\left(\frac{a_1^{\rm SP}}{k^{\rm P}a}h_1^{(2)}(k^{\rm P}a)+\frac{b_1^{\rm SS}}{k^{\rm S}a}h_1^{(2)}(k^{\rm S}a)\right)\vec e_x
\eeq
In the Rayleigh-scattering regime, $k^{\rm P}a\ll 1$ and $k^{\rm S}a\ll 1$, the scattering coefficients can be expanded according to
\beq
\begin{split}
a_1^{\rm SP}(a) &= -\frac{2\irm}{9}(k_{\rm P}a)^3\left[1-\frac{1}{18}\left(5(k_{\rm P}a)^2+6(k_{\rm S}a)^2\right)\right]\\
b_1^{\rm SS}(a) &= \frac{2\irm}{9}(k_{\rm S}a)^3\left[1-\frac{1}{40}\left(20(k_{\rm P}a)^2+87(k_{\rm S}a)^2\right)\right]
\end{split}
\eeq
Note that the explicit expressions for the scattering coefficients in the appendix of \cite{KoJo1996} are wrong. However, since the equations in the main part of the paper are correct, it is straight-forward to recalculate the scattering coefficients. Finally, we can write down the gravity perturbation in the Rayleigh-scattering regime
\beq
\delta\vec a(\vec 0,t) = \frac{4\pi}{9} G\rho_0\xi_0\e^{-\irm\omega t}\left(\frac{2}{3}(k^{\rm P}a)^2-\frac{7}{10}(k^{\rm S}a)^2\right)\vec e_x
\label{eq:scatterSNN}
\eeq
This completes our analysis of scattering effects on gravity perturbations. We found that waves scattered from a spherical cavity with incident P-waves and S-waves have negligible impact on gravity perturbations if the cavity radius is much smaller than the length of seismic waves. The gravity change according to Equations (\ref{eq:scatterPNN}) and (\ref{eq:scatterSNN}) is quadratic in the cavity radius $a$. In addition, the gravity perturbation from scattered waves is in phase with gravity perturbations of the incident wave (in the Rayleigh-scattering regime), which is beneficial for coherent noise cancellation, if necessary. 

\subsection{Gravity perturbations from seismic waves in a homogeneous half space}
\label{sec:halfspace} 
In this section, the gravity perturbation produced by plane seismic waves in a homogeneous half space will be calculated. The three types of waves that will be considered are compressional, shear, and Rayleigh waves. Reflection of seismic waves from the free surface will be taken into account. The purpose is to provide equations that can be used to estimate seismic Newtonian noise in GW detectors below and above surface. For underground detectors, corrections from the presence of a cavity will be neglected, but with the results of Section \ref{sec:scatterNN}, it is straight-forward to calculate the effect of a cavity also for the half-space problem.

\subsubsection{Gravity perturbations from body waves}
\label{sec:bodyhalf}
As a first step, we will calculate the gravity perturbation from plane shear and compressional waves without taking reflection from the free surface into account. The compressional wave has the form in Equation (\ref{eq:comprPW}), and the perturbation of the gravity potential above surface integrated over the half space and including the surface contribution is found to be
\beq
\delta\phi^{\rm P}(\vec r,t) = -2\pi G\rho_0 \xi_0^{\rm P}\e^{\irm(\vec k_\varrho\cdot\vec\varrho-\omega t)}\e^{-k_\varrho h}\frac{1}{\irm k^{\rm P}},
\eeq
with $h$ being the height of the point $\vec r$ above surface, $\vec\varrho$ being the projection of $\vec r$ onto the surface, and $\vec k_\varrho$ being the horizontal wave vector (omitting superscript 'P' to ease notation). The solution above surface can be understood as pure surface term characterized by an exponential suppression with increasing height. Also the phase term is solely a function of horizontal coordinates. These are typical characteristics for a surface gravity perturbation, and we will find similar results for gravity perturbations from Rayleigh waves. Below surface, $h$ reinterpreted as (positive valued) depth, the solution reads
\beq
\delta\phi^{\rm P}(\vec r,t) = -2\pi G\rho_0 \xi_0^{\rm P}\e^{-\irm\omega t}\frac{1}{\irm k^{\rm P}}\left(2\e^{\irm\vec k^{\rm P}\cdot\vec r}-\e^{-k_\varrho h}\e^{\irm\vec k_\varrho\cdot\vec\varrho}\right)
\label{eq:underP}
\eeq
It consists of a surface term with exponential suppression as a function of depth, and of the infinite-space solution of Equation (\ref{eq:gravP}). If the point $\vec r$ is at the surface ($h=0$), then the total half-space gravity perturbation is simply half of the infinite-space perturbation. 

The calculation is substantially easier for shear waves. Shear waves being transversal waves can have two different orthogonal polarizations. If the displacement is parallel to the free surface, then the polarization is called SH, otherwise it is called SV. An SH polarized wave cannot produce gravity perturbation, since shear waves do not produce density perturbations inside media, and SH waves also do not displace the surface along its normal. Gravity perturbations can be produced by SV waves through surface displacement. The result valid for gravity perturbations underground and above surface is 
\beq
\delta\phi^{\rm SV}(\vec r,t) = 2\pi G\rho_0 \xi_0^{\rm SV}\e^{\irm(\vec k_\varrho\cdot\vec\varrho-\omega t)}\e^{-k_\varrho h}\frac{1}{k^{\rm S}},
\eeq
where $h$ is the distance to the surface. These solutions can now be combined to calculate the gravity perturbation from an SV or P wave reflected from the surface. An incident compressional wave is partially converted into an SV wave and vice versa. Only waves with the same horizontal wave vector $\vec k_\varrho$ couple at reflection from a flat surface \cite{AkRi2009}. Therefore, the total gravity perturbation above surface in the case of an incident compressional wave can be written
\beq
\delta\phi^{\rm P}(\vec r,t) = -2\pi G\rho_0 \xi_0^{\rm P}\e^{\irm(\vec k_\varrho\cdot\vec\varrho-\omega t)}\e^{-k_\varrho h}\frac{1}{\irm k^{\rm P}}\left((1+{\rm PP}(k_\varrho))-\irm\frac{k^{\rm P}}{k^{\rm S}} {\rm PS}(k_\varrho)\right)
\eeq
The conversion of amplitudes is described by two reflection coefficients ${\rm PP}(k_\varrho),\,{\rm PS}(k_\varrho)$, as functions of the horizontal wave number. Their explicit form can for example be found in \cite{AkRi2009}, which leads to the gravity perturbation 
\beq
\begin{split}
\delta\phi^{\rm P}(\vec r,t) &= -2\pi G\rho_0 \xi_0^{\rm P}\e^{\irm(\vec k_\varrho\cdot\vec\varrho-\omega t)}\e^{-k_\varrho h}\frac{1}{\irm k^{\rm P}}\delta(\nu,k_\varrho)\\
\delta(\nu,k_\varrho) &\equiv \frac{8k_\varrho^2k^{\rm P}_zk^{\rm S}_z-\irm 4k_\varrho k^{\rm P}_z((k^{\rm S})^2-2k_\varrho^2)}{((k^{\rm S})^2-2k_\varrho^2)^2+4k_\varrho^2k^{\rm P}_zk^{\rm S}_z}
\end{split}
\label{eq:incidentP}
\eeq 
The gravity perturbation vanishes for horizontally and vertically propagating incident P-waves: the total P-wave contribution proportional to $1+{\rm PP}(k_\varrho)$ vanishes because of interference of the incident and reflected P-wave, while there is no conversion ${\rm PS}(k_\varrho)$ from P to S-waves for these two angles. The gravity amplitude $\delta(\cdot)$ depends on the Poisson's ratio, and the angle of incidence of the P-wave. Its absolute value is plotted in Figure \ref{fig:coeffNN} for three different angles of incidence $10^\circ,\,45^\circ,\,80^\circ$ of the P-wave with respect to the surface normal. Important to note is that above surface, the gravity perturbation produced by shear and body waves assumes the form of a surface density perturbation with exponential suppression as a function of height above ground, determined by the horizontal wavenumber. The expression for an incident S-wave can be constructed analogously. 

\subsubsection{Gravity perturbations from Rayleigh waves}
\label{sec:gravRayleigh}
The results for body waves can be compared with gravity perturbations from fundamental Rayleigh waves. There are two options to calculate the gravity perturbation. One is based on a representation of the Rayleigh wave in terms of seismic potentials (explicit expression can be found in \cite{Nov1999}), and using the last line in Equation (\ref{eq:gravHelm}). In the following, we choose to calculate gravity based on the displacement field since it is more intuitive, and not significantly more difficult. The Rayleigh displacement field can be written as \cite{HaNa1998}
\beq
\begin{split}
\vec\xi(\vec r,t) &= \xi_k(\vec r,t) \vec e_k+\xi_z(\vec r,t) \vec e_z\\
\xi_k(\vec r,t)  &= \irm\left(H_1\e^{h_1z}+H_2\e^{h_2z}\right)\e^{\irm (\vec k_\varrho\cdot\vec \varrho-\omega t)}\\
&= \irm H(z)\e^{\irm (\vec k_\varrho\cdot\vec \varrho-\omega t)}\\
\xi_z(\vec r,t)  &= \left(V_1\e^{v_1z}+V_2\e^{v_2z}\right)\e^{\irm (\vec k_\varrho\cdot\vec \varrho-\omega t)}\\
&= V(z)\e^{\irm (\vec k_\varrho\cdot\vec \varrho-\omega t)}
\end{split}
\label{eq:Rayleigh}
\eeq
The parameters $H_i,\,V_i,\,h_i,\,v_i$ are real numbers, see Equation (\ref{eq:Rayfield}), and so there is a constant $90^\circ$ phase difference between horizontal and vertical displacement leading to elliptical particle motion. The surface displacement and the density change inside the medium caused by the Rayleigh wave lead to gravity perturbations. The surface contribution valid below and above ground is given by
\beq
\begin{split}
\delta\phi_{\rm surf} (\vec r_0,t)&= -G\rho_0V(0)\e^{\irm(\vec k_\varrho\cdot\vec\varrho_0-\omega t)}\int\drm S\frac{\e^{\irm k_\varrho \varrho\cos(\phi)}}{\sqrt{\varrho^2+h^2}}\\
&= -2\pi G\rho_0(V_1+V_2)\e^{-hk_\varrho}\e^{\irm(\vec k_\varrho\cdot\vec\varrho_0-\omega t)}
\end{split}
\eeq
As before, the distance of the test mass to the surface is denoted by $h$. The density perturbations in the ground are calculated from the divergence of the Rayleigh-wave field:
\beq
\nabla\cdot\vec\xi(\vec r,t)=(-k_\varrho H(z)+V^\prime(z))\e^{\irm (\vec k_\varrho\cdot\vec \varrho-\omega t)},
\eeq
and therefore the bulk contribution to the gravity perturbation above surface reads:
\beq
\begin{split}
\delta\phi_{\rm bulk}(\vec r_0,t) &= G\rho_0\e^{\irm (\vec k_\varrho\cdot\vec \varrho_0-\omega t)}\int\drm V\frac{(-k_\varrho H(z)+V^\prime(z))\e^{\irm k \varrho\cos(\phi)}}{\sqrt{\varrho^2+(h-z)^2}}\\
&= 2\pi G\rho_0\e^{\irm (\vec k_\varrho\cdot\vec \varrho_0-\omega t)}\frac{1}{k_\varrho}\int\limits_{-\infty}^0\drm z(-k_\varrho H(z)+V^\prime(z))\e^{-(h-z)k_\varrho}
\end{split}
\label{eq:Surface}
\eeq
Inserting the definitions of $H(z),\,V(z)$ from Equation (\ref{eq:Rayleigh}) into the last equation, we finally obtain
\beq
\delta\phi_{\rm bulk}(\vec r_0,t) = 2\pi G\rho_0\e^{-hk_\varrho}\e^{\irm (\vec k_\varrho\cdot\vec \varrho_0-\omega t)}\frac{1}{k_\varrho}\left[-\frac{k_\varrho H_1}{h_1+k_\varrho}-\frac{k_\varrho H_2}{h_2+k_\varrho}+\frac{v_1V_1}{v_1+k_\varrho}+\frac{v_2V_2}{v_2+k_\varrho}\right]
\eeq
Adding bulk and surface contributions, one obtains the full gravity perturbation from a Rayleigh wave:
\beq
\begin{split}
\delta\phi_{\rm surf} (\vec r_0,t)&+\delta\phi_{\rm bulk}(\vec r_0,t)=\\
&\phantom{=}-2\pi G\rho_0 \e^{-hk_\varrho}\e^{\irm (\vec k_\varrho\cdot\vec \varrho_0-\omega t)}\left[\frac{H_1}{h_1+k_\varrho}+\frac{H_2}{h_2+k_\varrho}+\frac{V_1}{v_1+k_\varrho}+\frac{V_2}{v_2+k_\varrho}\right]\\
&=-2\pi G\rho_0 A\e^{-hk_\varrho}\e^{\irm (\vec k_\varrho\cdot\vec \varrho_0-\omega t)}(1-\zeta(k_\varrho)),
\end{split}
\label{eq:RayleighNN}
\eeq
where in the last line the parameters $H_i,\,V_i,\,h_i,\,v_i$ have been substituted by the expressions in Equation (\ref{eq:Rayfield}) for fundamental Rayleigh waves. The gravity perturbation underground contains an additional contribution from the compressional-wave content of the Rayleigh field:
\beq
\begin{split}
\delta\phi_{\rm surf} (\vec r_0,t)&+\delta\phi_{\rm bulk}(\vec r_0,t)=2\pi G\rho_0 A\e^{\irm (\vec k_\varrho\cdot\vec \varrho_0-\omega t)}\left(-2\e^{-hq_z^{\rm P}}+(1+\zeta(k_\varrho))\e^{-hk_\varrho}\right),
\end{split}
\label{eq:RayleighNNUG}
\eeq
where $q_z^{\rm P}$ is the vertical wavenumber of evanescent compressional waves defined in Section \ref{sec:seismic}, and $h$ is the depth of the test mass. Contributions from a cavity wall need to be added, which is straight-forward at least for a very small cavity, by using results from Section \ref{sec:bodynoscatt} and amplitudes of shear and compressional waves dependent on depth as given in Equation (\ref{eq:Rayfield}).

\epubtkImage{}{%
    \begin{figure}[htbp]
    \centerline{\includegraphics[width=0.6\textwidth]{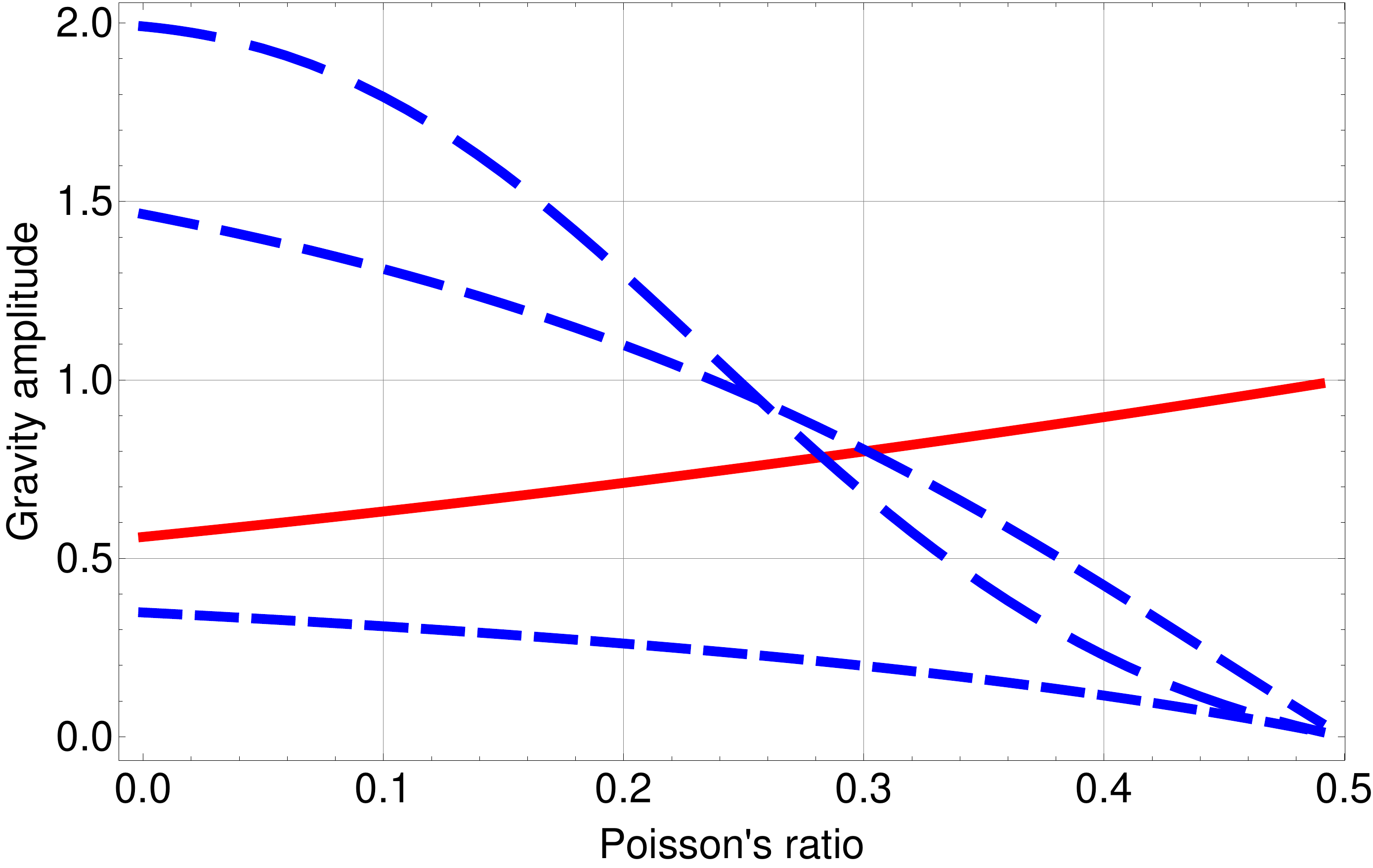}}
    \caption[Plane-wave Newtonian-noise amplitudes]{Gravity amplitudes for a medium with free, flat surface as functions of the Poisson's ratio. The solid line shows the gravity amplitude for Rayleigh waves, whereas the dashed lines show the gravity amplitudes for incident P-waves for three different angles of incidence: $10^\circ,\,45^\circ,\,80^\circ$ with increasing dash length.}
\label{fig:coeffNN}
    \end{figure}}   
Comparing with Equation (\ref{eq:incidentP}), one can see that the analytical expressions of gravity perturbations above ground produced by incident compressional waves or by Rayleigh waves are very similar. Only the wavenumber-dependent amplitude term, either in the form of wave-reflection coefficients or Rayleigh-wave amplitude coefficients, is different. In order to plot the results, it is convenient to substitute the amplitude $A$ by vertical surface displacement:
\beq
A=\frac{\xi_z(\vec 0,0)}{q_z^{\rm P}-k_\varrho\zeta(k_\varrho)}
\eeq
Inserting this expression into Equation (\ref{eq:RayleighNN}), and applying the gradient operator to both sides of the equation (which yields an expression for $\delta\vec a(\vec r_0,t)$), we obtain a unit-less factor that depends on the elastic properties of the half-space:
\beq
\gamma(\nu)=\frac{k_\varrho(1-\zeta(k_\varrho))}{q_z^{\rm P}-k_\varrho\zeta(k_\varrho)}.
\eeq
The wavenumbers of shear, compressional, and Rayleigh waves all have fixed proportions determined by the Poisson's ratio $\nu$ of the half-space medium (see Section \ref{sec:seismic}). Therefore, $\gamma$ itself is fully determined by $\nu$. A plot of $\gamma(\nu)$ is shown in Figure \ref{fig:coeffNN}. The maximum value of $\gamma(\nu)$ is equal to 1, which also corresponds to the case of gravity perturbations from pure surface displacement. This means that the density perturbations generated by the Rayleigh wave inside the medium partially cancel the surface contribution for $\nu<0.5$.

\subsection{Numerical simulations}
\label{sec:numsim}\index{numerical simulations}
Numerical simulations have become an important tool in seismic Newtonian-noise modelling. There are two types of numerical simulations. The first will be called ``kinematic'' simulation.\index{numerical simulations!kinematic} It is based on a finite-element model where each element is displaced according to an explicit, analytical expression of the seismic field. These can be easily obtained for individual seismic surface or body waves. The main work done by the kinematic simulation is to integrate gravity perturbations from a complex superposition of waves over the entire finite-element model according to Equation (\ref{eq:dipoleacc}). Today, we have explicit expressions for all types of seismic waves produced by all types of seismic sources, in infinite and half-space media. While this means in principle that many interesting kinematic simulations can be carried out, some effects are very hard to deal with. The kinematic simulation fails whenever it is impossible to provide analytical expressions for the seismic field. This is generally the case when heterogeneities of the ground play a role. Also deviations from a flat surface may make it impossible to run accurate kinematic simulations. In this case, a ``dynamical'' simulation needs to be employed.\index{numerical simulations!dynamical} 

A dynamical simulation only requires accurate analytical models of the seismic sources. The displacement field evolves from these sources governed by equations of motion that connect the displacement of neighboring finite elements. Even though the dynamical simulation can be considered more accurate since it does not rely on guessing solutions to the equations of motion, it is also true that not a single simulation of Newtonian noise has been carried out so far that could not have been done with a kinematic simulation. The reasons are that dynamical simulations are computationally very expensive, and constructing realistic models of the medium can be very challenging. It is clear though dynamical simulations will play an important role in future studies when effects from heterogeneities on gravity signals are investigated in detail.

Since kinematic simulations are easy to set up from scratch, we will focus on the discussion of dynamical simulations. Two tools have been used in the past for Newtonian-noise simulations. The first one is the commercial software Comsol. It interfaces with Matlab, which facilitates analyzing sometimes complex results. Simulation results for a seismic field produced by a point force at the origin are shown in Figure \ref{fig:comsolimp}.
\epubtkImage{}{%
    \begin{figure}[htbp]
    \centerline{\includegraphics[width=0.8\textwidth]{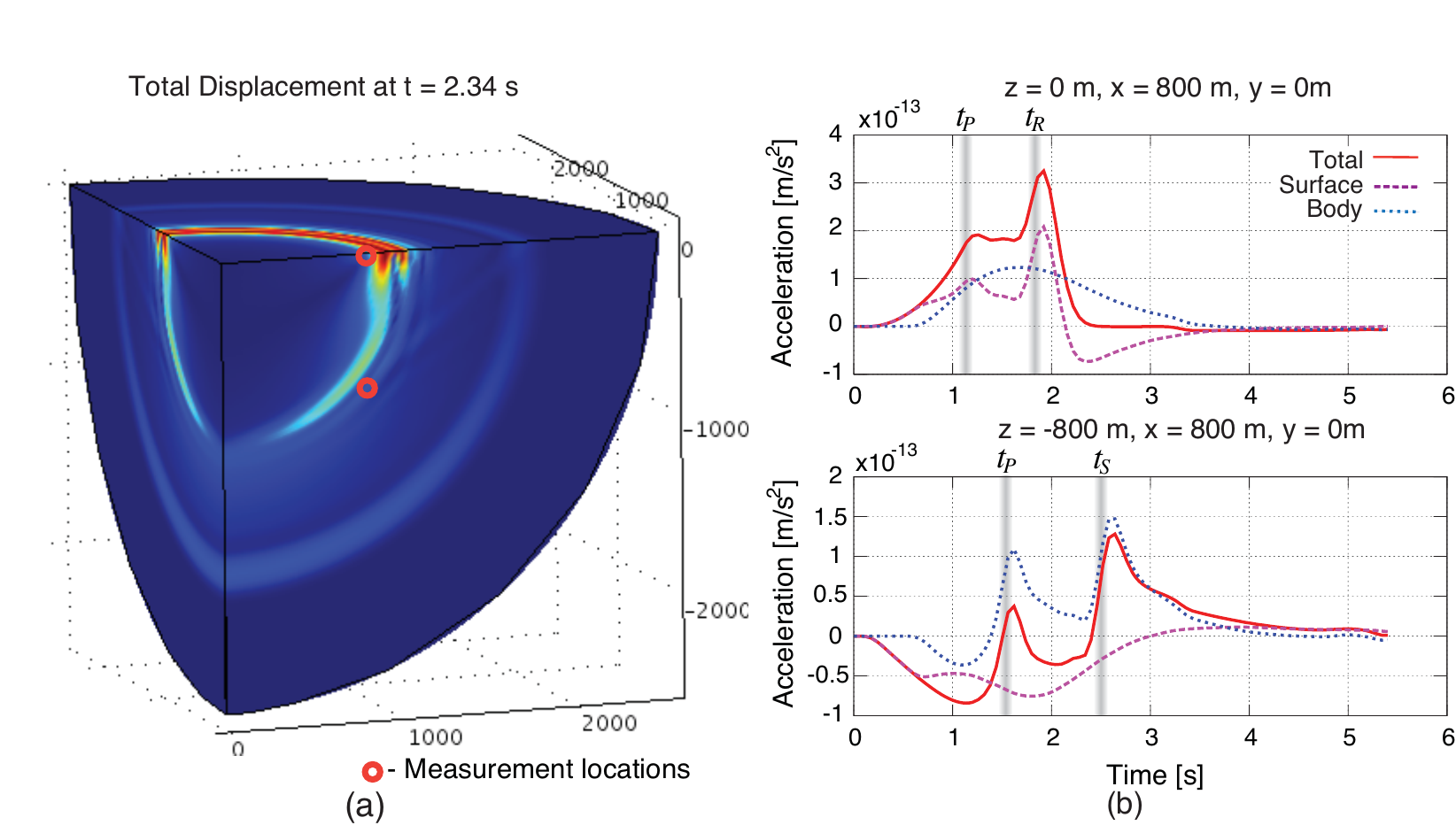}}
    \caption[Finite-element simulation of gravity perturbations from a surface impact]{Comsol simulation of gravity perturbation from seismic fields. The plot to the left shows a snapshot of the displacement field produced by a step-function point source at the origin. The plots to the right show the corresponding gravity perturbations evaluated at the two points marked in red in the left plot. Courtesy of Beker at al \cite{BeEA2010c}.}
\label{fig:comsolimp}
    \end{figure}}  
The results were presented in \cite{BeEA2010c}. A snapshot of the displacement field is plotted on the left. The P-wavefront is relatively weak and has already passed half the distance to the boundaries of the grid. Only a spherical octant of the entire finite-element grid is shown. The true surface in this simulation is the upper face of the octant. Consequently, a strong Rayleigh-wave front is produced by the point force. Slightly faster than the Rayleigh waves, an S-wavefront spreads underground. Its maximum is close to the red marker located underground. This seismic field represents a well-known problem in seismology, the so-called Lamb's problem, which has an explicit time-domain solution \cite{Ric1979}.\index{Lamb's problem} The plots on the right show the gravity perturbations evaluated at the two red markers. The P-wave, S-wave and Rf-wave arrival times are $t_{\rm P},\,t_{\rm S}$ and $t_{\rm R}$ respectively. The gravity perturbations are also divided into contributions from density perturbations inside the medium according to Equation (\ref{eq:bulkNN}) and surface contributions according to Equation (\ref{eq:surfNN}).

A second simulation package used in the past is SPECFEM3D. It is a free software that can be downloaded at \url{http://www.geodynamics.org/cig/software/specfem3d}. It is one of the standard simulation tools in seismology. It implements the spectral finite element method \cite{KoVi1998,KoTr1999}. Recently, Equation (\ref{eq:dipoleacc}) has been implemented for gravity calculations \cite{HaEA2015}. SPECFEM3D simulations typically run on computer clusters, but it is also possible to execute simple examples on a modern desktop. Simulations of wave propagation in heterogeneous ground and based on realistic source models such as shear dislocations are probably easier to carry out with SPECFEM3D than with commercial software. However, it should be noted that it is by no means trivial to run any type of simulation with SPECFEM3D, and a large amount of work goes into defining a realistic model of the ground for SPECFEM3D simulations. 
\epubtkImage{}{%
    \begin{figure}[htbp]
    \centerline{\includegraphics[width=0.33\textwidth]{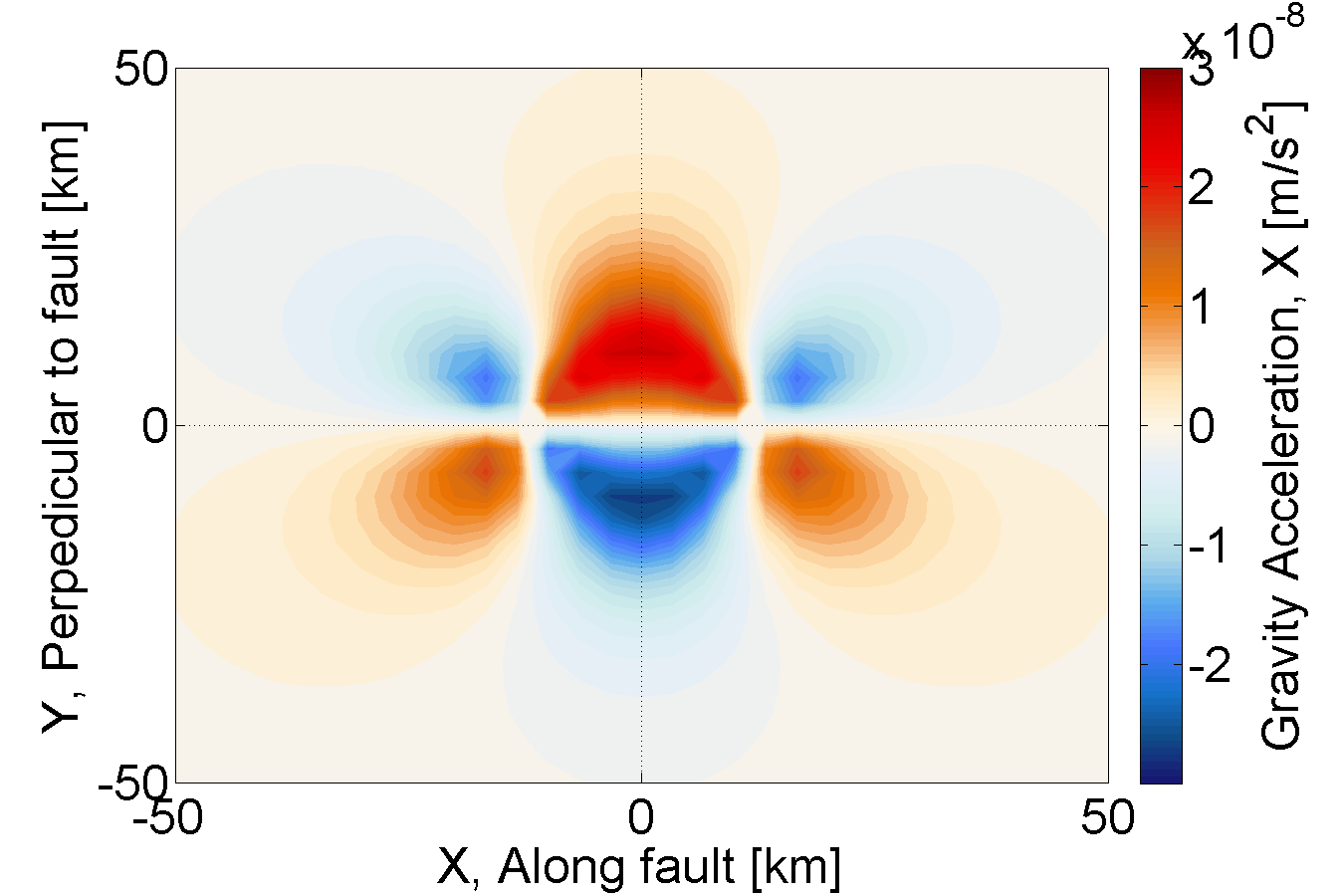}
                \includegraphics[width=0.33\textwidth]{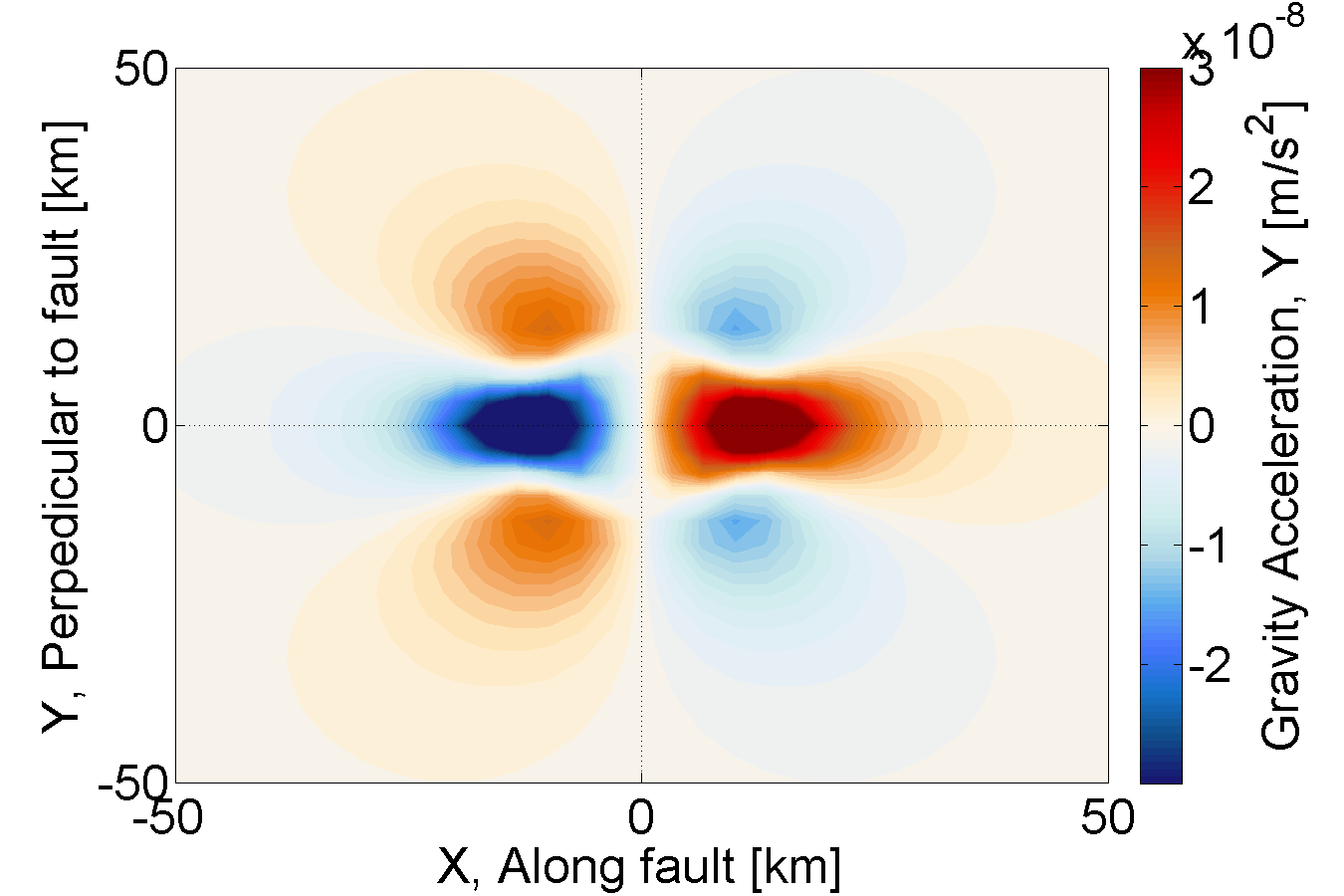}
                \includegraphics[width=0.33\textwidth]{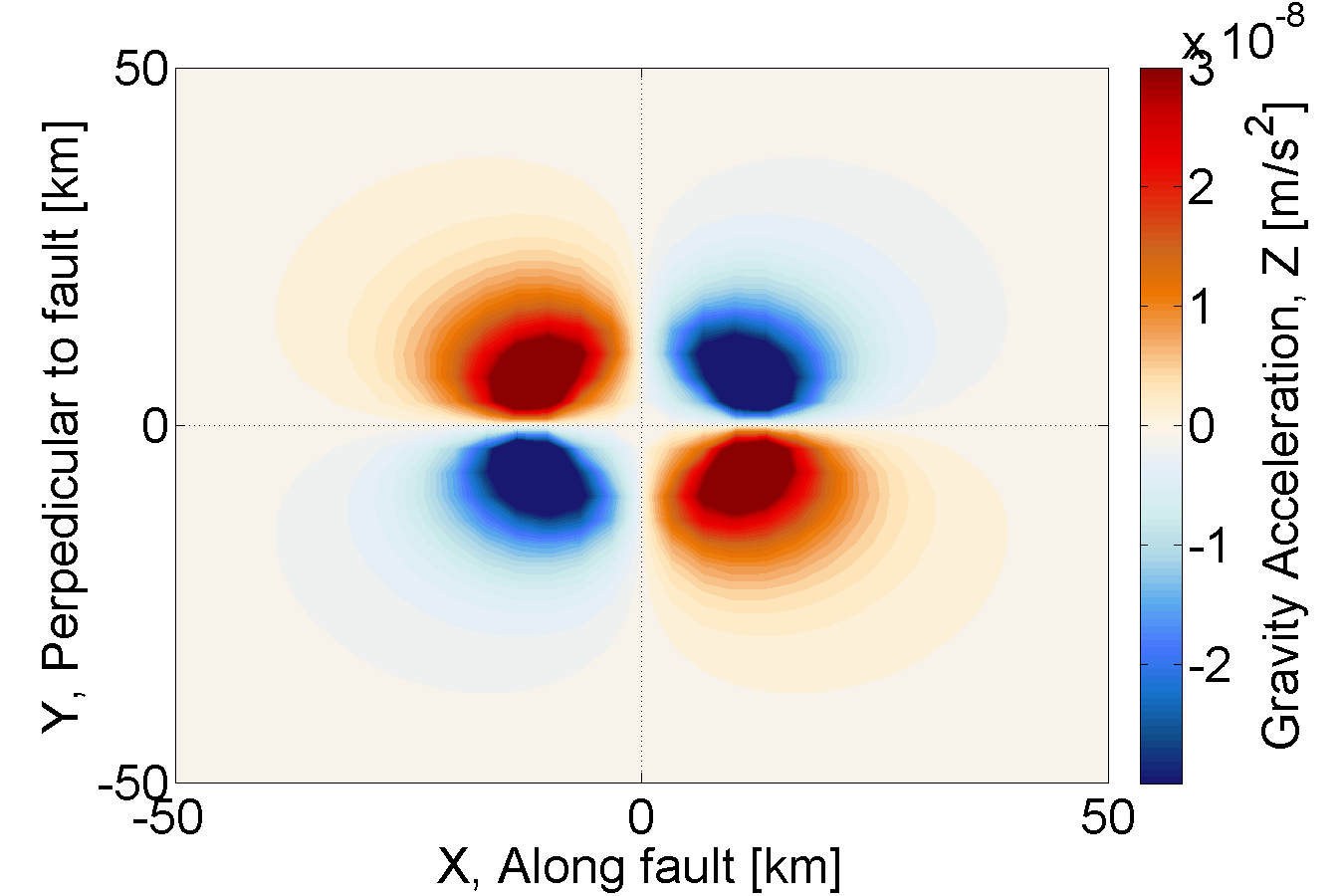}}
    \caption[Finite-element simulation of fault rupture]{SPECFEM3D simulation of a strike-slip fault rupture. The gravity is evaluated on a horizontal plane that includes the hypocenter.}
\label{fig:specfem}
    \end{figure}}  
Nonetheless, this is the realm of dynamical simulations, and simplifying geological models one should always check if a kinematic simulation can be used. An example of a gravity simulation using SPECFEM3D is shown in Figure \ref{fig:specfem}. The contour plots are snapshots of the gravity field after 5\,s of rupture on a strike-slip fault. The length of the vertical fault is 30\,km with hypocenter located 7.5\,km underground. The plots show the gravity perturbation on a horizontal plane that includes the hypocenter. Gravity perturbations in the vicinity of the fault are dominated by the lasting gravity change. The transient perturbation carried by seismic waves is invisible in these plots simply because of their small amplitudes compared to the lasting gravity change. An explicit time-domain expression of the gravity field does not exist, but it could be constructed with a kinematical simulation using the results of Section \ref{sec:sourcehalf}. In conclusion, while dynamical simulations are required to represent seismic fields in complex geologies and surface topographies, one should always favor kinematic simulations when possible. Kinematic simulations are faster by orders of magnitude facilitating systematic studies of gravity perturbations.

\subsection{Seismic Newtonian-noise estimates}
\label{sec:estSeismicNN}
The results of the analytical calculations can be used to estimate seismic Newtonian noise in GW detectors above surface and underground. The missing steps are to convert test-mass acceleration into gravity strain, and to understand the amplitudes of perturbation as random processes, which are described by spectral densities (see Section \ref{sec:noisefreq}). For a precise noise estimate, one needs to measure the spectrum of the seismic field, its two-point spatial correlation or anisotropy. These properties have to be known within a volume of the medium under or around the test masses, whose size depends on the lengths of seismic waves within the relevant frequency range. Practically, since all these quantities are then used in combination with a Newtonian-noise model, one can apply simplifications to the model, which means that some of these quantities do not have to be known very accurately or do not have to be known at all. For example, it is possible to obtain good Newtonian-noise estimates based on the seismic spectrum alone. All of the published Newtonian-noise estimates have been obtained in this way, and only a few conference presentations showed results using additional information such as the anisotropy measurement or two-point spatial correlation. In the following, the calculation of Newtonian-noise spectra is outlined in detail.

\subsubsection{Using seismic spectra}
We start with the simplest approach based on measured spectra of the ambient seismic field, all other quantities are represented by simple analytical models. At the LIGO Hanford site, it was found by array measurements that the main contribution to the vertical seismic spectrum at frequencies relevant to Newtonian noise comes from Rayleigh waves \cite{Dri2012}. Even if the wave composition of a seismic field at a surface site is unknown, then it would still be reasonable to assume that Rayleigh waves dominate the vertical spectrum since they couple most strongly to surface or near-surface sources \cite{Moo1976,BoEA2006}. We emphasize that this only holds for the vertical displacement since horizontal displacement can contain strong contributions from Love waves, which are shear waves with purely horizontal displacement trapped in surface layers\index{seismic waves!Love}. This means that we can use Equation (\ref{eq:RayleighNN}) to obtain an estimate of seismic Newtonian noise. We first rewrite it to give the Cartesian components of gravity acceleration:
\beq
\begin{split}
\delta\vec a(\vec r_0,t) &= 2\pi G\rho_0 \xi_z\e^{-hk_\varrho}\e^{\irm (\vec k_\varrho\cdot\vec \varrho_0-\omega t)}\gamma(\nu)\begin{pmatrix}\irm\cos(\phi)\\ \irm\sin(\phi)\\ -1\end{pmatrix}
\end{split}
\label{eq:RayleighS}
\eeq
where $\phi$ is the angle of propagation with respect to the $x$-axis. Note that all three components of acceleration are determined by vertical surface displacement. This is possible since vertical and horizontal displacements of Rayleigh waves are not independent. As we will argue in Section \ref{sec:mitigate}, expressing Newtonian noise in terms of vertical displacement is not only a convenient way to model Newtonian noise, but it is also recommended to design coherent cancellation schemes at the surface based on vertical sensor data, since horizontal sensor data can contain contributions from Love waves, which do not produce Newtonian noise. Hence, horizontal channels are expected to show lower coherence with Newtonian noise. Assuming that the Rayleigh-wave field is isotropic, one can simply average the last equation over all propagation directions. The noise spectral density of differential acceleration along a baseline of length $L$ parallel to the $x$-axis reads
\beq
S(\delta \vec a(L\vec e_x)-\delta \vec a(\vec 0);\omega) = \left(2\pi G\rho_0 \e^{-hk_\varrho}\gamma(\nu)\right)^2S(\xi_z;\omega)\begin{pmatrix}1-2J_0(k_\varrho L)+2J_1(k_\varrho L)/(k_\varrho L)\\ 1-2J_1(k_\varrho L)/(k_\varrho L)\\ 2-2J_0(k_\varrho L)\end{pmatrix}
\label{eq:RayNN}
\eeq
The vector contains the three direction-averaged response functions of horizontal and vertical gravity perturbations. Rayleigh Newtonian noise in one direction is uncorrelated with Newtonian noise in the other two directions independent of the value of $L$. Introducing $\lambda_{\rm R}\equiv 2\pi/k_\varrho$, the response functions, i.~e.~the square roots of the components of the vector in Equation (\ref{eq:RayNN}), divided by $L/\lambda_{\rm R}$ are shown in Figure \ref{fig:rayResp}. 
\epubtkImage{}{     
    \begin{figure}[htbp]
    \centerline{\includegraphics[width=0.6\textwidth]{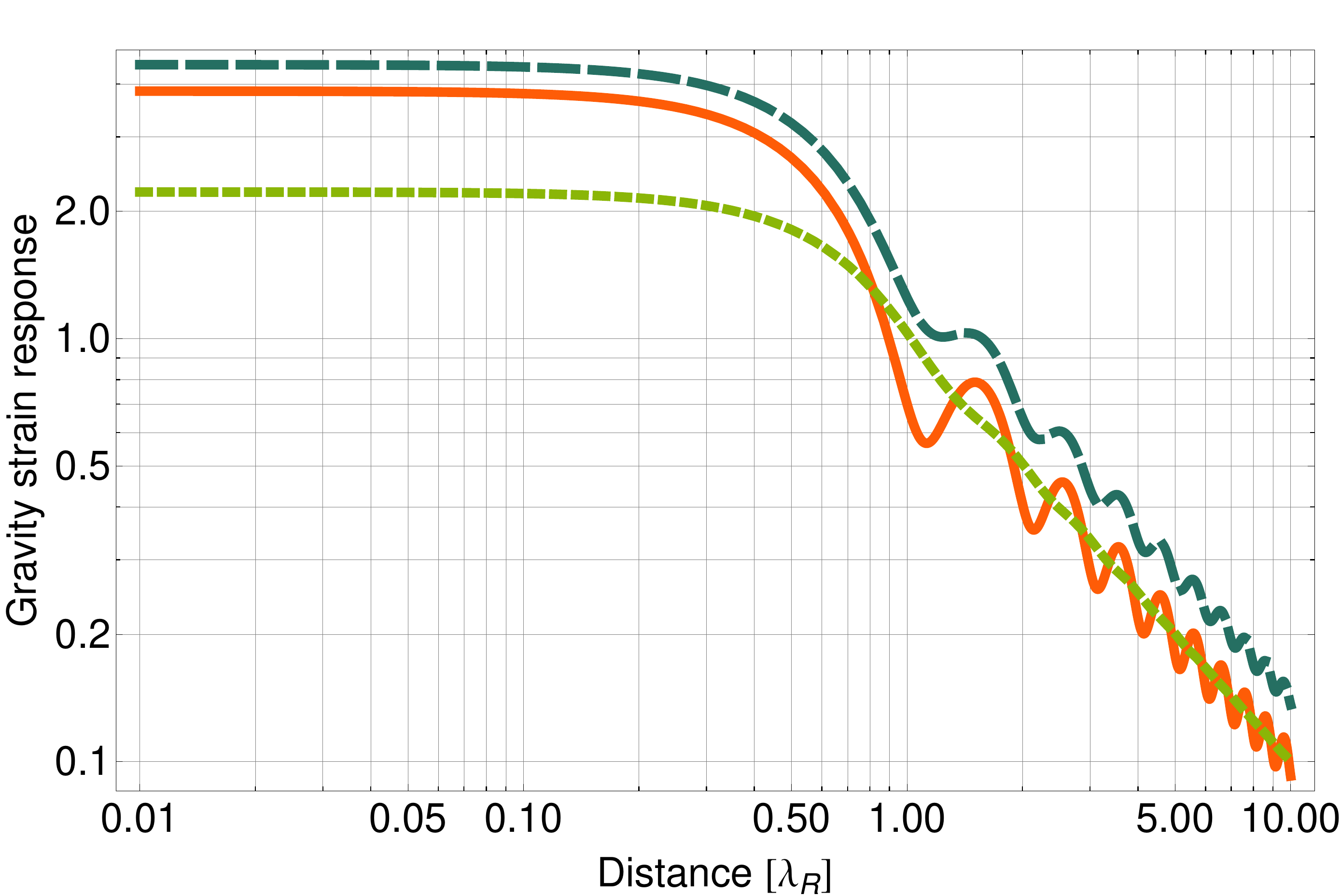}}
    \caption[Strain response to Rayleigh gravity perturbations]{Strain response to Rayleigh gravity perturbations. The solid curve shows the horizontal response for gravity perturbations along the line of separation, the dotted curve the horizontal response for perturbations perpendicular to the line of separation, and the dashed curve for perturbations in vertical direction.}
\label{fig:rayResp}
    \end{figure}}    
Gravity perturbations at the two locations $x=\{0,L\},\,y=z=0$ are uncorrelated for sufficiently large distances, and therefore the strain response decreases with increasing $L$. In other words, increasing the length of large-scale GW detectors would decrease Newtonian noise. Rayleigh Newtonian noise is independent of $L$ for short separations. This corresponds to the regime relevant to low-frequency GW detectors \cite{HaEA2013}. Equation (\ref{eq:RayNN}) is the simplest possible seismic surface Newtonian-noise estimate. Spatial correlation of the isotropic seismic field is fully determined by the fact that all seismic waves are assumed to be Rayleigh waves. Practically one just needs to measure the spectral density of vertical surface displacement, and also an estimate of the Poisson's ratio needs to be available (assuming a value of $\nu=0.27$ should be a good approximation in general \cite{ZaAm1995}). In GW detectors, the relevant noise component is along the $x$-axis. Taking the square-root of the expression in Equation (\ref{eq:RayNN}), and using a measured spectrum of vertical seismic motion, we obtain the Newtonian-noise estimate shown in Figure \ref{fig:VirgoNN} \footnotetext{Seismic data stem from channel SEBDCE06 between June 4, 2011, UTC 00:00 and September 3, 2011 UTC 00:00.}. Virgo's arm length is $L=3000\,$m, and the test masses are suspended at a height of about $h=1\,$m (although, it should be mentioned that the ground is partially hollow directly under the Virgo test masses). In order to take equal uncorrelated noise contributions from both arms into account, the single-arm strain noise needs to be multiplied by $\sqrt{2}$. The seismic spectrum falls approximately with $1/f$ in units of $\rm m/s/\sqrt{ Hz}$ within the displayed frequency range, which according to Equation (\ref{eq:RayNN}) means that the Newtonian-noise spectrum falls with $1/f^4$ (two additional divisions by $f$ from converting differential acceleration noise into differential displacement noise, and another division by $f$ from converting the seismic spectrum into a displacement spectrum). Note that the knee frequency of the response curve in Figure \ref{fig:rayResp} lies well below the frequency range of the spectral plots, and therefore does not influence the frequency dependence of the Newtonian-noise spectrum.
\epubtkImage{}{%
    \begin{figure}[htbp]
    \centerline{\includegraphics[width=0.49\textwidth]{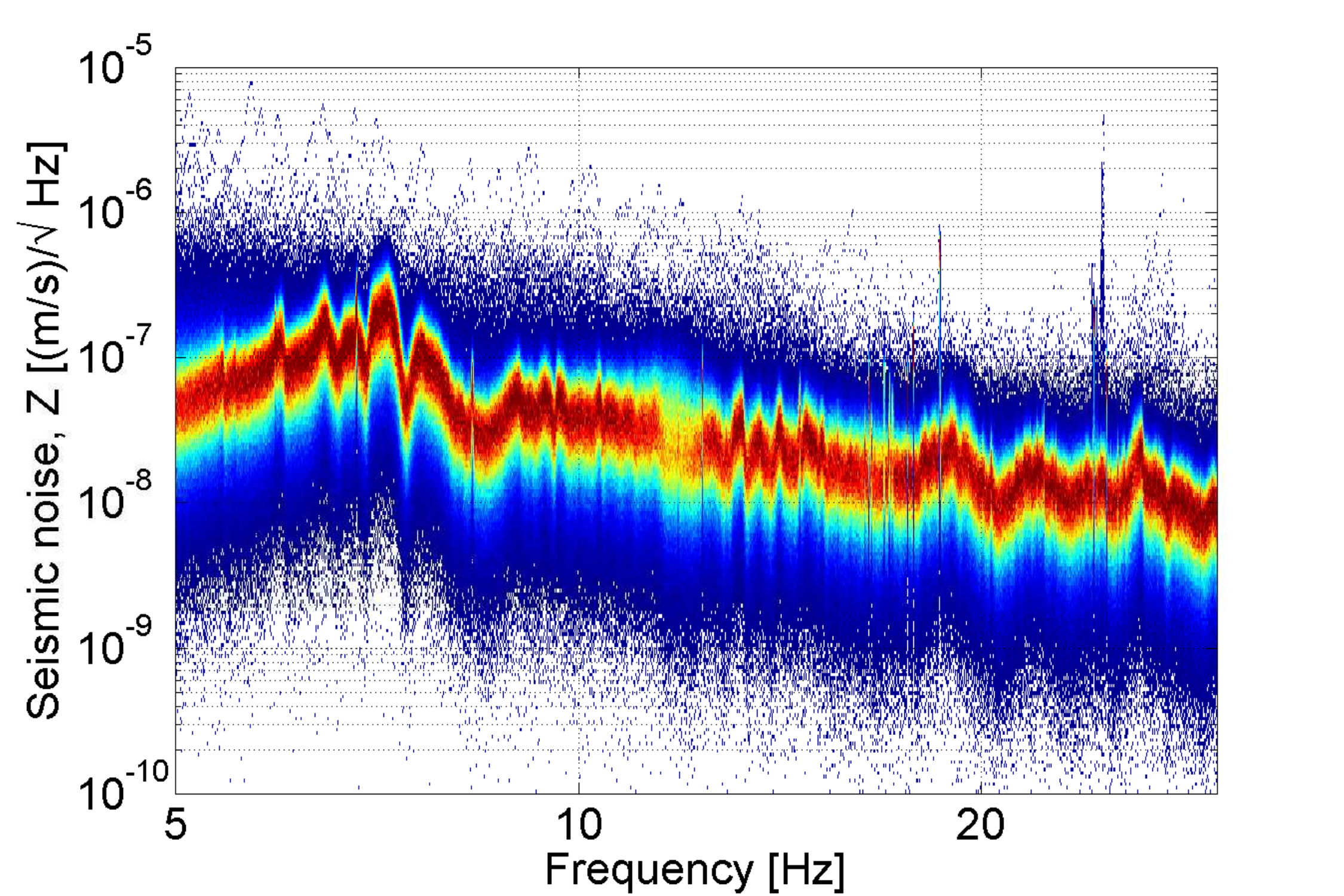}                 
                \includegraphics[width=0.49\textwidth]{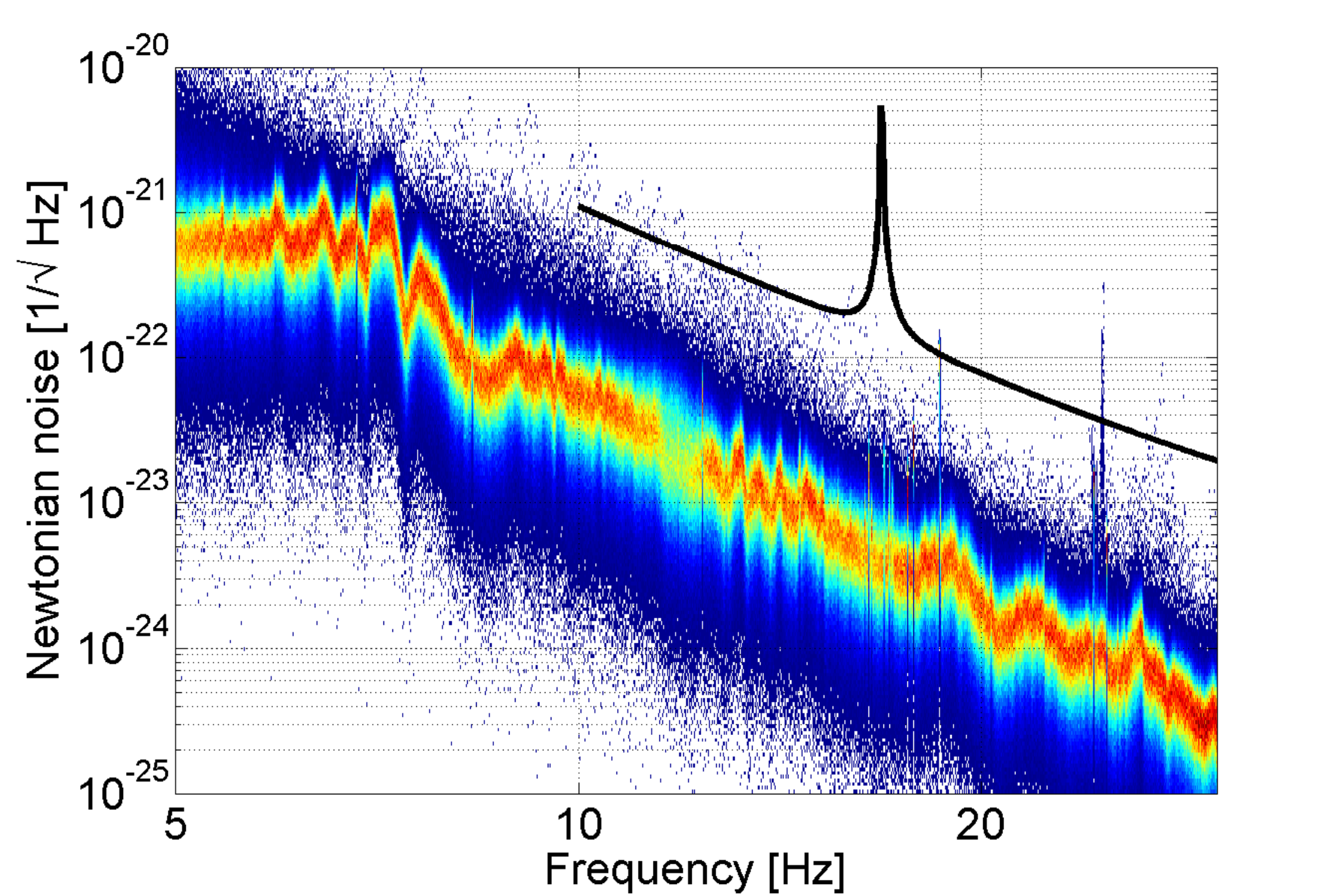}}
    \caption[Rayleigh Newtonian noise at Virgo site]{Histograms of seismic spectra at the central station of the Virgo detector and modelled Rayleigh-wave Newtonian noise \addtocounter{footnote}{-1}\footnotemark. A sensitivity model of the Advanced Virgo detector is plotted for comparison.}
\label{fig:VirgoNN}
    \end{figure}} 
Since seismic noise is non-stationary in general, and therefore can show relatively large variations in spectra, it is a wise idea to plot the seismic spectra as histograms instead of averaging over spectra. The plots can then be used to say for example that Newtonian noise stays below some level 90\% of the time (the corresponding level curve being called 90th percentile). In the shown example, a seismic spectrum was calculated each 128\,s for 7 days. Red colors mean that noise spectra often assume these values, blue colors mean that seismic spectra are rarely observed with these values. No colors mean that a seismic spectrum has never assumed these values within the 7 days of observation. Interesting information can be obtained in this way. For example, it can be seen that between about 11\,Hz -- 12\,Hz a persistent seismic disturbance increases the spectral variation, which causes the distribution to be wider and therefore the maximum value of the histogram to be smaller. \addtocounter{footnote}{-1}
\epubtkImage{}{%
    \begin{figure}[htbp]
    \centerline{\includegraphics[width=0.50\textwidth]{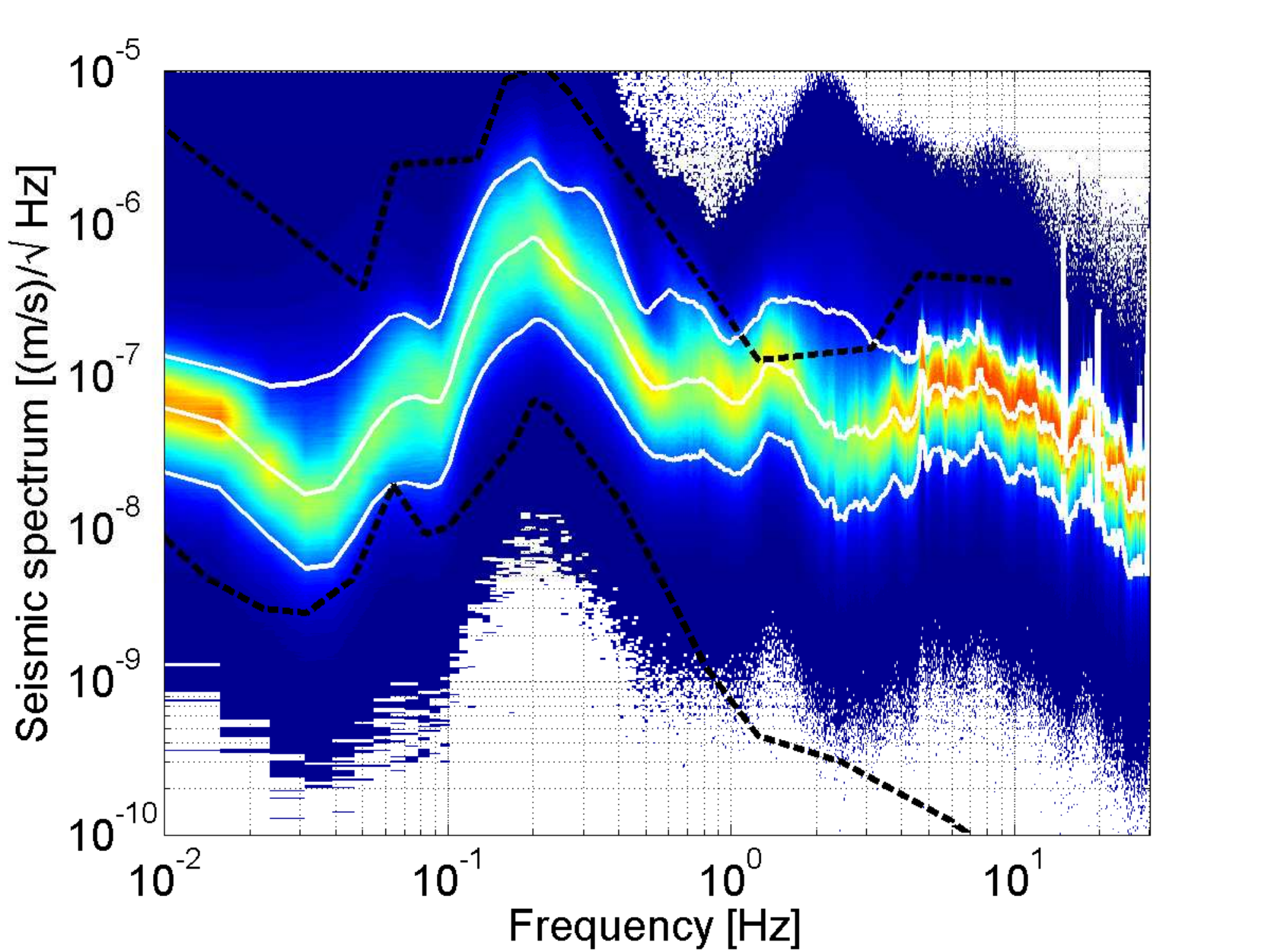}
                \includegraphics[width=0.50\textwidth]{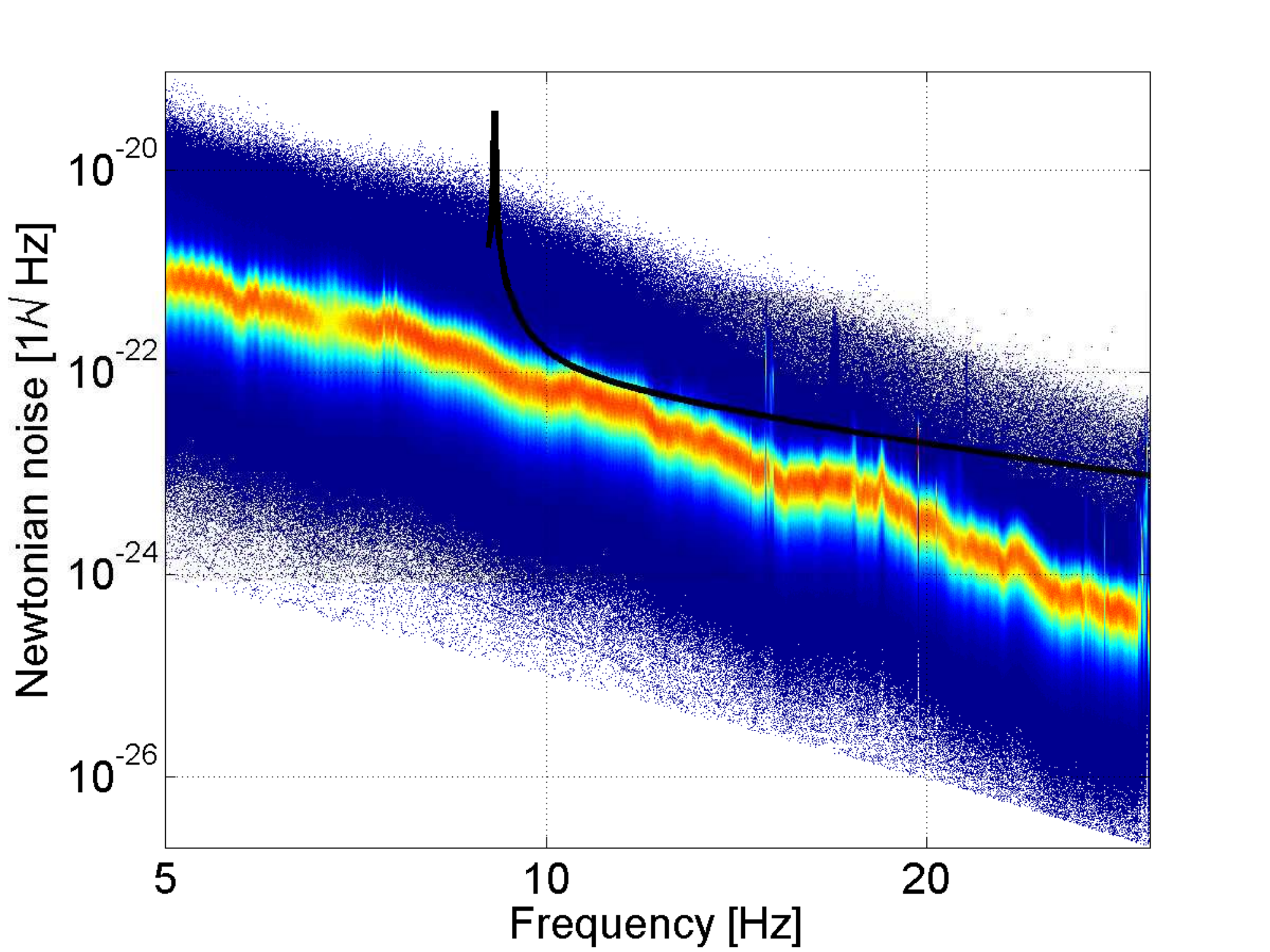}}
    \caption[Seismic noise at LIGO Livingston]{Histograms of seismic spectra at the central station of the LIGO Livingston detector and modelled Rayleigh-wave Newtonian noise \footnotemark. In the left plot, the dashed black curves are the global seismic high-noise and low-noise models. The white curves are the 10th, 50th, and 90th percentiles of the histogram. In the right plot, a sensitivity model of the Advanced LIGO detector is plotted for comparison.}
\label{fig:seismicLLO}
    \end{figure}} 
Generally, seismic spectra at the Virgo and LIGO sites show a higher grade of stationarity above 10\,Hz than at lower frequencies. For example, between 1\,Hz and 10\,Hz, seismic spectra have pronounced diurnal variation from anthropogenic activity, and between 0.05\,Hz and 1\,Hz seismic spectra follow weather conditions at the oceans. These features are shown in Figure \ref{fig:seismicLLO}. The white curves mark the 10th, 50th and 90th percentiles of the histogram. The histogram is based on a full year of 128\,s spectra. Strong disturbances during the summer months from logging operations near the site increase the width of the histogram in the anthropogenic band. In general, a 90th percentile curve exceeding the global high-noise model is almost certainly a sign of anthropogenic disturbances. At lowest frequencies, strong spectra far above the 90th percentile are frequently being observed due to earthquakes. Additional examples of Newtonian-noise spectra evaluated in this way can be found in \cite{DHA2012,BeEA2012}. \footnotetext{Seismic data stem from channel L0:PEM-LVEA$\_$SEISZ between August 1, 2009, UTC 00:00 and August 1, 2010 UTC 00:00.}\addtocounter{footnote}{-1}

\subsubsection{Corrections from anisotropy measurements}
Anisotropy of the seismic field can be an important factor in Newtonian-noise modelling. According to Equation (\ref{eq:RayleighNN}), Rayleigh waves that propagate perpendicularly to the relevant displacement direction of a test mass (which is along the arm of a GW detector), do not produce Newtonian noise. The chances of the Rayleigh-wave field to show significant anisotropy at Newtonian-noise frequencies are high since the dominant contribution to the field comes from nearby sources, probably part of the detector infrastructure. At one of the end stations of the LIGO Hanford detector, an array of 44 vertical seismometers was used to show that indeed the main seismic source of waves around 10\,Hz lies in the direction of an exhaust fan \cite{Har2013}. Coincidentally, this direction is almost perpendicular to the direction of the detector arm.
\epubtkImage{}{%
    \begin{figure}[htbp]
    \centerline{\includegraphics[width=0.49\textwidth]{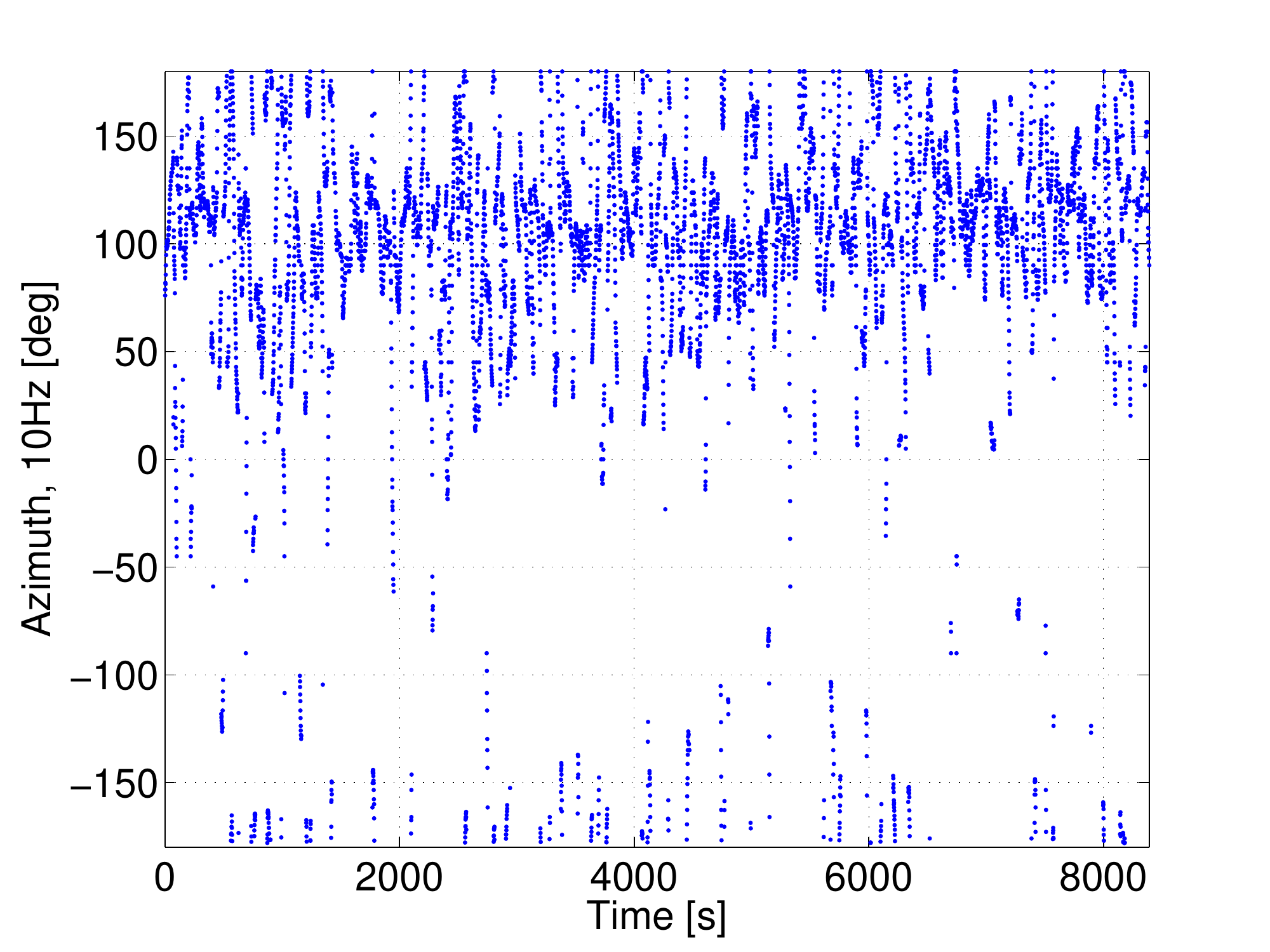}
                \includegraphics[width=0.49\textwidth]{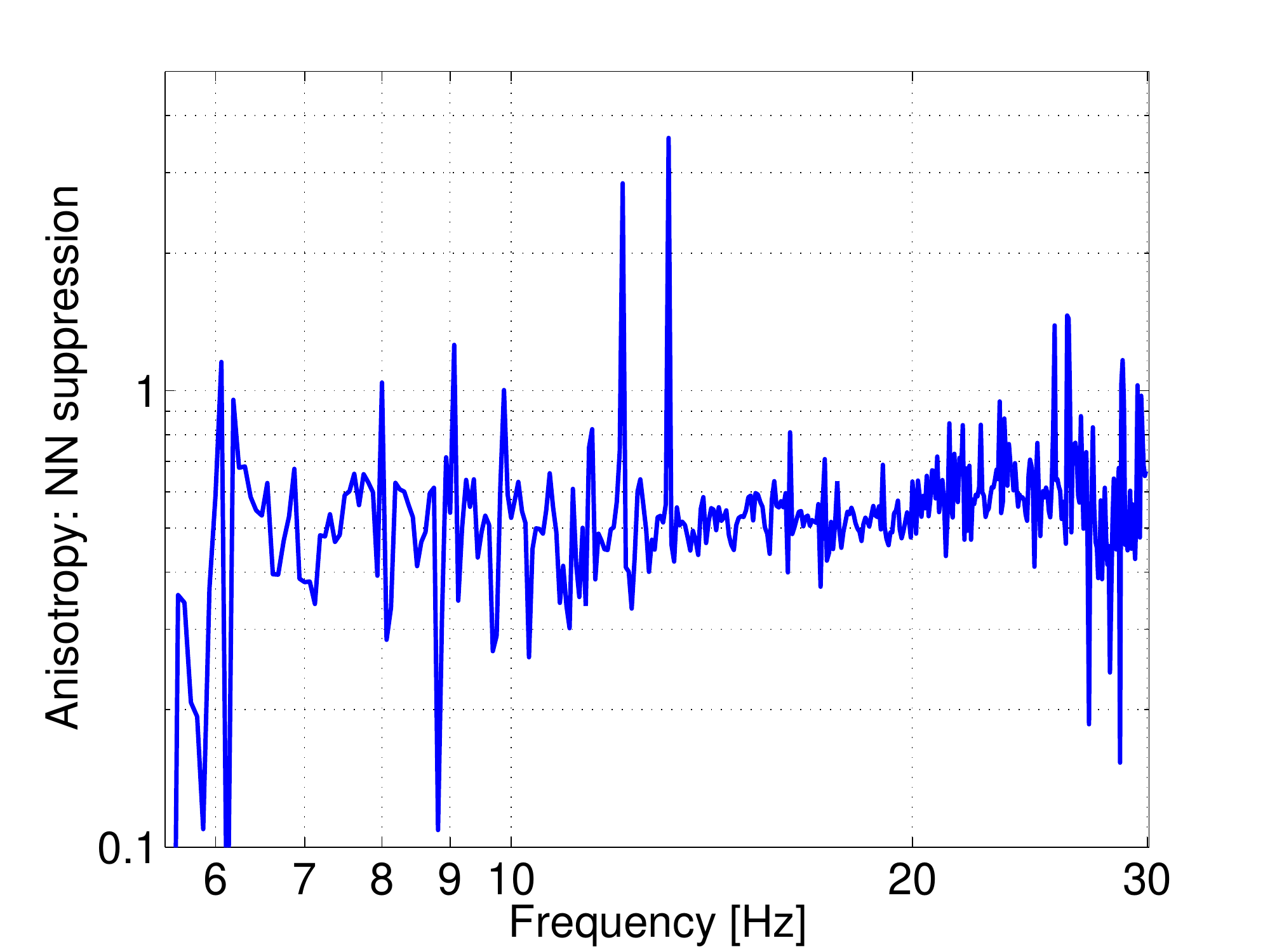}}
    \caption[Rayleigh-wave anisotropy at LIGO Hanford]{Anisotropy of the Rayleigh-wave field at 10\,Hz and Newtonian-noise suppression of a single test mass.}
\label{fig:anisoRayleigh}
    \end{figure}} 
Figure \ref{fig:anisoRayleigh} shows the anisotropy measurement at 10\,Hz and Newtonian-noise suppression of a single test mass obtained from anisotropy measurements over a range of frequencies. The seismic array was used to triangulate the source of dominant seismic waves over a period of a few hours. As shown in the left plot of Figure \ref{fig:anisoRayleigh}, the waves at 10\,Hz almost always come from a preferred direction approximately equal to 100$^\circ$. The same is true at almost all frequencies between 5\,Hz and 30\,Hz. Using the mean azimuth of waves within this range of frequencies, the Newtonian-noise suppression was calculated using Equation (\ref{eq:RayleighS}) inserting the mean azimuth at each frequency as direction of propagation $\phi$ of the Rayleigh waves. An azimuth of 90$^\circ$ corresponds to a direction perpendicular to the arm, which means that one expects Newtonian noise to be lower compared to the isotropic case. The suppression factor is plotted on the right of Figure \ref{fig:anisoRayleigh} with a typical value of about 2. If the situation is the same at the other end station at LIGO Hanford (which is a reasonable assumption, also for the Livingston site), and conservatively assuming that the field is isotropic in the central station, then Newtonian noise would be reduced overall by about a factor $\sqrt{2}$.

\subsubsection{Corrections from two-point spatial correlation measurements}
\label{sec:spatialcorr}
A calculation of Newtonian noise based on seismic two-point spatial correlation was first presented in \cite{BeEA2010}. In this section, we will outline the main part of the calculation focussing on gravity perturbations of a single test mass. The goal is to provide the analytical framework to make optimal use of array data in Newtonian-noise estimation. We will also restrict the analysis to surface arrays and Rayleigh waves. It is straightforward though to extend the analysis to 3D arrays, and as explained below, it is also in principle possible to integrate contributions from other wave types. Assuming that surface displacement is dominated by Rayleigh waves, the most general form of the single test-mass surface gravity perturbation is given by
\beq
S(\delta a_x;\vec\varrho,\omega) = \left(2\pi G\rho_0\gamma(\nu)\right)^2\int\frac{\drm^2k}{(2\pi)^2}\frac{\drm^2k'}{(2\pi)^2} S(\xi_z;\vec k_\varrho,\vec k_\varrho',\omega)\,\frac{k_x}{k_\varrho}\frac{k_x'}{k_\varrho'}\e^{-hk_\varrho}\e^{-hk_\varrho'}\e^{\irm\vec\varrho\cdot(\vec k_\varrho-\vec k_\varrho')}
\eeq
If the Rayleigh field is homogeneous, then the last equation can be simplified to
\beq
S(\delta a_x;\omega) = \left(2\pi G\rho_0\gamma(\nu)\right)^2\int\frac{\drm^2k}{(2\pi)^2}S(\xi_z;\vec k_\varrho,\omega)\frac{k_x^2}{k_\varrho^2}\e^{-2hk_\varrho}
\label{eq:specHomNN}
\eeq
If in addition the field is isotropic, one obtains
\beq
S(\delta a_x;\omega) = \left(2\pi G\rho_0\gamma(\nu)\right)^2\frac{1}{4\pi}\int\limits_0^\infty\drm k_\varrho\, k_\varrho S(\xi_z;k_\varrho,\omega)\e^{-2hk_\varrho}
\label{eq:specIsoNN}
\eeq
Equation (\ref{eq:specHomNN}) is probably the most useful variant since one should always expect that isotropy does not hold, and at the same time, it is practically unfeasible to characterize a seismic field that is inhomogeneous (corrections from inhomogeneities are probably minor as well). Nevertheless, the wavenumber spectra in all three equations can be measured in principle with seismic arrays as Fourier transforms of two-point spatial correlation measurements. In general, the correlation function and wavenumber spectrum are related via
\beq
S(\xi_z;\vec k_\varrho,\vec k_\varrho',\omega)=\int\drm^2\varrho\,\drm^2\varrho'\,C(\xi_z;\vec \varrho,\vec \varrho\,',\omega)\e^{-\irm(\vec\varrho\cdot\vec k_\varrho+\vec\varrho\,'\cdot\vec k_\varrho')}
\eeq
For a homogeneous field, we have
\beq
S(\xi_z;\vec k_\varrho,\omega)=\int\drm^2\varrho\,C(\xi_z;\vec \varrho,\omega) \e^{-\irm\vec\varrho\cdot\vec k_\varrho}
\label{eq:spec2D}
\eeq
We can first insert this expression into Equation (\ref{eq:specHomNN}), and integrate over wavenumbers to obtain the Newtonian noise spectrum in terms of the two-point spatial correlation $C(\xi_z;\vec\varrho,\omega)$ of the seismic field:
\beq
\begin{split}
S(\delta a_x;\omega) = \left(2\pi G\rho_0\gamma(\nu)\right)^2\frac{1}{2\pi}\int\drm^2 \varrho\, \Bigg[&\frac{x^2}{\varrho^2}\frac{2h}{\left((2h)^2+\varrho^2\right)^{3/2}}\\
&+\frac{y^2-x^2}{\varrho^4} \left(1-\frac{2h}{\left((2h)^2+\varrho^2\right)^{1/2}}\right)\Bigg] C(\xi_z;\vec\varrho,\omega)
\end{split}
\label{eq:homNNC}
\eeq
For isotropic and homogeneous fields, the wavenumber spectrum can be calculated as
\beq
S(\xi_z;k_\varrho,\omega)=2\pi\int\limits_0^\infty\drm\varrho\,\varrho J_0(k_\varrho\varrho)C(\xi_z;\varrho,\omega)
\eeq
Together with Equation (\ref{eq:specIsoNN}), we can express the gravity spectrum in terms of the isotropic two-point correlation:
\beq
S(\delta a_x;\omega) = \left(2\pi G\rho_0\gamma(\nu)\right)^2\frac{1}{2}\int\limits_0^\infty\drm \varrho\, \frac{2h\varrho}{\left((2h)^2+\varrho^2\right)^{3/2}} C(\xi_z;\varrho,\omega)
\label{eq:isoNNC}
\eeq
This result can also be obtained directly from Equation (\ref{eq:homNNC}) by integrating over the azimuth. The fraction inside the integral can be understood as the kernel of an integral transformation of the spatial correlation function with the two variables $\varrho,\,h$. \index{kernel} For vanishing test-mass height $h$, the kernel is to be substituted by the Delta-distribution $\delta(\varrho)$. This means that for negligible test-mass height, the gravity perturbation from a homogeneous and isotropic field is determined by the seismic spectral density $S(\xi_z;\omega)=C(\xi_z;0,\omega)$. 
\epubtkImage{}{%
    \begin{figure}[htb]
    \centerline{\includegraphics[width=0.5\textwidth]{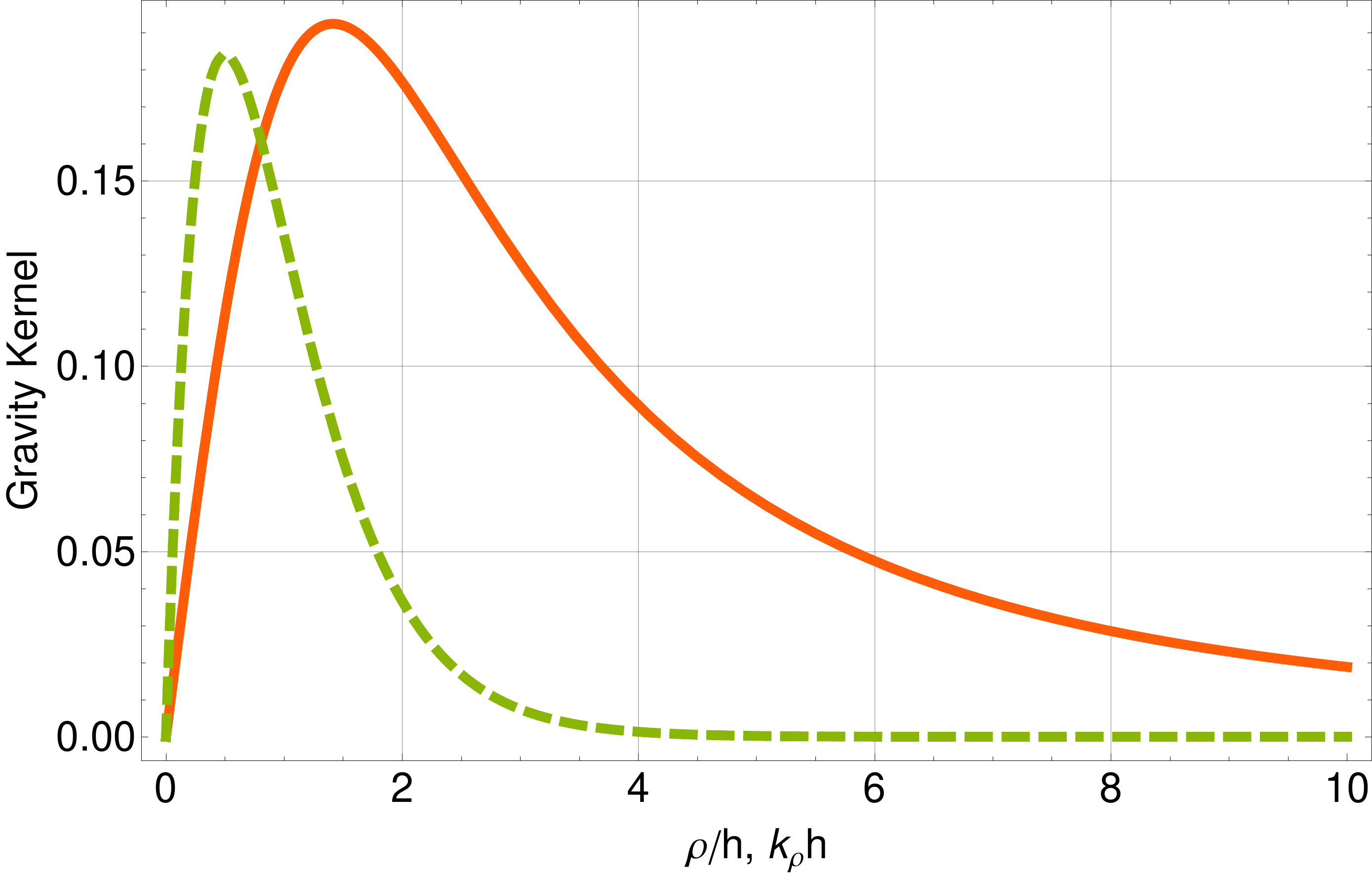}}
    \caption[Newtonian-noise kernel for isotropic, homogeneous Rayleigh fields]{Newtonian-noise kernel for isotropic, homogeneous Rayleigh fields. The dashed line is the kernel in wavenumber domain, Equation (\ref{eq:specIsoNN}), the solid line is the kernel in coordinate space, Equation (\ref{eq:isoNNC}).}
\label{fig:kernelRay}
    \end{figure}} 
Equation (\ref{eq:isoNNC}) also states that for a homogeneous, isotropic field, the values of $\varrho$ that are most relevant to the Newtonian-noise estimate lie around $\varrho=\sqrt{2}h$ where the kernel assumes its maximum. The kernel is plotted as solid curve in Figure \ref{fig:kernelRay}. For example, LIGO test masses are suspended 1.5\,m above ground. Spatial correlation over distances much longer than 5\,m are irrelevant to estimate Newtonian noise at the LIGO sites from homogeneous and isotropic fields. Consequently a seismic experiment designed to measure spatial correlations to improve Newtonian-noise estimates does not need to cover distances longer than this. Of course, in reality, fields are neither homogeneous nor isotropic, and seismic arrays should be designed conservatively so that all important features can be observed. The kernel of the integral transform in Equation (\ref{eq:specIsoNN}) is a function of the variables $k_\varrho,\,h$ with maximum at $k_\varrho=1/(2h)$. It is displayed in Figure \ref{fig:kernelRay} as dashed curve. The behavior of the two kernels with changing $h$ is intuitive. The higher the test mass above ground, the larger the scales of the seismic field that dominate the gravity perturbation, which means larger values of $\varrho$ and smaller values of $k_\varrho$. Kernels in higher dimension can also be calculated for homogeneous seismic fields, and for the general case. The calculation is straight-forward and will not be presented here. 

The isotropic, homogeneous case is further illustrated for the Rayleigh field. A homogeneous, isotropic Rayleigh wave field has a two-point spatial correlation given by \cite{DHA2012}
\beq
C(\xi_z;\varrho,\omega)=S(\xi_z;\omega)J_0(k^{\rm R}_\varrho\varrho),
\label{eq:corrRayiso}
\eeq
which gives rise to a wavenumber spectrum equal to
\beq
S(\xi_z;k_\varrho,\omega)=2\pi S(\xi_z;\omega)\frac{\delta(k^{\rm R}_\varrho-k_\varrho)}{k^{\rm R}_\varrho},
\label{eq:specRayIso}
\eeq
where we used the closure relation in Equation (\ref{eq:closureJ}). According to Equation (\ref{eq:specIsoNN}) or (\ref{eq:isoNNC}), the corresponding Newtonian noise of a single test mass is
\beq
S^{\rm R}(\delta a_x;\omega) = \left(2\pi G\rho_0 \e^{-hk_\varrho^{\rm R}}\gamma(\nu)\right)^2\frac{1}{2}S(\xi_z;\omega)
\label{eq:specNNiso}
\eeq
This result is consistent with our previous solution (the limit $L\rightarrow\infty$ of Equation (\ref{eq:RayNN}) is twice as high). As mentioned already, in the form given here with the numerical factor $\gamma(\nu)$, the results are strictly only valid for Rayleigh waves. Contributions from other types of waves to $C(\xi_z;\vec \varrho,\omega)$ could potentially be integrated separately with different numerical factors, but then one needs some prior information helping to distinguish wave types in these spectra (e.~g.~based on estimated seismic speeds). 
\epubtkImage{}{%
    \begin{figure}[htbp]
    \centerline{\includegraphics[width=1\textwidth]{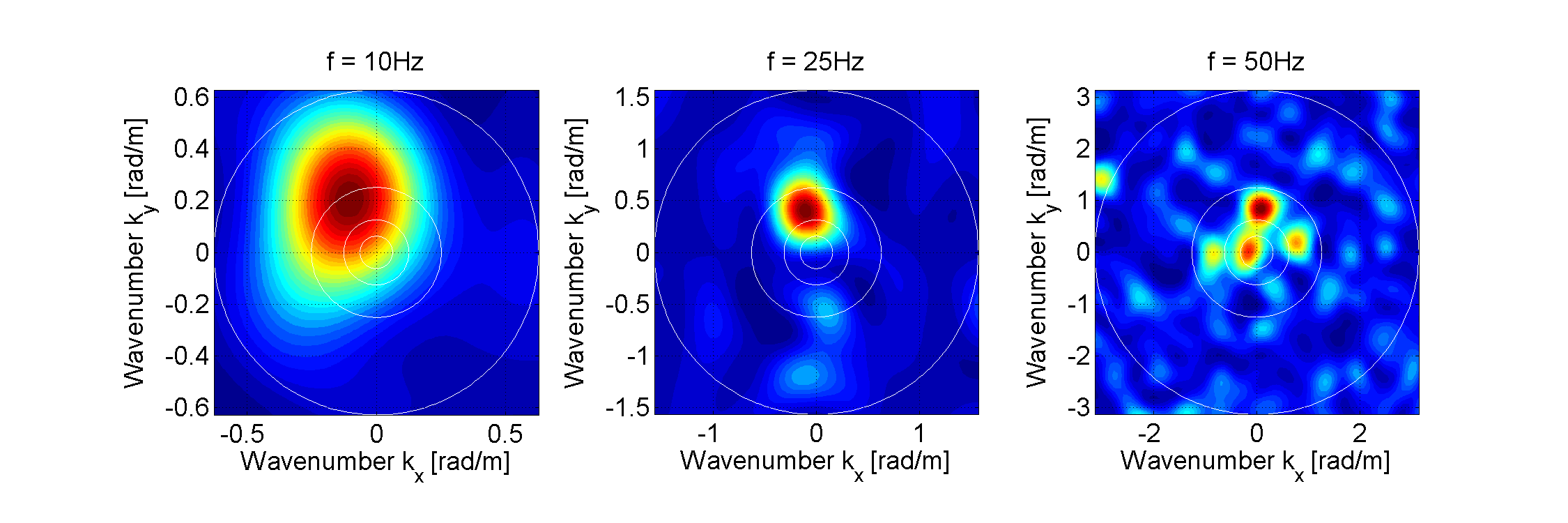}}
    \caption[Wavenumber spectrum at LIGO Hanford]{Wavenumber spectra measured at the LIGO Hanford site using a 44 seismometer array \footnotemark. The white circles with decreasing radius mark wave speeds of 100\,m/s, 250\,m/s, 500\,m/s and 1000\,m/s.}
\label{fig:specKLHO}
    \end{figure}} 
In Figure \ref{fig:specKLHO}, wavenumber spectra measured at the LIGO Hanford site are shown for three different frequencies. The maxima in all three spectra correspond to Rayleigh waves (since the corresponding speeds are known to be Rayleigh-wave speeds). However, the 50\,Hz spectrum contains a second mode with significant amplitude that lies much closer to the origin, which is therefore much faster than a Rayleigh wave. It can only be associated with a body wave. One can now split the integration of this map into two parts, one for the Rayleigh wave, and one for the body wave, using a different numerical factor in each case. This can work, but with the information that can be extracted from this spectrum alone, it is not possible to say what type of body wave it is. So one can either study particle motion with three-axes sensors to characterize the body wave further (which was not possible in this case since the array consisted of vertical sensors only), or instead of $\gamma(\nu)<1$ one can use the conservative numerical factor equal to 1 to calculate at least the corresponding gravity perturbation from pure surface displacement. The latter method would neglect sub-surface density perturbations produced by a P-wave. It should be noted that one can obtain a model independent estimate of Newtonian noise with a 3D array. The numerical factor $\gamma(\nu)$ came from a calculation of sub-surface gravity perturbations based on surface displacement. With information about the entire 3D displacement field, this step is not necessary and the noise estimate becomes model independent and does not require any other prior knowledge. An example of calculating Newtonian noise based on a 3D spatial correlation function is given in Section \ref{sec:quasitemp}. \footnotetext{The data of the LIGO Hanford array are stored in LIGO channels H2:PEM-EY$\_$AUX$\_$NNARRAY$\_$ACC$\_\{$1--44$\}\_$OUT$\_$DQ. The plot uses 16\,s starting from April 28, 2012 UTC 09:00.}

\subsubsection{Low-frequency Newtonian-noise estimates}
\label{sec:lowfNNRay}
There are qualitative differences between low- and high-frequency Newtonian noise that are worth being discussed more explicitly. First of all, we need to provide a definition of what should be considered low frequencies. There are two length scales relevant to Newtonian-noise estimates. The first is the size $L$ of the GW detector. The second is the depth $h$ of the detector. In this section, we will consider the scenario where both length scales are much shorter than the reduced length of seismic waves: $h,L\ll\lambda/(2\pi)$. This should typically be the case below about 1\,Hz. 

If the detector is much smaller than the reduced wavelength, then gravity perturbations along the same directions are significantly correlated over the extent of the detector. We can see this by expanding Equation (\ref{eq:RayNN}) rewritten into units of strain acceleration for small $L$:
\beq
S((\delta a_L-\delta a_0)/L;\omega) = \left(2\pi G\rho_0 \e^{-hk_\varrho}\gamma(\nu)\right)^2S(\xi_z;\omega)\frac{k_\varrho^2}{8}\begin{pmatrix}3\\ 1\\ 4\end{pmatrix}
\label{eq:RayNNlow}
\eeq
The next order is proportional to $L^2$, and we recall that the test masses are separated by $L$ along the $x$-coordinate. The first important observation is that the strain noise is independent of the detector size. The common-mode rejection of the differential acceleration, which is proportional to $L^2$ with respect to noise power, exactly compensates the $1/L^2$ from the conversion into strain. This also means that Newtonian-noise is significantly weaker at low frequencies consistent with Figure \ref{fig:rayResp}, which shows that gravity gradient response saturates below some test-mass distance. 

Next, we discuss the role of detector depths. It should be emphasized that Equation (\ref{eq:RayNNlow}) is valid only above surface. As we have seen in Equation (\ref{eq:underP}), density changes below surface give rise to additional contributions if the test mass is located underground. We have not explicitly calculated these contributions for Rayleigh waves in this article. The point that we want to make though is that if the length of the Rayleigh wave is much longer than the depth of the detector, then the surface model in Equation (\ref{eq:RayNNlow}) is sufficiently accurate. It can be used with the parameter $h$ set to 0. This is not only true for Newtonian noise from Rayleigh waves, but for all forms of seismic Newtonian noise. It should be noted though that these conclusions are not generally true in the context of coherent Newtonian-noise cancellation. If a factor 1000 noise reduction is required (as predicted for low-frequency GW detectors, see \cite{HaEA2013}), then much more detail has to be included into the noise models, to be able to predict cancellation performance. Here, not only the depth of the detector could matter, but also the finite thickness of the crust, the curvature of Earth, etc. 

Estimates of seismic Newtonian noise at low frequencies were presented with focus on atom-interferometric GW detectors in \cite{VeVi2013}. The interesting aspect here is that atom interferometers in general have a more complicated response to gravity perturbations. A list of gravity couplings for atom interferometers can be found in \cite{DiEA2008}. So while atom-interferometric GW detectors would also be sensitive to gravity strain only, the response function may be more complicated compared to laser interferometers depending on the detector design.

\subsection{Summary and open problems}
\label{sec:ambientsummary} 
In this section on Newtonian noise from ambient seismic fields, we reviewed basic analytical equations to calculate density perturbations in materials due to vibrations, to calculate the associated gravity perturbations, and to estimate Newtonian noise based on observations of the seismic field. Equations were given for gravity perturbations of seismic body waves in infinite and half spaces, and for Rayleigh waves propagating on a free surface. Newtonian noise above a half space can be fully characterized by surface displacement, even for body waves. It was found that analytical expressions for gravity perturbations from body and Rayleigh waves have the same form, just the numerical, material dependent conversion factor between seismic and gravity amplitudes has different values also depending on the propagation direction of a body wave with respect to the surface normal. In practice, this means that prior information such as seismic speeds of body waves is required to calculate gravity perturbations based on surface data alone. Another important difference between body and Rayleigh gravity perturbations is that the conversion factor has a material and propagation-direction dependent complex phase in the body-wave case. This has consequences on the design of a seismic surface array that one would use to coherently cancel the gravity perturbations, which will be discussed further in Section \ref{sec:mitigate}.
 
Scattering of body waves from spherical cavities was calculated concluding that gravity perturbations on a test mass inside a cavity are insignificantly affected by seismic scattering from the cavity. Here, ``insignificantly" is meant with respect to Newtonian-noise estimates. In coherent noise cancellation schemes, scattering could be significant if the subtraction goals are sufficiently high. An open problem is to understand the impact of seismic scattering on gravity perturbations in heterogeneous materials where scattering sources are continuously distributed. This problem was studied with respect to its influence on the seismic field \cite{Nor1986,LGZ2009}, but the effect on gravity perturbations has not been investigated yet. 

We also showed that the calculation of simple Newtonian-noise estimates can be based on seismic spectra alone, provided that one has confidence in prior information (e.~g.~that Rayleigh waves dominate seismic noise). In general, seismic arrays help to increase confidence in Newtonian-noise estimates. It was shown that either simple anisotropy measurements or measurements of 2D wavenumber spectra can be used to improve Newtonian-noise estimates. In this section, we did not discuss in detail the problem of estimating wavenumber spectra. Simply carrying out the Fourier transform in Equation (\ref{eq:spec2D}) is prone to aliasing. A review on this problem is given in \cite{KrVi1996}. Estimation of wavenumber spectra has also become an active field of research in GW groups, using data from the LIGO Hanford array deployed between April 2012 and February 2013, and the surface and underground arrays at the Sanford Underground Research Facility, which are currently being deployed with data to be expected in 2015. The problem of Newtonian-noise estimation using seismic arrays needs to be separated though from the problem of Newtonian-noise cancellation. The latter is based on Wiener filtering. From an information theory perspective, the Wiener filtering process is easier to understand than the noise estimation since Wiener filters are known to extract information from reference channels in an optimal way for the purpose of noise cancellation (under certain assumptions). There is no easy way to define a cost function for spectral estimation, which makes the optimal estimation of wavenumber spectra rather a philosophical problem than a mathematical or physical one. The optimal choice of analysis methods depends on which features of the seismic field are meant to be represented most accurately in a wavenumber spectrum. For example, some methods are based on the assumption that all measurement noise is stationary and effectively interpreted as isotropic seismic background. This does not have to be the case if the seismic field itself acts as a noise background for measurements of dominant features of the field. Nonetheless, designs of seismic arrays used for noise cancellation need to be based on information about wavenumber spectra. Initially, array data are certainly the only reliable sources of information, but also with Newtonian-noise observations, optimization of noise-mitigation schemes will be strongly guided by our understanding of the seismic field. 
\section{Gravity Perturbations from Seismic Point Sources}
\label{sec:pointsources}
In Section \ref{sec:ambient}, we have reviewed our understanding of how seismic fields produce gravity perturbations. We did however not pay attention to sources of the seismic field. In this section, gravity perturbations will be calculated based on models of seismic sources, instead of the seismic field itself. This can serve two purposes. First, a seismic source can be easier to characterize than the seismic field itself, since characterization of a seismic field requires many seismometers in general deployed in a 3D array configuration. Second, it is conceivable to obtain information about a seismic source based on observations of gravity perturbations. For example, it was suggested to promptly detect and characterize fault ruptures leading to earthquakes using low-frequency gravity strain meters \cite{HaEA2015}. In this case, the analysis of gravity data from high-precision gravity strain meters can be understood as a new development in the field of terrestrial gravimetry. Until today, observation of only very slow changes in the terrestrial gravity field (slower than about 1\,mHz) were possible using networks of ground-based gravimeters \cite{CrHi2010} or the satellite mission GRACE \cite{WaEA2004} with applications in hydrology, seismology and climate research. Also co-seismic gravity changes, i.~e.~changes following large earthquakes, were observed with gravimeters \index{gravimeter}\cite{ImEA2004} as well as with GRACE (see for example \cite{WSJ2012,CaSa2013}). These observations were predicted based on a theory of static gravity perturbations from fault rupture first developed by Okubo in \cite{Oku1991,Oku1992,Oku1993}. Only lasting gravity changes can be detected with these instruments, and it is to be expected that high-precision gravity strain meters will contribute significantly to this field by opening a window to gravity changes at higher frequencies. New models need to be constructed that describe time-varying gravity changes from seismic fields produced by various seismic sources. The first steps thereof are outlined in the following. We also want to point out that the same formalism can be applied to point sources of sound waves as shown in Section \ref{sec:shockNN}. We emphasize that all known time-domain models of gravity perturbations from seismic sources are for infinite media. The inclusion of surface effects, which is not always necessary, is one of the important calculations that needs to be done still. We will give some ideas how to approach this problem in Section \ref{sec:pointsummary}. According to the title of this section, the models presented here are for point sources only. It is however numerically trivial to combine point source solutions to represent an extended source. Also certain analytical calculations of gravity perturbations from extended sources should be feasible. 

\subsection{Gravity perturbations from a point force}
\label{sec:forcegrav}
Point forces can be a good model of various real sources such as vibrating engines or impacts of small objects on ground. A point force is modelled as force density according to
\beq
\vec f(\vec r,t) = F(t)\vec e_f\delta(\vec r\,)
\eeq
with source function \index{source function}$F(t)=0$ for $t<0$, and $\vec e_f$ being the normal vector pointing along the direction of the force. Such a force generates a complicated seismic field that is composed of a near field, and shear and compressional waves propagating in the intermediate and far field \cite{AkRi2009}, all components with different radiation patterns (explicit expressions for a point shear dislocation are given in Section \ref{sec:disdensity}). However, Equation (\ref{eq:gravP}) can be applied here, which means that we only need to know the potential of compressional waves in infinite media to simply write down the corresponding gravity perturbation. It is not too difficult to calculate the seismic potential, but one can also find the solution in standard text books \cite{AkRi2009}. The solution for the perturbed gravity acceleration reads
\beq
\delta \vec a(\vec r_0,t)=\frac{G}{r_0^3}(\vec e_f-3(\vec e_f\cdot\vec e_{r_0})\vec e_{r_0})\int\limits_0^{r_0/\alpha}\drm\tau\,\tau F(t-\tau)+\frac{G}{r_0\alpha^2}(\vec e_f\cdot\vec e_{r_0})\vec e_{r_0}F(t-r_0/\alpha)
\eeq
with the source being located at the origin. This perturbation is based on the full seismic field produced by the point force. The solution consists of a component proportional to an integral over the source function, and another component proportional to the retarded source function. At early times, when $t<r_0/\alpha$, i.~e.~when the seismic waves produced by the source have not yet reached the location $\vec r_0$, the second term vanishes while the integral can be rewritten as double time integral
\beq
\int\limits_0^t\drm\tau\,\tau F(t-\tau)=\int\limits_0^t\drm\tau\int\limits_0^\tau\drm\tau'\,F(\tau')\equiv \mathcal{I}_2[F](t)
\eeq
The acceleration simplifies to
\beq
\delta \vec a(\vec r_0,t)=\frac{G}{r_0^3}(\vec e_f-3(\vec e_f\cdot\vec e_{r_0})\vec e_{r_0})\mathcal{I}_2[F](t),\,\mbox{for}\;t<r_0/\alpha
\label{eq:earlyF}
\eeq
Interestingly, the early-time solution is independent of any geophysical parameters such as ground density and seismic speeds (assuming that the ground is homogeneous). The gravity perturbation from a point force can be used to model the contribution of local sources to Newtonian noise based on a measured source time function $F(t)$. In this section, we use it to present another interesting result. It has often been conjectured that a transient source of seismic vibrations would be a problem to coherent mitigation schemes since the gravity perturbation starts to be significant before any of the seismometers can sense the first ground motion produced by this source. Therefore, it would be impossible to coherently remove a significant contribution to Newtonian noise using seismic data. Some evidence speaking against this conjecture was already found in numerical simulations of approaching wavefronts from earthquakes \cite{HaEA2009b}, but there was no analytical explanation of the results. We can make up for this now.  Let us make the following Gedankenexperiment. Let us assume that all seismic noise is produced by a single source. Let us assume that this source is switched on at time $t=0$. Before this time, the entire seismic field is zero. 
\epubtkImage{}{%
    \begin{figure}[htbp]
    \centerline{\includegraphics[width=0.6\textwidth]{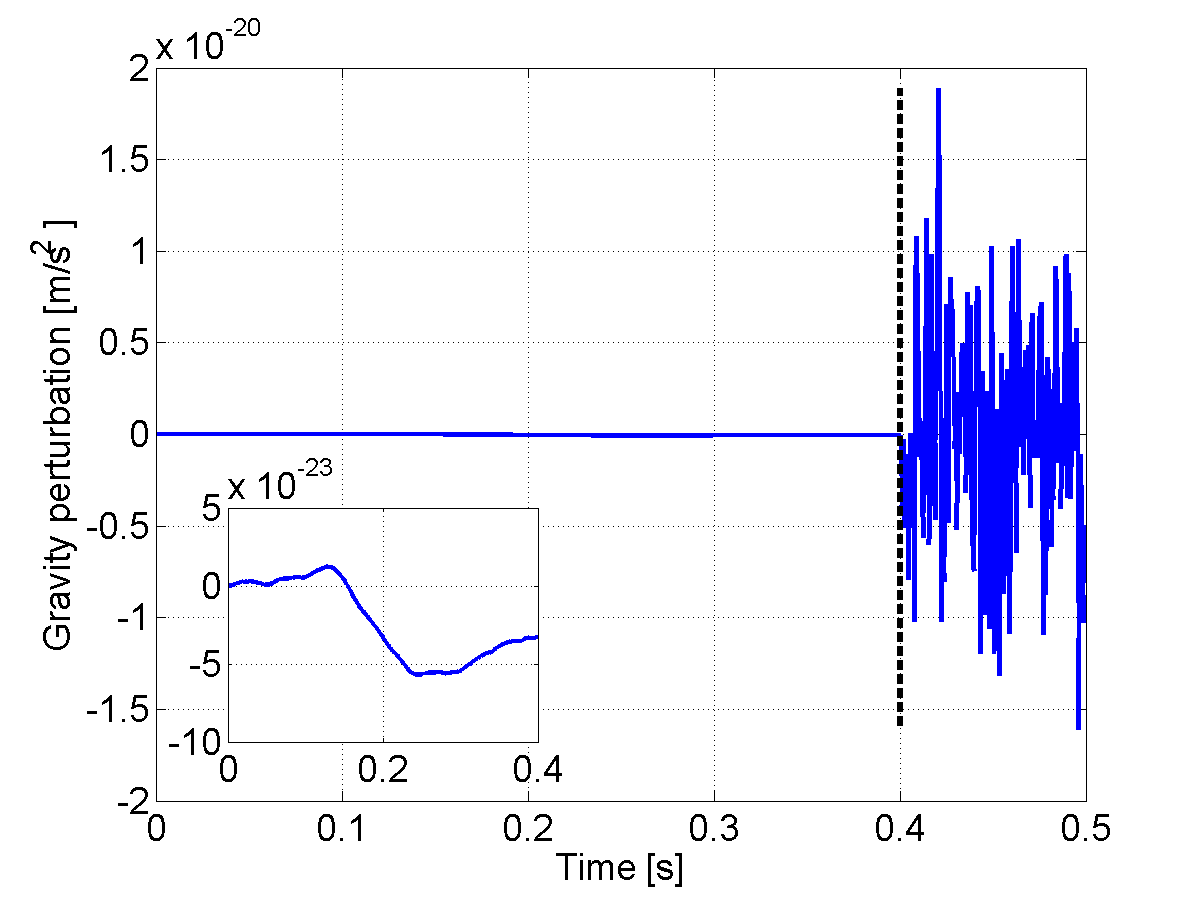}}
    \caption[Gravity perturbation point force]{Gravity perturbation from a point force assumed to be the only source of seismic noise, and switching on at $t=0$.}
\label{fig:forcepoint}
    \end{figure}}  
Now the source starts to irradiate seismic waves. The waves do not reach the test mass before $t=r_0/\alpha$, where $r_0$ is the distance between the source and the test mass. The situation is illustrated in Figure \ref{fig:forcepoint}. The dashed line marks the arrival of seismic waves. From that time on, we have the usual Newtonian noise from ambient seismic fields. Interesting however is what happens before arrival. The Newtonian noise is hardly visible. Therefore, an inset plot was added to show gravity perturbations before wave arrival. Not only is the rms of the gravity perturbation much lower, but as expected, it evolves much slower than the Newtonian noise from ambient seismic fields. Equation (\ref{eq:earlyF}) says that the source function is filtered by a double integrator to obtain the gravity acceleration. Another double integrator needs to be applied to convert gravity acceleration into test-mass displacement. Therefore, whatever the source function is, and the corresponding source spectrum $F(\omega)$\index{source spectrum}, gravity perturbations will be strongly suppressed at high frequencies. Due to the transient character of this effect, it is difficult to characterize the problem in terms of Newtonian-noise spectra, but it should be clear that a seismic source would have to be very peculiar (i.~e.~radiating very strongly at high frequencies and weakly at low frequencies), to cause a problem to coherent Newtonian-noise cancellation, without causing other problems to the detector such as a loss of cavity lock due to low-frequency ground disturbances.

\subsection{Density perturbation from a point shear dislocation in infinite homogeneous media}
\label{sec:disdensity}
In this subsection, we briefly review the known solution of the seismic field produced by a shear dislocation\index{shear dislocation}. The shear dislocation is modelled as a double couple\index{double couple}, which consists of two perpendicular pairs of forces pointing against each other with infinitesimal offset.  The coordinate system used in the following is shown in Figure \ref{fig:pointshear}. Its origin coincides with the location of the shear dislocation, with the $z$-axis being parallel to the slip direction, and the $x$-axis perpendicular to the fault plane. Spherical coordinates $r,\theta,\phi$ will be used in the following that are related to the Cartesian coordinates via $x=r\sin(\theta)\cos(\phi)$, $y=r\sin(\theta)\sin(\phi)$, $z=r\cos(\theta)$, with $0<\theta<\pi$, and $0<\phi<2\pi$.
\epubtkImage{}{%
    \begin{figure}[htbp]
    \centerline{\includegraphics[width=0.5\textwidth]{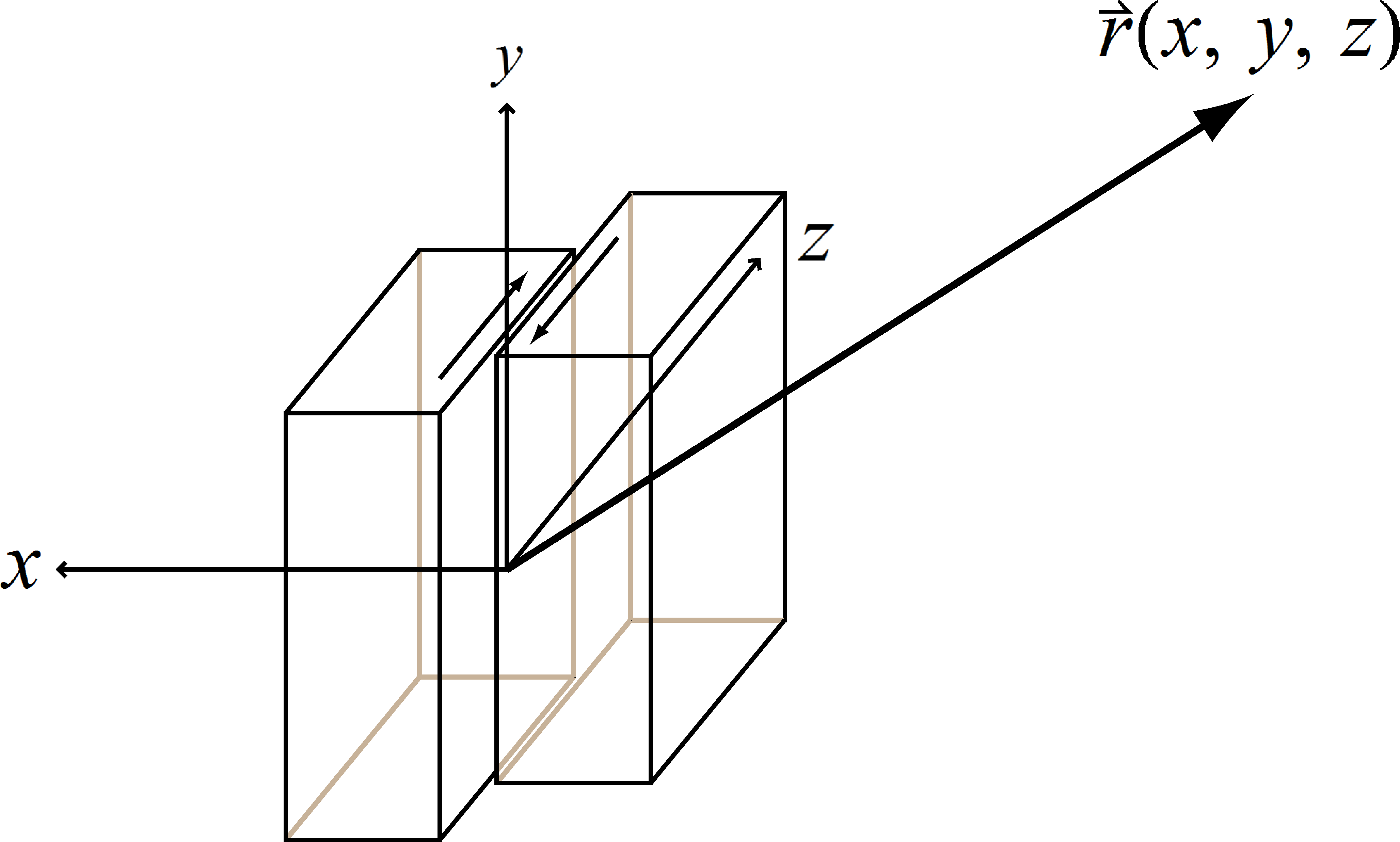}}
    \caption[Point shear dislocation]{Definition of the coordinate system used to describe a point shear dislocation.}
    \label{fig:pointshear}
    \end{figure}}
The double couple drives a displacement field that obeys conservation of linear and angular momenta. Its explicit form is given in Aki \& Richards \cite{AkRi2009}. It consists of a near-field component:
\beq
\begin{split}
\vec\xi_{\rm N}(\vec r\,,t) &= \frac{1}{4\pi\rho_0}\vec A_{\rm N}\frac{1}{r^4}\int\limits_{r/\alpha}^{r/\beta}\drm\tau\,\tau M_0(t-\tau),\\
\vec A_{\rm N} &\equiv 9\sin(2\theta)\cos(\phi)\vec e_r-6(\cos(2\theta)\cos(\phi)\vec e_\theta-\cos(\theta)\sin(\phi)\vec e_\phi),
\end{split}
\eeq
an intermediate-field component
\beq
\begin{split}
\vec\xi_{\rm I}(\vec r\,,t) &= \frac{1}{4\pi\rho_0\alpha^2}\vec A_{\rm IP}\frac{1}{r^2} M_0(t-r/\alpha)+\frac{1}{4\pi\rho_0\beta^2}\vec A_{\rm IS}\frac{1}{r^2} M_0(t-r/\beta),\\
\vec A_{\rm IP} &\equiv 4\sin(2\theta)\cos(\phi)\vec e_r-2(\cos(2\theta)\cos(\phi)\vec e_\theta-\cos(\theta)\sin(\phi)\vec e_\phi),\\
\vec A_{\rm IS} &\equiv -3\sin(2\theta)\cos(\phi)\vec e_r+3(\cos(2\theta)\cos(\phi)\vec e_\theta-\cos(\theta)\sin(\phi)\vec e_\phi),
\end{split}
\eeq
and a far-field component
\beq
\begin{split}
\vec\xi_{\rm F}(\vec r\,,t) &= \frac{1}{4\pi\rho_0\alpha^3}\vec A_{\rm FP}\frac{1}{r} \dot M_0(t-r/\alpha)+\frac{1}{4\pi\rho_0\beta^3}\vec A_{\rm FS}\frac{1}{r}\dot M_0(t-r/\beta),\\
\vec A_{\rm FP} &\equiv \sin(2\theta)\cos(\phi)\vec e_r,\\
\vec A_{\rm FS} &\equiv \cos(2\theta)\cos(\phi)\vec e_\theta-\cos(\theta)\sin(\phi)\vec e_\phi,
\end{split}
\eeq
which have to be added to give the total displacement field $\vec\xi(\vec r\,,t)$. The source function $M_0(t)$ of the double couple is called moment function. As for the point force, we assume again that the source function is zero for $t<0$. If a double couple is used to represent fault ruptures, than the source function increases continuously as long as the fault rupture lasts. 

In contrast to the intermediate and far-field terms, the near-field term does not describe a propagating seismic wave. The far field is the only component that generally vanishes for $t\rightarrow\infty$. According to Equation (\ref{eq:densinh}), density perturbations in infinite, homogeneous media can only be associated with compressional waves, since the divergence of the shear field is zero. This is confirmed by inserting the total displacement field into Equation (\ref{eq:densinh}). One obtains the density change
\beq
\begin{split}
\delta\rho(\vec r\,,t) &= -\rho_0\nabla\cdot\vec\xi(\vec r\,,t)\\
&= \frac{3\cos(\phi)\sin(2\theta)}{4\pi r^3\alpha^2}\left(M_0(t-r/\alpha)+\frac{r}{\alpha}\dot M_0(t-r/\alpha)+\frac{r^2}{3\alpha^2}\ddot M_0(t-r/\alpha)\right)\\
&\equiv \cos(\phi)\sin(2\theta)R(r,t)
\end{split}
\label{eq:densdis}
\eeq
The density perturbation assumes a much simpler form than the seismic field. The perturbation propagates with the speed of compressional waves, and has a quadrupole radiation pattern. A lasting density change is built up proportional to the final moment $M_0(t\rightarrow\infty)$ of the shear dislocation. 
\epubtkImage{}{
    \begin{figure}[htbp]
    \centerline{\includegraphics[width=0.7\textwidth]{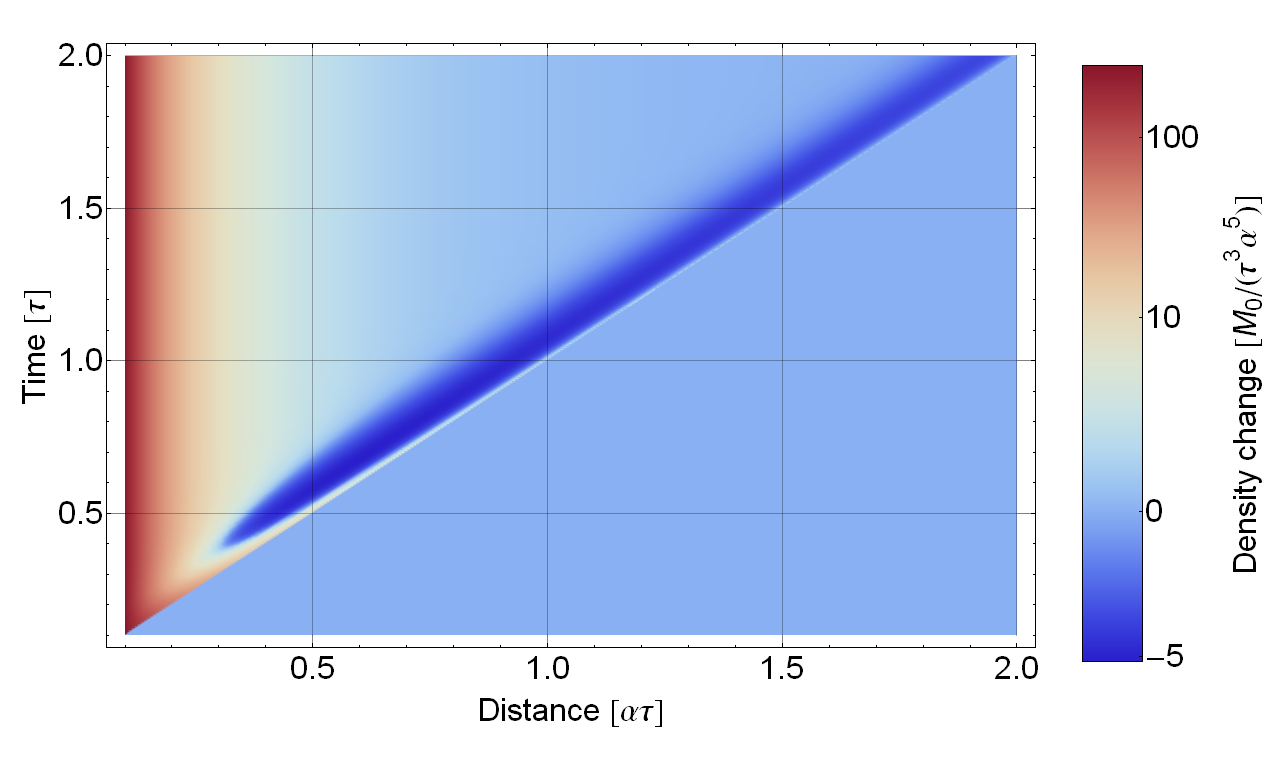}}
    \caption[Density perturbation of double couple]{Density perturbation produced by a double couple.}
    \label{fig:densdouble}
    \end{figure}}
In Figure \ref{fig:densdouble}, the gravity perturbation is shown for $\theta = \pi/4,\,\phi=0$. The source function is $M_0(t)=M_0\tanh(t/\tau)$ for $t>0$ and zero otherwise. A log-modulus transform is applied to the density field since its value varies over many orders of magnitude \cite{JoDr1980}. This transform preserves the sign of the function it is applied to. A transient perturbation carried by compressional waves propagates parallel to the line $t=r/\alpha$. A lasting density change, which quickly decreases with distance to the source, forms after the transient has passed. 

\subsection{Gravity perturbations from a point shear dislocation}
\label{sec:disgravity}
\index{earthquakes}
Fault slip generates elastodynamic deformation (static and transient), including compression and dilation that induce local perturbations of the material density. These in turn lead to global perturbations of the gravity field. In this section, we consider an elementary problem: we develop an analytical model of time-dependent gravity perturbations generated by a point-shear dislocation in an infinite, elastic, and homogeneous medium. We are interested in frequencies higher than $0.01$\,Hz, for which we can ignore the effects of self-gravitation \cite{DaTr1998}: we compute the gravity changes induced by mass redistribution caused by elastic deformation, but ignore the effect of gravity force fluctuations on the deformation. The results in this subsection were published in \cite{HaEA2015}. The gravity perturbation can either be obtained analogously to the case of a point force by seeking for a known solution of the P-wave potential and rewriting it as gravity potential, or by attempting a direct integration of density perturbations. First, we will show how to carry out the direct integration. 

The density perturbation $\delta\rho(\vec r\,,t)$ caused by the displacement field $\vec\xi(\vec r\,,t)$ was presented in Equation (\ref{eq:densdis}). The perturbation of the gravity potential at some point $\vec r_0$ is obtained by integrating over the density field according to
\beq
\delta\psi(\vec r_0\,,t)=-G\int{\rm d} V\,\frac{\delta\rho(\vec r\,,t)}{|\vec r-\vec r_0|}.
\label{eq:potential}
\eeq
The integration can be carried out using a multipole expansion of the gravity potential. This requires us to divide the integration over the radial coordinate $r$ into two intervals: $0<r<r_0$ and $r_0<r$. Over the first interval, one obtains the exterior multipole expansion:\index{multipole expansion}
\beq
\delta\psi_{\rm ext}(\vec r_0\,,t) 
  = \sum\limits_{l=0}^\infty\sum\limits_{m=-l}^l I_l^m(\vec r_0)^*
\cdot\int\limits_0^{r_0}{\rm d} r\,r^2\int{\rm d}\Omega\,\delta\rho(\vec r,t)R_l^m(\vec r)
\label{eq:exterior}
\eeq
The corresponding expression for the interior multipole expansion is given by
\beq
\delta\psi_{\rm int}(\vec r_0\,,t) 
= \sum\limits_{l=0}^\infty\sum\limits_{m=-l}^l R_l^m(\vec r_0)^*
\cdot\int\limits_{r_0}^\infty{\rm d} r\,r^2\int{\rm d}\Omega\,\delta\rho(\vec r,t)I_l^m(\vec r),
\label{eq:interior}
\eeq
where we used the solid spherical harmonics defined in Equation (\ref{eq:solidharm}). The two integrals in Equations (\ref{eq:exterior}) and (\ref{eq:interior}) are readily solved by expressing the radiation pattern in Equation (\ref{eq:densdis}) in terms of the surface spherical harmonics (see Table \ref{tab:sphereharm}),
\beq
 \sin(2\theta)\cos(\phi) =2\sqrt{2\pi/15}\,\left(Y_2^{-1}(\theta,\phi)-Y_2^1(\theta,\phi)\right),
\eeq
and subsequently making use of the orthogonality relation in Equation (\ref{eq:normalY}). For example, inserting the density perturbation into the exterior multipole expansion of the gravity potential we have
\beq
\begin{split}
\delta\psi_{\rm ext}(\vec r_0\,,t) 
  &= \sum\limits_{l=0}^\infty\sum\limits_{m=-l}^l I_l^m(\vec r_0)^*
\cdot\int\limits_0^{r_0}{\rm d} r\,r^2R(r,t)\int{\rm d}\Omega\,\sin(2\theta)\cos(\phi)R_l^m(\vec r)\\
 &= \frac{4\pi}{5}\frac{1}{r_0^3}\sum\limits_{m=-2}^2 Y_2^m(\theta_0,\phi_0)^*
\cdot\int\limits_0^{r_0}{\rm d} r\,r^4R(r,t)\int{\rm d}\Omega\,\sin(2\theta)\cos(\phi)Y_2^m(\theta,\phi)
\end{split}
\eeq
The integral over angles can be carried out using Equation (\ref{eq:conjugateY}). The result is again a quadrupole radiation pattern. The integral over the radius can be simplified considerably by integration by parts. The solution $\delta \psi(\vec r_0\,,t)=\delta \psi_{\rm ext}(\vec r_0\,,t)+\delta \psi_{\rm int}(\vec r_0\,,t)$ for the gravity potential perturbation can then be written in the form
\beq
\delta \psi(\vec r_0\,,t)=G  \sin(2\theta_0)\cos(\phi_0)\left[\frac{1}{r_0\alpha^2}M_0(t-r_0/\alpha)-\frac{3}{r_0^3}\int\limits_0^{r_0/\alpha}{\rm d} u\,u M_0(t-u)\right]
\label{eq:potentialext}
\eeq 

The second, more elegant approach to solve Equation\;(\ref{eq:potential}) is again based on Equation (\ref{eq:gravP}). Given the known solution for seismic potentials from a point force in infinite media \cite{AkRi2009}, one can derive the corresponding expression for a double-couple by applying derivatives to the gravity potential with respect to the source coordinates along two orthogonal directions, and rescale it according to Equation (\ref{eq:potential}) to obtain the expression of the perturbed gravity potential given in Equation (\ref{eq:potentialext}).

As for the point force, the gravity potential perturbation from a double couple has a particularly simple structure at early times, $t<r_0/\alpha$, i.e. before the arrival of P waves at $\vec r_0$:
\beq
\begin{split}
\delta \psi(\vec r_0\,,t) &= -\frac{3G}{r_0^3} \sin(2\theta_0)\cos(\phi_0)\mathcal{I}_2[M_0](t)\\
&= -\frac{6G}{r_0^3} (\vec e_{r_0}\cdot\vec e_x)(\vec e_{r_0}\cdot\vec e_z)\mathcal{I}_2[M_0](t)
\end{split}
\label{eq:psiearly}
\eeq
The early gravity potential perturbation appears to emerge from the acausal component of the P-wave potential, whose contribution to the seismic wavefield is cancelled out by a similar contribution from the S-wave potential. Finally, we also give the early-time solution for the gravity acceleration:
\beq
\delta \vec a(\vec r_0\,,t) = \frac{6G}{r_0^4} ((\vec e_{r_0}\cdot\vec e_z)\vec e_x+(\vec e_{r_0}\cdot\vec e_x)\vec e_z-5(\vec e_{r_0}\cdot\vec e_x)(\vec e_{r_0}\cdot\vec e_z)\vec e_{r_0})\mathcal{I}_2[M_0](t)
\label{eq:accearly}
\eeq
Written in this form, the expression becomes frame independent. The directions $\vec e_x,\vec e_z$ are physical directions denoting fault normal and slip direction. They can be reexpressed in any other coordinate system, as for example an Earth coordinate system. 

\subsubsection{Gravity-gradient tensor}
The gravity-gradient tensor $\ddot h(\vec r_0\,,t)$, whose components can be measured by torsion-bar antennas or atom interferometers, is obtained by calculating
\beq
\ddot h(\vec r_0\,,t)=-(\nabla\otimes\nabla)\delta \psi(\vec r_0\,,t)
\label{eq:hdddef}
\eeq
where '$\otimes$' denotes the Kronecker product (also known as dyadic or tensor product). For arbitrary $t$, the result is a symmetric tensor that can be divided into four distinct parts. The first part is proportional to the density perturbation at $\vec r_0$:
\beq
\ddot h_1(\vec r_0\,,t)=-4\pi G\,\delta\rho(\vec r_0\,,t)\vec e_r\otimes\vec e_r
\label{eq:hddpart1}
\eeq
It is the only contribution with non-vanishing trace. Using ${\rm Tr}(\vec a\otimes\vec b\,)=\vec a\cdot\vec b$, one obtains
\beq
{\rm Tr}(\ddot h_1(\vec r_0,t))=-4\pi G\,\delta\rho(\vec r_0,t),
\eeq
consistent with the Poisson equation. The second part can be cast into the form
\beq
\ddot h_2 (\vec r_0\,,t) = -\frac{6 G}{r_0^5}S(\theta,\phi)\int\limits_0^{r_0/\alpha}{\rm d} u\,u M_0(t-u)
\label{eq:hddpart2}
\eeq
where
\beq
\begin{split}
S(\theta,\phi) &=
5(\vec e_x\cdot\vec e_r)(\vec e_z\cdot\vec e_r)(3
{\hbox{$1\hskip -1.2pt\vrule depth 0pt height 1.6ex width 0.7pt \vrule depth 0pt height 0.3pt width 0.12em$}}
-7\vec e_r\otimes\vec e_r)\\
&\quad+4(\vec e_x\otimes\vec e_z)_{\rm sym}+5((\vec e_x\times\vec e_r)\otimes(\vec e_z\times\vec e_r))_{\rm sym}.
\end{split}
\eeq
Here $(\vec a\otimes\vec b)_{\rm sym}\equiv\vec a\otimes\vec b+\vec b\otimes\vec a$. The third part is given by
\beq
\ddot h_3(\vec r_0\,,t) = 
\frac{2 G}{5 r_0^3\alpha^2}\left(6M_0(t-r_0/\alpha)+\frac{r_0}{\alpha}\dot M_0(t-r_0/\alpha)\right)\left(S(\theta,\phi)+(\vec e_x\otimes\vec e_z)_{\rm sym}\right),
\label{eq:hddpart3}
\eeq
and the last part is proportional to the moment function
\beq
\ddot h_4(\vec r_0\,,t)=-\frac{2 G}{\alpha^2r_0^3} M_0(t-r_0/\alpha)
\cdot(\vec e_x\otimes\vec e_z)_{\rm sym},
\label{eq:hddpart4}
\eeq
Note that the unit vectors $\vec e_x,\,\vec e_z$ are not arbitrary coordinate axes, but have a physical interpretation being normal to the shear plane, and along shear direction respectively. The full gravity gradient is simply the sum of these four contributions.

The first and last two contributions vanish for $\alpha t<r_0$ since $M_0(t)=0$ for $t<0$, and the integral of the second contribution can be rewritten into 
\beq
\ddot h (\vec r_0\,,t)=-\frac{6 G}{r_0^5}S(\theta,\phi) I_2[M_0](t).
\label{eq:earlytime}
\eeq
None of the four contributions vanishes for $t\rightarrow\infty$. Instead the time derivatives of the moment function go to zero, and the moment function itself can be substituted by its final value $M_0(t\rightarrow\infty)$. The result is a gravity-gradient tensor whose components decrease with $1/r_0^3$. In addition, the gravity gradient for $t\rightarrow \infty$ is identical to the static gravity perturbation found by \cite{Oku1991} for shear dislocations in a half space, provided that his result is evaluated for an event far from the surface (so that surface effects are suppressed).

For small times $\alpha t<r_0$, the gravity-gradient perturbation is not delayed by $r_0/\alpha$. This delay only emerges once the P waves have reached the point $\vec r_0$. In other words, the point-shear dislocation behaves as a point source of gravity perturbations for $\alpha t<r_0$ even though the actual source is an expanding wavefront of seismic compressional waves. In this case, the effective (point) source function of gravity-gradient perturbations is the fourth time integral of the moment function, which also entails that contributions to the gravity gradient from higher-frequency components of the moment spectrum are strongly suppressed.

\subsection{Gravity perturbation from the Tohoku earthquake}
\label{sec:Tohoku}
\index{focal mechanism}
In this section, the specific example of the 2011 Tohoku earthquake will be used to estimate gravity perturbations. The Tohoku event had a magnitude of 9.0, and ruptured a fault of width and length of several hundred kilometers \cite{AmEA2011}. The hypocenter was located at latitude N37.52 and longitude E143.05. The estimate of the gravity perturbation will be based on the early-time approximation given in Equation (\ref{eq:psiearly}). The result should not be expected to be accurate, since the source cannot be approximated as a point source, and the influence of the surface on the gravity perturbation may be substantial, but nonetheless, it serves as an order-of-magnitude estimate.

With these simplifications, the main work is to understand specifications of focal mechanisms published by seismological institutions such as USGS \footnote{http://earthquake.usgs.gov/}, and translate them into the fault-slip oriented coordinate system of Section \ref{sec:disgravity}. The focal mechanisms can be specified by three angles: the strike angle $\gamma_{\rm s}$, the dip angle $\gamma_{\rm d}$, and the rake angle $\gamma_{\rm r}$. The strike angle is subtended by the intersection of the fault plane with the horizontal plane, and the North cardinal direction. The dip is the angle between the fault plane and the horizontal plane. Finally, the rake is subtended by the slip vector and the horizontal direction on the fault plane. The fault geometry is displayed in Figure \ref{fig:fault}.
\epubtkImage{}{%
    \begin{figure}[htbp]
    \centerline{\includegraphics[width=0.6\textwidth]{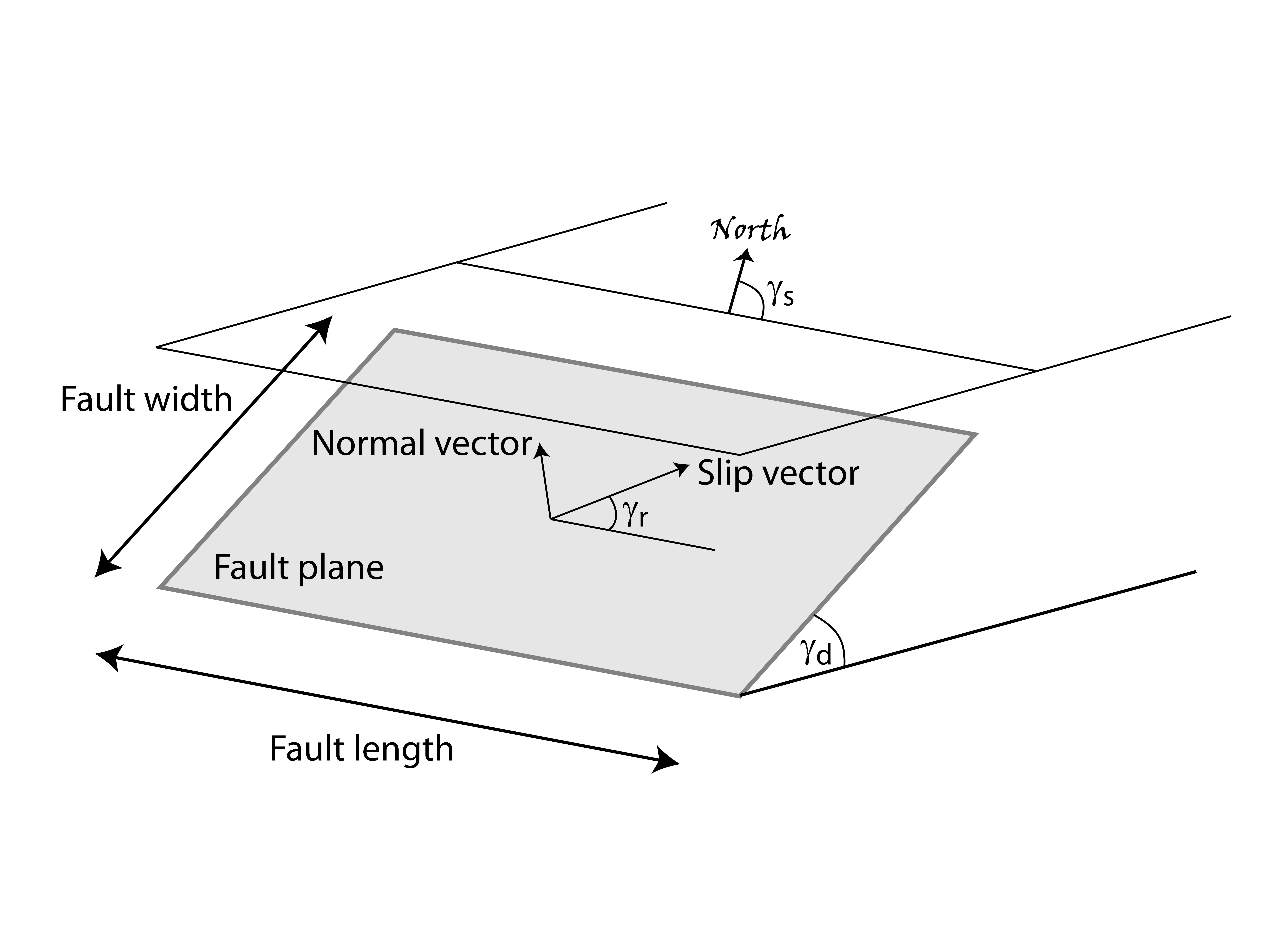}}
    \caption[Focal mechanism]{Focal mechanism.}
    \label{fig:fault}
    \end{figure}}
For the Tohoku earthquake, the angles are $\gamma_{\rm s}=3.54$, the dip angle $\gamma_{\rm d}=0.17$, and the rake angle $\gamma_{\rm r}=1.54$. In the coordinate system shown in Figure \ref{fig:pointshear}, the normal vector of the fault defines the direction of the $x$-axis, and the slip vector defines the direction of the $z$-axis. In this coordinate system, the gravity perturbation is given by:
\beq
\delta \vec a(\vec r_0\,,t) = -\frac{6G}{r_0^4} \left[(\vec e_x\cdot\vec e_{r_0})\vec e_z+(\vec e_z\cdot\vec e_{r_0})\vec e_x-5(\vec e_x\cdot\vec e_{r_0})(\vec e_z\cdot\vec e_{r_0})\vec e_{r_0}\right]\, I_2[M_0](t).
\eeq
Now vectors are to be expressed in a new coordinate system whose axes correspond to the cardinal directions $\vec e_{\rm E},\,\vec e_{\rm N}$, and the normal vector of Earth's surface $\vec e_{\rm V}$. Based on the geometry shown in Figure \ref{fig:fault}, the following relations can be found
\beq
\begin{split}
\vec e_x &= R(\vec e_{\rm V},-\gamma_{\rm s})\cdot R(\vec e_{\rm N},-\gamma_{\rm d})\cdot\vec e_{\rm V},\\
\vec e_z &= R(\vec e_{\rm V},-\gamma_{\rm s})\cdot R(\vec e_{\rm N},-\gamma_{\rm d})\cdot R(\vec e_{\rm V},\gamma_{\rm r})\cdot\vec e_{\rm N},
\end{split}
\eeq
and $\vec e_y=\vec e_z\times\vec e_x$. A matrix $R(\vec a,\alpha)$ describes a rotation around axis $\vec a$ by an angle $\alpha$. Sensors designed to monitor changes in gravity acceleration are called gravimeters \index{gravimeter}. For example, networks of gravimeters have been used in the past to detect coseismic gravity changes following large earthquakes \cite{ImEA2004}. However, these were pre-post event comparisons of DC gravity changes. A prompt detection of a coseismic gravity perturbation using gravimeters has not been achieved yet. 

\epubtkImage{}{
    \begin{figure}[htbp]
    \centerline{\includegraphics[width=0.49\textwidth]{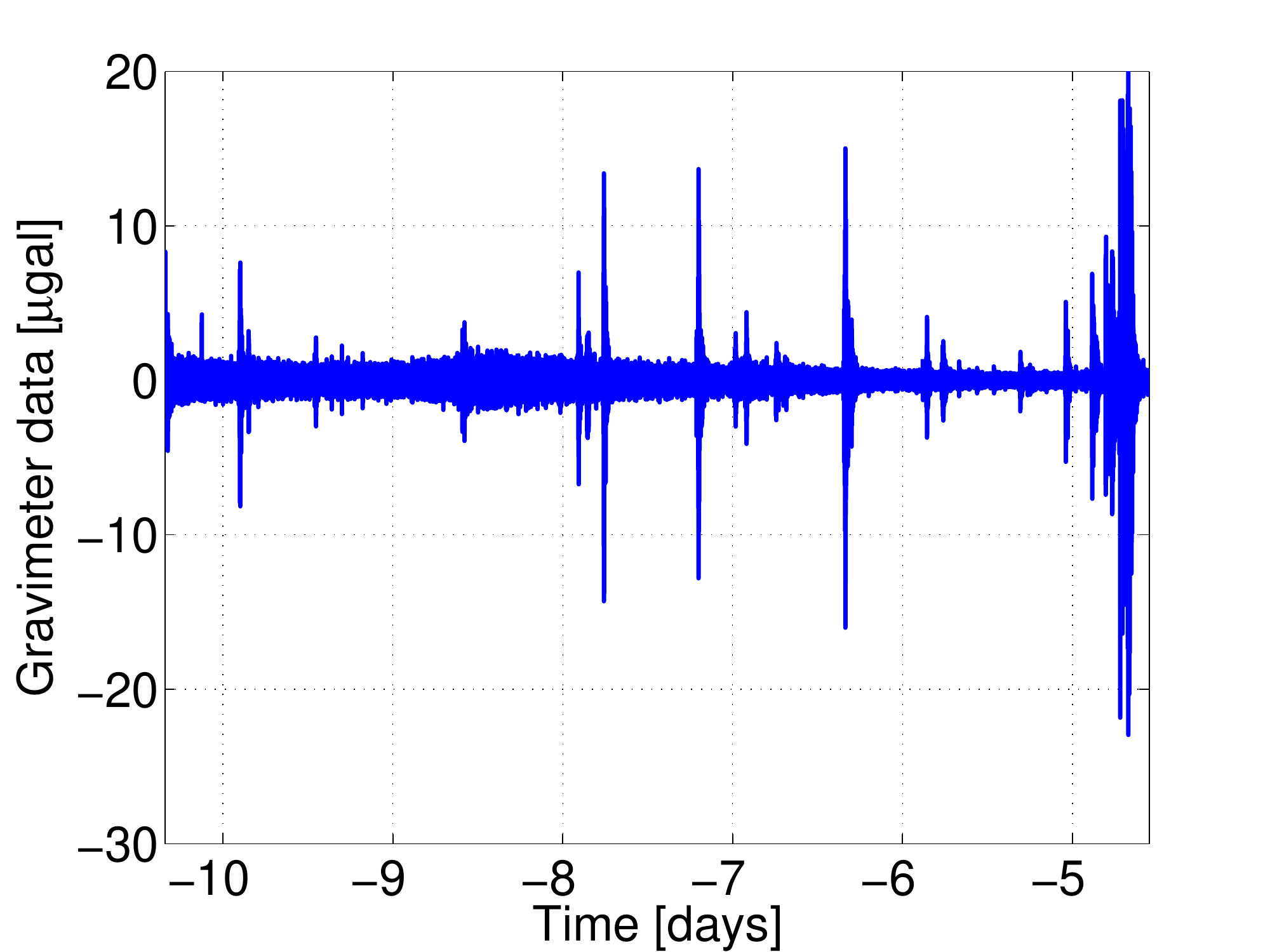} 
                \includegraphics[width=0.49\textwidth]{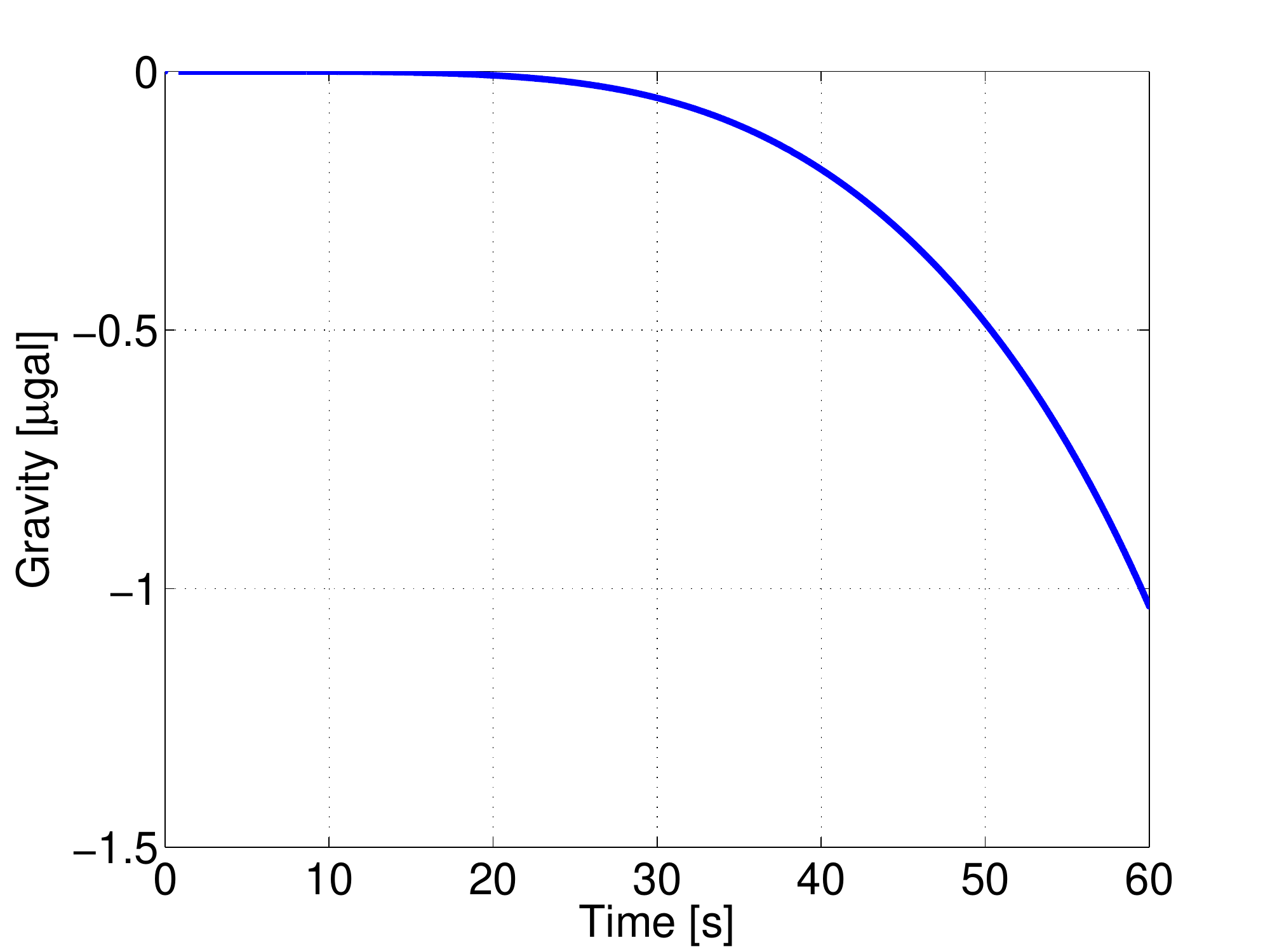}}
    \caption[Modelled Tohoku gravity perturbation]{The left plot shows gravimeter time series at the Kamioka site high-passed at 2\,mHz days before the Tohoku earthquake. The right plot shows the modelled gravity perturbation in vertical direction of the Tohoku earthquake at the Kamioka site.}
    \label{fig:Tohoku}
    \end{figure}}
The results of this and the previous subsection can be used to make quantitative predictions of gravity perturbations from earthquakes. Since the model of Equation (\ref{eq:potentialext}) is valid for fault ruptures in infinite space, a prediction of gravity perturbations for sources buried in half spaces can only be valid as long as the seismic waves have not reached the surface. In reality, this typically allows us to model up to a few seconds of time series of an earthquake, but it was shown in \cite{HaEA2015} using numerical simulations that the duration of the modelled gravity perturbation can be extended for some time without causing major deviations from the half-space signal. It would of course be useful to have the analytical half-space solution in hand. This said, it should nevertheless be true that the infinite space solution provides a useful order-of-magnitude estimate of the gravity perturbation, even beyond the duration validated by numerical simulations, at least as long as seismic waves have not yet reached the location of the gravity sensor. Figure \ref{fig:Tohoku} shows the result for the perturbation of gravity acceleration in vertical direction. The first 60\,s of the signal are simulated for a gravimeter at the Kamioka station at latitude N36.42 and longitude E137.31, about 500\,km away from the hypocenter of the earthquake. The curve uses the estimated source function of the Tohoku earthquake \footnote{\url{http://www.tectonics.caltech.edu/slip_history/2011_tohoku_joint/index.html}}, which had a total rupture duration of about 300\,s, with almost all of the total seismic moment, $5\times 10^{22}$\,Nm, already released after 120\,s. After 68\,s, the first seismic waves reach the gravimeter, which makes gravity measurements impossible for more than a day. A signal of about $-10^{-8}\,\rm m/s^2$ is substantial. The rms of the data between 2\,mHz and 0.5\,Hz is about $5\times10^{-9}\,\rm m/s^2$ during relatively quiet times, and gravimeter data are highly non-stationary (mostly due to direct seismic perturbation of the instrument). It may be possible to detect this signal before arrival of the seismic waves, based on a fit to the predicted gravity perturbation and integrating over the available 68\,s of data. 

\subsection{Seismic sources in a homogeneous half space}
\label{sec:sourcehalf}
We now consider the case of gravity perturbations above surface produced by seismic fields in a homogeneous half space. There are two major differences to the case of infinite space. First, the explicit solution of the seismic field produced by point sources contains an integral, which is impossible to solve except for the easiest source time functions. Sophisticated analytical techniques known as Cagniard -- de Hoop methods had to be invented to obtain these solutions \cite{dHo1961,AkRi2009}. Second, even when the seismic field is left unspecified, the explicit solution of the gravity perturbation involves other integrals, while the infinite space solution of Equation (\ref{eq:gravP}) was free of integrals. This is at least the conclusion of the preliminary investigation presented in the following. The purpose of this section is to introduce a suitable theoretical framework and to simplify the expression for the gravity perturbation as much as possible. The starting point is Equation (\ref{eq:gravHelm}) without the last term since it vanishes above surface. The $z$-axis is chosen as surface normal: $\vec n=(0,0,1)$. The goal is to calculate a gravity perturbation directly above the free surface at $z=0$. In general, the gravity potential above surface takes the form
\beq
\begin{split}
\delta\phi_{\rm surf}(\vec r_0,t)&=-G\rho_0\int\drm S\,\vec n\cdot\left[\vec\psi_{\rm s}(\vec r,t)\times\nabla\frac{1}{|\vec r-\vec r_0|}+\phi_{\rm s}(\vec r,t)\nabla\frac{1}{|\vec r-\vec r_0|}\right]\\
&=-G\rho_0\int\drm S\,(\psi_{\rm s}^x(\vec r,t)\partial_y-\psi_{\rm s}^y(\vec r,t)\partial_x+\phi_{\rm s}(\vec r,t)\partial_z)\frac{1}{|\vec r-\vec r_0|}\\
&= G\rho_0\left[\partial_{y_0}\int\drm S\,\frac{\psi_{\rm s}^x(\vec r,t)}{|\vec r-\vec r_0|}-\partial_{x_0}\int\drm S\,\frac{\psi_{\rm s}^y(\vec r,t)}{|\vec r-\vec r_0|}+\partial_{z_0}\int\drm S\,\frac{\phi_{\rm s}(\vec r,t)}{|\vec r-\vec r_0|}\right]
\end{split}
\label{eq:surfpot}
\eeq
Seismic fields in half spaces can be elegantly represented as expansion into cylindrical harmonics (see Section \ref{sec:cylindrical}). A good review of the theory with application to seismology can be found in \cite{Kim1989}. Looking at Equation (\ref{eq:surfpot}), we see that in addition to an expansion of the seismic potentials, one also needs an expansion of the inverse distance into cylindrical harmonics. Evaluating seismic fields on the surface $z=0$, and gravity above surface so that $z_0>0$, the inverse distance can be expanded according to
\beq
\frac{1}{|\vec r-\vec r_0|}=\sum\limits_{n=0}^\infty \int\limits_0^\infty\drm k\,(2-\delta_{n0})J_n(k\varrho_0)J_n(k\varrho)\cos(n (\phi-\phi_0))\e^{-kz_0},
\eeq
The expansion of the scalar potential components in Equation (\ref{eq:surfpot}) are given by \cite{Kim1989}
\beq
\begin{split}
\psi^x_{\rm s}(\rho,\phi,z)&=\e^{\irm\omega t}\sum\limits_{m=-\infty}^\infty\int\limits_0^\infty\drm p\,J_m(p\varrho)\e^{\irm m\phi}\left(a_m^{x,1}(p)\e^{zk_z^{\rm S}(p)}+a_m^{x,2}(p)\e^{-zk_z^{\rm S}(p)}\right)\\
\psi^y_{\rm s}(\rho,\phi,z)&=\e^{\irm\omega t}\sum\limits_{m=-\infty}^\infty\int\limits_0^\infty\drm p\,J_m(p\varrho)\e^{\irm m\phi}\left(a_m^{y,1}(p)\e^{zk_z^{\rm S}(p)}+a_m^{y,2}(p)\e^{-zk_z^{\rm S}(p)}\right)\\
\phi_{\rm s}(\rho,\phi,z)&=\e^{\irm\omega t}\sum\limits_{m=-\infty}^\infty\int\limits_0^\infty\drm p\,J_m(p\varrho)\e^{\irm m\phi}\left(b_m^{1}(p)\e^{zk_z^{\rm P}(p)}+b_m^{2}(p)\e^{-zk_z^{\rm P}(p)}\right)
\end{split}
\label{eq:cylexpand}
\eeq
The integration variable $p$ can be interpreted as horizontal wavenumber of the harmonics that constitute the seismic field, while the vertical wavenumbers $k_z^{\rm P}(p),\,k_z^{\rm S}(p)$ have the form in Equation (\ref{eq:wavek}). Explicit expressions of the amplitudes $a_m^x(p),\,a_m^y(p),\,b_m(p)$ depend on the nature of the seismic source, and can be found for a few important cases in \cite{Kim1989}. They also depend on the depth $z_{\rm s}$ of the seismic source. The evaluation of the surface integrals is analogous for the three potentials. We outline the calculation for the P-wave potential $\phi_{\rm s}$:
\beq
\begin{split}
\partial_{z_0}\int\drm S\,\frac{\phi_{\rm s}(\vec r,t)}{|\vec r-\vec r_0|} &= \partial_{z_0}\int\limits_0^{2\pi}\drm\phi\int\limits_0^\infty\drm\varrho\,\varrho\sum\limits_{n=0}^\infty \int\limits_0^\infty\drm k\,(2-\delta_{n0})J_n(k\varrho_0)J_n(k\varrho)\cos(n (\phi-\phi_0))\e^{-kz_0}\\
&\hspace{4cm}\cdot\e^{\irm\omega t}\sum\limits_{m=-\infty}^\infty\int\limits_0^\infty\drm p\,(b_m^1(p)+b_m^2(p))J_m(p\varrho)\e^{\irm m\phi}\\
&=2\pi\e^{\irm\omega t}\partial_{z_0}\sum\limits_{m=-\infty}^\infty \int\limits_0^\infty\drm k\,(2-\delta_{m0})J_m(k\varrho_0)\e^{-kz_0}\e^{\irm n\phi_0}\int\limits_0^\infty\drm p\,(b_m^1(p)+b_m^2(p))\\
&\hspace{4cm}\cdot\int\limits_0^\infty\drm\varrho\,\varrho J_m(p\varrho)J_m(k\varrho)\\
&=2\pi\e^{\irm\omega t}\partial_{z_0}\sum\limits_{m=-\infty}^\infty \int\limits_0^\infty\drm k\,(2-\delta_{m0})\frac{b_m^1(k)+b_m^2(k)}{k}J_m(k\varrho_0)\e^{-kz_0}\e^{\irm m\phi_0}\\
&=-2\pi\e^{\irm\omega t}\sum\limits_{m=-\infty}^\infty \int\limits_0^\infty\drm k\,(2-\delta_{m0})(b_m^1(k)+b_m^2(k))J_m(k\varrho_0)\e^{-kz_0}\e^{\irm m\phi_0}
\end{split}
\eeq
The last equation has strong similarity with the expansion in Equation (\ref{eq:cylexpand}) of the P-wave potential itself. The important difference is that the vertical wavenumber $k_z^{\rm P}(p)$ is now substituted by the horizontal wavenumber $k$. We have seen this already happening in the explicit solutions for Rayleigh and plane body waves, where the seismic amplitude changes with depth $z$ in terms of a vertical wavenumber, but the gravity perturbation changes with the horizontal wavenumber (see Sections \ref{sec:bodyhalf} \& \ref{sec:gravRayleigh}). For this reason, it is unfortunately impossible to express the P-wave contribution to the gravity potential directly in terms of the P-wave potential. However, the solution simplifies if the gravity field is to be calculated directly above surface at $z_0=0$:
\beq
\partial_{z_0}\int\drm S\,\frac{\phi_{\rm s}(\vec r,t)}{|\vec r-\vec r_0|}\bigg|_{z_0=0} = 2\pi\e^{\irm\omega t}\int\limits_0^\infty\drm k\,(b_0^1(k)+b_0^2(k))J_0(k\varrho_0)-4\pi\phi_{\rm s}(\rho_0,\phi_0,z_0=0,t)
\label{eq:halfgravP}
\eeq
The gravity perturbation contributed by the P-wave potential consists of a part that has the same form as the infinite-space solution in Equation (\ref{eq:gravP}), and a second part that is a simple integral involving only the zero-order Bessel function. It may be possible to carry out the remaining integral for specific seismic sources, and possibly also to carry out the inverse Fourier transform explicitly to obtain a full time-domain solution. It should be kept in mind also that the seismic potentials in half space are considerably more complicated than in infinite space, and therefore the solution in Equation (\ref{eq:halfgravP}) has only formal similarity with the infinite-space solution. We leave it to the reader to repeat the exercise for the components of the shear potential, which can be simplified by realizing that the horizontal components of the shear potential can be obtained from a single scalar potential using the identity $\vec\psi_{\rm s}(\vec r\,)=\nabla\times(0,0,\Lambda_{\rm s}(\vec r\,))+(0,0,\psi_{\rm s}^z(\varrho,\phi))$. As a final note, in order to translate the gravity model into a gravimeter signal, one also needs to take into account the self-gravity effect described in Section \ref{sec:gravground}, which means that gravity fluctuations induce surface motion. Whether corrections from self-gravity effects are significant depends on the distance of the gravimeter to the source as well as on the spectrum of the gravity fluctuations \cite{Run1980,Wan2005}.

\subsection{Summary and open problems}
\label{sec:pointsummary} 
We have shown how to calculate gravity perturbations based on models of seismic sources. The general expressions for these perturbations can be complicated, but especially when neglecting surface effects, the gravity perturbations assume a very simple form due to a fundamental equivalence between seismic and gravity potentials according to Equation (\ref{eq:gravP}). We have seen demonstrations of this principle in Section \ref{sec:forcegrav} for point forces, and in Section \ref{sec:disgravity} for point shear dislocations. 

The solution of the point force was used to highlight the difference between locally generated gravity perturbations, i.~e.~at the test mass, and perturbations from an incident seismic wavefront. It was shown that due to the strong low-pass filtering effect of gravity perturbations from distant seismic wavefronts, seismic sources need to have very peculiar properties to produce significant, instantaneous gravity perturbations at the test mass. Consequently, gravity perturbations from distant seismic wavefronts are more likely to play a role in sub-Hz GW detectors, and also there the seismic event producing the wavefront needs to be very strong. As an example, we have presented the formalism to estimate perturbations from earthquakes in Sections \ref{sec:disdensity} and \ref{sec:Tohoku}.

These results also have important implications for coherent Newtonian-noise cancellation schemes. It was argued in the past that seismic sensors deployed around the test mass can never provide information of gravity perturbations from incident seismic disturbances that have not yet reached the seismic array. Therefore, there would be a class of gravity perturbations that cannot be subtracted with seismic sensors. While the statement is generally correct, we now understand that the gravity perturbations are significant only well below the GW detection band (of any $>1\,$Hz GW detector), unless the source of the seismic wavefront has untypically strong high-frequency content. 

The theory of gravity perturbations from seismic point sources has just begun to be explored. Especially a thorough analysis of surface effects is essential for future developments. In Section \ref{sec:sourcehalf}, a first calculation of gravity perturbations from point sources in half spaces was outlined. The full solution still needs to be analyzed in detail. Open questions are how the Rayleigh waves generated in half spaces affect gravity perturbations at larger distances, and also how the contribution of body waves is altered by reflection from the surface. In light of the possible applications of low-frequency GW detectors in geophysics, further development of the theory may significantly influence future directions in this field.

\section{Atmospheric Gravity Perturbations}
\label{sec:atmos} 
\index{Newtonian noise!atmospheric}
The properties of the atmosphere give rise to many possible mechanisms to produce gravity perturbations. Sound fields are one of the major sources of gravity perturbations. Typically, sound is produced at boundaries between air and solid materials, but in general, one also needs to consider the \emph{internal} production of sound via the Lighthill process. The models of gravity perturbations from sound fields are very similar to perturbations from seismic compressional waves as given in Section \ref{sec:ambient}. The main difference in the models is related to the fact that the two fields are observed by different types of sensors. Additional mechanisms of producing atmospheric gravity noise are related to the fact that air can flow. This can lead to the formation of vortices or convection, and turbulence can always play a role in these phenomena. The Navier-Stokes equations directly predict density perturbations in these phenomena \cite{Dav2004}. Also static density perturbations produced by non-uniform temperature fields can be transported past a gravity sensor and cause gravity noise. One goal of Newtonian-noise modelling is to provide a strategy for noise mitigation. For this reason, it is important to understand the dependence of each noise contribution on distance between source and test mass, and also to calculate correlation functions. The former determines the efficiency of passive isolation schemes, such as constructing detectors underground, the latter determines the efficiency of coherent cancellation using sensor arrays. 

Atmospheric gravity perturbations have been known since long to produce noise in gravimeter data \cite{Neu2010}, where they can be observed below about 1\,mHz. At these frequencies, they are modelled accurately as a consequence of pressure fluctuations and loading of Earth's surface. Atmospheric gravity perturbations are generally expected to be the dominant contribution to ambient Newtonian noise below 1\,Hz \cite{HaEA2013}. In contrast, Creighton showed that atmospheric Newtonian noise can likely be neglected above 10\,Hz in large-scale GW detectors \cite{Cre2008}. His paper is until today the only detailed study of atmospheric Newtonian noise at frequencies above the sensitive band of gravimeters, and includes noise models for infrasound waves, quasi-static temperature fluctuations advected in various modes past test masses, and shockwaves. His results will be reviewed in the following with the exception that a new solution is given for gravity perturbations from shockwaves based on the point-source formalism of Section \ref{sec:pointsources}. Preliminary work on modelling gravity perturbations from turbulence was first published in \cite{CaAl2009}, and is reviewed and improved in Section \ref{sec:turbNN}. 

\subsection{Gravity perturbation from atmospheric sound waves}
\label{sec:soundNN}
Sound waves are typically understood as propagating perturbations of the atmosphere's mean pressure $p_0$. The pressure change can be translated into perturbation of the mean density $\rho_0$. The relation between pressure and density fluctuations depends on the adiabatic index $\gamma\approx 1.4$ of air \cite{Woo1955} \index{adiabatic index}
\beq
\gamma\frac{\delta \rho(\vec r,t)}{\rho_0}= \frac{\delta p(\vec r,t)}{p_0}
\eeq
The classical explanation for $\gamma>1$ is that the temperature increases when the sound wave compresses the gas sufficiently slowly, and this temperature increase causes an increase of the gas pressure beyond what is expected from compression at constant temperature. Note that in systems whose size is much smaller than the length of a sound wave, the statement needs to be reversed, i.~e.~fast pressure fluctuations describe an adiabatic process, not slow changes. An explanation of this counter-intuitive statement in terms of classical thermodynamics is given in \cite{Fle1974}. It can also be explained in terms of the degrees of freedom of gas molecules \cite{Hen1963}. At very high frequencies (several kHz or MHz depending on the gas molecule), vibrations and also rotations of the molecules cannot follow the fast sound oscillation, and their contribution to the specific heat freezes out (thereby lowering the adiabatic index). At low audio frequencies, sound propagation in air is adiabatic \footnote{Only at really low frequencies, below 10\,mHz, where the finite size of the atmosphere starts to matter, pressure oscillations can be isothermal again.}.

Let us return to the calculation of gravity perturbations from sound fields. Assuming that a sound wave incident on the surface is reflected without loss such that its horizontal wavenumber is preserved, one obtains the gravity perturbation as the following integral over the half space $z>0$:
\beq
\begin{split}
\delta\phi(\vec r_0,t) &= -\frac{G\rho_0}{\gamma\,p_0}\e^{\irm(\omega t-\vec k_\varrho\cdot\vec\varrho_0)}\delta p(\omega)\int\limits_{\mathcal H}\drm V\,\frac{(\e^{-\irm k_z z}+\e^{\irm k_z z})\e^{-\irm \vec k_\varrho\cdot\vec\varrho}}{(\varrho^2+(z-z_0)^2)^{1/2}}\\
&= 4\pi\frac{G\rho_0}{\gamma\,p_0}\e^{\irm(\omega t-\vec k_\varrho\cdot\vec\varrho_0)}(\e^{-k_\varrho|z_0|}(2\Theta(z_0)-1)-2\cos(k_zz_0)\Theta(z_0))\frac{\delta p(\omega)}{k^2}
\end{split}
\label{eq:gravinfra}
\eeq
Here, the Heaviside function $\Theta(\cdot)$ has the value 1 at $z_0=0$. The gravity potential and acceleration are continuous across the surface. We neglect the surface term here, but this is mostly to simplify the calculation and not fully justified. Part of the energy of a sound wave is transmitted into the ground in the form of seismic waves. Intuitively, one might be tempted to say that only a negligible amount of the energy is transmitted into the ground, but at the same time the density of the ground is higher, which amplifies the gravity perturbations. Let us analyze the case for a sound wave incident at a normal angle to the surface. In this case, the sound wave is transmitted as pure compressional wave into the ground. We denote the air medium by the index ``1'' and the ground medium by ``2''. Multiplying the seismic transmission coefficient (see \cite{AkRi2009}) by $\rho_2/\rho_1$, the relative amplitude of gravity perturbations is\index{sound!reflection}
\beq
\frac{\delta a_1}{\delta a_2}=2\frac{\rho_2\alpha_1}{\rho_1\alpha_1+\rho_2\alpha_2},
\eeq
where $\alpha_1$ is the speed of sound, and $\alpha_2$ the speed of compressional waves. The sum in the denominator can be approximated by $\rho_2\alpha_2$, which leaves $2\alpha_1/\alpha_2$ as gravity ratio. The ratio of wave speeds does not necessarily have to be small at the surface. We know that the Rayleigh-wave speed at the LIGO sites is about 250\,m/s \cite{HaOR2011}, which we can use to estimate the compressional-wave speed to be around 600\,m/s (by making a guess about the Poisson's ratio of the ground medium). This means that the effective transmissivity with respect to gravity perturbations could even exceed a value of 1! Therefore, it is clear that the physics of infrasound gravity perturbations is likely more complicated than outlined in this section. Nonetheless, we will keep this for future work and proceed with the simplified analysis assuming that sound waves are fully reflected by the ground. 

\epubtkImage{}{%
    \begin{figure}[htbp]
    \centerline{\includegraphics[width=0.49\textwidth]{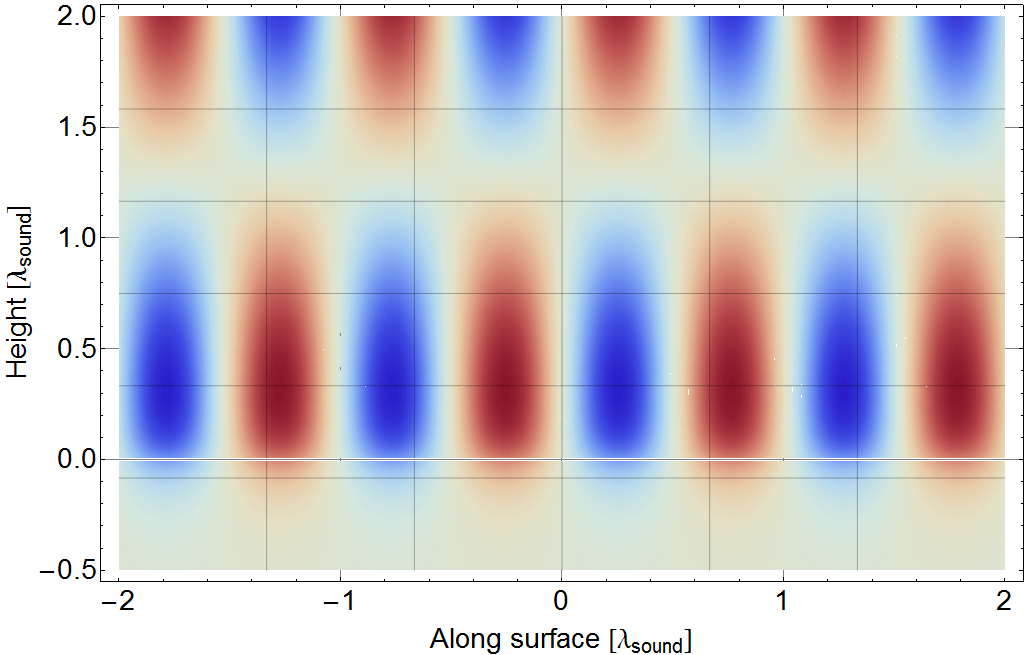}     
                \includegraphics[width=0.49\textwidth]{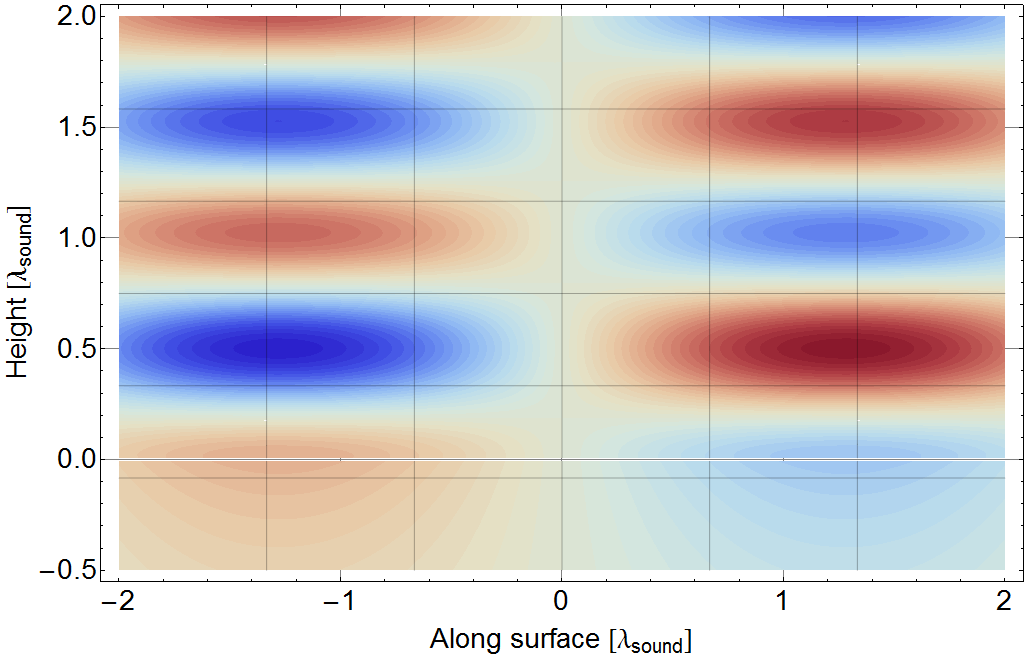}}
    \caption[Infrasound Newtonian noise]{Gravity acceleration along a horizontal direction produced by plane infrasound waves. The left plot shows the field for an angle of incidence of $7\pi/16$, the right plot for an angle of $\pi/16$ with respect to the surface normal.}
\label{fig:infraNN}
    \end{figure}}
The gravity acceleration caused by infrasound waves is shown in Figure \ref{fig:infraNN} for two different angles of incidence with respect to the surface normal. Note that the infrasound field modelled in Equation (\ref{eq:gravinfra}) consists of two plane waves propagating in opposite directions with respect to the normal, and along the same direction with respect to the horizontal. Therefore, the pressure and consequently gravity field have the form of a standing wave along the normal direction. Below surface, the gravity perturbation falls off exponentially. The decrease is faster when the infrasound wave propagates nearly horizontally. The length scale that determines the exponential fall off becomes infinite if the wave propagates vertically, but at the same time the projection of gravity acceleration onto a horizontal direction vanishes. This is why underground construction of GW detectors is an efficient means to mitigate infrasound Newtonian noise. Creighton also considered the case of a shield against infrasound disturbances around the test masses of surface detectors, which in its simplest form is already given by the buildings hosting the test masses \cite{Cre2008}. A detailed investigation of noise-reduction techniques is given in Section \ref{sec:mitigate}.

\subsection{Gravity perturbations from quasi-static atmospheric temperature perturbations}
\label{sec:quasitemp}
In this section, we review the rather complex calculation of gravity perturbations from a temperature field presented as appendix in \cite{Cre2008}. The calculation is also instructive to solve similar problems in the future. The basic idea is the following. Temperature fluctuations in the atmosphere lead to density changes. In terms of the mean temperature $T_0$ and density $\rho_0$ of the atmosphere, and according to the ideal gas law at constant pressure, small fluctuations in the temperature field cause perturbations of the density:
\beq
\delta \rho(\vec r,t)=-\frac{\rho_0}{T_0}\delta T(\vec r,t)
\eeq
Pressure fluctuations also cause density perturbations, but as we have seen in the previous section, they result in quickly propagating infrasound waves. The effect that we want to study here is the Newtonian noise from slowly changing density fields, transported past a test mass by air flow. These are predominantly associated with slow temperature fluctuations. The gravity perturbation produced by such a temperature field is given by
\beq
\delta\vec a(\vec r_0,t)=-\frac{G\rho_0}{T_0}\int\drm V\frac{\delta T(\vec r,t)}{|\vec r-\vec r_0|^3}(\vec r-\vec r_0)
\eeq
Trying to obtain an explicit expression of the temperature field, inserting it into this integral, and solving the integral is hopeless here. What one can do instead is to work with the statistical properties of the temperature field. If the temperature field is stationary, then we can calculate the spectral density as 
\beq
S(\delta a_x;\vec r_0,\omega)=2\left(\frac{G\rho_0}{T_0}\right)^2\int\drm\tau\int\drm V\int\drm V'\frac{xx'}{r^3(r')^3}\langle\delta T(\vec r,t)\delta T(\vec r\,',t+\tau)\rangle\e^{-\irm\omega\tau},
\label{eq:tempNNspec}
\eeq
where we have used Equation (\ref{eq:defspecdens}). The vectors $\vec r,\,\vec r\,'$ point from the test mass to temperature fluctuations in the atmosphere. The next step is to characterize the temperature field. Temperature fluctuations in the vicinity of Earth surface are distributed by turbulent mixing. As shown in \cite{KuNa2006}, temperature inhomogeneities of the surface play a minor role in the formation of the temperature field at frequencies above a few tens of a mHz. Therefore, at sufficiently high frequencies, one can approximate the temperature perturbations as homogeneous and isotropic. In this case, the second-order noise moments of $\delta T(\vec r,t)$ can be characterized by the \emph{temperature structure function} $D(\delta T;r)$:\index{temperature structure function}
\beq
\langle(\delta T(\vec r,t)-\delta T(\vec r+\Delta\vec r,t))^2\rangle=D(\delta T;\Delta r)
\label{eq:tempstruct}
\eeq
The structure function can typically be approximated as a power law
\beq
D(\delta T;|\Delta\vec r|)=c_T^2(\Delta r)^p
\eeq
provided that the distance $\Delta r$ is sufficiently small. This relation also breaks down at distances similar to and smaller than the Kolmogorov length scale, which is about 0.4\,mm for atmospheric surface layers \cite{AnAt1978}\index{Kolmogorov length scale}. Turbulent mixing enforces power laws with $p\sim 2/3$ \cite{AnAt1978}. Applying Taylor's hypothesis, the distance $\Delta r$ can be substituted by the product of wind speed $v$ with time $\tau$, and Equation (\ref{eq:tempstruct}) can be reformulated as
\beq
\langle\delta T(\vec r,t)\delta T(\vec r,t+\tau)\rangle=\sigma_T^2-\frac{c_T^2}{2}(v\tau)^p
\label{eq:eulercorr}
\eeq
The parameter $c_T$ depends on the dissipation rate of turbulent kinetic energy and the temperature diffusion rate, and $\sigma_T$ is the standard deviation of temperature fluctuations. Since Taylor's hypothesis is essential for the following calculations, we should make sure to understand it\index{Taylor's hypothesis}. Qualitatively it states that turbulence is transported as frozen pattern with the mean wind speed. More technically, it links measurements in Eulerian coordinates, i.~e.~at points fixed in space, with measurements in Lagrangian coordinates, i.~e.~that are connected to fluid particles. The practical importance is that two-point spatial correlation functions such as Equation (\ref{eq:tempstruct}) can be estimated based on a measurement at a single location when carried out over some duration $\tau$ as in Equation (\ref{eq:eulercorr}). In either case, the hypothesis can be expected to fail over sufficiently long periods $\tau$ or distances $\Delta r$, which are linked to the maximal scale of turbulent structures \cite{DaSo1997}. In any case, we assume that Taylor's hypothesis is sufficiently accurate for our purposes. The Fourier transform of Equation (\ref{eq:eulercorr}) yields the spectral density of temperature fluctuations
\beq
S(\delta T;\vec r_0,\omega)=c_T^2v^p(\vec r_0)\omega^{-(p+1)}\Gamma(p+1)\sin(\pi p/2)
\label{eq:spectemp}
\eeq
The Fourier transform cannot be calculated without employing an upper cutoff on the variable $\tau$. This means that the spectral density given here only holds at sufficiently high frequencies (at the same time not exceeding the Kolmogorov limit defined by the size $l$ of the smallest turbulence structures, $\omega<v/l$, which is of the order kHz). Technically, the Fourier transform can be calculated by multiplying an exponential term $\exp(-\epsilon \tau)$ to the integrand, and subsequently taking the limit $\epsilon\rightarrow 0$. 

The next step is to calculate the temperature correlation that appears in Equation (\ref{eq:tempNNspec}). Using Taylor's hypothesis to convert Equation (\ref{eq:eulercorr}) back into a two-point spatial correlation, we see that correlations over large distances are negligible. In terms of the frequency of temperature fluctuations, correlations are significant over distances of the order $v/\omega$ (which is shown in the following). Consider the scenario displayed in Figure \ref{fig:tempNN}. Two air pockets are shown at locations $\vec r,\,\vec r\,'$ and times $t,\,t'$ on two steam lines that we denote by $S$ and $S'$. 
\epubtkImage{}{%
    \begin{figure}[htbp]
    \centerline{\includegraphics[width=0.85\textwidth]{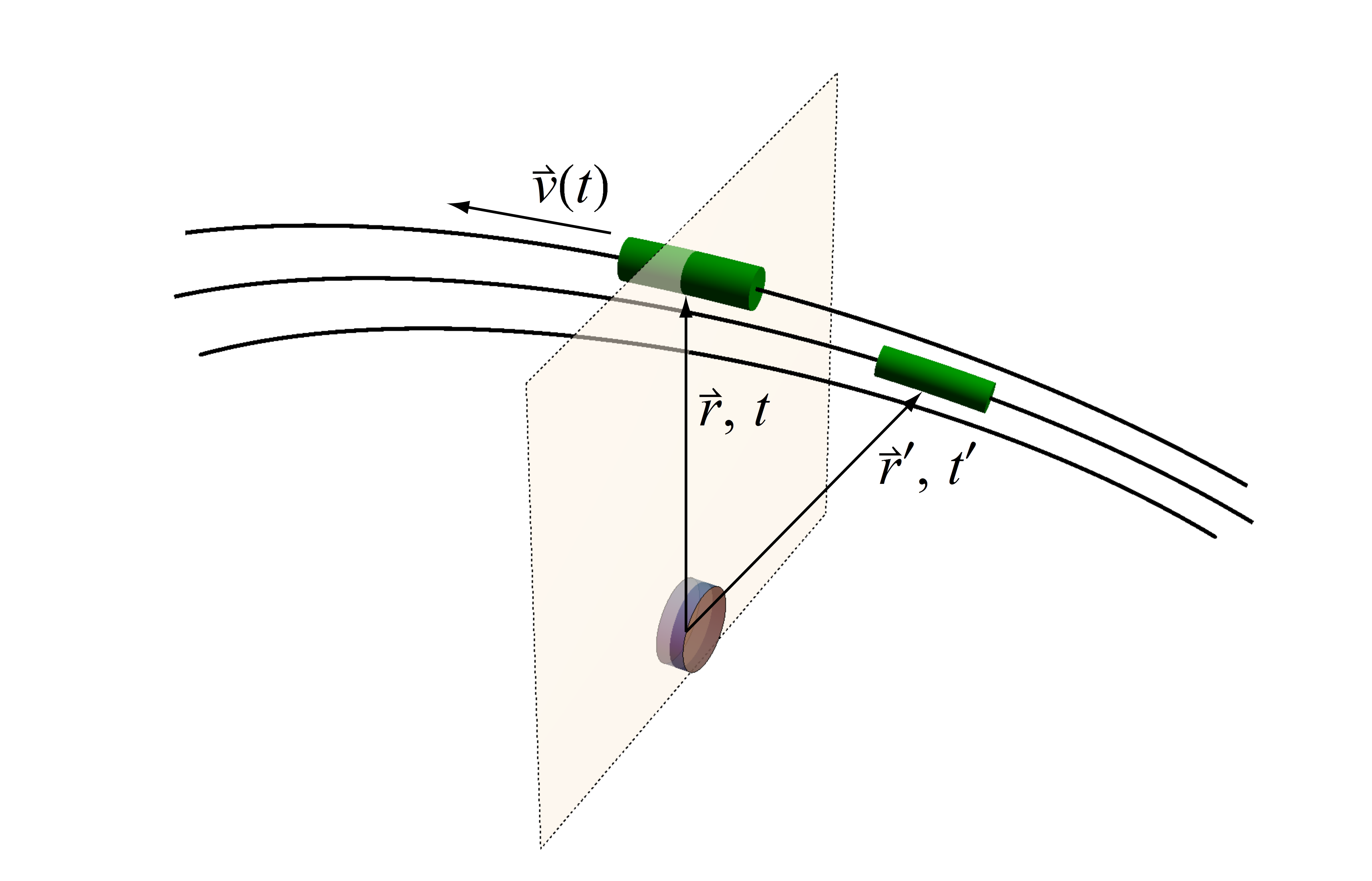}}
    \caption[Newtonian noise from temperature fields]{Sketch of laminar air flow past the test mass. Air volume is divided into cells that move along streamlines. Speed can change with time, and be different along a streamline.}
\label{fig:tempNN}
    \end{figure}}
If $\tau=t-t'$ is sufficiently small, then the separation of the two pockets can be written $(s^2+(v\tau)^2)^{1/2}$, where the distance $s$ of the two streamlines $S,\,S'$ and $v$ are evaluated at $\vec r$. Together with Taylor's hypothesis, temperature fluctuations between the two pockets are significant if $\tau$ is sufficiently close to the time $\tau_0$ it takes for the pocket at $\vec r\,'$ to reach the reference plane, and also $s$ must be sufficiently small. The temperature correlation can then be written as
\beq
\langle\delta T(\vec r\,',t')\delta T(\vec r,t'+\tau)\rangle=\sigma_T^2-\frac{c_T^2}{2}(s^2+v^2(\tau-\tau_0)^2)^{p/2}
\eeq
This allows us to carry out the integral over $\tau$ in Equation (\ref{eq:tempNNspec}):
\beq
\begin{split}
\int\drm\tau\langle&\delta T(\vec r\,',t')\delta T(\vec r,t'+\tau)\rangle\e^{-\irm\omega\tau} = \\
&\sqrt{\frac{2^{p+1}}{\pi}\frac{s}{v\omega}}\Gamma(p/2+1)c_T^2 \left(\frac{vs}{\omega}\right)^{p/2}\sin(\pi p/2)K_{(p+1)/2}(\omega s/v)\e^{-\irm\omega\tau_0},
\end{split}
\eeq
where $K_n(\cdot)$ is the modified Bessel function of the second kind, and $v,\,s$ are functions of $\vec r$. Again, the integral can only be evaluated if an exponential upper cutoff on the variable $\tau$ is multiplied to the integrand, which means that we neglect contributions from large-scale temperature perturbations. The correlation spectrum is plotted in Figure \ref{fig:corrTemp}. 
\epubtkImage{}{
    \begin{figure}[htbp]
    \centerline{\includegraphics[width=0.8\textwidth]{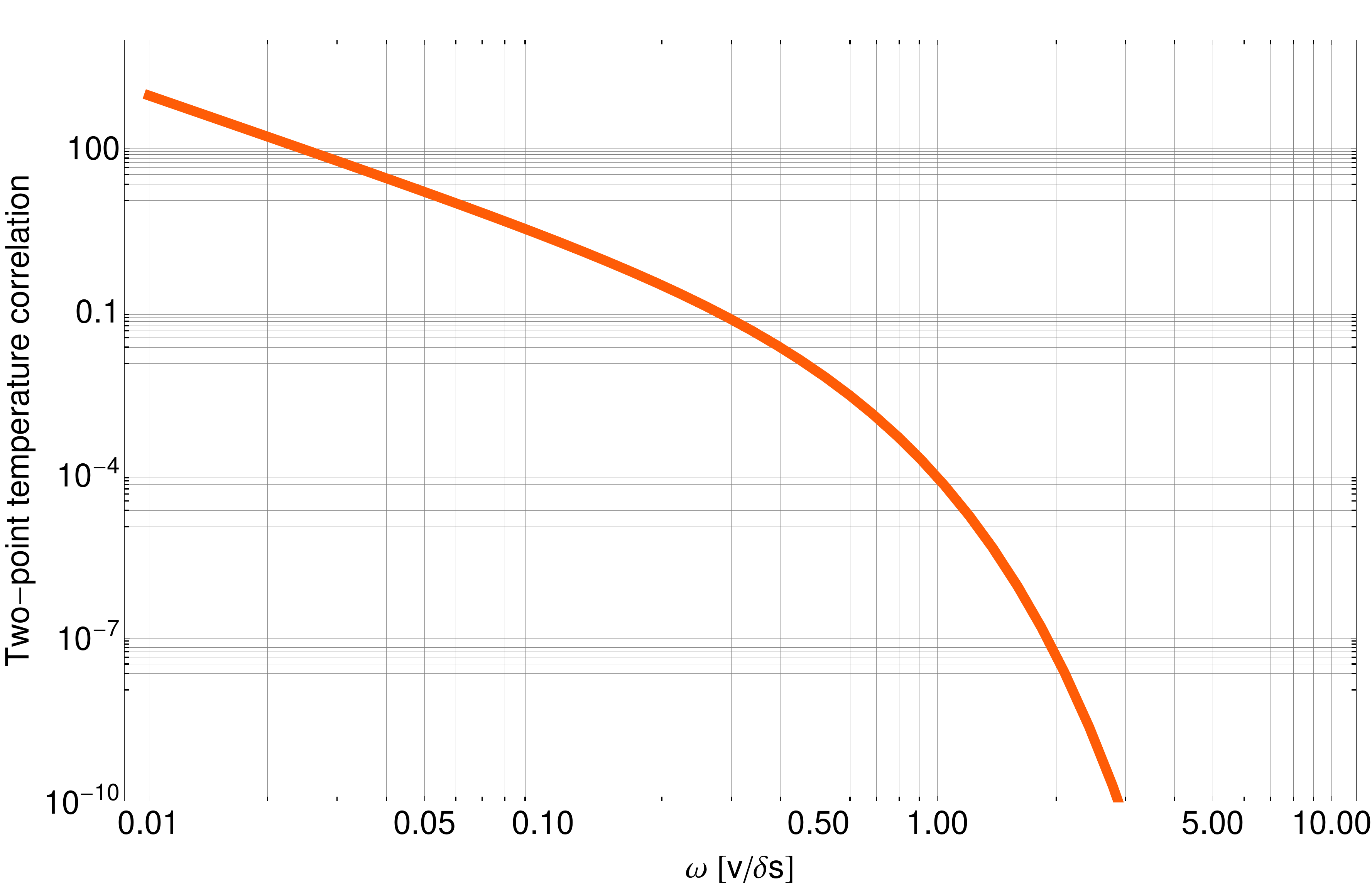}}
    \caption[Temperature correlation spectrum]{Two-point temperature correlation spectrum.}
\label{fig:corrTemp}
    \end{figure}}
At frequencies above $v/s$, the spectrum falls exponentially since $K_\nu(x)\rightarrow \sqrt{\pi/(2x)}\exp(-x)$ for $x\gg|\nu^2-1/4|$. This means that the distance between streamlines contributing to the two-point spatial correlation must be very small to push the exponential suppression above the detection band. The integral over $V'$ in Equation (\ref{eq:tempNNspec}) can be turned into an integral over streamlines $S'$ that lie within a bundle $s\lesssim v/\omega$ of streamline $S$, which allows us to approximate the volume element as cylindrical bundle $\drm V'=2\pi s \drm s\, \drm\tau_0v(\vec r\,)$. The form of the volume element is retained over the whole extent of the streamline since the air is nearly incompressible for all conceivable wind speeds, i.~e.~changes in the speed of the cylindrical pocket are compensated by changes in the radius of the pocket to leave the volume constant. Hence, the speed in the volume element can be evaluated at $\vec r$. With this notation, the integral can be carried out over $0<s<\infty$ since the modified Bessel function automatically enforces the long-distance cutoff necessary for our approximations, which yields
\beq
S(\delta a_x;\vec r_0,\omega)=\left(\frac{G\rho_0}{T_0}\right)^24\pi c_T^2\Gamma(2+p)\sin(\pi p/2) \int\drm V\int\drm \tau_0\frac{xx'}{r^3(r')^3} \left(\frac{v(\vec r\,)}{\omega}\right)^{p+3}\e^{-\irm\omega\tau_0},
\eeq
Here the vector $\vec r$ is  parameterized by $\tau_0$. This result can be interpreted as follows. We have two streamlines $S$, $S'$, whose contributions to this integral are evaluated in terms of the duration $\tau_0$ it takes for the pocket at $\vec r\,'$ to reach the reference plane that goes through all streamlines, and contains the test mass at $\vec r_0$ and location $\vec r$ (as indicated in Figure \ref{fig:tempNN}). Since we consider the pocket on streamline $S$ to be at the reference plane at time $t$, we can set $\tau_0=t-t'$, and integrating contributions from all streamlines over the reference plane with area element $\drm A$, with wind speed $v(\vec\varrho\,)$, and $\vec\varrho$ pointing from the test mass to streamlines on the reference plane, we can finally write
\beq
\begin{split}
S(\delta a_x;\vec r_0,\omega)&=\left(\frac{G\rho_0}{T_0}\right)^2 4\pi c_T^2\omega^{-(p+3)}\Gamma(2+p)\sin(\pi p/2)\\
& \hspace*{2cm}\cdot\int\limits_{\mathcal A(\vec r_0)}\drm A\, v(\vec\varrho\,)\int\drm t'\frac{x'}{(r')^3} \e^{-\irm\omega t'}\int\drm t\frac{x}{r^3}v(\vec r\,)^{p+3} \e^{\irm\omega t}\\
&=\left(\frac{G\rho_0}{T_0}\right)^2 \frac{4\pi}{\omega^2}(2+p)S(\delta T;\vec r_0,\omega)\\
& \hspace*{2cm}\cdot\int\limits_{\mathcal A(\vec r_0)}\drm A\, v(\vec\varrho\,)\int\drm t'\frac{x'}{(r')^3} \e^{-\irm\omega t'}\int\drm t\frac{x}{r^3}\left(\frac{v(\vec r\,)}{v(\vec r_0)}\right)^p v^3(\vec r\,)\e^{\irm\omega t}
\end{split}
\label{eq:advectNN}
\eeq
In this equation, $\vec r\,' = \vec r\,'(\vec\varrho,t')$, and $\vec r = \vec r(\vec\varrho,t)$ are the parameterized streamlines. For uniform airflow we have $v=\;$const, and the remaining integrals can be solved with the results given in Section \ref{sec:objectline}. Other examples have been calculated by Creighton \cite{Cre2008}.

\subsection{Gravity perturbations from shock waves}
\label{sec:shockNN}
In \cite{Cre2008}, an estimate of gravity perturbations from a shock wave produced in air was presented based on the infrasound perturbation in Equation (\ref{eq:gravinfra}). The goal was to estimate the transient gravity perturbation produced when the shock wave reaches the test mass. It did not address the question whether significant gravity perturbations can be produced before the arrival of the shock wave. A time-domain description may give further insight into this problem. As we have seen for seismic point sources, a time domain solution can reveal important characteristics of the gravity perturbation, such as the distinction between gravity perturbations from a distant wavefront, and from a wavefront that has reached the test mass. In the following, we will provide a full time-domain solution for an explosive point source of an atmospheric shock wave. It is assumed that the shock wave is produced sufficiently close to the test mass so that the pressure field can be approximated as spherical at the time the shock wave reaches the test mass. Reflections from the surface and upper atmospheric layers need to be considered for a more refined model applicable to distant sources. A shock wave from an explosive source is isotropic (which is rather a definition of what we mean by explosive source). The pressure change is built up over a brief amount of time initially involving an air mass $M=V\rho_0$ determined by the source volume $V$. In the theory of moment tensor sources, an explosion in air at $t_0=0$ can be represented by a diagonal moment tensor according to 
\beq
\mathbf{M}(t)=-\alpha^2\frac{M}{\gamma p_0}\Delta p(t)\mathbf{1}
\label{eq:explsource}
\eeq
where $\alpha$ is the speed of sound, $\gamma$ is the adiabatic coefficient of air, $p_0$ the mean air pressure, $\Delta p(t)$ the pressure change, and $\mathbf{1}$ the unit matrix. Since shock-wave generation is typically non-linear \cite{Whi1974}, the source volume should be chosen sufficiently large so that wave propagation is linear beyond its boundary. This entails that the pressure change $\Delta p$ is also to be evaluated on the boundary of the source volume. Note that in comparison to solitons, shock waves always show significant dissipation, which means that there should not be a fundamental problem with this definition of the source volume. Alternatively, if nonlinear wave propagation is significant over long distances, then one can attempt to linearize the shock-wave propagation by introducing a new nonlinear wave speed, which needs to be used instead of the speed of sound \cite{Whi1974}. In general, a sudden increase of atmospheric pressure by an explosive source must relax again in some way, which means that $\Delta p(t\rightarrow\infty)=0$.

Next, we need an expression to obtain the acoustic potential in terms of the moment tensor. The acoustic potential is analogous to the seismic P-wave potential for a medium with vanishing shear modulus, and we can calculate the corresponding perturbation of the gravity potential using Equation (\ref{eq:gravP}). The coupling of a tensor source to the acoustic field can be expressed in terms of the Green's matrix
\beq
\mathbf{\Phi}(\vec r,t)=\frac{1}{4\pi\rho_0}\left[-\frac{1}{\alpha^2r}\delta(t-r/\alpha)(\vec e_r\otimes\vec e_r)+\frac{1}{r^3}\left(3\vec e_r\otimes\vec e_r-\mathbf{1}\right)\int\limits_0^{r/\alpha}\drm\tau\,\tau\delta(t-\tau)\right],
\label{eq:coupletensor}
\eeq
where we assume that the shock wave is linear and propagates with the speed of sound outside the source volume, i.~e.~the amplitude of the shock wave has decreased to a level where non-linear propagation effects can be neglected. The acoustic potential can now be written
\beq
\phi_{\rm s}(\vec r,t) = \int\drm\tau\,{\rm Tr}(\mathbf{\Phi}(\vec r,t)\mathbf{M}(t-\tau))
\eeq
and together with Equation (\ref{eq:gravP}), we find the gravity potential perturbation
\beq
\delta\phi(\vec r_0,t) = -\frac{GM}{r_0}\frac{\Delta p(t-r_0/\alpha)}{\gamma p_0},
\label{eq:gravshock}
\eeq
where $\vec r_0$ points from the source to the test mass. This result is of very different nature compared to the gravity potentials for point forces and point shear dislocations presented in Section \ref{sec:pointsources}. Due to spherical symmetry of the source, the instantaneous gravity perturbation far away from the source vanishes. If the diagonal components of the source tensor had different values, then the integral contribution in Equation (\ref{eq:coupletensor}) would remain, which gives rise to instantaneous gravity perturbations at all distances. Source symmetry plays an important role. 

The corresponding perturbation of gravity acceleration reads
\beq
\delta\vec a(\vec r_0,t) = -\frac{GM}{r_0^2}\frac{1}{\gamma p_0}\left(\Delta p(t-r_0/\alpha)+\frac{r_0}{\alpha}\Delta p'(t-r_0/\alpha)\right)\vec e_{r_0}
\label{eq:gaccshock}
\eeq
The gravity perturbation in the far field is dominated by the derivative of the pressure change. One of the examples given in \cite{Cre2008} was a sonic boom from a supersonic aircraft. In this case, the source location changes with time along the trajectory of the aircraft. This amounts to an integral of Equation (\ref{eq:gaccshock}) over the trajectory. It is convenient in this case to introduce $r_0$ as distance at closest approach of the air craft to the test mass. The source volume is replaced by the rate $V\rightarrow A v$ ($A$ being the cross-sectional area of the ``source tube'' around the aircraft trajectory, and $v$ the speed of the aircraft). In the case of uniform motion of the aircraft, the calculation of the integral over the trajectory is straight-forward. 
\epubtkImage{}{
    \begin{figure}[htbp]
    \centerline{\includegraphics[width=0.45\textwidth]{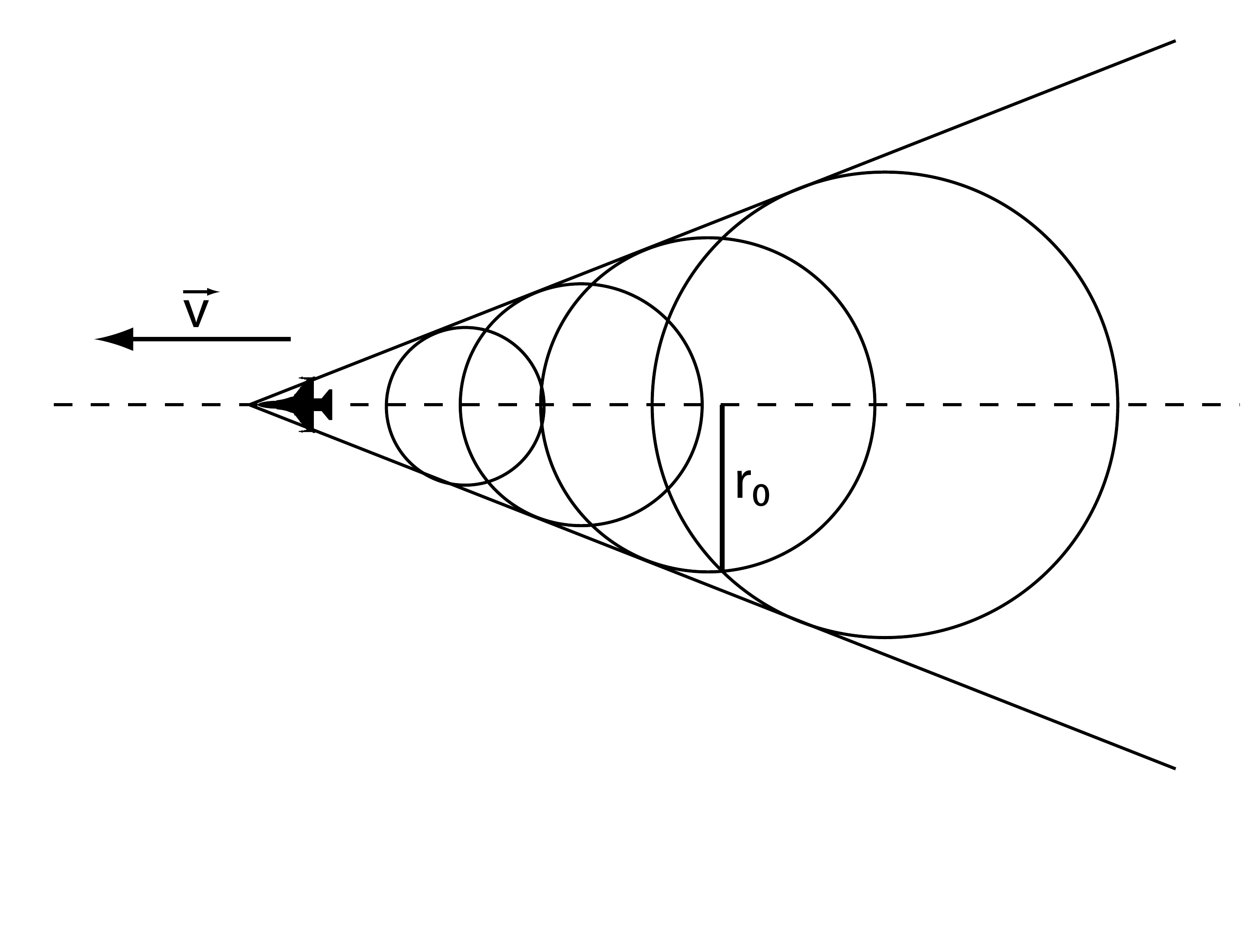}
                \includegraphics[width=0.55\textwidth]{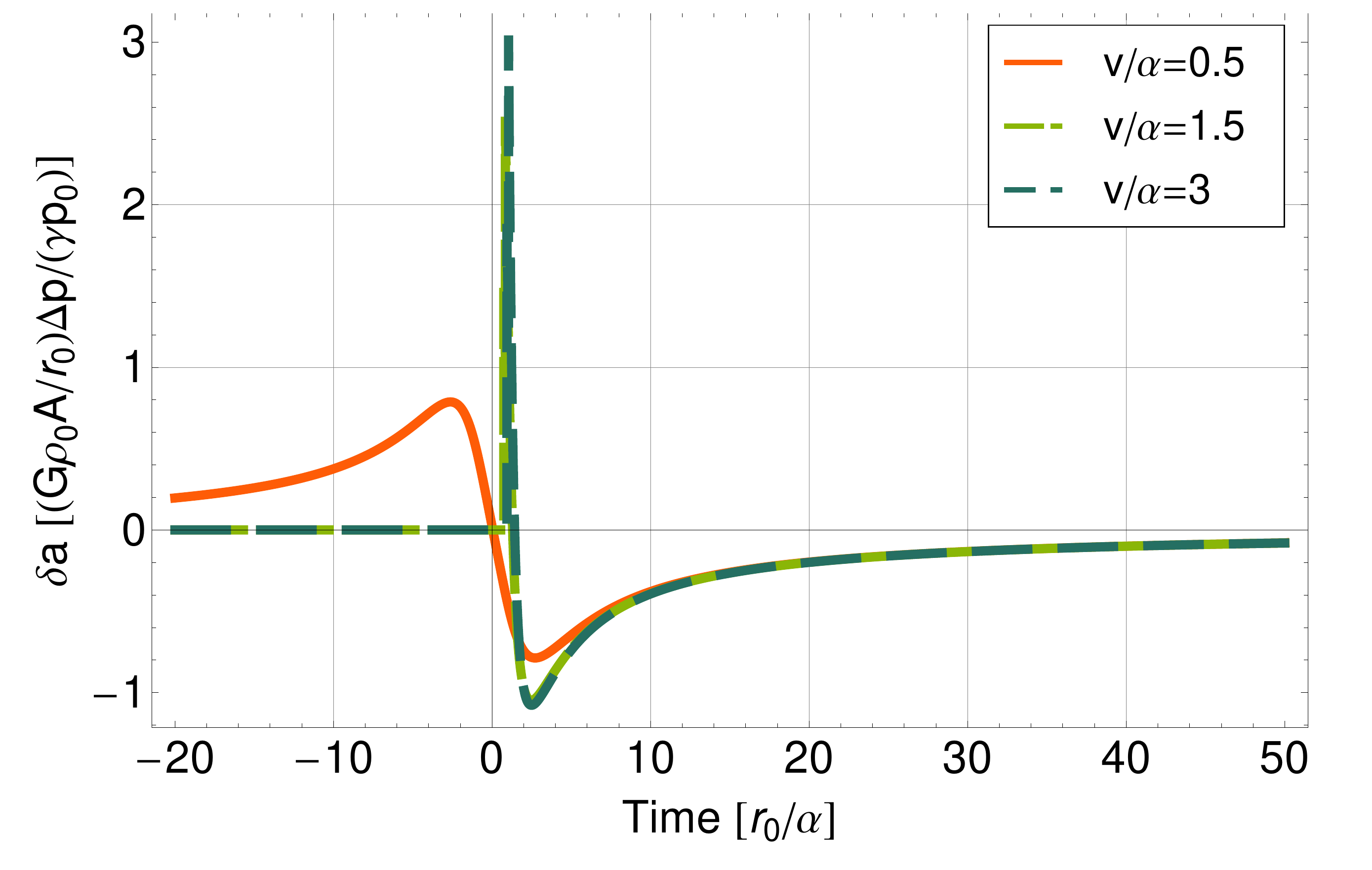}}
    \caption[Gravity perturbation from a sonic boom]{Gravity perturbation from a sonic boom produced by an aircraft. The curve with $v/\alpha=0.5$ is only for illustration purposes since a shockwave is not produced in sub-sonic flight. The direction of gravity perturbation plotted here is along the direction of the aircraft trajectory. The separation of the initial and final pressure change of a propagating wavefront is $\Delta t=0.1r_0/\alpha$.}
\label{fig:shockgrav}
    \end{figure}}
The result is shown in Figure \ref{fig:shockgrav} for three different ratios of aircraft speed over speed of sound. The pressure change is modelled as N-profile \cite{Cre2008}
\beq
\Delta p(t)=-\frac{2\Delta p}{\Delta t}(t-\Delta t/2)\theta(t)\theta(\Delta t-t),
\eeq
which consists of two positive pressure changes by $\Delta p$ at times $t=0$ and $t=\Delta t=0.1$, and a linear pressure fall between these two times. The aircraft trajectory is assumed to be horizontal and passing directly above the test mass. Time $t=0$ corresponds to the moment when the aircraft reaches the point of closest approach. If $v<\alpha$, then sound waves reach the test mass well before the aircraft reaches the closest point of approach. In the case of supersonic flight, $\alpha<v$, the first sound waves reach the test mass at $t=r_0/\alpha$. Inserting the pressure change into Equation (\ref{eq:gaccshock}), we see that the far-field gravity perturbation is characterized by two $\delta$-peaks. The derivative of the linear pressure change between the peaks cancels with a contribution of the near-field term. As can be understood from the left plot in Figure \ref{fig:shockgrav}, the gravity perturbation falls gradually after the initial peak since a test mass inside the cone still responds to pressure changes associated with two propagating wavefronts. 

\subsection{Gravity perturbations in turbulent flow}
\label{sec:turbNN}
In this section, we review the calculation of gravity perturbations from turbulent flow \cite{CaAl2009}. While in Section \ref{sec:quasitemp}, the problem was to calculate gravity perturbations from an advected temperature field whose spectrum is determined by turbulent mixing, we are now interested in the gravity perturbations from pressure fluctuations produced in turbulent flow. Generation of pressure fluctuations (sound) in air is a non-linear phenomenon known as Lighthill process \cite{Lig1952,Lig1954}. Lighthill found that the Navier-Stokes equations can be rearranged into equations for the propagation of sound,
\beq
\left(\Delta-\frac{1}{c_{\rm s}^2}\partial_t^2\right)\rho(\vec r,t)=-\frac{1}{c_{\rm s}^2}(\nabla\otimes\nabla):\boldsymbol\tau(\vec r,t),
\label{eq:Lighthill}
\eeq
where $\tau_{ij}=\rho v_iv_j+\sigma_{ij}-c_{\rm s}^2\rho\delta_{ij}$ is an effective stress field, and $c_{\rm s}$ is the speed of sound in the uniform medium. The terms in the effective stress tensor are the fluctuating Reynolds stress $\rho v_iv_j$, the compressional stress tensor $\sigma_{ij}$, and the stress $c_{\rm s}^2\rho\delta_{ij}$ of a uniform acoustic medium at rest. In other words, the effective stress tensor acting as a source term of sound is the difference between the stresses in the real flow and the stress of a uniform medium at rest. Equation (\ref{eq:Lighthill}) is exact.

In order to calculate the associated gravity perturbations, we introduce some approximations. First, we consider viscous stress contributions to $\sigma_{ij}$ unimportant (we neglect viscous damping in sound propagation), and therefore the temperature field can be assumed to be approximately uniform. This means that the difference $\sigma_{ij}-c_{\rm s}^2\rho\delta_{ij}$ is negligible with respect to the fluctuating Reynolds stress. Furthermore, we will assume that the root mean square of the velocities $v_i$ are much smaller than the speed of sound $c_{\rm s}$ (i.~e.~the turbulence has a small Mach number), and consequently the relative pressure fluctuations $\delta p(\vec r,t)/p_0$ produced by the Reynolds stress is much smaller than 1. In this case, we can rewrite the Lighthill equation into the approximate form
\beq
\left(\Delta-\frac{1}{c_{\rm s}^2}\partial_t^2\right)\frac{\delta p(\vec r,t)}{p_0}=-\frac{1}{c_{\rm s}^2}(\nabla\otimes\nabla):(\vec v(\vec r,t)\otimes\vec v(\vec r,t)),
\label{eq:lightLight}
\eeq
with $(\nabla\otimes\nabla):(\vec v\otimes\vec v\,)\equiv\partial_{x_i}\partial_{x_j}v_iv_j$ (summing over indices $i,\,j$). This equation serves as a starting point for the calculation of the pressure field. It describes the production of sound in turbulent flow through conversion of shear motion into longitudinal motion. The Reynolds stress represents a quadrupole source, which means that sound production is less efficient in turbulent flow than for example at vibrating boundaries where the source has dipole form. The remaining task is to characterize the velocity fluctuations in terms of spatial correlation functions, translate these into a two-point correlation function of the pressure field using Equation (\ref{eq:lightLight}), and finally obtain the spectrum of gravity fluctuations from these correlations. The last step is analogous to the calculation carried out in Section \ref{sec:quasitemp}, specifically Equation (\ref{eq:tempNNspec}), for the perturbed temperature field. The calculation of gravity perturbations will be further simplified by assuming that the velocity field is stationary, isotropic, and homogeneous. These conditions can certainly be contested, but they are necessary to obtain an explicit solution to the problem (at least, solutions for a more general velocity field are unknown to the author). 

Since the source term is quadratic in the velocity field, it is clear that the problem of this section is rather complicated. For example, the relation between temperature perturbations and gravity fluctuations was linear. For this reason, the authors of \cite{CaAl2009} decided to carry out the calculation in Fourier space (with respect to time and space). From Equation (\ref{eq:lightLight}), we can calculate the Fourier transform of the auto-correlation of the pressure field, which yields (see Section \ref{sec:noisefreq})
\beq
\begin{split}
S(\delta p;\vec k,\omega) &= \frac{1}{(2\pi)^4}\frac{p_0^2}{(\omega^2-c_{\rm s}^2 k^2)^2}\\
&\cdot\int\drm \tau\,\e^{-\irm \omega \tau}\int\drm V\,\e^{\irm \vec k\cdot\vec r}\langle (\vec k\cdot \vec v(\vec r_0,t))(\vec k\cdot \vec v(\vec r_0,t))(\vec k\cdot\vec  v(\vec r_0+\vec r,t+\tau))(\vec k\cdot \vec v(\vec r_0+\vec r,t+\tau))\rangle
\end{split}
\label{eq:turbpress}
\eeq
Note that the convention in turbulence theory used here to normalize the Fourier transform by $1/(2\pi)^4$ is different from the convention used elsewhere in this article, where the inverse Fourier transform obtains this factor. The fact that noise amplitudes at different wave vectors and frequencies do not couple is a consequence of homogeneity and stationarity of the velocity field. Once the spectral density of pressure fluctuations is known, we can use it to calculate the gravity perturbation according to
\beq
S(\delta\vec a;\vec k,\omega)=\left(\frac{4\pi G}{c_{\rm s}^2}\right)^2\frac{\vec k\otimes\vec k}{k^4}S(\delta p;\vec k,\omega),
\label{eq:accpress}
\eeq
which is given in tensor form to describe spectral densities of the three acceleration components including their cross-spectral densities. This equation is obtained by taking the negative gradient of the first line in Equation (\ref{eq:totalNNinh}), and subsequently calculating its spatial Fourier transform. 

We can now focus on the calculation of the source spectrum. According to Isserlis' theorem\index{Isserlis' theorem}, the ensemble average in Equation (\ref{eq:turbpress}) can be converted into a product of second-order moments in case that the velocity fluctuations are Gaussian. We assume this to be the case (one of the less disputable assumptions), and write:
\beq
\langle v_iv_jv'_lv'_m\rangle
=\langle v_iv_j\rangle\langle v'_lv'_m\rangle+\langle v_iv'_l\rangle\langle v_jv'_m\rangle+\langle v_iv'_m\rangle\langle v_jv'_l\rangle
\label{eq:Isserlis}
\eeq
The second-order moments are determined by turbulence theory. An isotropic turbulence has the wavenumber spectrum \cite{Dav2004}
\beq
\begin{split}
\langle (\vec k\cdot \vec v(\vec r_0,t))(\vec k\cdot \vec v(\vec r_0,t))\rangle &= \frac{2}{3}k^2\int\limits_{k_0}^{k_\nu}\drm k' \mathcal E(k')\\
\langle (\vec k\cdot \vec v(\vec r_0,t))(\vec k\cdot \vec v(\vec r_0+\vec r,t))\rangle &= k^2 \int\limits_{\mathcal I}\drm^3k'\,\e^{-\irm\vec k'\cdot\vec r}\left(1-\frac{(\vec k\cdot\vec k\,')^2}{k^2k'^2}\right)\frac{\mathcal E(k')}{4\pi k'^2}\\
&=\frac{2}{3}k^2\int\limits_{k_0}^{k_\nu}\drm k' \mathcal E(k')\left((j_0(k'r)-\frac{1}{2}j_2(k'r))+\frac{3}{2}\frac{(\vec k\cdot\vec r)^2}{k^2r^2}j_2(k'r)\right)
\end{split}
\label{eq:corrvel}
\eeq
where $\mathcal E(k)=\mathcal K_0\epsilon^{2/3}k^{-5/3}$ is the Kolmogorov energy spectrum, $\mathcal K_0$ the Kolmogorov number, and $\epsilon$ the total (specific) energy dissipated by viscous forces
\beq
\epsilon=2\nu\int\limits_0^\infty\drm k'\,k'^2\mathcal E(k')
\label{eq:dissturb}
\eeq
Here, $\nu$ is the fluid's viscosity. The Kolmogorov energy spectrum holds for the inertial regime $\mathcal I$ (viscous forces are negligible), i.~e.~for wavenumbers between $k_0=2\pi/\mathcal R$ and $k_\nu=(\epsilon/\nu^3)^{1/4}$, where $\mathcal R$ is the linear dimension of the largest eddy in the turbulent flow. In Equation (\ref{eq:corrvel}), we have only written the equal-time correlations (the first following from the second equation). The velocities in the second equation should however be evaluated at two different times $t,\,t+\tau$. In \cite{Kan1993}, we find that for $k\gg k_0$
\beq
\int\drm V\e^{\irm\vec k\cdot\vec r}\langle v_i(\vec r_0,t)v_j(\vec r_0+\vec r,t+\tau)\rangle=\exp\left(-\frac{1}{2}\frac{\tau^2}{\tau_0^2(k)}\right)\int\drm V\e^{\irm\vec k\cdot\vec r}\langle v_i(\vec r_0,t)v_j(\vec r_0+\vec r,t)\rangle
\label{eq:tcorrturb}
\eeq
with $\tau_0^2(k)=1/(k^2\langle v_i^2\rangle)$, where $v_i$ is any of the components of the velocity vector. The first term in Equation (\ref{eq:Isserlis}) is independent of time for a stationary velocity field (both expectation values are equal-time). Therefore, its energy only contributes to frequency $\omega=0$, and we can neglect it. The Fourier transform in Equation (\ref{eq:turbpress}) of the second and third terms in Equation (\ref{eq:Isserlis}) with respect to $\tau$ can be carried out easily using Equation (\ref{eq:tcorrturb}). Also integrating over the angular coordinates of the spatial Fourier transform in Equation (\ref{eq:turbpress}), the gravity spectrum can be written
\beq
\begin{split}
S(\delta\vec a;&\vec k,\omega) = \left(\frac{4\pi G}{c_{\rm s}^2}\right)^2\frac{\vec k\otimes\vec k}{k^4}\frac{1}{(2\pi)^3}\frac{p_0^2}{(\omega^2-c_{\rm s}^2 k^2)^2}\frac{\tau_0(k)}{2\sqrt{\pi}}\exp\left(-\frac{\tau_0^2(k)\omega^2}{4}\right)\\
&\cdot 2\int\drm V\,\e^{\irm \vec k\cdot\vec r}\left[\frac{2}{3}k^2\int\limits_{k_0}^{k_\nu}\drm k' \mathcal E(k')\left((j_0(k'r)-\frac{1}{2}j_2(k'r))+\frac{3}{2}\frac{(\vec k\cdot\vec r)^2}{k^2r^2}j_2(k'r)\right)\right]^2\\
&=\left(\frac{2 G p_0}{c_{\rm s}^2}\right)^2\frac{\vec k\otimes\vec k}{(\omega^2-c_{\rm s}^2 k^2)^2}\frac{\tau_0(k)}{\pi^{3/2}}\exp\left(-\frac{\tau_0^2(k)\omega^2}{4}\right)\\
&\cdot \Bigg\{\frac{4}{9}\int\limits_0^\infty\drm r\,r^2j_0(k r)\left[\,\int\limits_{k_0}^{k_\nu}\drm k' \mathcal E(k')\left(j_0(k'r)-\frac{1}{2}j_2(k'r)\right)\right]^2\\
&\quad+\frac{4}{9}\int\limits_0^\infty\drm r\,r^2\left(j_0(kr)-2j_2(kr)\right)\left[\,\int\limits_{k_0}^{k_\nu}\drm k' \mathcal E(k')\left(j_0(k'r)-\frac{1}{2}j_2(k'r)\right)\right]\left[\,\int\limits_{k_0}^{k_\nu}\drm k' \mathcal E(k')j_2(k'r)\right]\\
&\quad+ \int\limits_0^\infty\drm r\,r^2\left(\frac{1}{5}j_0(kr)-\frac{4}{7}j_2(kr)+\frac{8}{35}j_4(kr)\right)\left[\,\int\limits_{k_0}^{k_\nu}\drm k' \mathcal E(k')j_2(k'r)\right]^2\Bigg\}
\end{split}
\eeq
Probably the best way to proceed is to carry out the integral over the radius $r$. The integrands are products of three spherical Bessel functions. An analytic solution for this type of integral was presented in \cite{MLM1991} where we find that the integral is non-zero only if the three wavenumbers fulfill the triangular relation $|k'-k''|\leq k\leq k'+k''$ (i.~e.~the sum of the three corresponding wave vectors needs to vanish), and the orders of the spherical Bessel functions must fulfill $|n'-n''|\leq n\leq n'+n''$. Especially the last relation is useful since many products can be recognized by eye to have zero value. In each case, the result of the integration is a rational function of the three wavenumbers if the triangular condition is fulfilled, and zero otherwise. While it may be possible to solve the integral analytically, we will stop the calculation at this point. Numerical integration as suggested in \cite{CaAl2009} is a valuable option. The square-roots of the noise spectra normalized to units of strain, $S(\delta\vec a;\vec k,\omega)(2/(L\omega^2)^2$, are shown in Figure \ref{fig:lightspec} for $k = \rm 0.1\,m^{-1},\,0.67\,m^{-1},\,1.58\,m^{-1},\,3.0\,m^{-1}$, where $L=3000\,$m is the length of an interferometer arm.
\epubtkImage{}{
    \begin{figure}[htbp]
    \centerline{\includegraphics[width=0.55\textwidth]{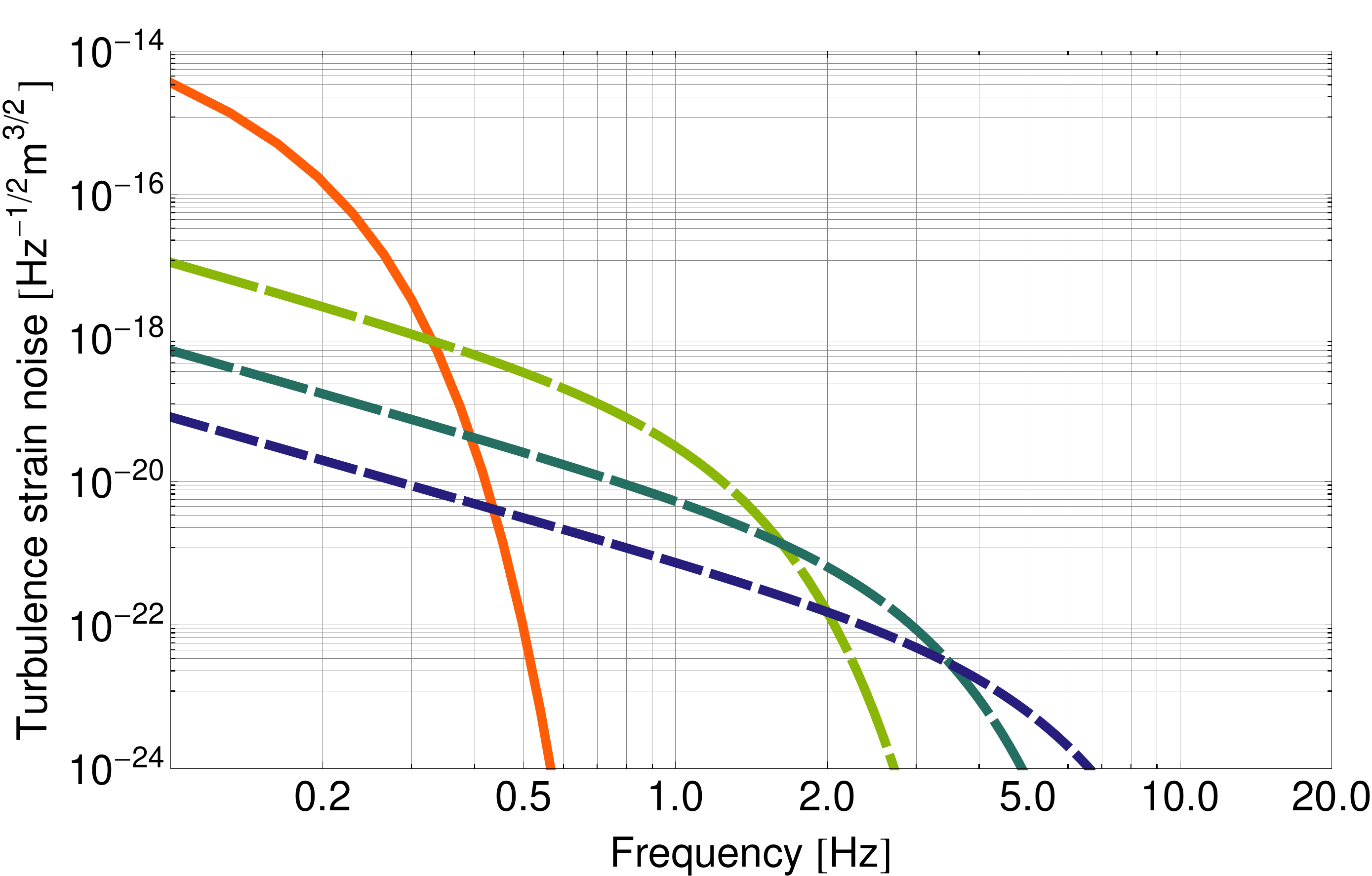}}
    \caption[Gravity perturbations from Lighthill acoustic noise]{Newtonian-noise spectra from Lighthill acoustic noise. The four curves are plotted for $k = \rm 0.1\,m^{-1},\,0.67\,m^{-1},\,1.58\,m^{-1},\,3.0\,m^{-1}$ (with decreasing dash length).}
\label{fig:lightspec}
    \end{figure}}
Each spectrum is exponentially suppressed above the corner frequency $1/\tau_0(k)$ with $\tau_0=\rm 3.5\,s,\,0.52\,s,\,0.22\,s,\,0.12\,s$. Below the corner frequency, the spectrum is proportional to $1/\omega^2$. In order to calculate the dissipation rate $\epsilon$, a measured spectrum was used \cite{AlEA1997}, which has a value of about $1\,\rm m^3 s^{-2}$ at $k=1\,\rm m^{-1}$, and wavenumber dependence approximately equal to the Kolmogorov spectrum. In this way, we avoid the implicit relation of the dissipation rate in Equation (\ref{eq:dissturb}), since $\epsilon$ also determines the Kolmogorov energy spectrum. Solving the implicit relation for $\epsilon$ gave poor numerical results, and also required us to extend the energy spectrum (valid in the inertial regime) to higher wavenumbers (the viscous regime). It is also worth noting that the energy spectrum and the scale $\mathcal R$ (we used a value of $150\,$m) are the only required model inputs related to properties of turbulence. Any other turbulence parameter in this calculation can be calculated from these two (and a few standard parameters such as air viscosity, air pressure, $\ldots$). The resulting spectra show that Newtonian noise from the Lighthill process is negligible above 5\,Hz, but it can be a potential source of noise in low-frequency detectors. In the future, it should be studied how strongly the Lighthill gravity perturbation is suppressed when the detector is built underground. 

\subsection{Atmospheric Newtonian-noise estimates}
\label{sec:estAtmNN}
In the following, we present the strain-noise forms of gravity perturbations from infrasound fields and uniformly advected temperature fluctuations. While the results of the previous sections allow us in principle to estimate noise at the surface as well as underground, we will only calculate the surface noise spectra here. Newtonian noise from advected temperature perturbations decreases strongly with depth and should not play a role in underground detectors. Suppression of infrasound gravity noise with depth depends strongly on the isotropy of the infrasound field. Using Equation (\ref{eq:gravinfra}), it is straight-forward to modify the results of this section to include noise suppression with depth once the infrasound field is characterized.

We start with the infrasound Newtonian noise. According to Equation (\ref{eq:gravinfra}), the gravity acceleration of a single test mass at $z_0=0$ due to an infrasound wave is given by 
\beq
\delta a_x(\vec \varrho_0,\omega)=-4\pi\irm\frac{G\rho_0}{\gamma p_0}\e^{-\irm\vec k_\varrho\cdot\varrho_0}\frac{\vec e_x\cdot\vec k}{k^2}\delta p(\omega)
\eeq
Averaging over all propagation directions, the strain noise measured between two test masses separated by a distance $L$ along $\vec e_x$ reads
\beq
S(h;\omega)=\frac{2}{3}\left(\frac{4\pi}{kL\omega^2}\frac{G\rho_0}{\gamma p_0}\right)^2S(\delta p;\omega)(1-j_0(kL)+2j_2(kL))
\label{eq:atmstrainNN}
\eeq
The gravity-strain amplitude response is plotted in Figure \ref{fig:infraresp} expressing the distance $L$ between the two test masses in units of sound wavelength $\lambda_{\rm IS}=2\pi/k$.
\epubtkImage{}{
    \begin{figure}[htbp]
    \centerline{\includegraphics[width=0.7\textwidth]{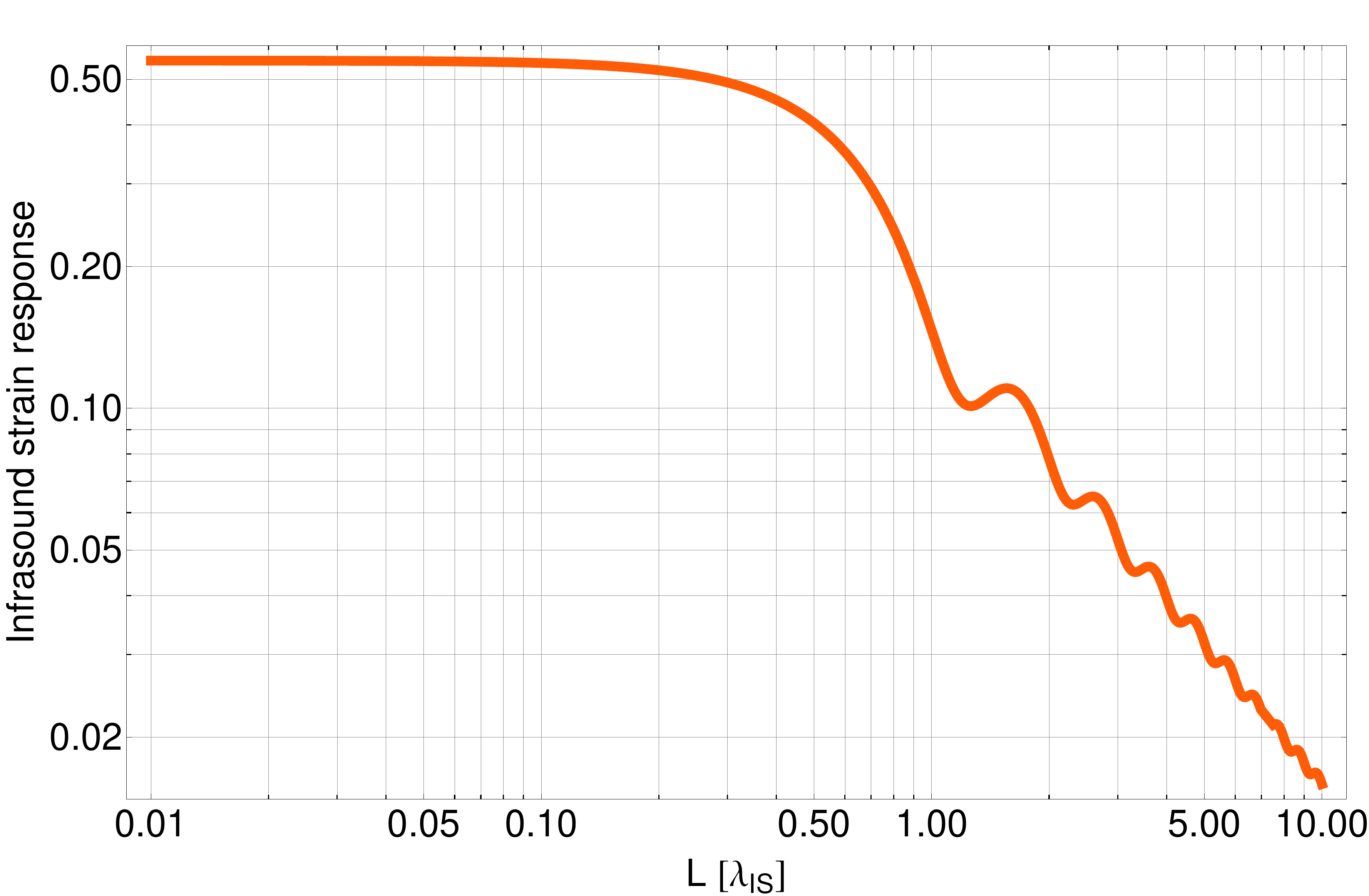}}
    \caption[Gravity-strain response to infrasound fields]{Gravity-strain amplitude response to infrasound fields.}
\label{fig:infraresp}
    \end{figure}}
For short distances between the test masses, the response is independent of $L$, and at large distances, the response falls with $1/L$. The long-distance response follows from the fact that gravity noise is uncorrelated between the two test masses, while the small-distance response corresponds to the regime where the two test masses sense gravity-gradient fluctuations. 

The strain noise spectrum from uniformly advected temperature fluctuations is calculated from Equation (\ref{eq:advectNN}) using the solution of the integrals given in Section \ref{sec:objectline}:
\beq
S(h;\omega)=(2\pi)^3\left(\frac{G\rho_0c_T}{LT_0}\right)^2\omega^{-(p+7)}\Gamma(2+p)\sin(\pi p/2)\e^{-2r_{\rm min}\omega/v}v^{p+2},
\label{eq:atmtempNN}
\eeq
where the modified Bessel functions were approximated according to Equation (\ref{eq:approxmov}). We assume that both test masses experience gravity perturbations characterized by the same spectral densities. The integral over stream lines in Equation (\ref{eq:advectNN}) was carried out over a semi-infinite disk with a disk-shaped excision of radius $r_{\rm min}$ around the test mass. The excision enforces a minimum distance between stream lines and test masses, for example because of buildings hosting the test masses. Due to the exponential suppression, this noise contribution can be expected to be insignificant deep underground. The integration includes an average over streamline directions. If the expression is to be converted into a GW strain sensitivity of a two-arm interferometer, then it is not fully accurate to simply multiply the strain noise by 2 due to gravity correlations between the two inner test masses of the two arms. Nonetheless, for the noise budget presented in this section, we will use the factor 2 conversion.

Figure \ref{fig:atmNN} shows the Newtonian-noise spectra together with a reference sensitivity of the Advanced Virgo detector.
\epubtkImage{}{
    \begin{figure}[htbp]
    \centerline{\includegraphics[width=0.7\textwidth]{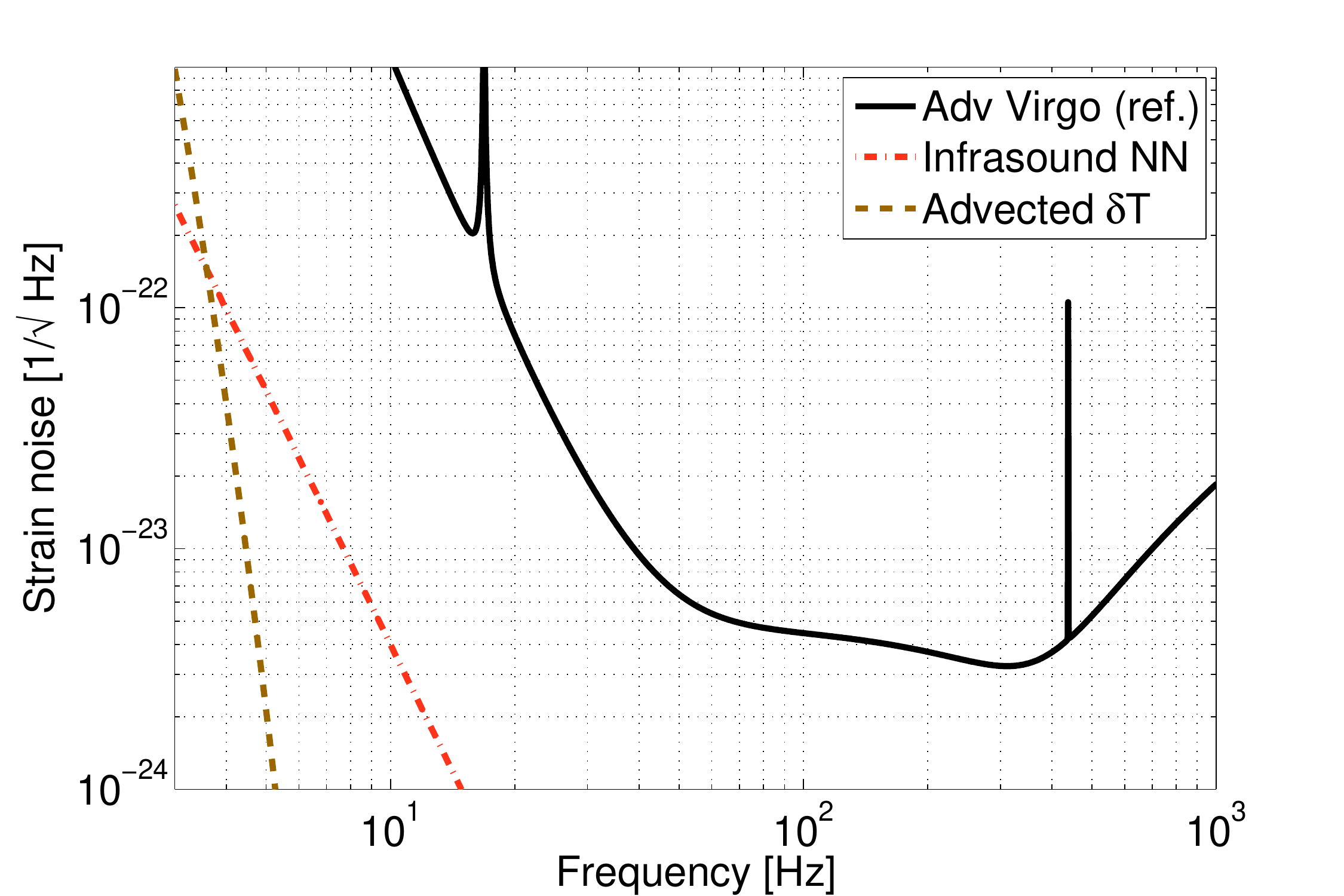}}
    \caption[Atmospheric Newtonian noise]{Atmospheric Newtonian noise from sources that were first discussed in \cite{Cre2008} in comparison with a reference sensitivity curve of Advanced Virgo. Temperature fluctuations are advected at speed 15\,m/s. The sound spectrum represents typical background noise inside laboratory buildings of large-scale GW detectors.}
\label{fig:atmNN}
    \end{figure}}
The Newtonian noise from advected temperature fields is evaluated using a wind speed of $v=15\,$m/s, and a minimum distance of $5\,$m to the test masses. With respect to the advanced GW detectors LIGO/Virgo, atmospheric Newtonian noise will be insignificant according to these results. 

The slope of infrasound Newtonian noise is steeper than of seismic Newtonian noise (see Figure \ref{fig:LIGONN}), which can be taken as an indication that there may be a frequency below which atmospheric Newtonian noise dominates over seismic Newtonian noise. This has in fact been predicted in \cite{HaEA2013}. Using measured spectra of atmospheric pressure fluctuations and seismic noise, the intersection between seismic and infrasound Newtonian noise happens at about 1\,Hz for a test mass at the surface. From Section \ref{sec:gravimeterNN}, we also know that Newtonian noise from atmospheric pressure fluctuations is the dominant ambient noise background around 1\,mHz. One might be tempted to conclude that gravity perturbations from advected temperature fields may be an even stronger contribution at low frequencies. However, one has to be careful since the noise prediction cannot be extended to much below a few Hz without modifying the model. The quasi-static approximation of the temperature field will fail at sufficiently low frequencies, and the temperature field cannot be characterized anymore as a result of turbulent mixing \cite{KuNa2006}. Also, the part of the model shown in Figure \ref{fig:atmNN} is characterized by an exponential suppression (effective above 3\,Hz). 

\subsection{Summary and open problems}
\label{sec:atmsummary} 
In this section, we reviewed models of atmospheric gravity perturbations that are either associated with infrasound waves, or with quasi-stationary temperature fields advected by wind. We have seen that atmospheric Newtonian noise will very likely be insignificant in GW detectors of the advanced generation. For surface detectors, atmospheric Newtonian noise starts to be significant below 10\,Hz according to these models. 

According to Equations (\ref{eq:atmstrainNN}) and (\ref{eq:atmtempNN}), and comparing with seismic Newtonian noise (see Figure \ref{fig:LIGONN}), we see that atmospheric spectra are steeper and therefore potentially the dominating gravity perturbation in low-frequency detectors. However, both models are based on approximations that may not hold at frequencies below a few Hz. A summary of approximations applied to the infrasound Newtonian noise model can be found in \cite{HaEA2013}, including modelling of a half-space atmosphere, neglecting wind, etc. Also the noise model of advected temperature fluctuations likely does not hold at low frequencies since it is based on the assumption that the temperature field is quasi-stationary. In addition, at low frequencies, near-surface temperature spectra can be affected by variations of ground temperature in addition to turbulent mixing. 

As we have seen, few time-varying atmospheric noise models have been developed so far, which leaves plenty of space for future work in this field. For example, convection may produce atmospheric gravity perturbations, and only very simple models of gravity perturbations from turbulence have been calculated so far. While these yet poorly modelled forms of atmospheric noise are likely insignificant in GW detectors sensitive above 10\,Hz, they may become important in low-frequency detectors. Another open problem is to study systematically the decrease in atmospheric Newtonian noise with depth in the case of underground GW detectors. Especially, it is unclear how much atmospheric noise is suppressed in sub-Hz underground detectors. We have argued that seismic Newtonian noise does not vary significantly with detector depth in low-frequency GW detectors, but some forms of atmospheric Newtonian noise depend strongly on the minimal distance between source and test mass. So the conclusion might be different for atmospheric noise. Finally, the question should be addressed whether atmospheric disturbances transmitted in the form of seismic waves into the ground can be neglected in Newtonian-noise models. As we outlined briefly in Section \ref{sec:soundNN}, even though transmission coefficients of sound waves into the ground are negligible with respect to their effect on seismic and infrasound fields, it seems that they may be relevant with respect to their effect on the gravity field.

\section{Gravity Perturbations from Objects}
\label{sec:objects}\index{Newtonian noise!objects}
In the previous sections, Newtonian-noise models were developed for density perturbations described by fields in infinite or half-infinite media. The equations of motion that govern the propagation of disturbances play an important role since they determine the spatial correlation functions of the density field. In addition, gravity perturbations can also be produced by objects of finite size, which is the focus of this section. Typically, the objects can be approximated as sufficiently small, so that excitation of internal modes do not play a role in calculations of gravity perturbations. The formalism that is presented can in principle also be used to calculate gravity perturbations from objects that experience deformations, but this scenario is not considered here. In the case of deformations, it is advisable to make use of a numerical simulation. For example, to calculate gravity perturbations from vibrations of vacuum chambers that surround the test masses in GW detectors, Pepper used a numerical simulation of chamber deformations \cite{Pep2007}. A first analytical study of gravity perturbations from objects was performed by Thorne and Winstein who investigated disturbances of anthropogenic origin \cite{ThWi1999}. The paper of Creighton has a section on gravity perturbations from moving tumbleweeds, which was considered potentially relevant to the LIGO Hanford detector \cite{Cre2008}. Interesting results were also presented by Lockerbie \cite{Loc2012}, who investigated corrections to gravity perturbations related to the fact that the test masses are cylindrical and not, as typically approximated, point masses. 

Section \ref{sec:thumb} presents rules of thumb that make it possible to estimate the relevance of perturbations from an object ``by eye'' before carrying out any calculation. Sections \ref{sec:objectline} and \ref{sec:oscobj} review well-known results on gravity perturbations from objects in uniform motion, and oscillating objects. A generic analytical method to calculate gravity perturbations from oscillating and rotating objects based on multipole expansions is presented in Sections \ref{sec:vibobj} and \ref{sec:rotobj}.

\subsection{Rules of thumb for gravity perturbations}
\label{sec:thumb}
From our modelling effort so far, we conclude that seismic fields produce the dominant contribution to Newtonian noise above a few Hz. In terms of test-mass acceleration, seismic Newtonian noise is proportional to the displacement $\xi$, and ground density $\rho$:
\beq
\delta a\sim G\rho\xi
\label{eq:scaleg}
\eeq
This relation is true for any type of seismic field, underground and above surface, with or without scattering, and to make this equation exact, a numerical factor needs to be multiplied, which, in the cases studied so far, should realistically lie within the range 1 -- 10. This was one of the results of Section \ref{sec:ambient}. Other forms of Newtonian noise would be deemed relevant if they lay within a factor 10 to seismic Newtonian noise (this number can increase in the future with improving noise-cancellation performance). The question that we want to answer now is under which circumstances an object would produce gravity perturbations comparable to perturbations from seismic fields. Intuitively, one might think that an object only needs to be close enough to the test mass, but this is insufficient unless the object almost touches the test mass, as will be shown in the following.

Let us consider the gravity perturbation from a small mass of volume $\delta V$ and density $\rho_0$ at distance $r$ to the test mass that oscillates with amplitude $\xi(t)\ll r$. We can use the dipole form in Equation (\ref{eq:dipoleacc}) to calculate the gravity perturbation at $\vec r_0=\vec 0$:
\beq
\delta\vec a(t) = G\rho_0\frac{\delta V}{r^3}\left(\vec\xi(t)-3(\vec e_r\cdot\vec\xi(t))\vec e_r\right)
\eeq
Acceleration produced by a point mass scales similarly to acceleration from seismic fields according to Equation (\ref{eq:scaleg}), but the amplitude is reduced by $\delta V/r^3$. In numbers, a solid object with 1\,m diameter at a distance of 5\,m oscillating with amplitude equal to seismic amplitudes, and equal density to the ground would produce Newtonian noise, which is about a factor 100 weaker than seismic Newtonian noise. Infrastructure at GW detectors near test masses include neighboring chambers, which can have diameters of several meters, but the effective density is low since the mass is concentrated in the chamber walls. 

If the distance $r$ is decreased to its minimum when the test mass and the perturbing mass almost touch, then the factor $\delta V/r^3$ is of order unity. It is an interesting question if there exist geometries of disturbing mass and test mass that minimize or maximize the gravitational coupling of small oscillations. An example of a minimization problem that was first studied by Lockerbie \cite{Loc2012} is presented in Section \ref{sec:momentinter}. The maximization of gravitational coupling by varying object and test-mass geometries could be interesting in some experiments. Maybe it is possible to base a general theorem on the multipole formalism for small oscillations introduced in Section \ref{sec:vibobj}.

One mechanism that could potentially boost gravity perturbations from objects are internal resonances. It is conceivable that vibration amplitudes are amplified by factors up to a few hundred on resonance, and therefore it is important to investigate carefully the infrastructure close to the test mass. There is ongoing work on this for the Virgo detector where handles attached to the ground are located within half a meter to the test masses. While the rule of thumb advocated in this section rules out any significant perturbation from the handles, handle resonances may boost the gravity perturbations to a relevant level. Finally, we want to emphasize that the rule of thumb only applies to perturbative motion of objects. An object that changes location, or rotating objects do not fall under this category.

\subsection{Objects moving with constant speed}
\label{sec:objectline}
Objects moving at constant speed produce gravity perturbations through changes in distance from a test mass. It is straight-forward to write down the gravitational attraction between test mass and object as a function of time. The interesting question is rather what the perturbation is as a function of frequency. While gravity fluctuations from random seismic or infrasound fields are characterized by their spectral densities, gravity changes from moving objects need to be expressed in terms of their Fourier amplitudes, which are calculated in this section. Since the results should also be applicable to low-frequency detectors where the test masses can be relatively close to each other, the final result will be presented as strain amplitudes.

We consider the case of an object of mass $m$ that moves at constant speed $v$ along a straight line that has distance $r_1,\,r_2$ to two test masses of an arm at closest approach. The vectors $\vec r_1,\,\vec r_2$ pointing from the test mass to the points of closest approach are perpendicular to the velocity $\vec v$. The closest approach to the first test mass occurs at time $t_1$, and at $t_2$ to the second test mass.

As a function of time, the acceleration of test mass 1 caused by the uniformly moving object reads
\beq
\delta \vec a_1(t)=-\frac{Gm}{\left(r_1^2+v^2(t-t_1)^2\right)^{3/2}}\left(\vec r_1+\vec v(t-t_1)\right)
\label{eq:movatime}
\eeq
The Fourier transform of $\delta \vec a_1(t)$ can be directly calculated with the result
\beq
\delta \vec a_1(\omega)=-\frac{2Gm\omega}{v^2}\Big(K_1(r_1\omega/v)\vec r_1/r_1+\irm K_0(r_1\omega/v)\,\vec v/v\Big)\e^{\irm\omega t_1}
\label{eq:movsingle}
\eeq
with $K_n(x)$ being the modified Bessel function of the second kind. This equation already captures the most important properties of the perturbation in frequency domain. The ratio $v/r_1$ marks a threshold frequency. Above this frequency, the argument of the modified Bessel functions is large and we can apply the approximation
\beq
K_n(x)\approx \sqrt{\frac{\pi}{2x}}\e^{-x}\left(1+\frac{4n^2-1}{8x}+\ldots\right),
\label{eq:approxmov}
\eeq
which is valid for $x\gg|n^2-1/4|$. We see that the Fourier amplitudes are exponentially suppressed above $v/r_1$. The expression in Equation (\ref{eq:movsingle}) has the same form for the second test mass. We can however eliminate $t_2$ in this equation since the distance travelled by the object between $t_1$ and $t_2$ is $L(\vec e_{12}\cdot\vec v)/v$, where $L$ is the distance between the test masses, and $\vec e_{12}$ is the unit vector pointing from test mass 1 to test mass 2, and so $t_2=t_1+L(\vec e_{12}\cdot\vec v)/v^2$. Another substitution that can be made is 
\beq
\vec r_2=\vec r_1-L\vec e_{12}+L(\vec e_{12}\cdot\vec v)\vec v/v^2
\eeq
The strain amplitude is then simply given by
\beq
h(\omega)=-\vec e_{12}\cdot(\delta \vec a_2(\omega)-\delta \vec a_1(\omega))/(\omega^2L)
\eeq
Let us consider a simplified scenario. The test masses are assumed to be underground at depth $D$, and a car is driving directly above the test masses with $\vec v$ parallel $\vec e_{12}$ and perpendicular to $\vec r_1$. Therefore, $\vec r_1=\vec r_2$, and $t_2-t_1=L/v$. The corresponding strain amplitude is
\beq
h(\omega)=\frac{2Gm}{v^2\omega L}\irm K_0(\omega D/v)\Big(e^{\irm\omega L/v}-1\Big)e^{\irm\omega t_1}
\eeq
Notice that the strain amplitude is independent of the test mass separation $L$ at frequencies $\omega\ll v/L$. The plots in Figure \ref{fig:movobj} show the strain amplitudes with varying speeds $v$ and arm lengths $L$. In the former case, the arm length is kept constant at $L=500\,$m, in the latter case, the speed is kept constant at $v=20\,$m/s. The mass of the car is 1000\,kg, and the depth of the test masses is 300\,m.
\epubtkImage{}{
    \begin{figure}[htbp]
    \centerline{\includegraphics[width=0.5\textwidth]{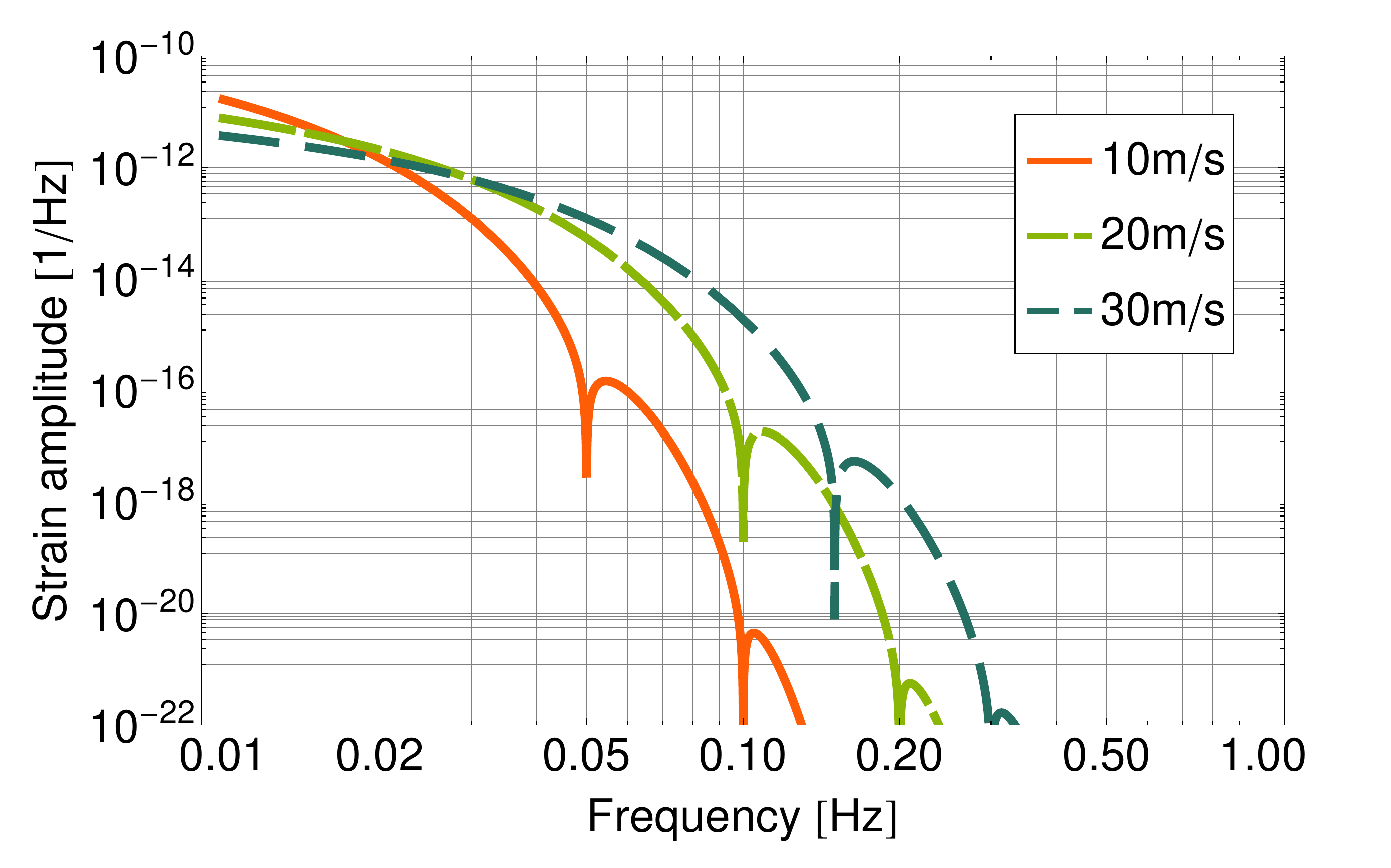}
                \includegraphics[width=0.5\textwidth]{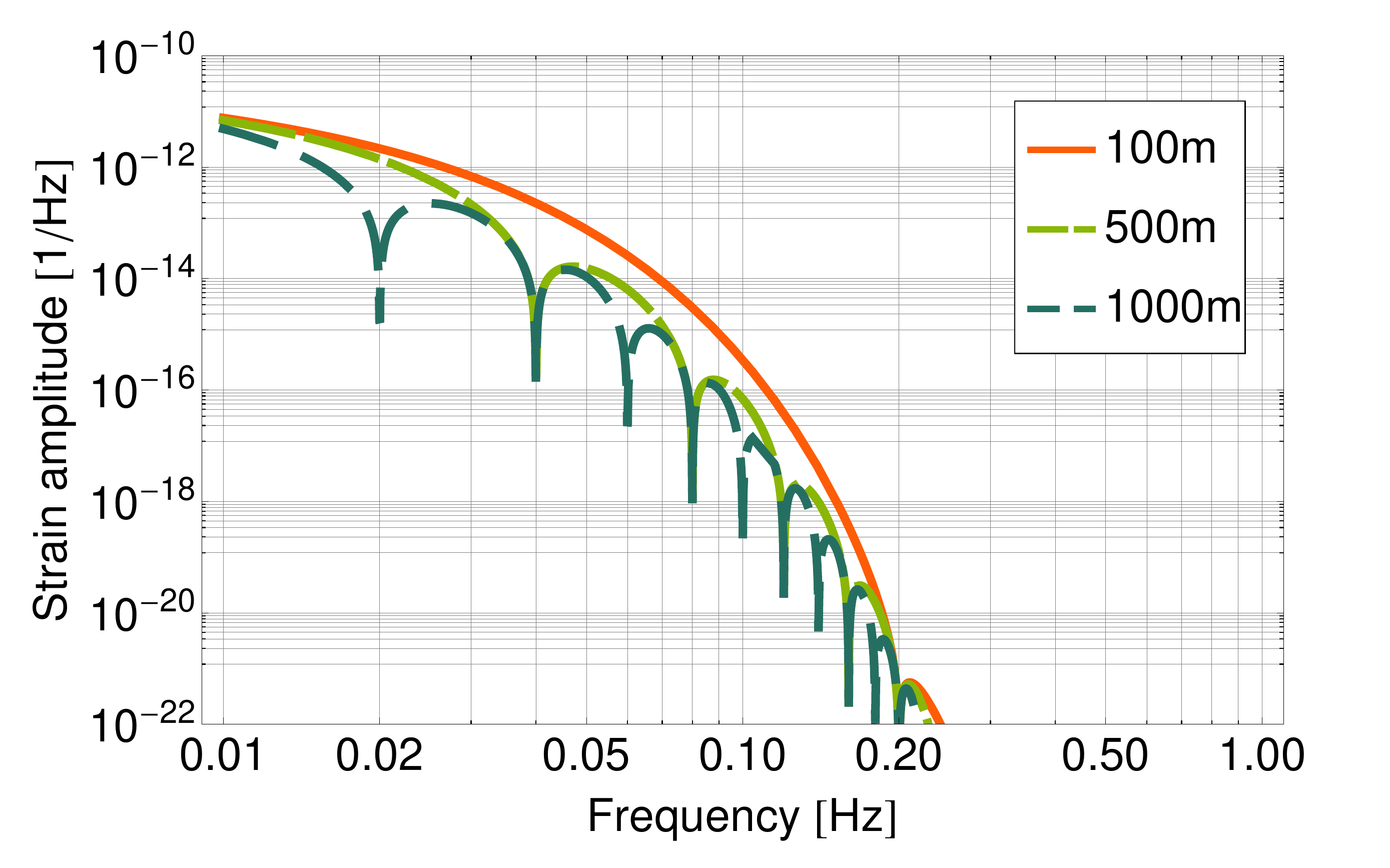}}
    \caption[Gravity perturbations from uniformly moving point mass]{Gravity perturbations from a uniformly moving point mass. In the left plot, distance between test masses is kept constant at $L=500\,$m/s, while in the right plot speed is kept constant at $20\,$m/s.}
    \label{fig:movobj}
    \end{figure}}
While this form of noise is irrelevant to large-scale GW detectors sensitive above 10\,Hz, low-frequency detectors could be strongly affected. According to the left plot, one should better enforce a speed limit on cars to below 10\,m/s if the goal is to have good sensitivity around 0.1\,Hz. Another application of these results is to calculate Newtonian noise from uniformly advected atmospheric temperature fields as discussed in Section \ref{sec:quasitemp}. For uniform airflow, the remaining integrals in Equation (\ref{eq:advectNN}) are the Fourier transform of Equation (\ref{eq:movatime}), whose solution was given in this section. 

\subsection{Oscillating point masses}
\label{sec:oscobj}
Oscillating masses can be a source of gravity perturbations, where we understand oscillation as a periodic change in the position of the center of mass. As we have seen in Section \ref{sec:thumb}, it is unlikely that these perturbations are dominant contributions to Newtonian noise, but in the case of strongly reduced seismic Newtonian noise (for example, due to coherent noise cancellation), perturbations from oscillating objects may become significant. For an accurate calculation, one also needs to model disturbances resulting from the reaction force on the body that supports the oscillation. In this section, we neglect the reaction force. Oscillation is only one of many possible modes of object motion that can potentially change the gravity field. A formalism that can treat all types of object vibrations and other forms of motion is presented in Section \ref{sec:momentinter}. 

The goal is to calculate the gravity perturbation as strain noise between two identical point masses $m$ at distance $L$ to each other separated along the direction of the unit vector $\vec e_{12}$. While the direction of oscillation is assumed to be constant, the amplitude is random and therefore characterized by a spectral density. As usual, we will denote the amplitudes of oscillation by $\xi(\omega)$ keeping in mind that these only have symbolic meaning and need to be translated into spectral densities. We only allow for small oscillations, i.~e.~with $\xi$ being much smaller than the distance of the object to the two test masses. The acceleration of the first test mass has the well-known dipole form
\beq
\delta \vec a_1(\omega)=\frac{Gm}{r_1^3}\,\left(\vec\xi(\omega)-3(\vec\xi(\omega)\cdot\vec e_{r_1})\vec e_{r_1}\right)
\eeq
where $\vec e_{r_1}$ is the unit vector pointing from the first test mass to the object, and $r_1$ is the distance between them. The acceleration of the second test mass has the same form, and we can substitute $\vec e_{r_2}=(\vec e_{r_1}-\lambda\vec e_{12})/\delta$ and $r_2=r_1\delta$ with $\delta\equiv(1+\lambda^2-2\lambda(\vec e_{r_1}\cdot\vec e_{12}))^{1/2}$ and $\lambda\equiv L/r_1$. 

Let us consider the case of an object oscillating along the direction $\vec e_{12}$, and $\vec e_{12}$ being perpendicular to $\vec e_{r_1}$. Then we can write for the strain noise
\beq
h_{\|}(\omega)=\vec e_{12}\cdot(\delta \vec a_2(\omega)-\delta \vec a_1(\omega))/(L\omega^2)=\frac{Gm\xi(\omega)}{r_1^4\omega^2}\frac{1}{\lambda}\left(\frac{1-2\lambda^2}{(1+\lambda^2)^{5/2}}-1\right)
\label{eq:oscpara}
\eeq
Changing the direction of oscillation from $\vec e_{12}$ to $\vec e_{r_1}$, the strain noise reads
\beq
h_{\perp}(\omega)=\frac{Gm\xi(\omega)}{r_1^4\omega^2}\frac{1}{\lambda}\frac{3\lambda}{(1+\lambda^2)^{5/2}}
\label{eq:oscperp}
\eeq
While $h_{\|}(\omega)$ becomes arbitrarily small with decreasing $\lambda$, $h_{\perp}(\omega)$ approaches a constant value. Towards high frequencies, $h_{\perp}$ falls rapidly since there is no force along $\vec e_{12}$ on the first test mass, and the distance of the object to the second test mass increases with growing $\lambda$, and also the projection of the gravity perturbation at the second test mass onto $\vec e_{12}$ becomes smaller.
\epubtkImage{}{
    \begin{figure}[htbp]
    \centerline{\includegraphics[width=0.5\textwidth]{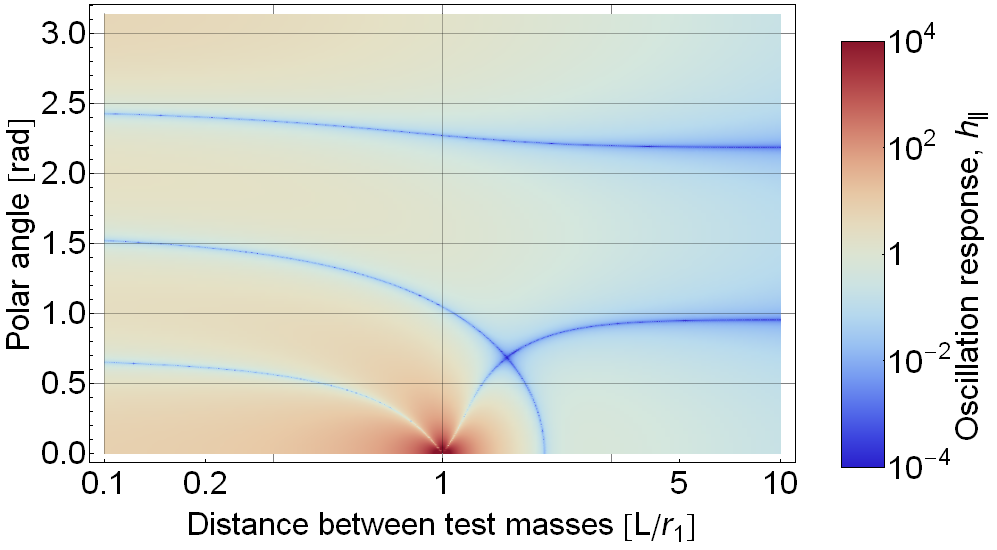}
                \includegraphics[width=0.5\textwidth]{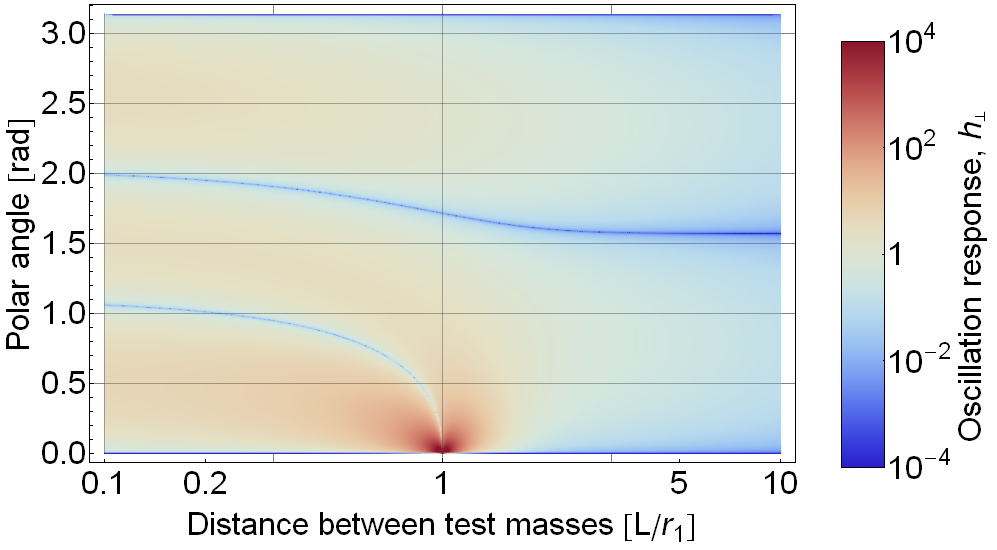}}
    \caption[Strain response to oscillating objects]{Strain response to gravity perturbations from oscillating objects.}
    \label{fig:oscresp}
    \end{figure}}
Figure \ref{fig:oscresp} shows the gravity strain response to an oscillating object for oscillations parallel to $\vec e_{12}$ (left) and perpendicular to $\vec e_{12}$ (right). The position of the object is parameterized by the angle $\phi={\rm arccos}(\vec e_{12}\cdot\vec e_{r_1})$ (which we call polar angle). Equations (\ref{eq:oscpara}) and (\ref{eq:oscperp}) correspond to the response on the lines $\phi=\pi/2$. The response grows to infinity for $\lambda=1$ and polar angle $\phi=0$ since the object is collocated with the second test mass. Note that $\lambda>1$ and $\phi=0$ means that the object lies between the two test masses.

\subsection{Interaction between mass distributions}
\label{sec:momentinter}
In the following, we discuss gravitational interaction between two compact mass distributions. We consider the case where the distance $R_{AB}$ between the two centers of mass is greater than the object diameters at largest extent. The so-called bipolar expansion allows us to express the gravitational force in terms of mass multipole moments\index{bipolar expansion}. The idea is to split the problem into three separate terms. One term depends on the vector $\vec R_{\rm AB}$ that points from the center of mass $A$ to the center of mass $B$. Each individual mass is expanded into its multipoles according to Equation (\ref{eq:multiext}) calculated in identically oriented coordinate systems, but with their origins corresponding to the two centers of mass. The situation is depicted in Figure \ref{fig:twocenter}. 
\epubtkImage{}{
    \begin{figure}[htbp]
    \centerline{\includegraphics[width=0.6\textwidth]{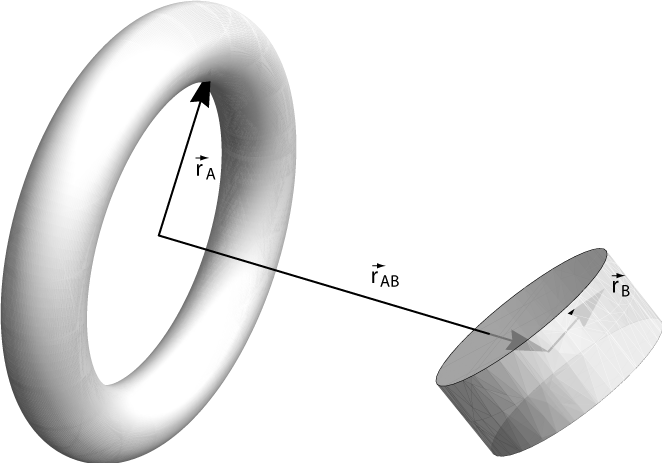}}
    \caption[Bipolar multipole expansion]{Bipolar multipole expansion.}
    \label{fig:twocenter}
    \end{figure}}
Technically, the origins do not have to be the centers of mass, but in many cases it is certainly the preferred choice. The calculation of the bipolar expansion of the interaction energy between two charge distributions is outlined \cite{SoEA2007}. The result can either be expressed in terms of the Wigner 3-j symbols or Clebsch-Gordan coefficients. We will use Clebsch-Gordan coefficients (see Appendix \ref{sec:clebsch}):    
\beq
\begin{split}
U_{AB}(\vec R_{AB}\,)=-G\sum\limits_{l_1=0}^\infty&\sum\limits_{l_2=0}^\infty(-1)^{l_2}{2L\choose 2l_1}^{1/2}\\
&\times\sum\limits_{m_1=-l_1}^{l_1}\sum\limits_{m_2=-l_2}^{l_2}\left(I_L^M(\vec R_{AB}\,)\right)^* X_{l_1}^{m_1,\,A}X_{l_2}^{m_2,\,B}\langle l_1,m_1;l_2,m_2|L,M\rangle,
\end{split}
\label{eq:multipotential}
\eeq
where $I_L^M(\cdot)$ are the interior solid spherical harmonics defined in Equation (\ref{eq:solidharm}), and $L\equiv l_1+l_2,\,M\equiv m_1+m_2$. It is not very difficult to generalize this equation for arbitrary mass distributions (one object inside another hollow object, etc), but we will leave this for the reader. The method is essentially an exchange of irregular and regular solid spherical harmonics in Equation (\ref{eq:multipotential}) together with Equations (\ref{eq:multiext}) and (\ref{eq:multiint}). Also, in general it may be necessary to divide the multipole integral in Equation (\ref{eq:multiext}) into several integrals over regular and irregular harmonics. A practical method to calculate the Clebsch-Gordan coefficients $\langle l_1,m_1;l_2,m_2|L,M\rangle$ is outlined in Section \ref{sec:clebsch}. 

As a first example, we apply the formalism to calculate the gravity force between a point mass and a cylindrical mass. This scenario has been first considered by Lockerbie \cite{Loc2012} to investigate whether the typical approximation of the test mass as a point mass is valid. The only non-vanishing multipole moment of a point mass $M$ in a coordinate system centered on its position is $X_0^0=M$. Therefore the interaction energy can be written
\beq
U_{AB}(\vec R_{AB}\,)=-GM\sum\limits_{l=0}^\infty\sum\limits_{m=-l}^{l}\left(I_l^m(\vec R_{AB}\,)\right)^* X_l^{m\,B}
\label{eq:pointinter}
\eeq
Let us consider the specific example of a point mass interacting with the quadrupole moment of a cylindrical mass of uniform density (the dipole moment of the cylinder is zero). The cylinder of mass $M$ has a radius $R$ and a height $H$. Aligning the $z$-axis of the coordinate system with the symmetry axis of the cylinder, the only non-vanishing moments of the cylinder have $m=0$ due to axial symmetry. Therefore, the relevant solid spherical harmonic expressed in cylindrical coordinates is given by
\beq
R_2^0=\frac{1}{2}(2z^2-\rho^2)
\eeq
According to Equation (\ref{eq:multiext}), the corresponding quadrupole moment with respect to the center of mass is
\beq
X_2^0=\frac{M_{\rm c}}{12}\left(H^2-3R^2\right)
\eeq
Since the $z$-axis is defined parallel to the symmetry axis of the cylinder, the spherical angular coordinate $\theta$ in $I_2^0(\vec R_{AB}\,)$ represents the angle between the symmetry axis and the separation vector $\vec R_{AB}$. The interaction energy of the quadrupole term can be written
\beq
U_{AB}(\vec R_{AB}\,)=-\frac{GMM_{\rm c}}{24R_{AB}^3}\left(H^2-3R^2\right)(3\cos^2(\theta)-1)
\label{eq:exquadpoint}
\eeq
Next we outline briefly how to calculate a gravitational force between two bodies based on the bipolar expansion involving one point mass $m$. The gravitational force is the negative gradient of the interaction energy, which can be calculated using Equation (\ref{eq:gradientscalar})
\beq
\begin{split}
\vec F(\vec R_{AB}\,)&=GM\sum\limits_{l=0}^\infty\sum\limits_{m=-l}^{l}\left(\nabla I_l^m(\vec R_{AB}\,)\right)^* X_{l}^{m,\,A}\\
&=GM\sum\limits_{l=0}^\infty\sqrt{\dfrac{4\pi}{2l+1}}\dfrac{1}{R_{AB}^{l+2}}\sum\limits_{m=-l}^{l}\left(-(l+1)\vec Y_l^m(\vec R_{AB}\,)+\vec \Psi_l^m(\vec R_{AB}\,)\right)^* X_{l}^{m,\,A},
\end{split}
\eeq
which involves the vector spherical harmonics defined in Equation (\ref{eq:vectharm}). Interestingly, the quadrupole moment of the cylinder, and the associated interaction energy and force, are zero when $H=\sqrt{3}R$. Since the quadrupole moment can be considered describing the lowest-order correction of the monopole gravitational force, a cylinder with vanishing quadrupole moment behaves very much like a point mass in interactions with nearby point masses. An interesting application of this result is presented in \cite{Loc2002}.

The interaction of the point mass with the monopole and quadrupole moments of the cylinder can also lead to a cancellation of certain components of the force. For example, calculating the sum of the monopole and quadrupole terms of the last equation in radial direction, we have
\beq
\begin{split}
0&=-\frac{GMM_{\rm c}}{R_{AB}^2}-\dfrac{3GMM_{\rm c}}{R_{AB}^4}\dfrac{1}{2}(3\cos^2(\theta)-1) \frac{1}{12}\left(H^2-3R^2\right)\\
R_{AB}^2&=\dfrac{\left(H^2-3R^2\right)}{8}(1-3\cos^2(\theta)) 
\end{split}
\eeq
Obviously, cancellation is impossible if $H=\sqrt{3}R$. Conversely, the quadrupole moment can also lead to an enhancement of components of the gravitational force relative to the monopole term. 

\subsection{Oscillating objects}
\label{sec:vibobj}\index{object!oscillation}
In the previous section, we introduced the formalism of bipolar expansion to calculate gravitational interactions between two bodies. However, what we typically want is something more specific such as the change in gravity produced by translations and rotations of bodies. Translations in the form of small oscillations will be studied in this section, rotations in the following section. We emphasize that the same formalism can also be used to describe changes in the gravity field due to arbitrary vibrations of bodies by treating these as changes in the coefficients of a multipole expansion.

In this section, we assume that the orientation of the two bodies does not change, while the separation between them changes. This can either be incorporated into the formalism as a change of $\vec R_{\rm AB}$, which, according to Equation (\ref{eq:multipotential}), requires a transformation rule for the irregular solid spherical harmonics under translation. Alternatively, we could also treat $\vec R_{\rm AB}$ as constant, but translate one of the bodies inside its own coordinate system leading to changes in its multipole moments. In either case, the effect of translation can be accounted for using the transformation rules on the regular or irregular solid spherical harmonics in the form of addition theorems. In the case of regular solid spherical harmonics, the result can be written in terms of Clebsch-Gordan coefficients
\beq
\begin{split}
R_l^m(\vec r_1+\vec r_2\,) &= \sum\limits_{l'=0}^l\sum\limits_{m'=-l'}^{l'}\mathcal{R}_{lm}^{l'm'}R_{l'}^{m'}(\vec r_1\,)R_{l-l'}^{m-m'}(\vec r_2\,)\\
\mathcal{R}_{lm}^{l'm'} &\equiv {2l\choose 2l'}^{1/2}\langle l',m';l-l',m-m'|lm\rangle
\end{split}
\label{eq:transregharm}
\eeq
The addition theorem for the irregular solid spherical harmonics reads
\beq
\begin{split}
I_l^m(\vec r_1+\vec r_2\,) &= \sum\limits_{l'=0}^\infty\sum\limits_{m'=-l'}^{l'}\mathcal{I}_{lm}^{l'm'}R_{l'}^{m'}(\vec r_<\,)I_{l+l'}^{m-m'}(\vec r_>\,)\\
\mathcal{I}_{lm}^{l'm'} &\equiv {2l+2l'+1\choose 2l'}^{1/2}\langle l',m';l+l',m-m'|lm\rangle
\end{split}
\label{eq:transirregharm}
\eeq
where $\vec r_>$ is the longer of the two vectors $\vec r_1,\,\vec r_2$, and $\vec r_<$ is the shorter one. These addition theorems can be found in different forms \cite{StRu1973,Cao1978}. We chose the Clebsch-Gordan variant since it bears some similarity to the bipolar expansion. 

In the following, we describe oscillations of a body as a small change in $\vec R_{\rm AB}$, which means that we need to apply the addition theorem of irregular harmonics. In Equation (\ref{eq:transirregharm}), we set $\vec r_1=\vec R_{\rm AB}$ and $\vec r_2=\vec \xi$. Since the oscillation amplitude $\xi$ is assumed to be small, we only keep terms up to linear order in $\xi$: 
\beq
\begin{split}
I_L^M(\vec R_{AB}+\vec \xi\,) &= \sum\limits_{l'=0}^\infty\sum\limits_{m'=-l'}^{l'}\mathcal{I}_{LM}^{l'm'}R_{l'}^{m'}(\vec \xi\,)I_{L+l'}^{M-m'}(\vec R_{AB}\,)\\
&\approx \sum\limits_{l'=0}^1\sum\limits_{m'=-l'}^{l'}\mathcal{I}_{LM}^{l'm'}R_{l'}^{m'}(\vec \xi\,)I_{L+l'}^{M-m'}(\vec R_{AB}\,)\\
 &= I_{L}^{M}(\vec R_{AB}\,)+{\mathcal I}_{L,M}^{1,-1}I_{L+1}^{M+1}(\vec R_{AB}\,)R_{1}^{-1}(\vec \xi\,)\\
&\qquad+{\mathcal I}_{L,M}^{1,0}I_{L+1}^{M}(\vec R_{AB}\,)R_{1}^{0}(\vec \xi\,)+{\mathcal I}_{L,M}^{1,1}I_{L+1}^{M-1}(\vec R_{AB}\,)R_{1}^{1}(\vec \xi\,)
\end{split}
\eeq
Let us illustrate this result with an example. We apply the linearized addition theorem to the case of an interaction between a quadrupole moment of a cylinder and an oscillating point mass. The cylinder is meant to represent a test mass of a GW detector, the point mass can represent part of a larger vibrating object in the vicinity of the test mass. Using the notation of the example in the previous section, the perturbed interaction energy can be written as
\beq
\begin{split}
U_{AB}(\vec R_{AB}+\vec \xi\,) &=-GM\left(I_2^0(\vec R_{AB}+\vec \xi\,)\right)^* X_2^{0\,B}\\
&= -\frac{GMM_c}{12}(H^2-3R^2)\\
&\qquad\cdot\left[I_{2}^{0}(\vec R_{AB}\,)+2\sqrt{6}\Re[I_{3}^{1}(\vec R_{AB}\,)R_{1}^{-1}(\vec \xi\,)]-3I_{3}^{0}(\vec R_{AB}\,)R_{1}^{0}(\vec \xi\,)\right] \\
&=-\frac{GMM_c}{24R_{AB}^3}(H^2-3R^2)\\
&\qquad\cdot\left[3(\vec e_{\rm AB}\cdot\vec e_z)^2-1-\frac{3\xi}{R_{\rm AB}}((5(\vec e_{\rm AB}\cdot\vec e_z)^2-1)(\vec e_{\rm AB}\cdot\vec e_\xi)-2(\vec e_{\rm AB}\cdot\vec e_z)(\vec e_\xi\cdot\vec e_z))\right]
\end{split}
\eeq
where $\vec e_z$ is the symmetry axis of the cylinder, $\vec e_{\rm AB}\equiv \vec R_{AB}/R_{AB}$, and $\vec e_\xi\equiv\vec \xi/\xi$. In the case of a point mass being displaced along the radial direction, parallel to $\vec R_{AB}$, the perturbed interaction potential simplifies to
\beq
\begin{split}
U_{AB}(\vec R_{AB}+\vec \xi\,)&=-\frac{GMM_c}{24R_{AB}^3}(H^2-3R^2)(3(\vec e_{\rm AB}\cdot\vec e_z)^2-1)\left(1-\frac{3\xi}{R_{AB}}\right)
\end{split}
\eeq
This result can also be derived directly from Equation (\ref{eq:exquadpoint}). For displacements perpendicular to the radial direction $\vec R_{AB}$, the interaction potential simplifies to
\beq
\begin{split}
U_{AB}(\vec R_{AB}+\vec \xi\,) &= -\frac{GMM_c}{24R_{AB}^3}(H^2-3R^2)\bigg[3(\vec e_{\rm AB}\cdot\vec e_z)^2-1+\frac{6\xi}{R_{AB}}(\vec e_{\rm AB}\cdot\vec e_z)(\vec e_\xi\cdot\vec e_z)\bigg]
\end{split}
\eeq
In this scenario, the displacement $\vec \xi$ of the point mass should be considered a function of time. The result describes the lowest order correction of the monopole-monopole time-varying interaction between a point mass and a cylinder. We see that the quadrupole contribution is suppressed by a factor $\xi/R_{\rm AB}$ (the vibration amplitude is at most a few millimeters). Therefore it is clear that corrections from higher-order moments only matter if gravitational interaction is measured very precisely, or the vibrating point mass is very close to the cylinder. 

\subsection{Rotating objects}
\label{sec:rotobj}\index{object!rotation}
Gravity perturbations can be generated by rotating objects such as exhaust fans or motors. We will again use the formalism of the bipolar expansion to calculate the gravitational interaction. In analogy to the previous section, a transformation rule is required for solid spherical harmonics under rotations. For this, we need to work in two coordinate systems. One coordinate system is body fixed. When the body rotates, this coordinate system rotates with it. For the bipolar expansion, we also need to define a coordinate system of the ``laboratory frame'', and the purpose of the rotation transformation is to describe the relative orientation of a body-fixed coordinate system to the laboratory frame. Rotations are easier to describe since we have chosen to work with spherical multipole expansions in this article. According to Equation (\ref{eq:solidharm}), if we understand the transformation of scalar surface spherical harmonics under rotations, then we automatically have the transformation of solid spherical harmonics. The transformation of surface spherical harmonics under rotations can be written in terms of the Wigner D-matrices \cite{RoKr2007,StRu1973}:
\beq
Y_l^m(\theta,\phi)=\sum\limits_{m'=-l}^lY_l^{m'}(\theta',\phi')D_{m',m}^{(l)}(\alpha,\beta,\gamma)
\eeq
and since the transformation is unitary:
\beq
Y_l^m(\theta',\phi')=\sum\limits_{m'=-l}^lY_l^{m'}(\theta,\phi)D_{m,m'}^{(l)^*}(\alpha,\beta,\gamma)
\eeq
Primed coordinates stand for the body-fixed system, while coordinates without prime belong to the laboratory frame. Rotations preserve the degree $l$ of spherical harmonics.

The rotation is defined in terms of the Euler angles $\alpha,\,\beta,\,\gamma$ around three axes derived from the body-fixed system. The first rotation is by $\alpha$ around the $z$-axis of the body-fixed system, then by $\beta$ around the $y'$-axis of the once rotated coordinate system (following the convention in \cite{StRu1973}), and finally by $\gamma$ around the $z''$-axis of the twice rotated coordinate system. Rotations around the $z$-axes lead to simple complex phases being multiplied to the spherical harmonics. The rotation around the $y'$-axis is more complicated, and the general, explicit expressions for the components $D_{m,m'}^{(l)^*}(\alpha,\beta,\gamma)$ of the rotation matrix are given by\cite{StRu1973}:
\beq
\begin{split}
D_{m',m}^{(l)}(\alpha,\beta,\gamma) &= \e^{-\irm m'\alpha}d_{m',m}^{(l)}(\beta)\e^{-\irm m\gamma}\\
d_{m',m}^{(l)}(\beta) &= \sqrt{\frac{(l+m')!(l-m')!}{(l+m)!(l-m)!}}(-1)^{m'-m}\\
& \qquad\cdot\sum\limits_k(-1)^k{l+m\choose k}{l-m\choose l-m'-k}\\
& \qquad\qquad\cdot(\cos(\beta/2))^{2l-m'+m-2k}(\sin(\beta/2))^{m'-m+2k}
\end{split}
\eeq
where the sum is carried out over all values of $k$ that give non-negative factorials in the two binomial coefficients: $\max(0,m-m')\leq k\leq\min(l-m',l+m)$. In the remainder of this section, we apply the rotation transformation to the simple case of a rotating ring of $N$ point masses. Its multipole moments have been calculated in Section \ref{sec:multipole}. The goal is to calculate the gravity perturbation produced by the rotating ring, assumed to have its symmetry axis pointing towards the test mass that is now modelled as a point mass. In this case, we can take Equation (\ref{eq:pointinter}) as starting point. The rotation transforms the exterior multipole moments $X_l^{m,\rm B}$. We have seen that multipole moments of the ring vanish unless $m=0,N,2N,\ldots$ and $l+m$ must be even. Only the first (or last) Euler rotation by an angle $\alpha=\omega t$ is required, which yields
\beq
U_{\rm AB}(\vec R_{\rm AB},t)=-GM\sum\limits_{l=0}^\infty\sum\limits_{m=-l}^{l}\left(I_l^m(\vec R_{\rm AB}\,)\right)^* X_l^{m,\rm B}\e^{-\irm m\omega t}
\eeq
This result is obtained immediately using $d_{m',m}^{(l)}(0)=\delta_{m',m}$. This result could have been guessed directly by noticing that the azimuthal angle $\phi_k$ that determines the position of a point mass on the ring appears in the spherical harmonics as phase factor $\exp(\irm m\phi_k)$. When the ring rotates, all azimuthal angles change according to $\phi_k(t)=\phi_k(0)-\omega t$. 

Since $X_l^m$ vanishes unless $m=0,N,2N,\ldots$, only specific multiples of the rotation frequency $\omega$ can be found in the time-varying gravity field. The number $N$ of point masses on the ring quantifies the level of symmetry of the ring, and acts as an up-conversion factor of the rotation frequency. Therefore, if gravity perturbations are to be estimated from rotating bodies such as a rotor, then the level of symmetry is important. However, the higher the up-conversion, the stronger is the decrease of the perturbation with distance from the ring. It would of course be interesting to study the effect of asymmetries of the ring on gravity perturbations. For example, the point masses can be slightly different, and their distance may not be equal among them. It is not a major effort to generalize the symmetric ring study to be able to calculate the effect of these deviations. 

\subsection{Summary and open problems}
\label{sec:objectsummary} 
In this section, we reviewed the theoretical framework to calculate gravity perturbations produced by finite-size objects. Models have been constructed for uniformly moving objects, oscillating objects, as well as rotating objects. In all examples, the object was assumed to be rigid, but expanding a mass distribution into multipole moments can also facilitate simple estimates of gravity perturbations from excited internal vibration modes. An``external'' vibration in the sense of an isolated oscillation does not exist strictly speaking since there must always be a physical link to another object to compensate the momentum change, but it is often possible to identify a part of a larger object as main source of gravity perturbations and to apply the formalism for oscillating masses. 

Many forms of object Newtonian noise have been estimated \cite{Pep2007,DHA2012}. So far, none of the potential sources turned out to be relevant. In Section \ref{sec:thumb}, we learned why it is unlikely that object Newtonian noise dominates over seismic Newtonian noise. Still, one should not take these rules of thumb as a guarantee. Strong vibration, i.~e.~with amplitudes much larger than ground motion, can in principle lead to significant noise contributions, especially if the vibration is enhanced by internal resonances of the objects. Any form of macroscopic motion including rotations (in contrast to small-amplitude vibrations) should of course be avoided in the vicinity of the test masses. 

A Newtonian-noise budget based on an extensive study of potential sources at the LIGO sites was published in \cite{DHA2012}. The result is shown in Figure \ref{fig:LIGONN}.  
\epubtkImage{}{
    \begin{figure}[htbp]
    \centerline{\includegraphics[width=0.75\textwidth]{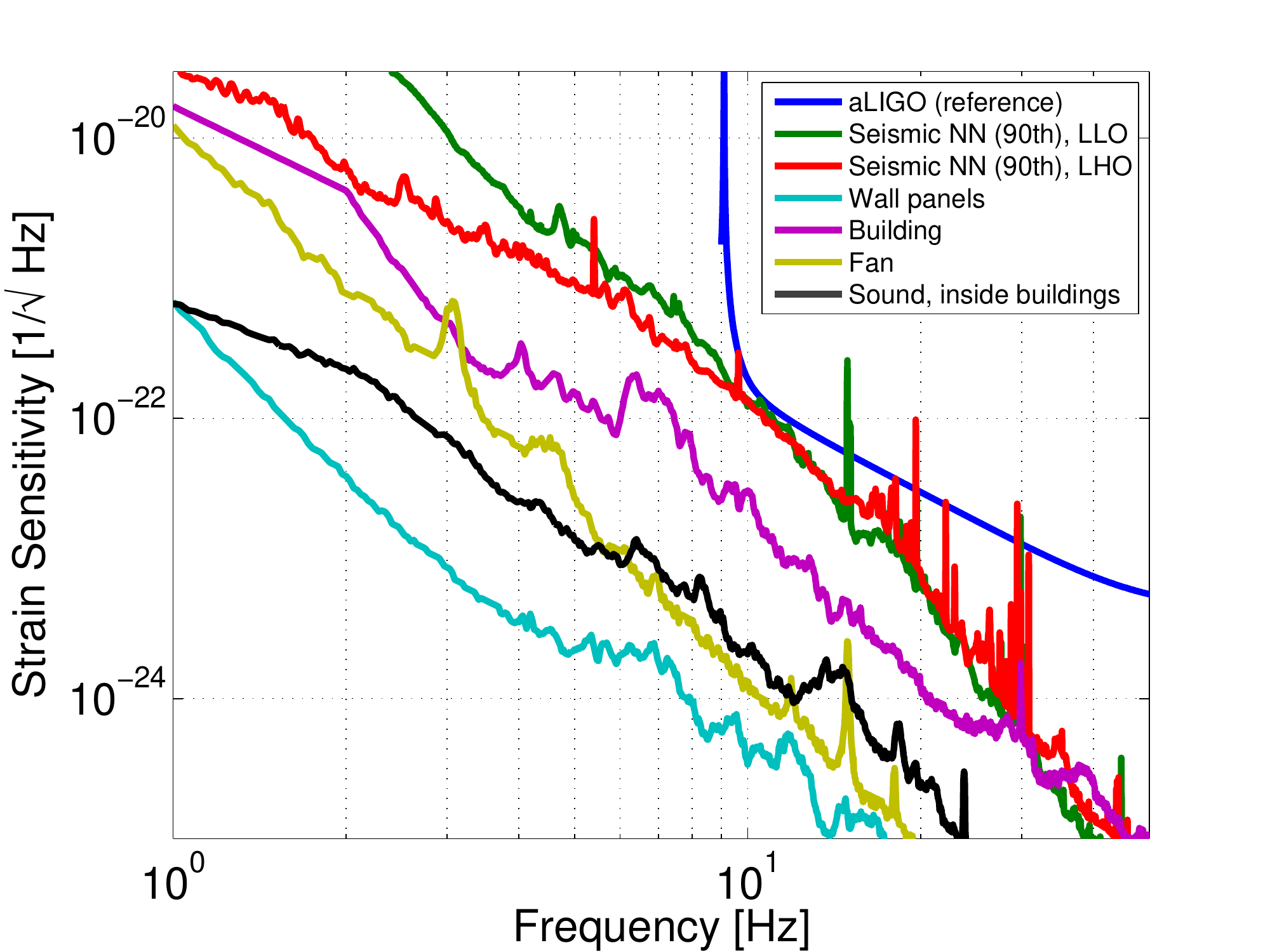}}
    \caption[Newtonian-noise budget for the LIGO sites]{Newtonian-noise budget for the LIGO sites as published in \cite{DHA2012}. Gravity perturbations from the wall panels, building, and fan were estimated based on equations from this section.}
    \label{fig:LIGONN}
    \end{figure}}
The curves are based on seismic, sound, and vibration measurements. The seismic Newtonian noise curves are modelled using Equation (\ref{eq:RayleighS}), the sound Newtonian noise using Equation (\ref{eq:atmstrainNN}), and estimates of gravity perturbations from wall panels, the buildings, and fans are modelled using equations from this section. Gravity perturbations from the buildings assume a rocking motion of walls and roof. The exhaust fan strongly vibrates due to asymmetries of the rotating parts, which was taken as source of gravity perturbations. Finally, panels attached to the structure of the buildings show relatively high amplitudes of a membrane like vibration. Nonetheless, these sources, even though very massive, do not contribute significantly to the noise budget. 

Greater care is required when designing future GW detectors with target frequencies well below 10\,Hz. These will rely on some form of Newtonian-noise mitigation (passive or active), which increases the relative contribution of other forms of gravity perturbations. Also, in some cases, as for the uniform motion discussed in Section \ref{sec:objectline}, there is a link between the shape of the gravity perturbation spectrum and the distance between object and test mass. These classes of gravity perturbations (and we have identified only one of them so far), can be much stronger at lower frequencies. 

Future work on object Newtonian noise certainly includes a careful study of this problem for low-frequency GW detectors. In general, it would be beneficial to set up a catalogue of potential sources and corresponding gravity models to facilitate the process of estimating object Newtonian noise in new detector designs. Another interesting application of the presented formalism could be in the context of experiments carried out with the intention to be sensitive to gravity perturbations produced by an object (such as the quantum-gravity experiment proposed by Feynman \cite{Zeh2011}). The formalism presented in this section may help to optimize the geometrical design of such an experiment.

\section{Newtonian-Noise Mitigation}
\label{sec:mitigate} \index{mitigation!Newtonian noise}
In early sensitivity plots of GW detectors, Newtonian noise was sometimes included as infrastructure noise. It means that it was considered a form of noise that cannot be mitigated in a straight-forward manner, except maybe by changing the detector site or applying other major changes to the infrastructure. Today however, some form of Newtonian-noise mitigation is part of every design study and planning for future generations of GW detectors, and it is clear that mitigation techniques will have a major impact on the future direction of ground-based GW detection. The first to mention strategies of seismic Newtonian-noise mitigation ``by modest amounts'' were Hughes \& Thorne \cite{HuTh1998}. Their first idea was to use arrays of dilatometers in boreholes, and seismometers at the surface to monitor the seismic field and use the sensor data for a coherent subtraction of Newtonian noise. The seismic channels serve as input to a linear filter, whose output is then subtracted from the target channel (i.~e.~the data of a GW detector). The output of an optimal filter can be interpreted as the best possible linear estimation of gravity perturbations based on seismic data. This method will be discussed in Section \ref{sec:cohcancel}. The second idea was to construct narrow moats around the test masses that reflect incoming Rf waves and therefore reduce seismic disturbances and associated gravity perturbations. As they already recognized in their paper, and as will be discussed in detail in Section \ref{sec:shieldseismic}, moats must be very deep (about 10\,m for the LIGO and Virgo sites). They are also less effective to reduce Newtonian noise from body waves.

The idea of coherent cancellation of seismic Newtonian noise has gained popularity in the GW community, probably because it is based on techniques that have already been implemented successfully in GW detectors to mitigate other forms of noise \cite{GiEA2003,DrEA2012,DeEA2012}. These techniques are known as \emph{active} noise mitigation\index{mitigation!active}. It is mostly considered as a means to reduce seismic Newtonian noise, but the same scheme may also be applied to atmospheric Newtonian noise (see especially Section \ref{sec:gravimeterNN}) and possibly also other forms of gravity perturbations. While for example active seismic isolation cancels seismic disturbances before they reach the final suspension stages of a test mass, gravity perturbations have to be cancelled in the data of the GW detector. Coherent cancellation comes without (known) ultimate limitations, which means that in principle any level of noise reduction can be achieved provided that the environmental sensors are sufficiently sensitive, and one can deploy as many senors as required. The prediction by Hughes \& Thorne of a modest noise reduction rather follows from a vision of a practicable solution at the time the paper was written. The first detailed study of coherent Newtonian-noise cancellation was carried out by Cella \cite{Cel2000}. He studied the Wiener-filter scheme. Wiener filters are based on observed mutual correlation between environmental sensors and the target channel. The Wiener filter is the optimal linear solution to reduce variance in a target channel as explained in Section \ref{sec:Wiener}. The goal of a cancellation scheme can be different though, e.~g.~reduction of a stationary noise background in non-stationary data. The focus in Section \ref{sec:cohcancel} will also lie on Wiener filters, but limitations will be demonstrated, and the creation of optimal filters using real data is mostly an open problem. 

Techniques to mitigate Newtonian noise without using environmental data are summarized under the category of \emph{passive} Newtonian-noise mitigation\index{mitigation!passive}. Site selection is the best understood passive mitigation strategy. The idea is to identify the quietest detector site in terms of seismic noise and possibly atmospheric noise, which obviously needs to precede the construction of the detector as part of a site-selection process. The first systematic study was carried out for the Einstein Telescope \cite{BeEA2012} with European underground sites. Other important factors play a role in site selection, and therefore one should not expect that future detector sites will be chosen to minimize Newtonian noise, but rather to reduce it to an acceptable level. Current understanding of site selection for Newtonian-noise reduction is reviewed in Section \ref{sec:siteselect}. Other passive noise-reduction techniques are based on building shields against disturbances that cause density fluctuations near the test masses, such as moats and recess structures against seismic Newtonian noise, which are investigated in Section \ref{sec:shieldseismic}.

\subsection{Coherent noise cancellation}
\label{sec:cohcancel}
\index{coherent noise cancellation}
Coherent noise cancellation, also known as \emph{active noise cancellation}, is based on the idea that the information required to model noise in data can be obtained from auxiliary sensors that monitor the sources of the noise. The noise model can then be subtracted from the data in real time or during post processing with the goal to minimize the noise. In practice, cancellation performance is limited for various reasons. Depending on the specific implementation, non-stationarity of data, sensor noise, and also signal and other noise in the target channel can limit the performance. Furthermore, the filter that represents the noise model, which in the context of Newtonian-noise cancellation is a multiple-input-single-output (MISO) filter with reference channels providing the inputs, and the noise model being the output, also maps sensor noise into the noise model, which means that sensor noise is added to the target channel. It follows that the auxiliary sensors must provide information about the sources with sufficiently high signal-to-noise ratio. 

The best way to understand the noise-cancellation problem is to think of it as an optimization of extraction of information, subject to constraints. Constraints can exist for the maximum number of auxiliary sensors, for the possible array configurations, and for the amount of data that can be used to calculate the optimal filter. Also the type of filter and the algorithm used to calculate it can enforce constraints on information extraction. There is little understanding of how most of these constraints limit the performance. A well-explored cancellation scheme is based on \emph{Wiener filters} \cite{Vas2001}. Wiener filters are linear filters calculated from correlation measurements between reference and target channels. They are introduced in Section \ref{sec:Wiener}. In the context of seismic or atmospheric Newtonian-noise cancellation, the auxiliary sensors monitor a field of density perturbations, which means that correlation between auxiliary sensors is to be expected. In this case, if the field is wide-sense stationary (defined in Section \ref{sec:Wiener}), if the target channel is wide-sense stationary, and if all forms of noise are additive, then the Wiener filter is known to be the optimal linear filter for a given configuration of the sensor array \cite{RVRe2013}. In Sections \ref{sec:arrayNNRay} to \ref{sec:arrayNNatm}, the problem is described for seismic and infrasound Newtonian noise. The focus lies on gravity perturbations from fluctuating density fields. Noise cancellation from finite-size sources is mostly a practical problem, and trivial from the theory perspective. The optimization of array configurations for noise cancellation is a separate problem, which is discussed in Section \ref{sec:optimarray}. 

\subsubsection{Wiener filtering}
\label{sec:Wiener} 
\index{Wiener filter} 
A linear, time-invariant filter that produces an estimate of a random stationary (target) process minimizing the deviation between target and estimation is known as Wiener filter \cite{BSH2008}. It is based on the idea that data from reference channels exhibit some form of correlation to the target channel, which can therefore be used to provide a coherent estimate of certain contributions to the target channel. Strictly speaking, the random processes only need to be wide-sense stationary\index{stationary!wide sense}, which means that noise moments are independent of time up to second order (i.~e.~variances and correlations). Without prior knowledge of the random processes, the Wiener filter itself needs to be estimated. In this section, we briefly review Wiener filtering, and discuss some of its limitations.

Two main modes of Wiener filtering exist: filtering in time domain (real-valued) or frequency domain (complex-valued). Let us start with the time-domain filter. Wiener filter require random processes as inputs that are assumed to be correlated with the target process. We will call these reference channels, and collect them as components of a vector $\vec x_n$. The subindex $n$ represents time $t_n=t_0+n\Delta t$, where $\Delta t$ is the common sampling time of the random processes. With discretely sampled data, a straight-forward filter implementation is the convolution with a finite-impulse response filter (FIR). These filters are characterized by a filter order $N$. Assuming that we have $M$ reference channels, the FIR filter $\mathbf w$ is a $(N+1)\times M$ matrix with components $w_{nm}$. The convolution assumes the form
\beq
\begin{split}
\hat y_n &= \sum\limits_{k=0}^N \vec w_k\cdot\vec x_{n-k}\\
&\equiv \mathbf w\circ\vec x_n
\end{split}
\eeq
where the dot-product is with respect to the $M$ reference channels. This equation implies that there is only one target channel $y_n$, in which case the FIR filter is also known as multiple-input-single-output (MISO) filter. We have marked the filter output with a hat to indicate that it should be interpreted as an estimate of the actual target channel. The coefficients of the Wiener filter can be calculated by demanding that the mean-square deviation $\langle (y_n-\hat y_n)^2\rangle$ between the target channel and filter output is minimized, which directly leads to the Wiener-Hopf equations\index{Wiener-Hopf equations}:
\beq
\mathbf C_{xx}\cdot\vec w(:)=\vec C_{xy}
\label{eq:wienerhopf}
\eeq
The Wiener-Hopf equations are a linear system of equations that determine the filter coefficients. Here, $\vec w(:)$ is the $NM$-dimensional vector that is obtained by concatenating the $M$ columns of the matrix $\mathbf w$. The $(N+1)M\times (N+1)M$ matrix $\mathbf C_{xx}$ is the cross-correlation matrix between reference channels. Correlations must be evaluated between all samples of all reference channels where sample times differ at most by $N\Delta t$. It contains the autocorrelations of each reference channel as $(N+1)\times (N+1)$ blocks on its diagonal:
\beq
\mathbf C_{xx}^{\rm auto}=
\left(\begin{matrix}
c_0 & c_1 & \cdots & c_N \\
c_1 & c_0 & \cdots & c_{N-1} \\
\vdots & \vdots & \ddots & \vdots \\
c_N & c_{N-1} & \cdots & c_0
\end{matrix}\right)
\eeq
with $c_k\equiv\langle x_n x_{n+k}\rangle$ for each of the $M$ reference channels. In this form it is a symmetric Toeplitz matrix. The $(N+1)M$-dimensional vector $\vec C_{xy}$ is a concatenation of correlations between each reference channel and the target channel. The components contributed by a single reference channel are
\beq
\vec C_{xy}^{\,\rm sgl}=\left(s_0,s_1,\ldots,s_N\right)
\eeq
with $s_k\equiv\langle x_n y_{n+k}\rangle$. Note that we do not assume independence of noise between different reference channels. This is important since there can be forms of noise correlated between reference channels, but uncorrelated with the target channel (e.~g.~shear waves in Newtonian-noise cancellation, see Section \ref{sec:arrayNNP}). In general, the correlations that determine the Wiener-Hopf equations are unknown and need to be estimated from measurements using data from reference and target channels. An elegant implementation of the code that provides these estimates and solves the Wiener-Hopf equations can be found in \cite{Pep2007}.

The residual of the target channel after subtraction of $\hat y_n$ is given by
\beq
r_n = y_n-\mathbf w\circ\vec x_n
\label{eq:reswiener}
\eeq
This equation summarizes the concept of coherent noise cancellation. In the context of Newtonian noise subtraction, the target channel $y_n$ corresponds to the GW strain signal contaminated by Newtonian noise, and $\hat y_n$ is the estimate of Newtonian noise provided by the Wiener filter using reference data from seismometers or other sensors. Time-domain Wiener filters were successfully implemented in GW detectors for the purpose of noise reduction \cite{DrEA2012,DeEA2012}. Results from a time-domain simulation of Newtonian-noise cancellation using Wiener filters was presented in \cite{DHA2012}. Not all coherent cancellation schemes are necessarily implemented as Wiener filters. For example in \cite{GiEA2003}, noise cancellation was optimized by solving a system-identification problem. 

A frequency-domain version of the Wiener filter can be obtained straight-forwardly by dividing the data into segments and calculating their Fourier transforms. Equation (\ref{eq:reswiener}) translates into a segment-wise noise cancellation where $n$ stands for a double index to specify the segment and the discrete frequency (also known as frequency bin). For stationary random processes, correlations between noise amplitudes at different frequencies are zero (keep in mind that amplitudes of stationary, random processes do not exist as Fourier amplitudes, and therefore this statement needs a suitable definition of these amplitudes, see Section \ref{sec:noisefreq} and \cite{RoPi1955}). This means that coherent noise cancellation in frequency domain can be done on each frequency bin separately, which is numerically much less demanding, and more accurate since the dimensionality of the system of equations in Equation (\ref{eq:wienerhopf}) is reduced from $NM$ to $M$ (for $N$ different frequency bins). In contrast, time-domain correlations $c_k,\,s_k$ can be large for small values of $k$. This can cause significant numerical problems to solve the Wiener-Hopf equations, and as observed in \cite{CoEA2014}, FIR filters of lower order can be more effective (even though theoretically, increasing the filter order should not make the cancellation performance worse).

It should be noted that coherence between channels needs to be very high even for ``modest'' noise cancellation. The ideal suppression factor $s(\omega)$ as a function of frequency in the case of a single reference channel is related to the reference-target coherence $c(\omega)$ via
\beq
s(\omega)=\frac{1}{\sqrt{1-c(\omega)^2}}
\label{eq:singlechcoh}
\eeq
where the coherence is defined in terms of the spectral densities (see Section \ref{sec:noisefreq}):
\beq
c(\omega) = \frac{S(x,y;\omega)}{(S(x;\omega)S(y;\omega))^{1/2}}
\eeq
If coherence between reference and target channels at some frequency is $\{0.9, 0.99, 0.999\}$, then the residual amplitude spectrum at that frequency will ideally be reduced by factors $\{2.3, 7.1, 22\}$, respectively. These numbers clearly do not pose a limit to cancellation with multiple reference sensors. Trivially, cancellation using $M$ collocated, identical sensors leads at least to a $\sqrt{M}$ reduction of the sensor noise limit. If the $M$ sensors monitor a field whose values at nearby points are dynamically correlated (i.~e.~the two-point spatial correlation is not just a $\delta$-peak), then further gain is to be expected for example by being able to distinguish between modes of the field that produce correlation with the target channel, and modes that do not. This will be discussed in detail in Section \ref{sec:arrayNNP} and Section \ref{sec:optimarray}.

\subsubsection{Cancellation of Newtonian noise from Rayleigh waves}
\label{sec:arrayNNRay}\index{active noise cancellation!Rayleigh waves}
As we have seen in Section \ref{sec:Wiener}, the correlations between reference channels and the target channel determine the Wiener filter. For seismic fields, correlations between reference channels (seismometers) can be measured, but we still need a model consistent with the seismic correlations that provides the correlation with the gravity channel. Obviously, as long as 2D arrays are used for the characterization of the seismic field, the predicted correlation with the gravity channel can be subject to systematic errors. For example, we will have to guess the types of seismic waves that contribute to the seismic correlations. In this section, the problem will be solved assuming that all seismic waves are Rayleigh waves. Here, we will also discuss the cancellation problem in low-frequency detectors explicitly since it is qualitatively different. 
 
It is assumed that the seismic correlation is known, and the problem is to calculate the correlation with the gravity channel. We will consider the case of a homogeneous, but not necessarily isotropic seismic field. In this case, we can choose to evaluate the gravity acceleration at $\vec\varrho_0=\vec 0$, and its correlation with the vertical surface displacement at $\vec \varrho=(x,y)$. For the gravity acceleration along the $x$-axis measured at height $h$ above surface, the correlation is given by (see Section \ref{sec:spatialcorr} for spectral representation of noise)
\beq
\begin{split}
\langle \delta a_x(\vec 0,\omega), \xi_z(\vec \varrho\,,\omega)\rangle &= -2\pi\irm G\rho_0\gamma(\nu)\int\frac{\drm^2k}{(2\pi)^2}S(\xi_z;\vec k_\varrho,\omega)\frac{k_x}{k_\varrho}\e^{-hk_\varrho}\e^{\irm\vec k_\varrho\vec\varrho}\\
&=G\rho_0\gamma(\nu)\int\drm^2\varrho'\,C(\xi_z;\vec \varrho\,',\omega) \frac{x-x'}{(h^2+|\vec\varrho-\vec\varrho\,'|^2)^{3/2}}
\end{split}
\label{eq:kernelRayaxi}
\eeq
The two dimensional kernel of the integral in the second line is plotted in Figure \ref{fig:kernelRayaxi}. The important coordinate range of the correlation function lies around the two extrema at $x'=x\pm h/\sqrt{2}$ and $y'=y$.
\epubtkImage{}{
    \begin{figure}[htbp]
    \centerline{\includegraphics[width=0.8\textwidth]{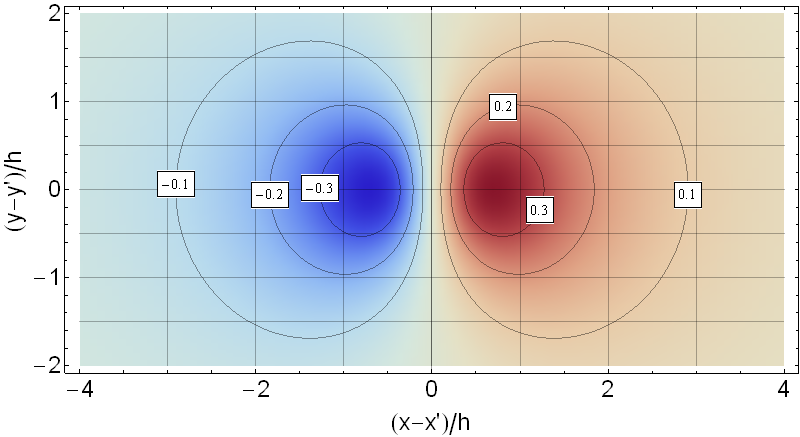}}
    \caption[Homogeneous displacement-gravity kernel for Rayleigh fields]{Homogeneous displacement-gravity kernel for Rayleigh fields according to Equation (\ref{eq:kernelRayaxi}).}
    \label{fig:kernelRayaxi}
    \end{figure}}
Next, we will consider the explicit example of an isotropic Rayleigh wave field. The easiest way to obtain the result is to insert the known solution of the wavenumber spectrum, Equation (\ref{eq:specRayIso}), into the first line in Equation (\ref{eq:kernelRayaxi}), which gives:
\beq
\langle \delta a_x(\vec 0,\omega), \xi_z(\vec \varrho\,,\omega)\rangle =2\pi G\rho_0\gamma(\nu) S(\xi_z;\omega)\e^{-hk_\varrho^{\rm R}}\cos(\phi)J_1(k_\varrho^{\rm R}\varrho)
\label{eq:corraxiRayiso}
\eeq
Interestingly, the correlation between vertical displacement and the gravity perturbation vanishes for $\varrho = 0$. This is a consequence of the fact that any elastic perturbation of the ground must fulfill the wave equation. If instead the ground were considered as a collection of infinitely many point masses without causal link, then the correlation of displacement of point masses nearest to the test mass with the gravity perturbation would be strongest.

Since the purpose of this section is to evaluate and design a coherent noise cancellation of gravity perturbations in $x$-direction, one may wonder why the correlation with the vertical surface displacement is used, and not the displacement along the direction of the $x$-axis. The reason is that in general horizontal seismic motion of a flat surface correlates weakly with gravity perturbations produced at the surface. Other waves such as horizontal shear waves can produce horizontal surface displacement without perturbing gravity. Vertical surface displacement always perturbs gravity, no matter by what type of seismic wave it is produced. The situation is different underground as we will see in Section \ref{sec:arrayNNP}. 

Notice that the results so far can only be applied to the case where Newtonian noise is uncorrelated between different test masses. In future GW detectors that measure signals below 1\,Hz, correlation of seismic Newtonian noise between two test masses can be very high since the seismic wavelength is much larger than the dimension of the detector. Since the only position dependence in Equation (\ref{eq:RayleighNN}) is the phase term $\exp(\irm \vec k_\varrho\cdot\vec\varrho_0)$, the differential acceleration between two test masses is governed by the difference of phase terms at the two test masses, which simplifies to $\irm \vec k_\varrho\cdot\vec L$ when the distance $L$ between the test masses is much smaller than the length of the Rayleigh wave. Considering the case that direction of acceleration and direction of separation are the same, the correlation is given by
\beq
\begin{split}
\langle(\delta a_x(L\vec e_x,\omega)-&\delta a_x(\vec 0,\omega))/L, \xi_z(\vec \varrho\,,\omega)\rangle_{\rm low-f} \\
&=2\pi G\rho_0\gamma(\nu)\int\frac{\drm^2k}{(2\pi)^2}S(\xi_z;\vec k_\varrho,\omega)\frac{k_x^2}{k_\varrho}\e^{-hk_\varrho}\e^{\irm\vec k_\varrho\vec\varrho}\\
&=G\rho_0\gamma(\nu)\int\drm^2\varrho'\,C(\xi_z;\vec \varrho\,',\omega) \frac{h^2-3(x-x')^2+|\vec\varrho-\vec\varrho\,'|^2}{(h^2+|\vec\varrho-\vec\varrho\,'|^2)^{5/2}}
\end{split}
\label{eq:kernelRayaxiLow}
\eeq
The maximum of the kernel lies at the origin $x'=x$, $y'=y$ independent of test-mass height. Now, for the homogeneous and isotropic field, the solution with respect to the strain acceleration reads
\beq
\begin{split}
\langle(\delta a_x(L\vec e_x,\omega)-&\delta a_x(\vec 0,\omega))/L, \xi_z(\vec \varrho\,,\omega)\rangle_{\rm low-f} \\
&=\pi G\rho_0\gamma(\nu) S(\xi_z;\omega)\e^{-hk_\varrho^{\rm R}}(J_0(k_\varrho^{\rm R}\varrho)-\cos(2\phi)J_2(k_\varrho^{\rm R}\varrho))
\end{split}
\label{eq:corraxiRayisoLow}
\eeq
Here, correlation does not vanish in the limit $\varrho\rightarrow 0$. Also notice that the result is independent of the distance $L$. This is the typical situation for strain quantities at low frequencies since the differential signal is proportional to the distance, which then cancels in the strain variable when dividing by $L$. We have seen this already in Section \ref{sec:lowfNNRay}. 

At this point, we have the required analytical expressions to evaluate the performance of Wiener filters. The goal is to derive equations that allow us to calculate the performance of the Wiener filter, given a specific array configuration and seismometer self noise. We also want to know whether it is possible to use the results to design optimal array configurations based on seismic correlation measurements alone. First, we continue with the specific example of a homogeneous and isotropic field, and a single test mass. Since the Wiener filter is based on measured correlations between seismometers and a gravity channel, we need to introduce the seismometer self noise. It is convenient to express the noise in terms of the signal-to-noise ratio $\sigma(\omega)$ with respect to measurements of seismic displacement. According to Equation (\ref{eq:corrRayiso}), the correlation between two seismometers at locations $\vec\varrho_i,\,\vec\varrho_j$ can then be written
\beq
C_{\rm SS}^{ij}(\xi_z;\omega)=S(\xi_z;\omega)\left(J_0(k_\varrho^{\rm R}|\vec\varrho_i-\vec\varrho_j|)+\frac{\delta_{ij}}{\sigma_i^2(\omega)}\right)
\eeq
Seismometer self noise is assumed to be uncorrelated among different seismometers and with the gravity channel. The correlations between all seismometers form (a frequency-domain version of) the correlation matrix $\mathbf C_{xx}$ in Equation (\ref{eq:wienerhopf}). The correlation of each seismometer with the gravity perturbation will be denoted as 
\beq
C_{\rm SN}^i(\xi_z,\delta a_x;\omega)\equiv 2\pi G\rho_0\gamma(\nu) S(\xi_z;\omega)\e^{-hk_\varrho^{\rm R}}\cos(\phi_i)J_1(k_\varrho^{\rm R}\varrho_i)
\eeq
Subtracting the output of a Wiener filter leaves a residual, whose spectrum relative to the original gravity spectrum $C_{\rm NN}(\omega)=S(\delta a_x;\omega)$ is \cite{Cel2000}
\beq
R(\omega)=1-\frac{\vec C_{\rm SN}^{\,\rm T}(\omega)\cdot(C_{\rm SS}(\omega))^{-1}\cdot\vec C_{\rm SN} (\omega)}{C_{\rm NN}(\omega)}
\label{eq:residualNN}
\eeq
A simple question to answer is where a single seismometer should be placed to minimize the residual. In this case, the residual spectrum is given by
\beq
R_1(\omega) = 1-\frac{2\cos^2(\phi_1)J_1^2(k_\varrho^{\rm R}\varrho_1)}{1+1/\sigma_1^2(\omega)}
\label{eq:residualNNRay1}
\eeq
Since the fraction is always positive and smaller than 1, it needs to be maximized. This means that $\phi_1=0$ or $\pi$, and $\varrho_1$ is chosen to maximize the value of the Bessel function. In the presence of $N>1$ seismometers, the optimization problem is non-trivial. The optimal array configuration fulfills the relation
\beq
\nabla^N(\vec C_{\rm SN}^{\,\rm T}(\omega)\cdot(C_{\rm SS}(\omega))^{-1}\cdot\vec C_{\rm SN} (\omega))=\vec 0,
\eeq
where $\nabla^N$ contains $2N$ derivatives with respect to the two horizontal coordinates of $N$ seismometers. Already with a few seismometers, it becomes very challenging to find numerical solutions to this equation (see Section \ref{sec:optimarray}). An easier procedure that we want to illustrate now is to perform a step-wise optimal placement of seismometers. In other words, one after the other, seismometers are added at the best locations, with all previous seismometers having fixed positions. 
\epubtkImage{}{
    \begin{figure}[htbp]
    \centerline{\includegraphics[width=0.3\textwidth]{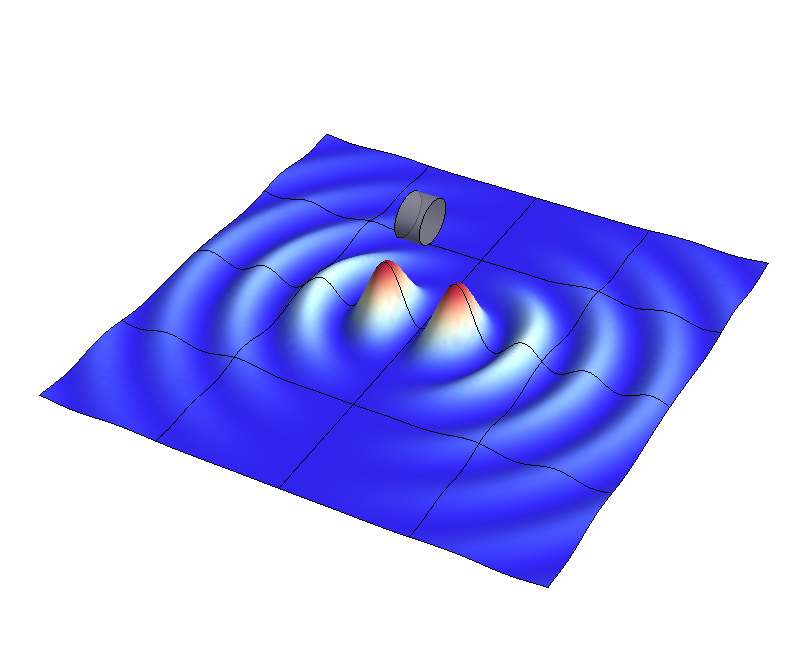}
                \includegraphics[width=0.3\textwidth]{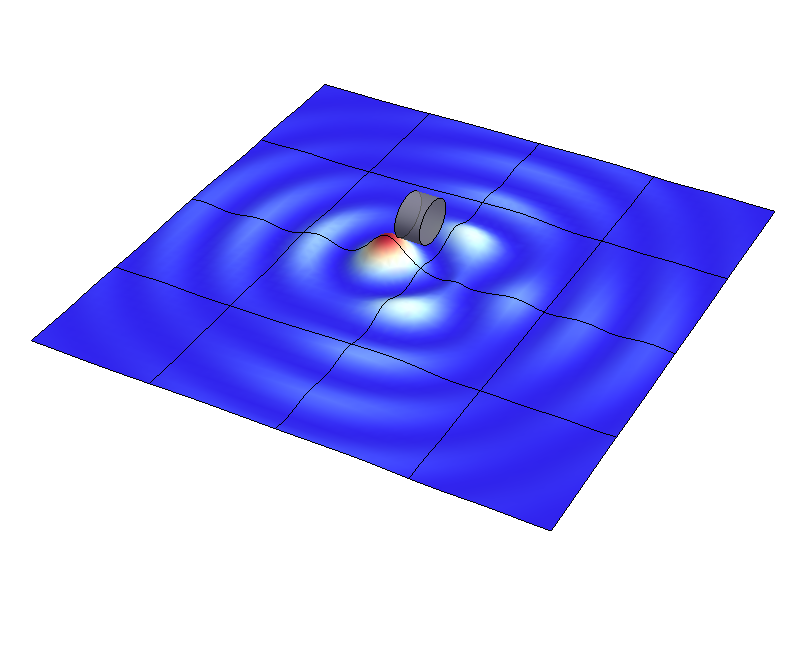}
                \includegraphics[width=0.3\textwidth]{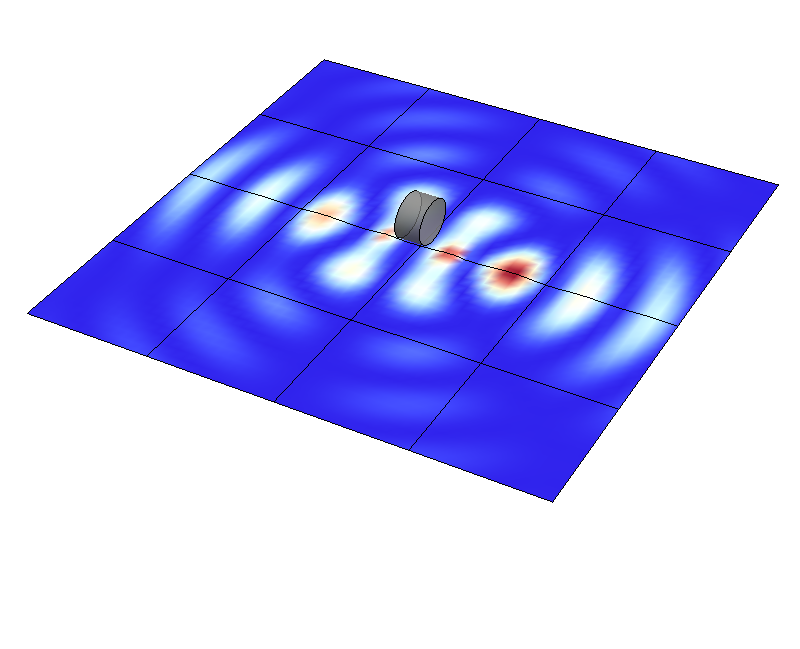}}
    \caption[Step-wise optimal array for Rayleigh Newtonian noise]{Step-wise optimal placement of seismometers for Wiener filtering of Rayleigh Newtonian noise. Maxima indicate best placement. Left: optimal location(s) of first seismometer. Middle: optimal location of second seismometer. Right: optimal location of third seismometer.}
    \label{fig:arrayRay}
    \end{figure}}
The procedure can be seen in Figure \ref{fig:arrayRay}. The first seismometer must be placed at $x_1=\pm 0.3\lambda^{\rm R}$ and $y_1=0$. We choose the side with positive $x$-coordinate. Assuming a signal-to-noise ratio of $\sigma=10$, the single seismometer residual would be 0.38. The second seismometer needs to be placed at $x_2=-0.28\lambda^{\rm R}$ and $y_2=0$, with residual 0.09. The third seismometer at $x_3=0.75\lambda^{\rm R}$ and $y_3=0$, with residual 0.07. The step-wise optimization described here works for a single frequency since the optimal locations depend on the length $\lambda^{\rm R}$ of a Rayleigh wave. In reality, the goal is to subtract over a band of frequencies, and the seismometer placement should be optimized for the entire band. The result is shown in the left of Figure \ref{fig:residualsRay} for a sub-optimal spiral array, and seismometers with frequency-independent $\sigma=100$. Rayleigh-wave speed is constant $c_{\rm R}=250\,$m/s. There are three noteworthy features. First, the minimal relative residual lies slightly below the value of the inverse seismometer signal-to-noise ratio. It is a result of averaging of self noise from different seismometers. Second, residuals increasing with $1/\omega$ at low frequencies is a consequence of the finite array diameter. An array cannot analyze waves much longer than its diameter. Third, the residuals grow sharply towards higher frequencies. The explanation is that the array has a finite seismometer density, and therefore, waves shorter than the typical distance between seismometers cannot be analyzed. If the seismic speed is known, then the array diameter and number of seismometers can be adjusted in this way to meet a subtraction goal in a certain frequency range.
\epubtkImage{}{
    \begin{figure}[htbp]
    \centerline{\includegraphics[width=0.45\textwidth]{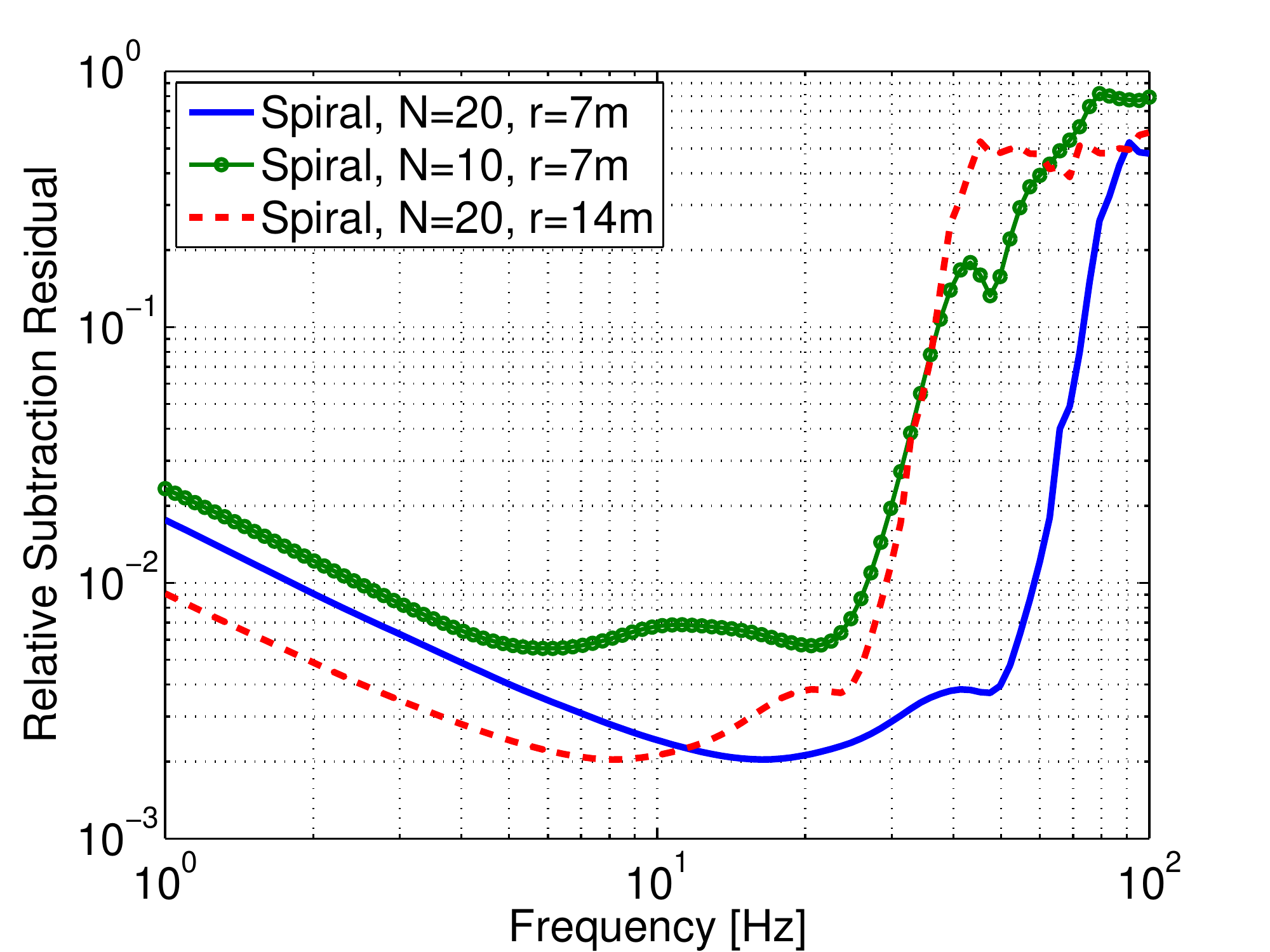}
                \includegraphics[width=0.45\textwidth]{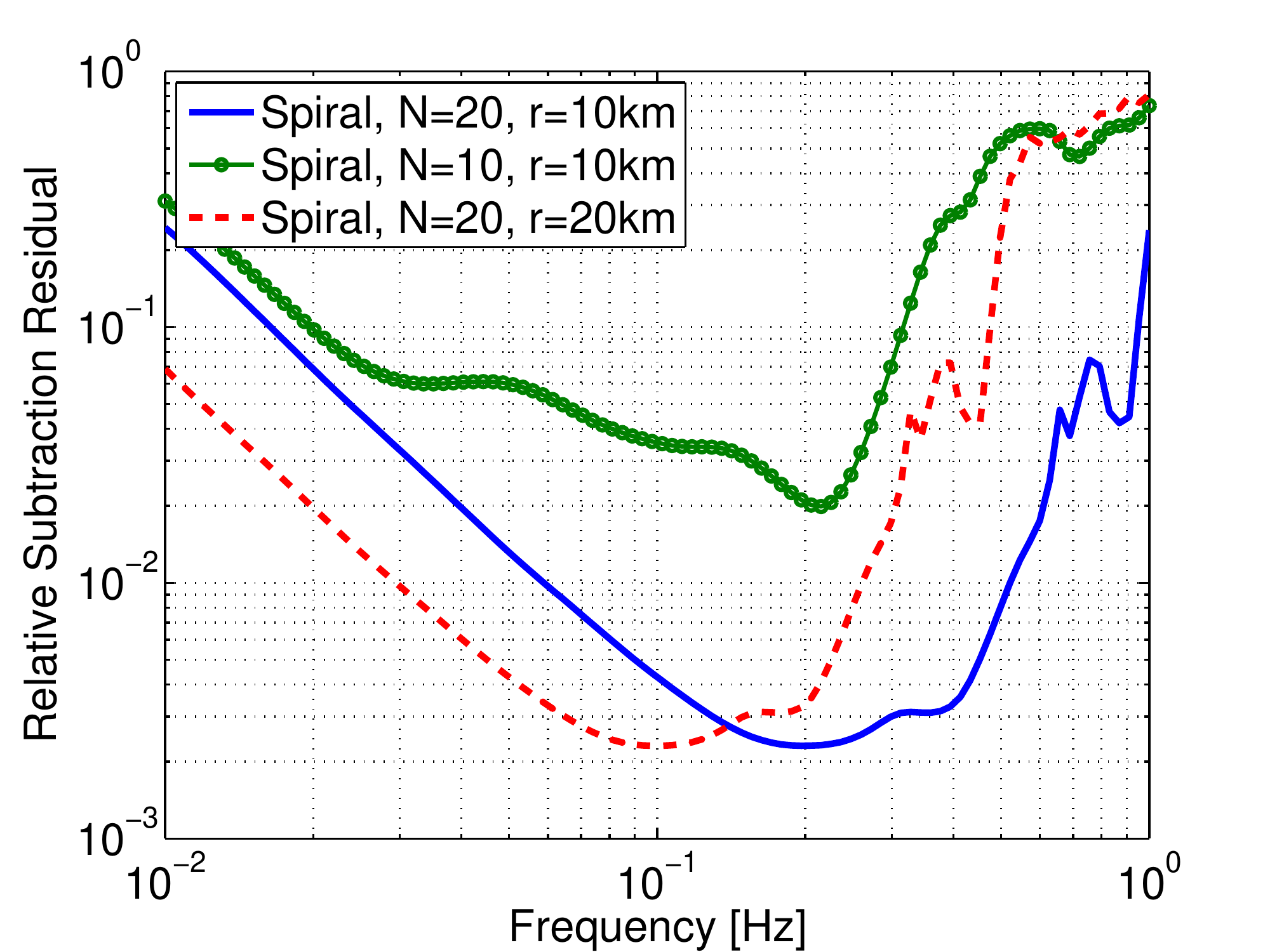}}
    \caption[Broadband Rayleigh Newtonian-noise residuals]{Residuals from Wiener noise cancellation using spiral seismometer arrays. The curves show the residuals for different numbers $N$ of seismometers, and also different spiral radii. In all cases, the spirals have two full windings. Left: subtraction of uncorrelated test-mass noise. Right: subtraction of gravity-gradient noise typical for low-frequency detectors.}
    \label{fig:residualsRay}
    \end{figure}}

Residuals are also shown in the right plot of Figure \ref{fig:residualsRay} after subtraction of gravity-gradient noise (i.~e.~the low-frequency case). The sensor signal-to-noise ratio is the same as before. As we have seen in Section \ref{sec:lowfNNRay}, Newtonian noise is suppressed at low frequencies in gravity strainmeters due to common-mode rejection of correlated gravity perturbations between two test masses. However, as soon as coherent cancellation is required, one has to pay the price for this gain. Each seismometer measures seismic displacement that is similarly correlated with gravity perturbations at both test masses. This means that the dominant part of the seismic data is useless since the corresponding gravity perturbations are rejected as common mode. Therefore, the data provided by the seismic array must make it possible to distinguish between the common-mode and differential noise. The Wiener filter needs to cancel the common-mode noise in the seismic data by combining data from different seismometers. An underlying weak correlation with the differential gravity signal then needs to be sufficient to optimize the Wiener filter for noise cancellation. It can be seen that the common-mode rejection causes the residuals to be higher, but only if the number of seismometers lies below a critical value. With the $N=20$ arrays it is possible to distinguish the common-mode noise from differential noise, and subtraction residuals are similar to the standard Newtonian-noise cancellation. However, in all cases, suppression of common-mode noise becomes less efficient at long wavelengths. For this reason, the low-frequency slope of the residual spectra has an additional $1/\omega$, which causes the cancellation to be less broadband. Further results from this analysis can be found in \cite{Har2013}. In the future, it should be analyzed if an inherently differential seismic sensor, such as a seismic strainmeter, naturally provides the required common-mode rejection of seismic data, leading to more efficient noise subtraction.

\subsubsection{Cancellation of Newtonian noise from body waves}
\label{sec:arrayNNP}\index{active noise cancellation!compressional waves}
In this section, the focus lies on noise subtraction in infinite media. As we have seen in Sections \ref{eq;basicsgrav} and \ref{sec:halfspace}, any gravity perturbation can be divided into two parts, one that has the form of gravity perturbations from seismic fields in infinite space, and another that is produced by the surface. Subtraction of the surface part follows the scheme outlined in Section \ref{sec:arrayNNRay} using surface arrays. The additional challenge is that body waves can have a wide range of angles of incidence leading to a continuous range of apparent horizontal speeds, which could affect the array design. In this section, we will investigate the properties of coherent noise cancellation of the bulk contribution. Therefore, this section is without purpose to low-frequency GW detectors. The reason is that a low-frequency detector (i.~e.~sub-Hz detector) can always be considered to be located at the surface with respect to seismic Newtonian noise, since feasible detector depths are only a small fraction of the length of seismic waves. In other words, surface perturbations will always vastly dominate bulk contributions. Since we consider the high-frequency case, we can assume here that Newtonian noise is uncorrelated between test masses. In order to simplify the analysis, only homogeneous and isotropic body-wave fields are considered, without contributions from surface waves.

The evaluation of Wiener-filter performance requires the calculation of two-point spatial correlation functions between seismic measurements and gravity measurements. Since gravity perturbations are assumed to be uncorrelated between test masses, we can focus on gravity perturbations at a single test mass. The test mass is assumed to be located underground inside a cavity. We know from Section \ref{sec:bodynoscatt} that gravity perturbations are produced by compressional waves through density perturbations of the medium, and by shear and compressional waves due to displacement of cavity walls. From the theory perspective, cancellation of noise from cavity walls is straight-forward and will not be discussed here. More interesting is the cancellation of noise from density perturbations in the medium. A seismic measurement is represented by the projection $\vec e_n\cdot\vec \xi(\vec r,\omega)$, where $\vec e_n$ is the direction of the axis of the seismometer, and the gravity measurement by a similar projection $\vec e_n\cdot\delta\vec a(\vec r,\omega)$. Therefore, the general two-point correlation function depends on the directions $\vec e_1,\,\vec e_2$ of two measurements, and the unit vector $\vec e_{12}$ that points from one measurement location at $\vec r_1$ to the other at $\vec r_2$. 

The correlation functions are calculated using the formalism presented in \cite{Fla1993,AlRo1999}, developed for correlations between measurements of strain tensors representing gravitational waves. The first step is to obtain an expression of correlations from single plane waves characterized by a certain polarization and direction of propagation, and then to average over all directions. Here our goal is to calculate separate solutions for compressional and shear waves, which means that we only average over the two transversal polarizations for the case of shear waves. We first calculate the two-point spatial correlation between two seismic measurements of a field composed of P-waves: 
\beq
\begin{split}
\langle (\vec e_1\cdot\vec \xi^{\,\rm P}(\vec r_1,\omega)), (\vec e_2\cdot\vec \xi^{\,\rm P}(\vec r_2,\omega))\rangle &= \frac{3S(\xi_n^{\rm P};\omega)}{4\pi}\int\drm\Omega_k(\vec e_1\cdot\vec e_k)(\vec e_2\cdot\vec e_k)\e^{\irm \vec k^{\rm P}\cdot(\vec r_2-\vec r_1)}\\
&= \frac{3S(\xi_n^{\rm P};\omega)}{4\pi}\vec e_1\cdot\left(\int\drm\Omega_k(\vec e_k\otimes\vec e_k)\e^{\irm k^{\rm P}|\vec r_2-\vec r_1|(\vec e_k\cdot\vec e_{12})}\right)\cdot\vec e_2,
\end{split}
\eeq
where $\vec e_k$ is the direction of propagation of a P-wave. The factor 3 accounts for the isotropic distribution of P-wave energy among the three displacement directions: $S(\xi_n^{\rm P};\omega)=S(\xi^{\rm P};\omega)/3$. The integral is carried out easily in spherical coordinates $\theta,\,\phi$ by choosing the $z$-axis parallel to $\vec e_{12}$ so that $\vec e_k\cdot\vec e_{12}=\cos(\theta)$. Instead of writing down the explicit expression of $\vec e_k\otimes\vec e_k$ and evaluating the integral over all of its independent components, one can reduce the problem to two integrals only. The point is that the matrix that results from the integration can in general be expressed in terms of two ``basis'' matrices $\mathbf 1$ and $\vec e_{12}\otimes \vec e_{12}$. For symmetry reasons, it cannot depend explicitly on any other combination of the coordinate basis vectors $\vec e_x\otimes\vec e_x$, $\vec e_x\otimes\vec e_y$, $\ldots$ Expressing the integral as linear combination of basis vectors, $P_1(\Phi_{12})\mathbf 1+P_2(\Phi_{12})(\vec e_{12}\otimes \vec e_{12})$ with $\Phi_{12}\equiv k^{\rm P}|\vec r_2-\vec r_1|$, solutions for $P_1(\Phi_{12}),P_2(\Phi_{12})$ can be calculated as outlined in \cite{Fla1993,AlRo1999}, and the correlation function finally reads
\beq
\begin{split}
\langle (\vec e_1\cdot\vec \xi^{\,\rm P}(\vec r_1,\omega)), (\vec e_2\cdot\vec \xi^{\,\rm P}(\vec r_2,\omega))\rangle &= S(\xi_n^{\rm P};\omega)\left(P_1(\Phi_{12})(\vec e_1\cdot\vec e_2)+P_2(\Phi_{12})(\vec e_1\cdot\vec e_{12})(\vec e_2\cdot\vec e_{12})\right)\\[0.3cm]
P_1(\Phi_{12})&=j_0(\Phi_{12})+j_2(\Phi_{12})\\
P_2(\Phi_{12})&=-3j_2(\Phi_{12})
\end{split}
\label{eq:corrPP}
\eeq
A great advantage of this expression is that it is coordinate independent. P-wave correlation is zero only if the two measurement directions are orthogonal to each other and to the separation vector. For small distances of the two seismometers, correlation is significant only when the two measurement directions are similar. 

For compressional waves, one also needs to calculate the correlation with gravity perturbations. For the bulk contribution, we can use the gradient of Equation (\ref{eq:gravP}). The analytic form of the correlation is identical to Equation (\ref{eq:corrPP}) since the gravity acceleration is simply a multiple of the seismic displacement. Therefore we can immediately write
\beq
\langle (\vec e_1\cdot\vec \xi^{\,\rm P}(\vec r_1,\omega)), (\vec e_2\cdot\delta\vec a(\vec r_2,\omega))\rangle = 4\pi G\rho_0\langle (\vec e_1\cdot\vec \xi^{\,\rm P}(\vec r_1,\omega)), (\vec e_2\cdot\vec \xi^{\,\rm P}(\vec r_2,\omega))\rangle
\label{eq:corrGravP}
\eeq
Consequently, also here, correlation does not necessarily vanish if gravity acceleration is measured in orthogonal direction to the seismic displacement. In contrast to the Rayleigh-wave correlation in Equation (\ref{eq:corraxiRayiso}), correlation between gravity perturbation and compressional wave displacement is maximal when $\vec r_1=\vec r_2$ assuming that $\vec e_1=\vec e_2$. Nonetheless, due to the more complex form of correlation functions of body fields, there are more choices to make when optimizing array configurations. For example, should single-axis seismometers all measure along the relevant direction of gravity perturbations? What do we gain from multi-axis seismometers? These questions still need to be investigated in detail.

Next, the S-wave correlation is calculated. Since shear waves can be polarized in two orthogonal transverse directions, we form two polarization matrices in terms of basis vectors of the spherical coordinate system, $\vec e_\theta\otimes\vec e_\theta$, $\vec e_\phi\otimes\vec e_\phi$, and average the integrals over these two matrices. The result is
\beq
\begin{split}
\langle (\vec e_1\cdot\vec \xi^{\,\rm S}(\vec r_1,\omega)), (\vec e_2\cdot\vec \xi^{\,\rm S}(\vec r_2,\omega))\rangle &= S(\xi_n^{\rm S};\omega)\left(S_1(\Phi_{12})(\vec e_1\cdot\vec e_2)+S_2(\Phi_{12})(\vec e_1\cdot\vec e_{12})(\vec e_2\cdot\vec e_{12})\right)\\[0.3cm]
S_1(\Phi_{12})&=j_0(\Phi_{12})-\frac{1}{2}j_2(\Phi_{12})\\
S_2(\Phi_{12})&=\frac{3}{2}j_2(\Phi_{12})
\end{split}
\label{eq:corrSS}
\eeq
Since shear waves do not produce gravity perturbations, they act as noise contribution correlated between seismometers. A mixing ratio $\mathpzc{p}$ needs to be introduced that parameterizes the ratio of energy in the P-wave field over the total energy in P- and S-waves. The correlation between seismometers depends on $\mathpzc{p}$:
\beq
\begin{split}
\langle (\vec e_1\cdot\vec \xi(\vec r_1,\omega)), &(\vec e_2\cdot\vec \xi(\vec r_2,\omega))\rangle= \\
&S(\xi_n;\omega)\bigg(\frac{\mathpzc{p}}{S(\xi_n^{\rm P};\omega)}\langle (\vec e_1\cdot\vec \xi^{\,\rm P}(\vec r_1,\omega)), (\vec e_2\cdot\vec \xi^{\,\rm P}(\vec r_2,\omega))\rangle\\
&\hspace*{1.5cm}+\frac{1-\mathpzc{p}}{S(\xi_n^{\rm S};\omega)}\langle (\vec e_1\cdot\vec \xi^{\,\rm S}(\vec r_1,\omega)), (\vec e_2\cdot\vec \xi^{\,\rm S}(\vec r_2,\omega))\rangle\bigg),
\end{split}
\eeq
with $S(\xi_n;\omega)=S(\xi_n^{\rm P};\omega)+S(\xi_n^{\rm S};\omega)$ 
All required quantities are calculated now to evaluate the Wiener filter. In the case of a single seismometer, the residual spectrum defined in Equation (\ref{eq:residualNN}) is given by
\beq
R_1(\omega) = 1-\frac{\mathpzc{p}}{1+1/\sigma_1^2(\omega)}\left(P_1(\Phi_{12})(\vec e_1\cdot\vec e_2)+P_2(\Phi_{12})(\vec e_1\cdot\vec e_{12})(\vec e_2\cdot\vec e_{12})\right)^2
\eeq
The optimal placement of a single seismometer is independent of the mixing ratio. The minimal residual is achieved for $\Phi_{12}=0$, i.~e.~when the seismometer is placed at the test mass. The residual is solely limited by the mixing ratio and signal-to-noise ratio. The case was different for Rayleigh waves, see Equation (\ref{eq:residualNNRay1}), where a limitation was also enforced by the correlation pattern of the seismic field. This is a great advantage of underground detectors. In fact, if the mixing ratio is $\mathpzc{p}=1$ (only P-waves), then it can be shown that the optimal placement of all seismometers would be at the test mass. With a single seismometer, a residual of $\approx 1/\sigma^2$ would be achieved over all frequencies (assuming that $\sigma$ is constant). However, the case is different for mixing ratios smaller than 1. Assuming a conservative mixing ratio of $\mathpzc{p}=1/3$ (P-waves are one out of three possible body-wave polarizations), the single-seismometer residual is about $2/3$ provided that $\sigma\gg 1$. 

As in Section \ref{sec:arrayNNRay}, we consider the step-wise optimized array configuration, which is illustrated in Figure \ref{fig:residualMixPS}.
\epubtkImage{}{
    \begin{figure}[htbp]
    \centerline{\includegraphics[width=0.3\textwidth]{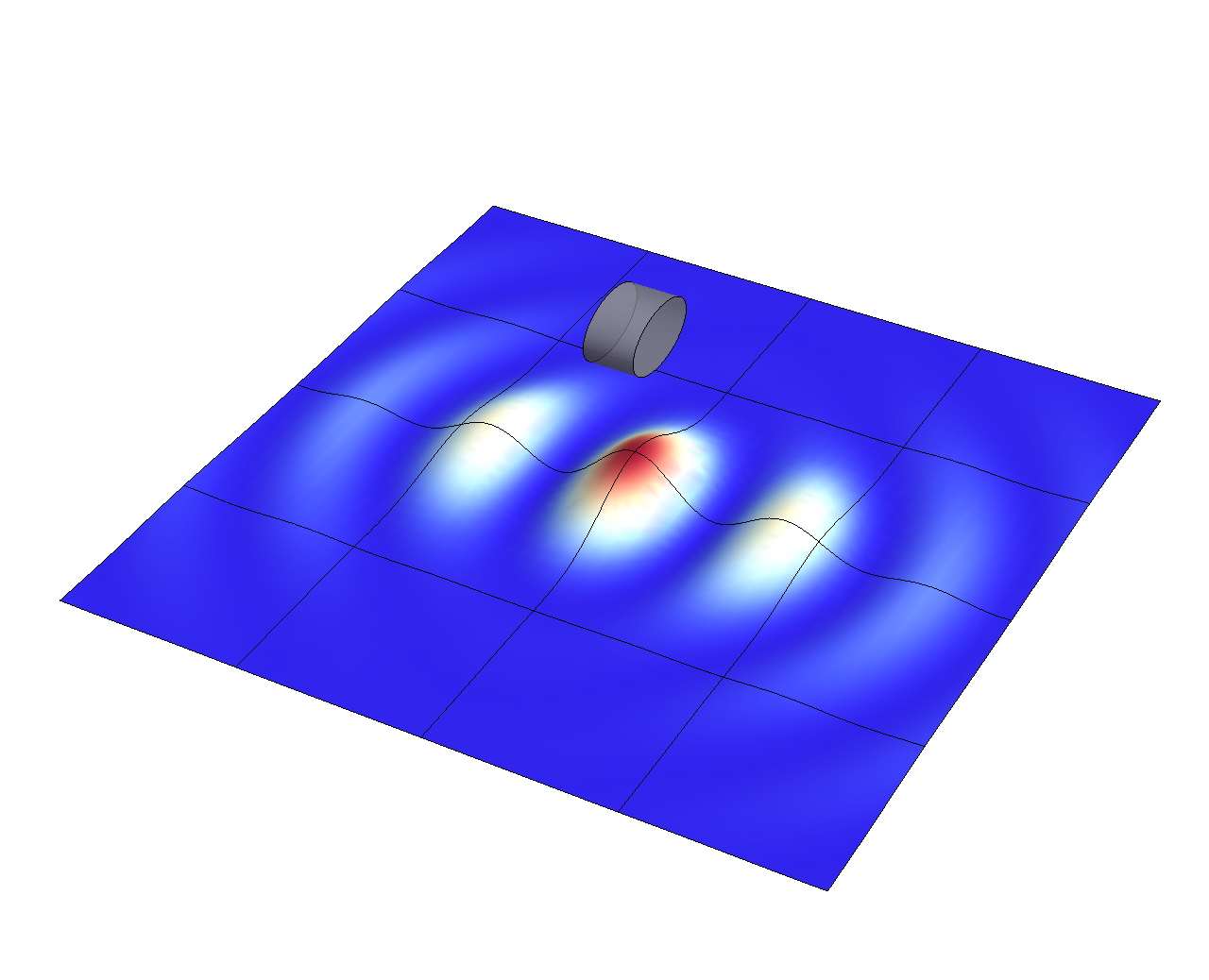}
                \includegraphics[width=0.3\textwidth]{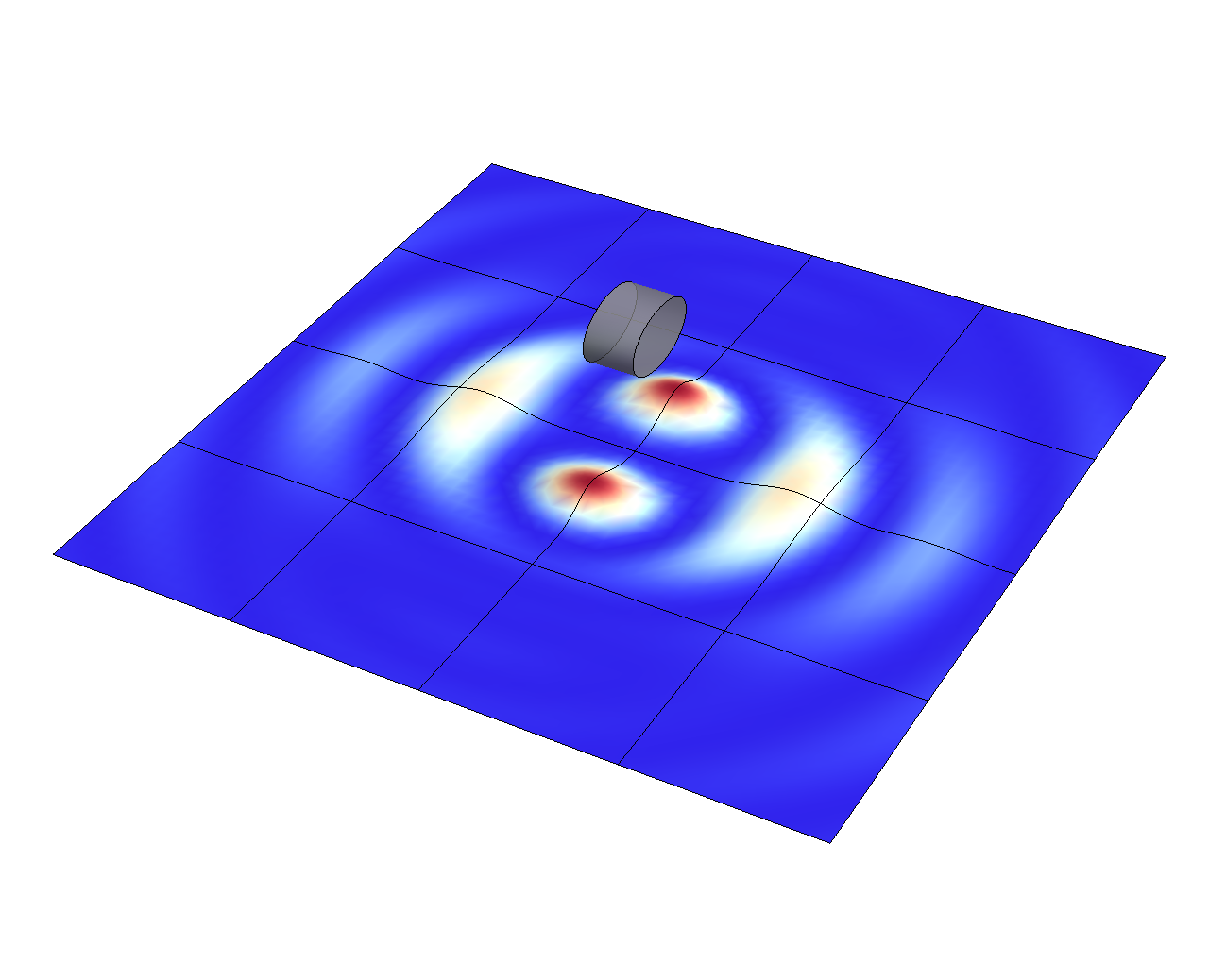}
                \includegraphics[width=0.3\textwidth]{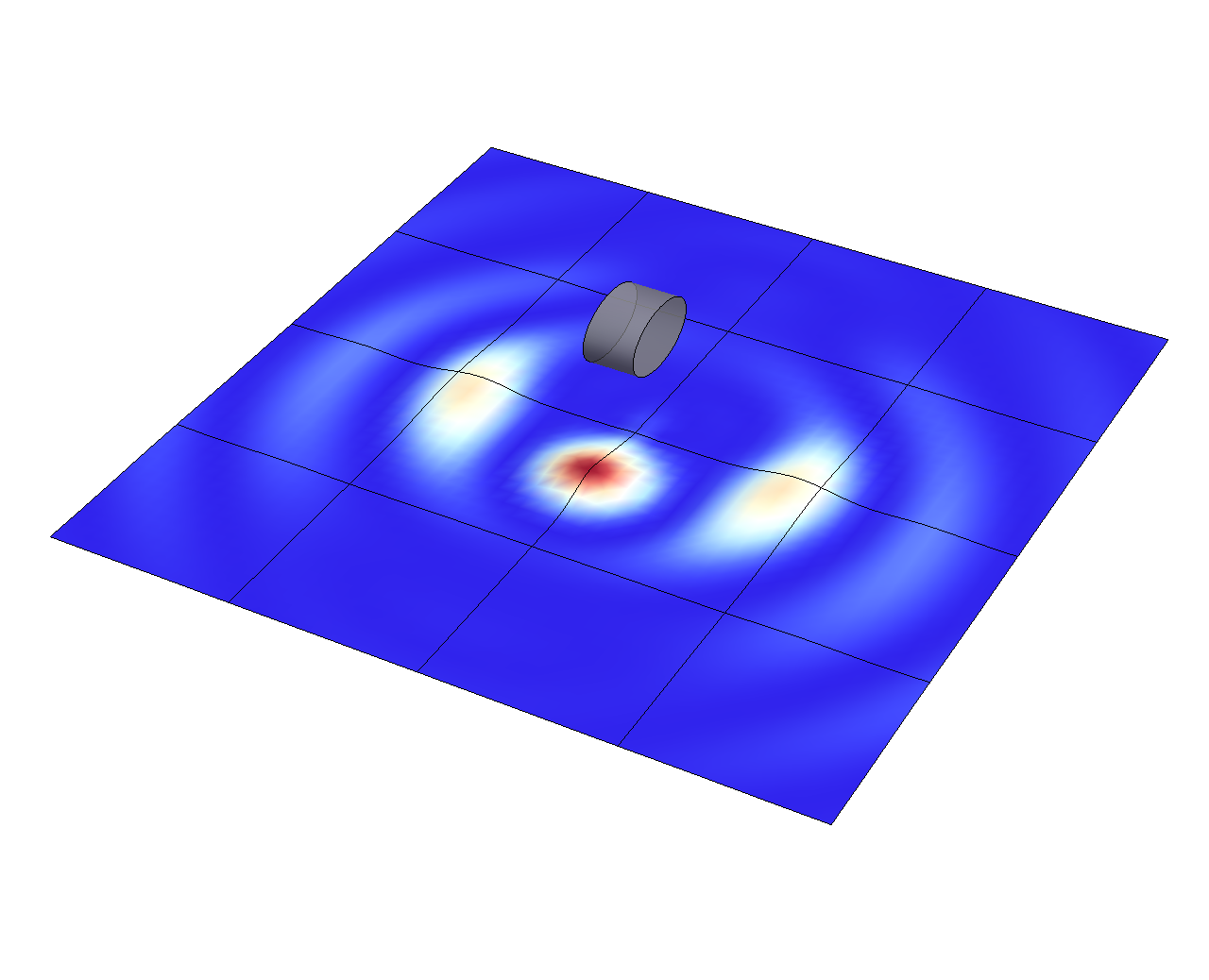}}
    \caption[Body-wave step-wise array optimization for noise cancellation]{Step-wise array optimization for noise cancellation of bulk Newtonian noise. The red maxima mark the optimal location of the next seismometer to be placed. Left: placement of first seismometer with residual 0.67. Middle: placement of second seismometer with residual 0.54. Right: placement of third seismometer with residual 0.44.}
    \label{fig:residualMixPS}
    \end{figure}}
The array is designed for cancellation of gravity perturbations along the $x$-axis. The plot only shows a plane of possible seismometer placement, and all seismometers measure along the relevant direction of gravity acceleration. Ideally, optimization should be done in three dimensions, but for the first three seismometers, the 2D representation is sufficient. In theses calculations, the P-wave speed is assumed to be a factor 1.8 higher than the S-wave speed. The mixing ratio is $1/3$, and the signal-to-noise ratio is 100. The optimal location of the second seismometer lies in orthogonal direction at $x_2=z_2=0$ and $y_2=\pm 0.33\lambda^{\rm P}$. We choose the positive $y$-coordinate. In this case, the third seismometer needs to be placed at $x_3=z_3=0$ and $y_3=- 0.33\lambda^{\rm P}$. With three seismometers, a residual of $0.44$ can be achieved. 

The left plot in Figure \ref{fig:residualXiStrain} shows the subtraction residuals of bulk Newtonian noise using a 3D spiral array with all seismometers measuring along the relevant direction of gravity acceleration. The mixing ratio is $1/3$. The ultimate limit enforced by seismometer self noise, $1/(\sigma\sqrt{N})$, is not reached. Nonetheless, residuals are strongly reduced over a wide range of frequencies. Note that residuals do not approach 1 at highest and lowest frequencies, since a single seismometer at the test mass already reduces residuals to $0.67$ at all frequencies assuming constant $\sigma=100$. 

Another idea is to use seismic strainmeters instead of seismometers\index{strainmeter!seismic}. Seismic strainmeters are instruments that measure the diagonal components of the seismic strain tensor \cite{Agn1986}. Off-diagonal components are measured by seismic tiltmeters\index{tiltmeter}. Strainmeters are also to be distinguished from dilatometers\index{dilatometer}, which are volumetric strainmeters measuring the trace of the seismic strain tensor. The advantage would be that strainmeters are ideally insensitive to shear waves. This means that the optimization of the array is independent of the mixing ratio $\mathpzc{p}$. The seismic strain field $\mathbf{h}(\vec r\,,t)$ produced by a compressional wave can be written as
\beq
\mathbf{h}(\vec r\,,t)=-\irm k^{\rm P}(\vec e_k\otimes\vec e_k)\xi^{\rm P}\e^{\irm(\omega t-\vec k^{\rm P}\cdot\vec r)},
\eeq
which is a $3\times 3$ strain tensor. A seismic strainmeter measures differential displacement along a direction that coincides with the orientation of the strainmeter, which rules out any type of rotational measurements. In this case, the correlation between two seismic strainmeters measuring strain along direction $\vec e_1,\,\vec e_2$ is given by
\beq
\begin{split}
\langle (\vec e_1\cdot\mathbf{h}(\vec r_1,\omega)\cdot\vec e_1),& (\vec e_2\cdot\mathbf{h}(\vec r_2,\omega)\cdot\vec e_2)\rangle\\
&=\frac{5S(h_n^{\rm P};\omega)}{4\pi}\int\drm\Omega_k(\vec e_1\cdot\vec e_k)^2(\vec e_2\cdot\vec e_k)^2\e^{\irm k^{\rm P}|\vec r_2-\vec r_1|\vec e_{12}\cdot \vec e_k}
\end{split}
\eeq
The factor 5 accounts for the isotropic distribution of strain-wave energy among the five strain degrees of freedom: $S(h_n^{\rm P};\omega)=S(h^{\rm P};\omega)/5$, with $h^{\rm P}\equiv k^{\rm P}\xi^{\rm P}$. There are five degrees of freedom since the strain tensor is symmetric and its trace is a constant (note that any symmetric tensor with constant trace can be diagonalized, in which case the resulting tensor only has two independent components, but here we also need to include the three independent rotations).

This integral can be solved fully analogously to the tensor calculation given in \cite{Fla1993,AlRo1999}, or more specifically, using the generalized result in \cite{CoHa2014b}. The required steps are to define a 4D polarization tensor $\vec e_k\otimes\vec e_k\otimes\vec e_k\otimes\vec e_k$ so that the projection along directions $\vec e_1,\,\vec e_2$ can be applied outside the integral, solve the integral, and then project the solution. As for the vector fields, due to symmetry, we can express the 4D matrix resulting from the integral as a sum over a relatively small number of basis matrices (in this case 5), and solve for the five expansion coefficients. It turns out (in the case of seismic strain measurements) that only 3 expansion coefficients are different, which means that the final solution can be expressed as a linear combination of three coefficients $T_1(\Phi_{12}),\,T_2(\Phi_{12}),\,T_3(\Phi_{12})$. The result is the following:
\beq
\begin{split}
\langle (\vec e_1\cdot\mathbf{h}(\vec r_1,&\omega)\cdot\vec e_1),(\vec e_2\cdot\mathbf{h}(\vec r_2,\omega)\cdot\vec e_2)\rangle\\
&=S(h_n^{\rm P};\omega)\bigg(T_1(\Phi_{12})(1+2(\vec e_1\cdot\vec e_2)^2) +T_2(\Phi_{12})((\vec e_1\cdot\vec e_{12})^2+(\vec e_2\cdot\vec e_{12})^2 \\
&\hspace*{2cm}+4(\vec e_1\cdot\vec e_2)(\vec e_{12}\cdot\vec e_1)(\vec e_{12}\cdot\vec e_2))+T_3(\Phi_{12})(\vec e_{12}\cdot\vec e_1)^2(\vec e_{12}\cdot\vec e_2)^2\bigg)\\[0.3cm]
T_1(\Phi_{12})&= \frac{1}{21}(7j_0(\Phi_{\rm 12})+10j_2(\Phi_{\rm 12})+3j_4(\Phi_{\rm 12}))\\
T_2(\Phi_{12})&= -\frac{5}{7}(j_2(\Phi_{\rm 12})+j_4(\Phi_{\rm 12}))\\
T_3(\Phi_{12})&= 5j_4(\Phi_{\rm 12})
\end{split}
\label{eq:straincorr}
\eeq
Even though this expression looks rather complicated, it is numerically straight-forward to implement it in Wiener-filter calculations. 
\epubtkImage{}{
    \begin{figure}[htbp]
    \centerline{\includegraphics[width=0.99\textwidth]{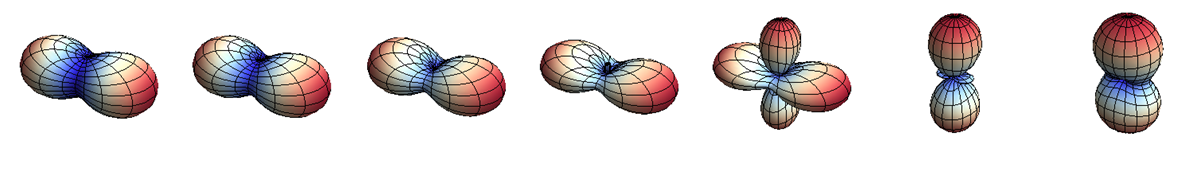}}
    \centerline{\includegraphics[width=0.99\textwidth]{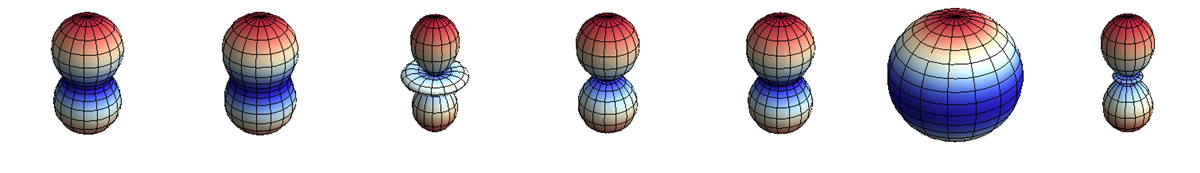}}
    \caption[Two-point correlation of seismic strain measurements]{Two-point correlation between seismic strain measurements. The direction $\vec e_1$ is kept constant. The components of $\vec e_2$ are represented in angular spherical coordinates, with $z$-axis parallel to $\vec e_{12}$ (i.~e.~the ``vertical'' direction in these plots parallel to the symmetry axes in the lower row). From left to right, the value of $\Phi_{12}$ changes from 0 to $2\pi$ in equidistant steps. Upper row: $\vec e_1\cdot\vec e_{12}=0$. Lower row: $\vec e_1\cdot\vec e_{12}=1$.}
    \label{fig:strainpatterns}
    \end{figure}}
Spherical plots of the two-point spatial correlation are shown in Figure \ref{fig:strainpatterns}. The vector $\vec e_1$ is kept constant, while the vector $\vec e_2$ is expressed in spherical coordinates $\theta,\,\phi$. For each value of these two angles, the resulting correlation between the two strainmeters corresponds to the radial coordinate of the plotted surfaces. Since the focus lies on the angular pattern of the correlation function, each surface is scaled to the same maximal radius. It can be seen that there is a rich variety of angular correlation patterns, which even includes near spherically symmetric patterns (which means that the orientation of the second strainmeter weakly affects correlation). 

A similar calculation yields the correlation between seismic strainmeter and gravity perturbation:
\beq
\begin{split}
\langle (\vec e_1\cdot\mathbf{h}(\vec r_1,&\omega)\cdot\vec e_1),(\vec e_2\cdot\delta\vec a(\vec r_2,\omega))\rangle
=4\pi G\rho_0S(h_n^{\rm P};\omega)\frac{1}{k^{\rm P}}\\
&\cdot\bigg(T_1(\Phi_{12})((\vec e_2\cdot\vec e_{12}) +2(\vec e_1\cdot\vec e_{12})(\vec e_1\cdot\vec e_2)) +T_2(\Phi_{12})(\vec e_{12}\cdot\vec e_1)^2(\vec e_{12}\cdot\vec e_2)\bigg)\\[0.3cm]
T_1(\Phi_{12})&= j_1(\Phi_{\rm 12})+j_3(\Phi_{\rm 12})\\
T_2(\Phi_{12})&= -5j_3(\Phi_{\rm 12})
\end{split}
\label{eq:corrNNstrain}
\eeq
The Wiener-filter cancellation using seismic strainmeters is independent of the mixing ratio. However, in contrast to the seismometer case, a strainmeter located at the test mass has zero correlation with the gravity perturbation. Therefore, a strainmeter located near the test mass can only have an indirect effect on the Wiener filter, such as improving the ability of a sensor array to disentangle shear and compressional waves. Without other seismic sensors, a strainmeter near the test mass is fully useless for the purpose of Newtonian-noise cancellation.
\epubtkImage{}{
    \begin{figure}[htbp]
    \centerline{
        \includegraphics[width=0.45\textwidth]{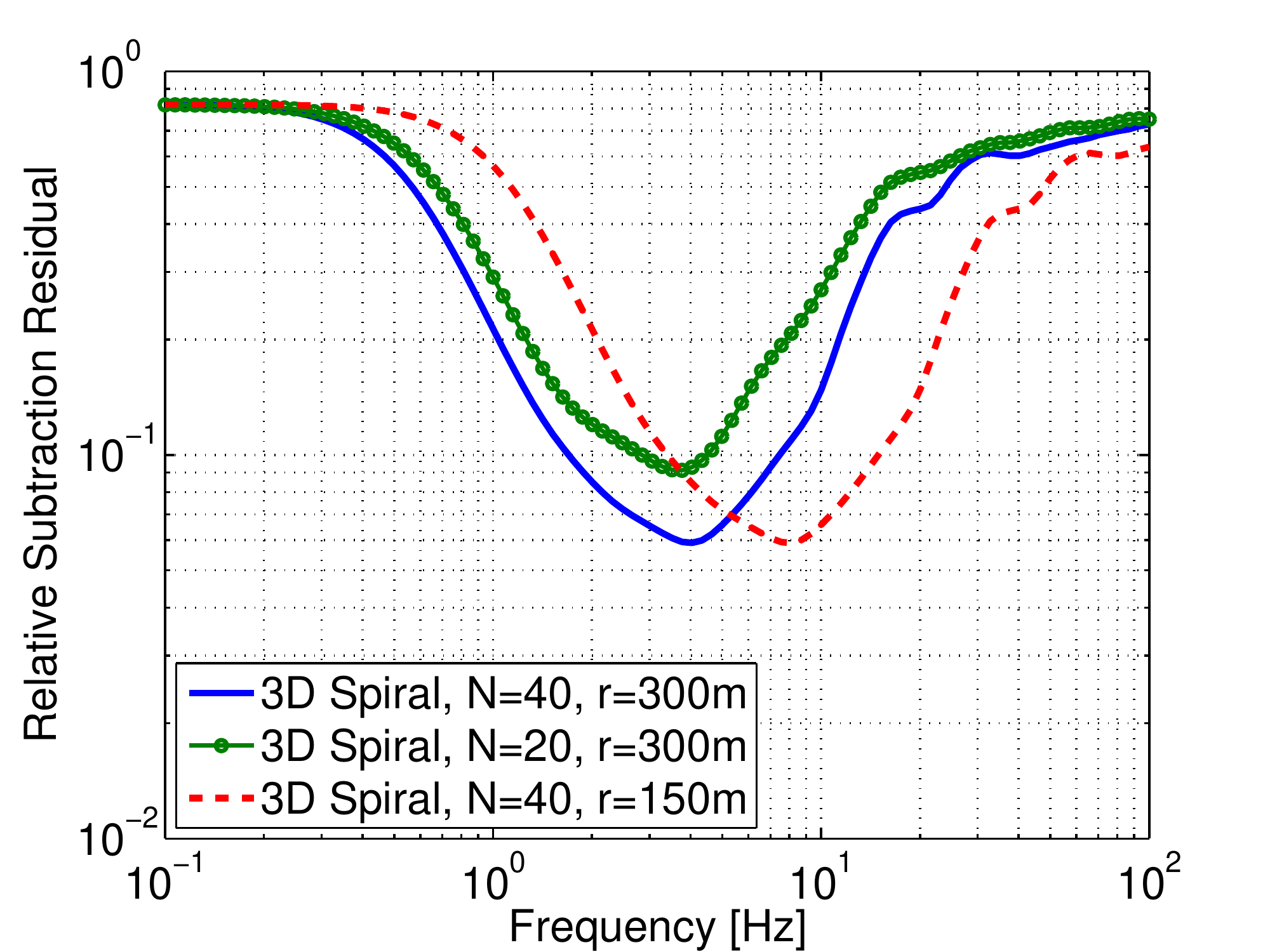}
        \includegraphics[width=0.45\textwidth]{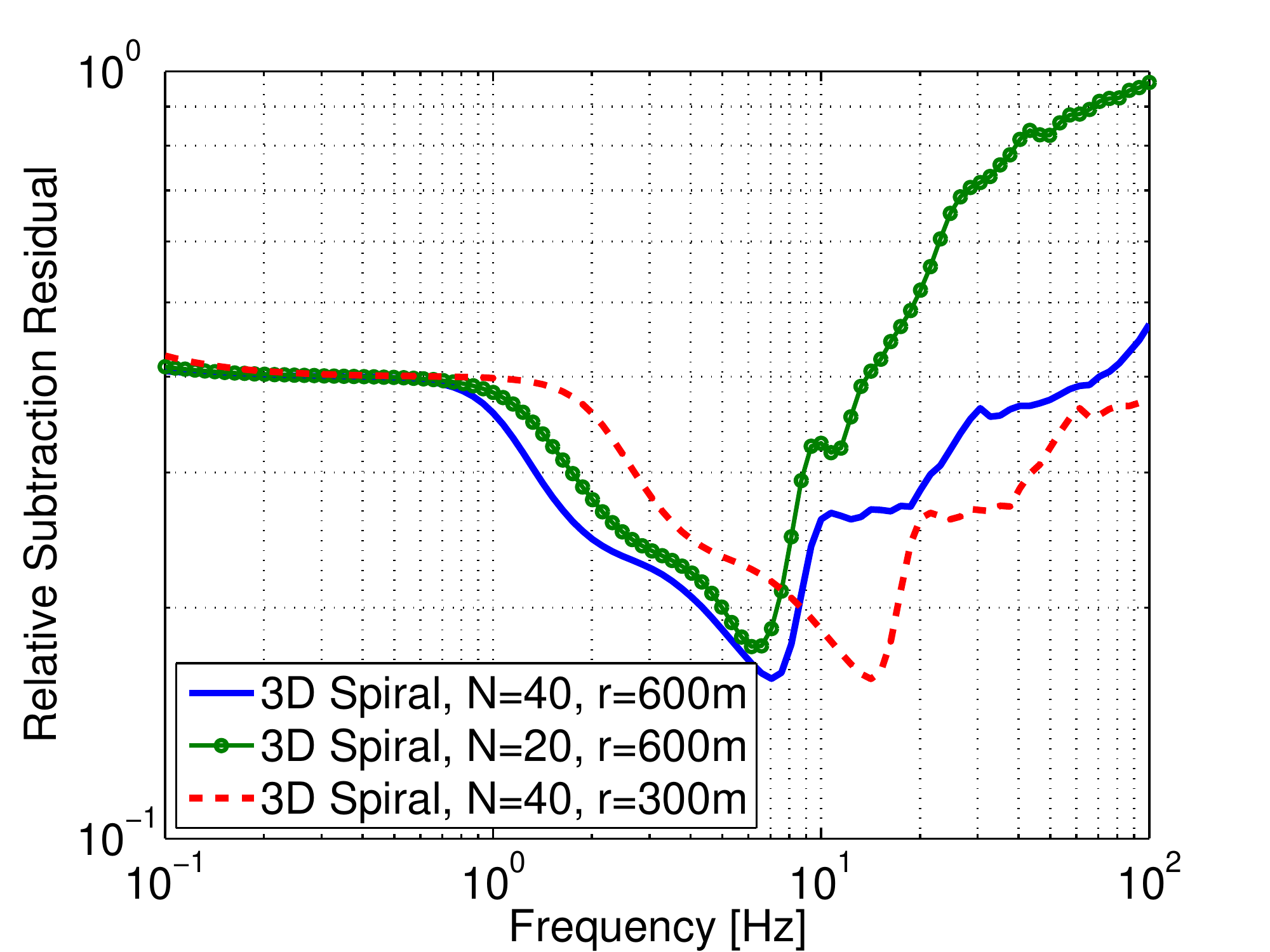}}
    \caption[Residual spectra using seismic displacement and strain sensors]{Residual spectra using seismic displacement and strain sensors in a 3D spiral array configuration. Left: seismometer array. Right: strain-meter array.}
    \label{fig:residualXiStrain}
    \end{figure}}
The subtraction residuals from a strainmeter array are shown in the right of Figure \ref{fig:residualXiStrain}. The array configuration is the same 3D spiral array used for the seismometer array in the left plot, but with twice as large extent to have peak performance at similar frequencies. All strainmeters are oriented parallel to the relevant direction of gravity perturbations. Apparently, there is no advantage in using strain-meter arrays even though the subtraction performance is independent of shear-wave content. It should be emphasized though that subtraction performance of 3D arrays depends strongly on the array configuration. Therefore, optimized array configurations may perform substantially better, and it is also possible that orienting sensors along different directions, and combining strainmeters with seismometers leads to lower subtraction residuals. This needs to be investigated in the future.

\subsubsection{Cancellation of Newtonian noise from infrasound}
\label{sec:arrayNNatm}
Coherent cancellation of Newtonian noise from infrasound is substantially different from the seismic case. Seismic sensors are substituted by microphones, which have more complicated antenna patterns. Here we will assume that a microphone measures the pressure fluctuations at a point without being able to distinguish directions. This is an important difference to seismic sensing. Furthermore, it is unfeasible to deploy a 3D array of microphones in the atmosphere. There may be other methods of sensing pressure fluctuations (e.~g.~some type of light/radar tomography of the atmosphere around the test masses), but it is unclear if they can be used to resolve the fast, relatively small-scale fluctuations produced by infrasound. So for now, we assume that pressure fluctuations can only be measured on surface. We also want to stress that we have not succeeded yet to calculate the correlation functions for the Wiener filtering in the case of a test mass underground. One can probably make progress in this direction starting with the scalar plane-wave expansion in Equation (\ref{eq:pwscalar}) and using the half-space integral in Equation (\ref{eq:intrelY}), but we will leave this as future work. In the following, we consider the test mass and all microphones to be located on the surface. In this case, the two-point spatial correlation is found to be
\beq
\begin{split}
\langle \delta p(\vec \varrho_1,\omega),\delta p(\vec \varrho_2,\omega)\rangle&= \frac{S(\delta p;\omega)}{4\pi}\int\drm\Omega_k\e^{\irm\vec k^{\rm P}\cdot(\vec \varrho_2-\vec \varrho_1)}\\
&= S(\delta p;\omega) j_0(k^{\rm P}|\vec \varrho_2-\vec \varrho_1|)
\end{split}
\label{eq:corrpress}
\eeq
This can be calculated starting with the plane-wave expansion in Equation (\ref{eq:pwscalar}), and using Equations (\ref{eq:spheraddition}) and (\ref{eq:normalY}). Note that it makes no difference for microphones at $z_0=0$ that sound waves are reflected from the surface (apart from a doubling of the amplitude). This also means that the direction average can be carried out over the full solid angle. For $z_0>0$, one has to be more careful, explicitly include the reflection of sound waves, and only average over propagation directions incident ``from the sky'' (assuming also that there are no sources of infrasound on the surface). 

The correlation between pressure fluctuations and resulting gravity perturbations at the surface can be calculated using the negative gradient of Equation (\ref{eq:gravinfra}). Since the projection of $\delta \vec a$ onto the $x$-coordinate can be technically obtained by calculating the derivative $\partial_x$, with $x\equiv x_2-x_1$ and Equation (\ref{eq:corrpress}), we find
\beq
\begin{split}
\langle \delta a_x(\vec \varrho_1,\omega),\delta p(\vec \varrho_2,\omega)\rangle&= -\frac{S(\delta p;\omega)}{(k^{\rm P})^2}\frac{G\rho_0}{\gamma\,p_0}\partial_x\int\drm\Omega_k\e^{\irm\vec k^{\rm P}_\varrho\cdot(\vec \varrho_2-\vec \varrho_1)}\\
&= 4\pi\frac{S(\delta p;\omega)}{k^{\rm P}}\frac{G\rho_0}{\gamma\,p_0}\frac{x_2-x_1}{|\vec \varrho_2-\vec \varrho_1|}j_1(k^{\rm P}|\vec \varrho_2-\vec \varrho_1|)
\end{split}
\eeq
The correlation vanishes for microphones collocated with the test mass. For this reason, the optimization of a microphone array is similar to the optimization of a surface seismometer array for Rayleigh-wave Newtonian-noise cancellation. Formally, the difference is that spherical Bessel functions determine correlations of the infrasound field instead of ordinary Bessel functions since the infrasound field is three dimensional. This results in a slightly weaker correlation of microphones near the test mass with gravity perturbations. 
\epubtkImage{}{
    \begin{figure}[htbp]
    \centerline{\includegraphics[width=0.6\textwidth]{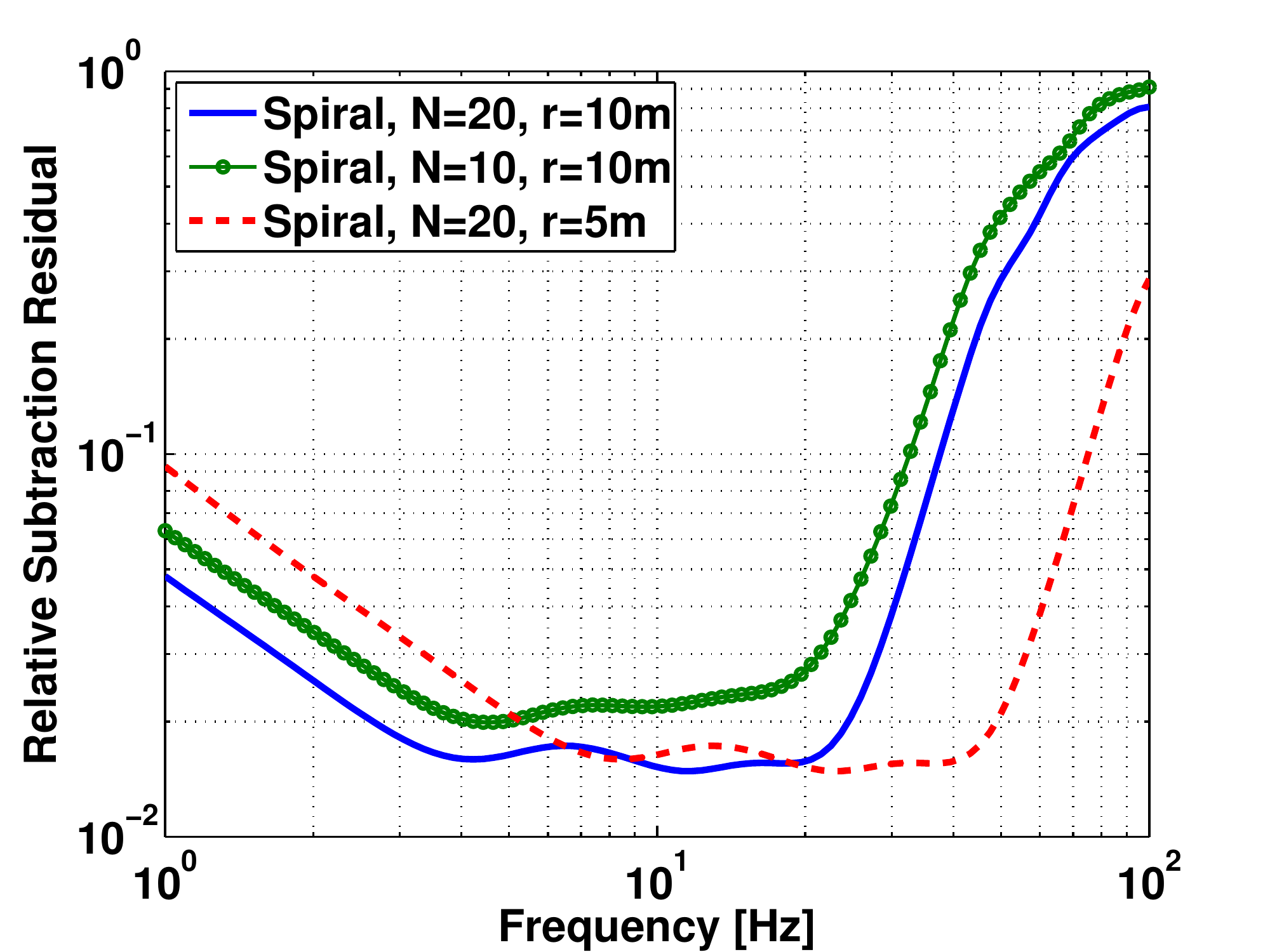}}
    \caption[Residuals of infrasound Newtonian-noise cancellation]{Residual spectra after coherent subtraction of infrasound Newtonian-noise. Signal-to-noise ratio of microphones is assumed to be 100 over all frequencies. The spirals have two full windings around the test mass.}
    \label{fig:residualSound}
    \end{figure}}
The residual spectra using spiral surface arrays of microphones can be seen in Figure \ref{fig:residualSound}. The sensors have a signal-to-noise ratio of 100. Important to realize is that the arrays are very small, and therefore located completely or partially inside the buildings hosting the test masses. In this case the assumptions of an isotropic and homogeneous infrasound field may not be fulfilled. Nonetheless, based on detailed studies of infrasound correlation, it is always be possible to achieve similar noise residuals, potentially with a somewhat increased number of microphones. 

As a final remark, infrasound waves have properties that are very similar to compressional seismic waves, and the result of Section \ref{sec:arrayNNRay} was that broadband cancellation fo Newtonian noise from compressional waves can be achieved with primitive array designs, provided that the field is not mixed with shear waves. Air does not support the propagation of shear waves, so one might wonder why subtraction of infrasound Newtonian noise does not have these nice properties. The reason lies in the sensors. Microphones provide different information. In a way, they are more similar in their response to seismic strainmeters. According to Equation (\ref{eq:corrNNstrain}), correlations between a strainmeter and gravity perturbations also vanishes if the strainmeter is located at the test mass. What this means though is that a different method to monitor infrasound waves may make a big difference. It is a ``game with gradients''. One could either monitor pressure gradients, or the displacement of air particles due to pressure fluctuations. Both would restore correlations of sensors at the test mass with gravity perturbations.

\subsubsection{Demonstration: Newtonian noise in gravimeters}
\label{sec:gravimeterNN}
The problem of coherent cancellation of Newtonian noise as described in the previous sections is not entirely new. Gravimeters are sensitive to gravity perturbations caused by redistribution of air mass in the atmosphere \cite{Neu2010}. These changes can be monitored through their effect on atmospheric pressure. For this reason, pressure sensors are deployed together with gravimeters for a coherent cancellation of atmospheric Newtonian noise \cite{BaCr1999}. In light of the results presented in Section \ref{sec:arrayNNatm}, it should be emphasized that the cancellation is significantly less challenging in gravimeters since the pressure field is not a complicated average over many sound waves propagating in all directions. This does not mean though that modelling these perturbations is less challenging. Accurate calculations based on Green's functions are based on spherical Earth models, and the model has to include the additional effect that a change in the mass of an air column changes the load on the surface, and thereby produces additional correlations with the gravimeter signal \cite{GuEA2004}. Nonetheless, from a practical point of view, the full result is more similar to the coherent relations such as Equation (\ref{eq:gravinfra}), which means that local sensing of pressure fluctuations should yield good cancellation performance. 

This is indeed the case as shown in Figure \ref{fig:gravimeterNN}. The original median of gravity spectra is shown as red line. Using a very simple filter, which is based on direct proportionality of local pressure and gravity fluctuations, gravity noise can be reduced by about a factor 5 at 0.1\,mHz.
\epubtkImage{}{
    \begin{figure}[htbp]
    \centerline{\includegraphics[width=0.8\textwidth]{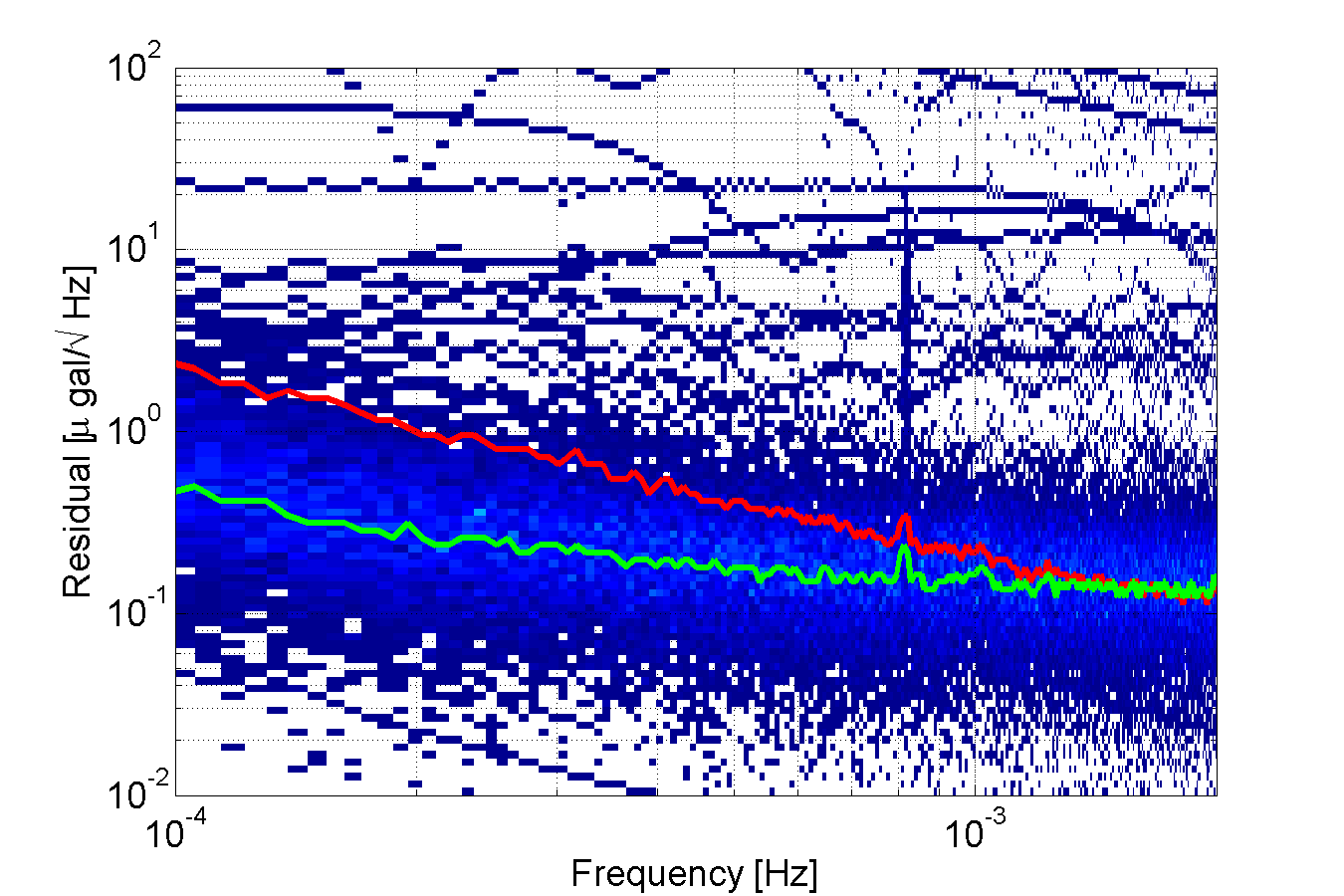}}
    \caption[Noise cancellation in gravimeters]{Spectral histogram of subtraction residuals using a local pressure sensor as reference channel. The two solid curves correspond to the medians of spectral histograms before (red) and after (green) subtraction.}
    \label{fig:gravimeterNN}
    \end{figure}}
The subtraction residuals are close to the instrumental noise of the gravimeters, which means that the simple scheme based on proportionality of the data is already very effective at these frequencies. Especially at lower frequencies, the filter design needs to be more complicated to achieve good broadband cancellation performance. Typically, a frequency domain version of Wiener filtering is applied in standard subtraction procedures \cite{Neu2010}. Due to non-Gaussianity and non-stationarity of the data, time-domain FIR Wiener filters as discussed in Section \ref{sec:Wiener} are less effective. We want to stress though that cancellation results are not this good in all gravimeters. Sometimes it can be explained by data quality of the pressure sensors, but often it is not clear what the reasons are. It may well be that detailed knowledge of the gravimeter sites can provide ideas for explanations. 

\subsubsection{Optimizing sensor arrays for noise cancellation}
\label{sec:optimarray}\index{array!optimization}
In the previous sections, we focussed on the design and performance evaluation of an optimal noise-cancellation filter for a given set of reference sensors. In this section, we address the problem of calculating the array configuration that minimizes noise residuals given sensor noise of a fixed number of sensors. The analysis will be restricted to homogeneous fields of density perturbations. The optimization can be based on a model or measured two-point spatial correlations $C(\delta\rho;\vec r,\omega)$. We start with a general discussion and later present results for the isotropic Rayleigh-wave field. 

The optimization problem will be formulated as a minimization of the noise residual $R$ defined in Equation (\ref{eq:residualNN}) as a function of sensor locations $\vec r_i$. Accordingly, the optimal sensor locations fulfill the equation
\beq
\nabla_k R=\vec 0,
\label{eq:noisemin}
\eeq
where the derivatives are calculated with respect to the coordinates of each of the $M$ sensors, i.~e.~$k\in 1,\ldots,M$. In homogeneous fields, the Newtonian-noise spectrum and seismic spectrum are independent of sensor location,
\beq
\nabla_k C_{\rm NN}=\vec 0,\quad \nabla_k C_{\rm SS}^{kk}=\vec 0,
\eeq
which allows us to simplify Equation (\ref{eq:noisemin}) into
\beq
\begin{split}
&\nabla_k\vec C^{\,\rm T}_{\rm SN}\cdot \vec w^{\,\rm T}+\vec w\cdot \nabla_k \vec C_{\rm SN}-\vec w\cdot \nabla_k\mathbf C_{\rm SS}\cdot\vec w^{\,\rm T}=\vec 0, \\
&2\vec w\cdot \nabla_k\vec C_{\rm SN}=\vec w\cdot \nabla_k\mathbf C_{\rm SS}\cdot\vec w^{\,\rm T},
\end{split}
\eeq
where we have introduced the Wiener filter $\vec w=\vec C^{\,\rm T}_{\rm SN}\cdot {\mathbf C}^{-1}_{\rm SS}$. For the following steps, let us use a slightly different notation. We will write the sensor cross-correlation matrix $\mathbf C_{\rm SS}=\mathbf C(\vec s;\vec s)$, and the correlations between sensor and target channels as $\vec C_{\rm SN}=\vec C(\vec s;n)$. Only the component $k$ of the vector $\vec C(\vec s;n)$ and the $k$th row and column of $\mathbf C(\vec s;\vec s)$ depend on the coordinates of the sensor $k$. This means that the derivative $\nabla_k$ produces many zeros in the last equation, which allows us to simplify it into the following form:
\beq
\nabla_kC(s_k;n)-\vec w\cdot \nabla_k\vec C(\vec s;s_k)=0.
\label{eq:optimal}
\eeq
The optimal array fulfills this equation for derivatives with respect to the coordinates of all $M$ sensors. Solutions to this equation need to be calculated numerically. Optimization of arrays using Equation (\ref{eq:optimal}) produces accurate solutions more quickly than traditional optimization methods, which directly attempt to find the global minimum of the residual $R$. Traditional codes (nested sampling, particle swarm optimization) produce solutions that converge to the ones obtained by solving Equation (\ref{eq:optimal}).

In the following, we will we present optimization results for a homogeneous and isotropic Rayleigh-wave field. The correlation functions are given in Equation (\ref{eq:corrRayiso}) and (\ref{eq:corraxiRayiso}). The filled contour plot in Figure \ref{fig:optRayNN} shows the residual $R$ as a function of sensor coordinates for a total of 1 to 3 sensors, from left to right. In the case of a single sensor, the axes represent its $x$ and $y$ coordinates. For more than one sensor, the axes correspond to the $x$ coordinates of two sensors. All coordinates not shown in these plots assume their optimal values. 
\epubtkImage{}{
    \begin{figure}[htbp]
    \centerline{
        \includegraphics[width=0.32\textwidth]{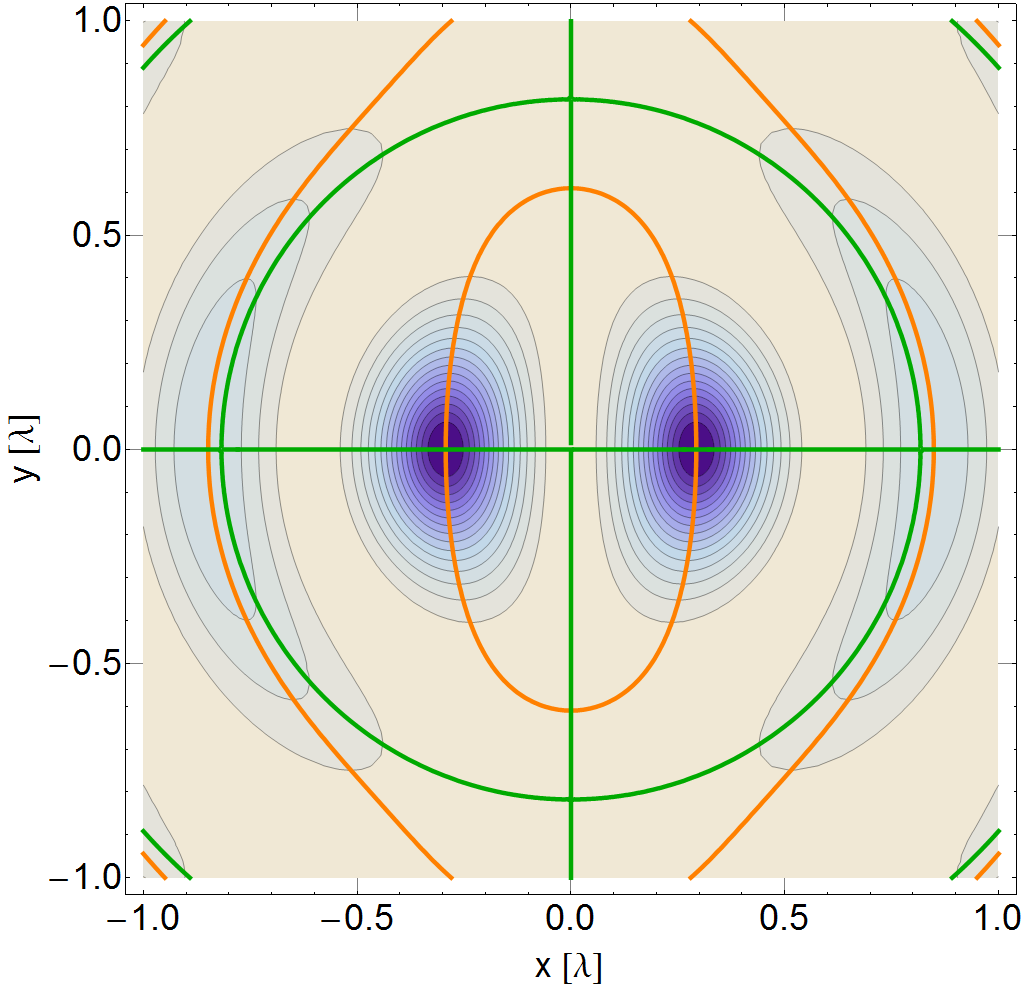}
        \includegraphics[width=0.32\textwidth]{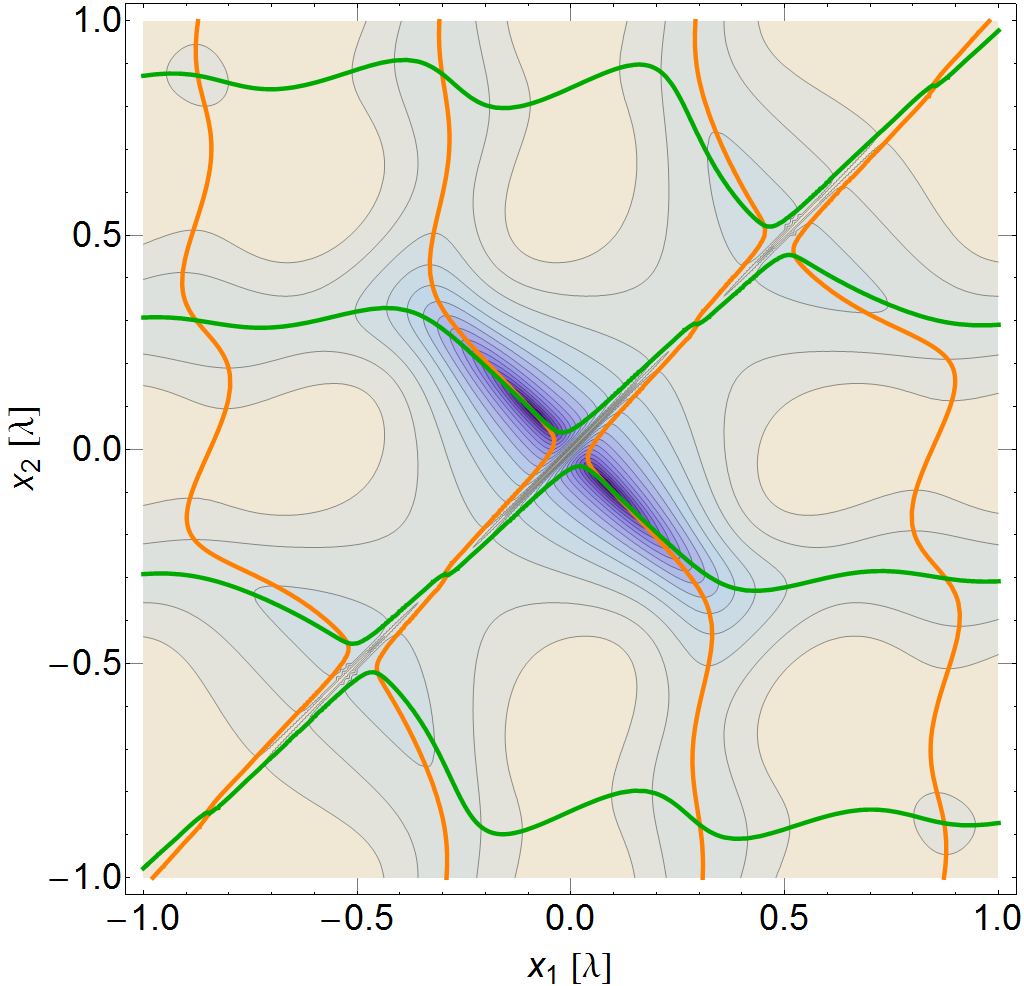}         
        \includegraphics[width=0.32\textwidth]{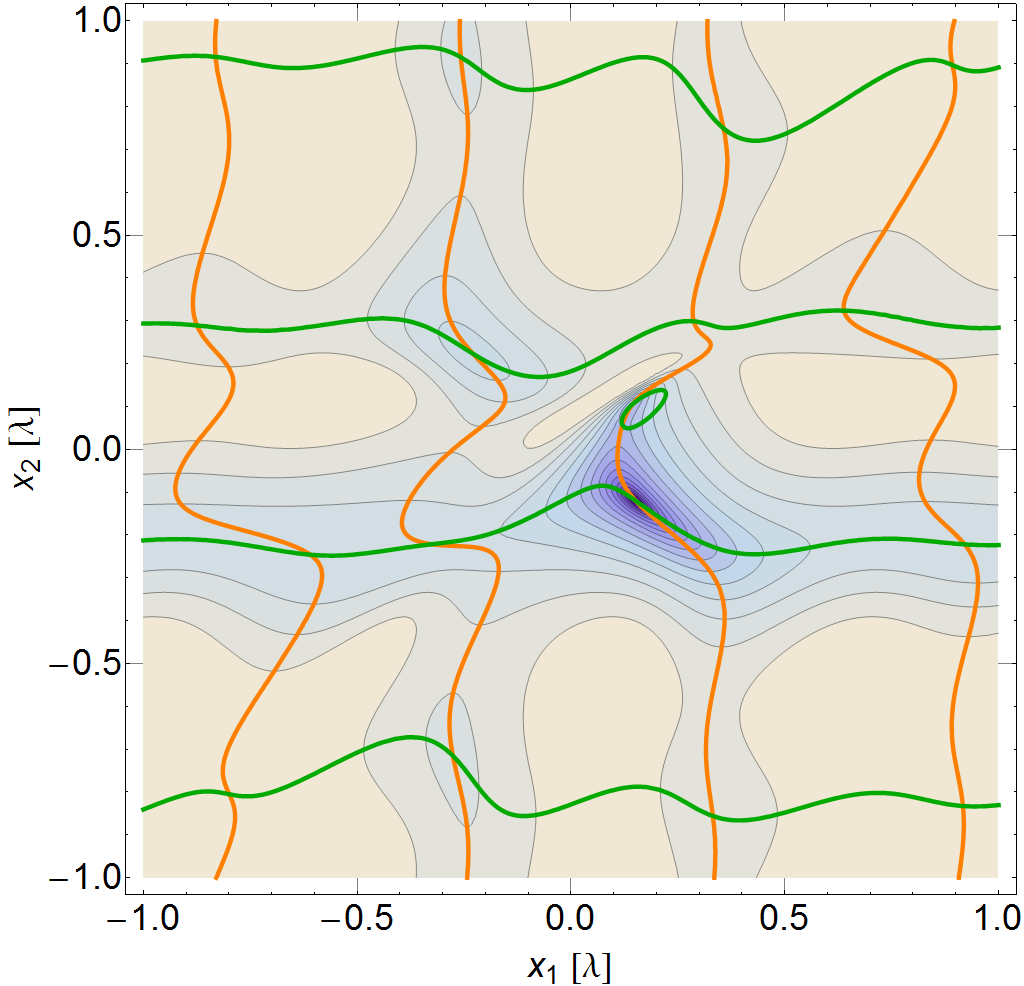}}
    \caption[Array optimization for cancellation of Rayleigh Newtonian noise]{Array optimization for cancellation of Rayleigh Newtonian noise. The array is optimized for cancellation at a single frequency using 1 to 3 sensors (left to right) with $\rm SNR=100$. The curves represent Equation (\ref{eq:optimal}) for the coordinates in the axis labels. The filled contour plots show the noise residual $R$. All coordinates not shown assume their optimal values.}
    \label{fig:optRayNN}
    \end{figure}}

The green and orange curves represent Equation (\ref{eq:optimal}) either for the derivatives $\partial_x,\,\partial_y$ or $\partial_{x_1},\,\partial_{x_2}$. These curves need to intersect at the optimal coordinates. It can be seen that they intersect multiple times. The numerical search for the optimal array needs to find the intersection that belongs to the minimum value of $R$. For the isotropic case, it is not difficult though to tune the numerical search such that the global minimum is found quickly. The optimal intersection is always the one closest to the test mass at the origin. While it is unclear if this holds for all homogeneous seismic fields, it seems intuitive at least that one should search intersections close to the test mass in general. 

In order to find optimal arrays with many sensors, it is recommended to build these solutions gradually from optimal solutions with one less sensor. In other words, for the initial placement, one should use locations of the $M-1$ optimal array, and then add another sensor randomly nearby the test mass. The search relocates all sensors, but it turns out that sensors of an optimal array with a total of $M-1$ sensors only move by a bit to take their optimal positions in an optimal array with $M$ sensors. So choosing initial positions in the numerical search wisely significantly decreases computation time, and greatly reduces the risk to get trapped in local minima.
\epubtkImage{}{
    \begin{figure}[htbp]
    \centerline{
        \includegraphics[width=0.6\textwidth]{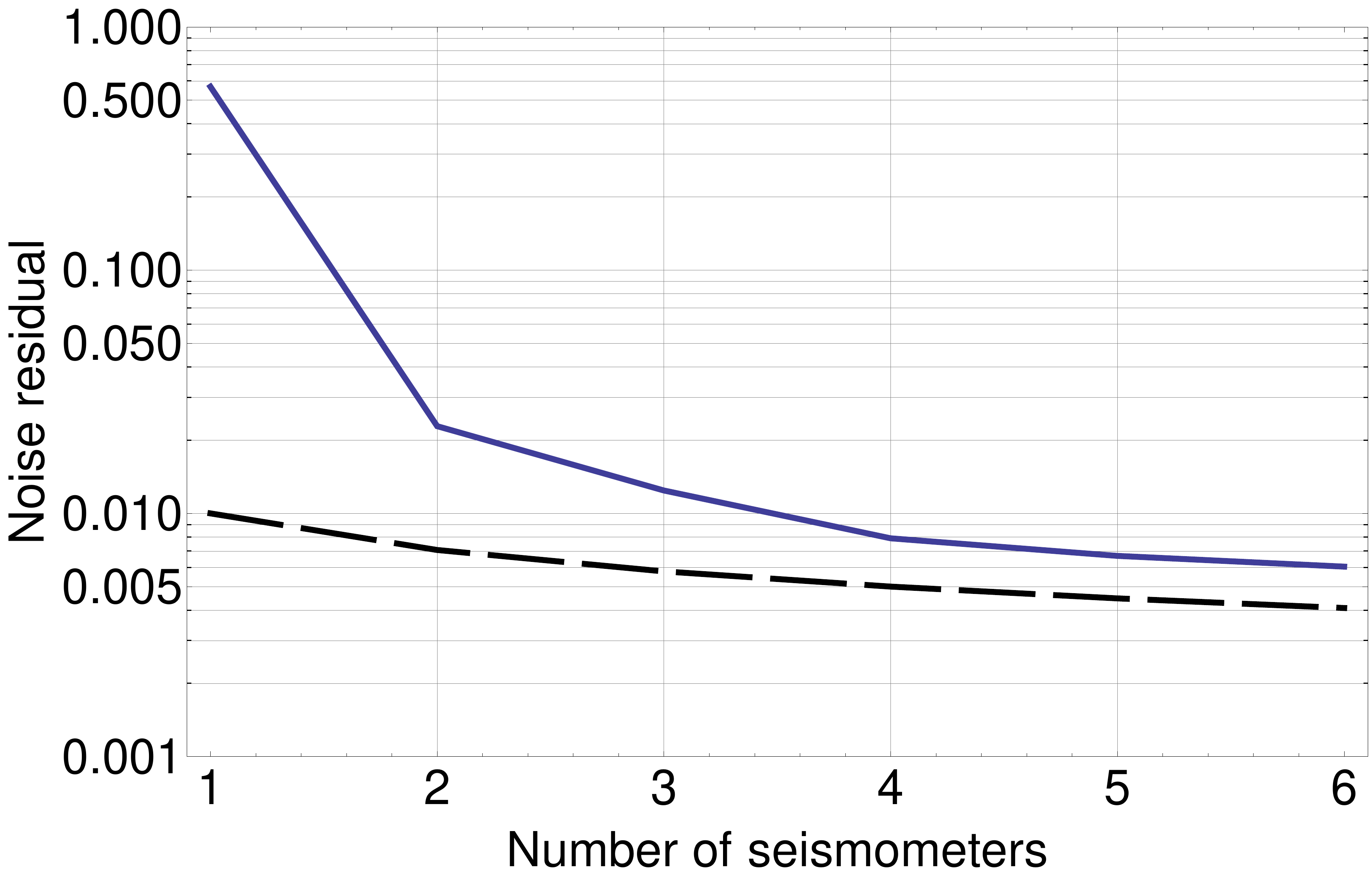}}
    \caption[Minimized noise residuals from Rayleigh Newtonian-noise cancellation]{Minimized noise residuals from Rayleigh Newtonian-noise cancellation. The dashed line marks the sensor-noise limit.}
    \label{fig:optResiduals}
    \end{figure}}
    
Figure \ref{fig:optResiduals} shows the noise residuals of Newtonian noise from an isotropic Rayleigh-wave field using optimal arrays with 1 to 6 sensors and sensor $\rm SNR=100$. The residuals are compared with the sensor-noise limit ${\rm 1/SNR}/\sqrt{M}$ (dashed curve). Arrays with $M>3$ yield residuals that are close to a factor $\sqrt{2}$ above the sensor-noise limit. The origin of the factor $\sqrt{2}$ has not been explained yet. It does not appear in all noise residuals, for example, the noise residual of a Wiener filter using a single reference channel perfectly correlated with the target channel, see Equation (\ref{eq:singlechcoh}), is given by 1/SNR. 

In many situations, it will not be possible to model the correlations $\mathbf C_{\rm SS}$ and $\vec C_{\rm SN}$. In this case, observations of seismic correlations $\mathbf C_{\rm SS}$ can be used to calculate $\vec C_{\rm SN}$, see Equation (\ref{eq:kernelRayaxi}), and also $C_{\rm NN}$, see Equation (\ref{eq:homNNC}). Seismic correlations are observed with seismometer arrays. It is recommended to choose a number of seismometers for this measurement that is significantly higher than the number of seismometers foreseen for the noise cancellation. Otherwise, aliasing effects and resolution limits can severely impact the correlation estimates. Various array-processing algorithms are discussed in \cite{KrVi1996}.

Table \ref{tab:residualRf} summarizes the noise residuals from optimized arrays of 1 to 6 sensors with SNR = 100, which may serve as reference values for alternative optimization methods. The $N=7$ array is the first optimal array that requires two seismometers placed on top of each other. Consequently, the broadband performance of the $N=6$ array is similar to the $N=7$ array. Residuals of optimal arrays can be compared with the stepwise optimized arrays as discussed in Section \ref{sec:arrayNNRay}, taking into account that $\rm SNR = 10$ was used in Section \ref{sec:arrayNNRay}.
\begin{table}[htbp]
\caption{Cancellation of Newtonian noise from isotropic Rayleigh-wave fields at wavelength $\lambda$. Shown are the optimal arrays for 1 to 6 sensors with SNR = 100.}
\label{tab:residualRf}
\renewcommand{\arraystretch}{1.5}
\centerline{
\begin{tabularx}{0.7\textwidth}{|X|l|}
\hline
Sensor coordinates $[\lambda]$ & Noise residual $\sqrt{R}$ \\
\hline
(0.293,0) & 0.568\\
(0.087,0), (-0.087,0) & $2.28\times 10^{-2}$\\
(0.152,-0.103), (0.152,0.103), (-0.120,0) & $1.24\times 10^{-2}$\\
(0.194,0.112), (0.194,-0.112), (-0.194,0.112), \newline(-0.194,-0.112) & $7.90\times 10^{-3}$\\
(0.191,0.215), (0.299,0), (0.191,-0.215), \newline(-0.226,0.116), (-0.226,-0.116) & $6.69\times 10^{-3}$\\
(0.206,0.196), (0.295,0), (0.206,-0.196), \newline(-0.206,0.196), (-0.295,0), (-0.206,-0.196) & $6.04\times 10^{-3}$\\
\hline
\end{tabularx}}
\end{table}
The noise residuals of the stepwise optimization were $R=0.38$, 0.09, and 0.07 for the first three seismometers, while the fully optimized residuals are $R=0.38$, 0.014 and 0.0074, i.~e.~much lower for $N\geq2$.

\subsubsection{Newtonian noise cancellation using gravity sensors}
\label{sec:gravsub}
In the previous sections, we have investigated Newtonian-noise cancellation using auxiliary sensors that monitor density fluctuations near the test masses. An alternative that has been discussed in the past is to use gravity sensors instead. One general concern about this scheme is that a device able to subtract gravity noise can also cancel GW signals. This fact indeed limits the possible realizations of such a scheme, but it is shown in the following that at least Newtonian noise in large-scale GW detectors from a Rayleigh-wave field can be cancelled using auxiliary gravity sensors. However, it will become clear as well that it will be extremely challenging to build a gravity sensor with the required sensitivity.

In the following discussion, we will focus on cancellation of gravity noise from isotropic Rayleigh-wave fields. Most of the results can be obtained from the two-point spatial correlation of gravity fluctuations, \index{correlation!gravity-gravity}
\beq
\langle \delta a_x(\vec 0,\omega), \delta a_x(\vec \varrho,\omega)\rangle =\big(2\pi G\rho_0 \gamma \e^{-hk_\varrho}\big)^2\frac{1}{2}S(\xi_z;\omega)\cdot\left[J_0(k_\varrho \rho)-\cos(2\phi)J_2(k_\varrho \rho)\right],
\label{eq:corrNNacc}
\eeq
evaluated at a specific frequency. Here, $\vec\varrho=\varrho(\cos(\phi),\sin(\phi))$, and $k_\varrho$ is the wavenumber of a Rayleigh wave. This result turns into Equation (\ref{eq:specNNiso}) for $\varrho\rightarrow 0$.

The only (conventional) type of gravity sensor that can be used to cancel Newtonian noise in GW detectors is the gravity strainmeter or gravity gradiometer \footnote{Here, we do not consider using seismic data from a gravimeter for Newtonian-noise cancellation.}. As we have discussed in Section \ref{sec:superg}, the sensitivity of gravimeters is fundamentally limited by seismic noise, and any attempt to mitigate seismic noise in gravimeters inevitably transforms its response into a gravity gradiometer type. So in the following, we will only consider gravity strainmeters/gradiometers as auxiliary sensors. 

Let us first discuss a few scenarios where noise cancellation cannot be achieved. If two identical large-scale GW detectors are side-by-side, i.~e.~with test masses approximately at the same locations, then Newtonian-noise cancellation by subtracting their data inevitably means that GW signals are also cancelled. Let us make the arms of one of the two detectors shorter, with both detectors' test masses at the corner station staying collocated. Already one detector being shorter than the other by a few meters reduces Newtonian-noise correlation between the two detectors substantially. The reason is that correlation of gravity fluctuations between the end test masses falls rapidly with distance according to Equation (\ref{eq:corrNNacc}). It can be verified that subtracting data of these two detectors to cancel at least gravity perturbations of the inner test masses does not lead to sensitivity improvements. Instead, it effectively changes the arm length of the combined detector to $\Delta L$, where $\Delta L$ is the difference of arm lengths of the two detectors, and correspondingly increases Newtonian noise.

If Newtonian noise is uncorrelated between two test masses of one arm, then decreasing arm length increases Newtonian strain noise. However, as shown in Figure \ref{fig:rayResp}, if the detector becomes shorter than a seismic wavelength and Newtonian noise starts to be correlated between test masses, Newtonian strain noise does not increase further. Compared to the Newtonian noise in a large-scale detector with arm length $L$, Newtonian noise in the short detector is greater by (up to) a factor $k_\varrho L$. In this regime, the small gravity strainmeter is better described as gravity gradiometer. The common-mode suppression of Newtonian noise in the gradiometer due to correlation between test masses greatly reduces Newtonian-noise correlation between gradiometer and the inner test masses of the large-scale detector. Consequently, a gravity gradiometer cannot be used for noise cancellation in this specific configuration.

It turns out though that there is a class of gravity gradiometers, known as full-tensor gradiometers, that can be used for cancellation of Newtonian noise from Rayleigh waves. The key is to understand that gravity gradients $\partial_z\delta a_x = \partial_x\delta a_z$, where $\delta\vec a$ are the fluctuations of gravity acceleration, and $x$ points along the arm of the large-scale detector, are perfectly correlated with $\delta a_x$. This can be seen from Equation (\ref{eq:RayleighS}), since derivatives of the acceleration $\delta a_x$ with respect to $z$, i.~e.~the vertical direction, does not change the dependence on directions $\phi$. The coherence (normalized correlation) between $\delta a_x$ and $\partial_z\delta a_x$ is shown in the left of Figure \ref{fig:gradLIGOCancel} making use of $\langle\delta a_x(\vec 0\,),\partial_z\delta a_x(\vec \varrho\,)\rangle_{\rm norm}=\langle\delta a_x(\vec 0\,),\delta a_x(\vec \varrho\,)\rangle_{\rm norm}$.

\epubtkImage{}{
    \begin{figure}[htbp]
    \centerline{
        \includegraphics[width=0.48\textwidth]{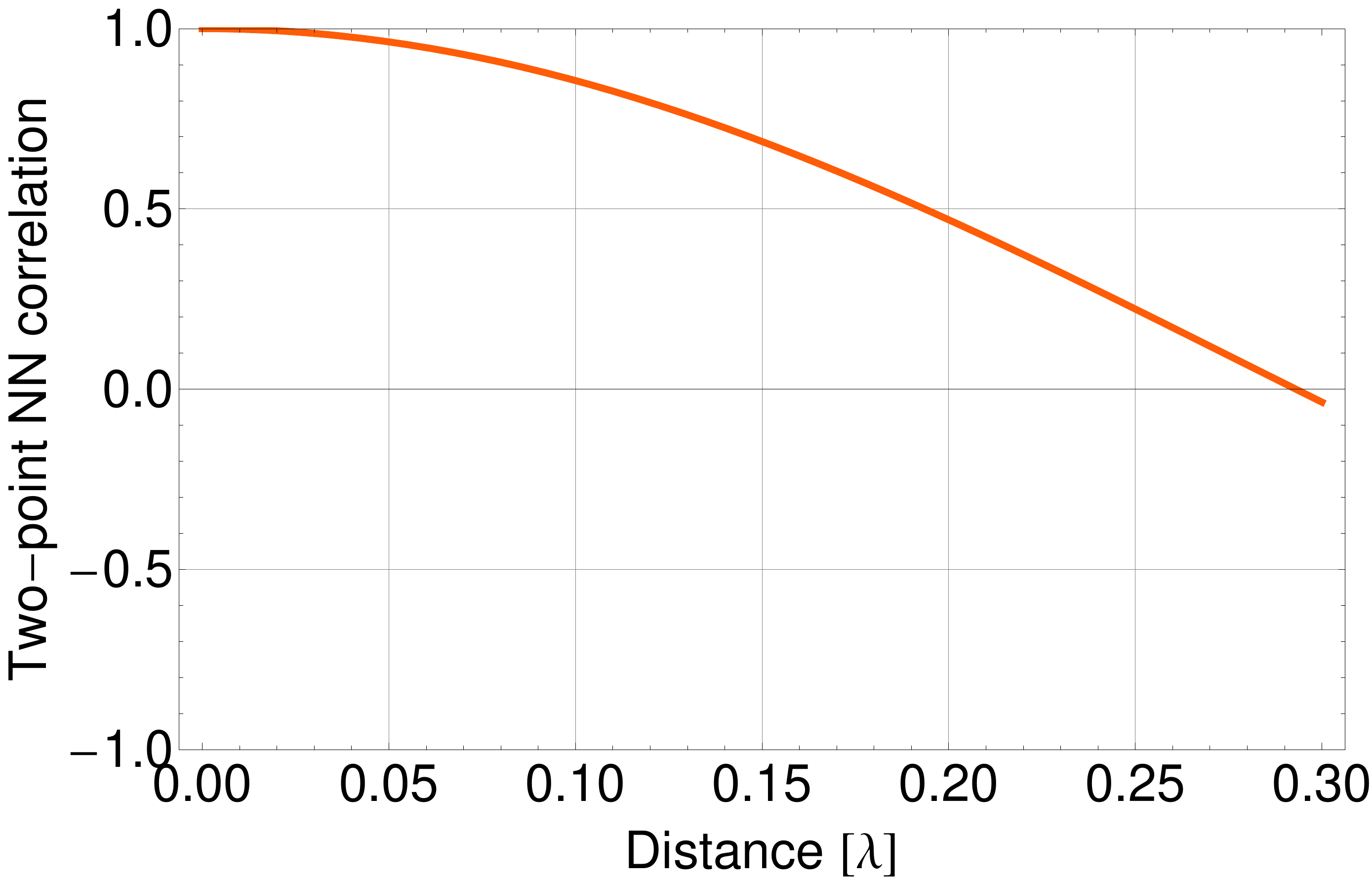}
        \includegraphics[width=0.48\textwidth]{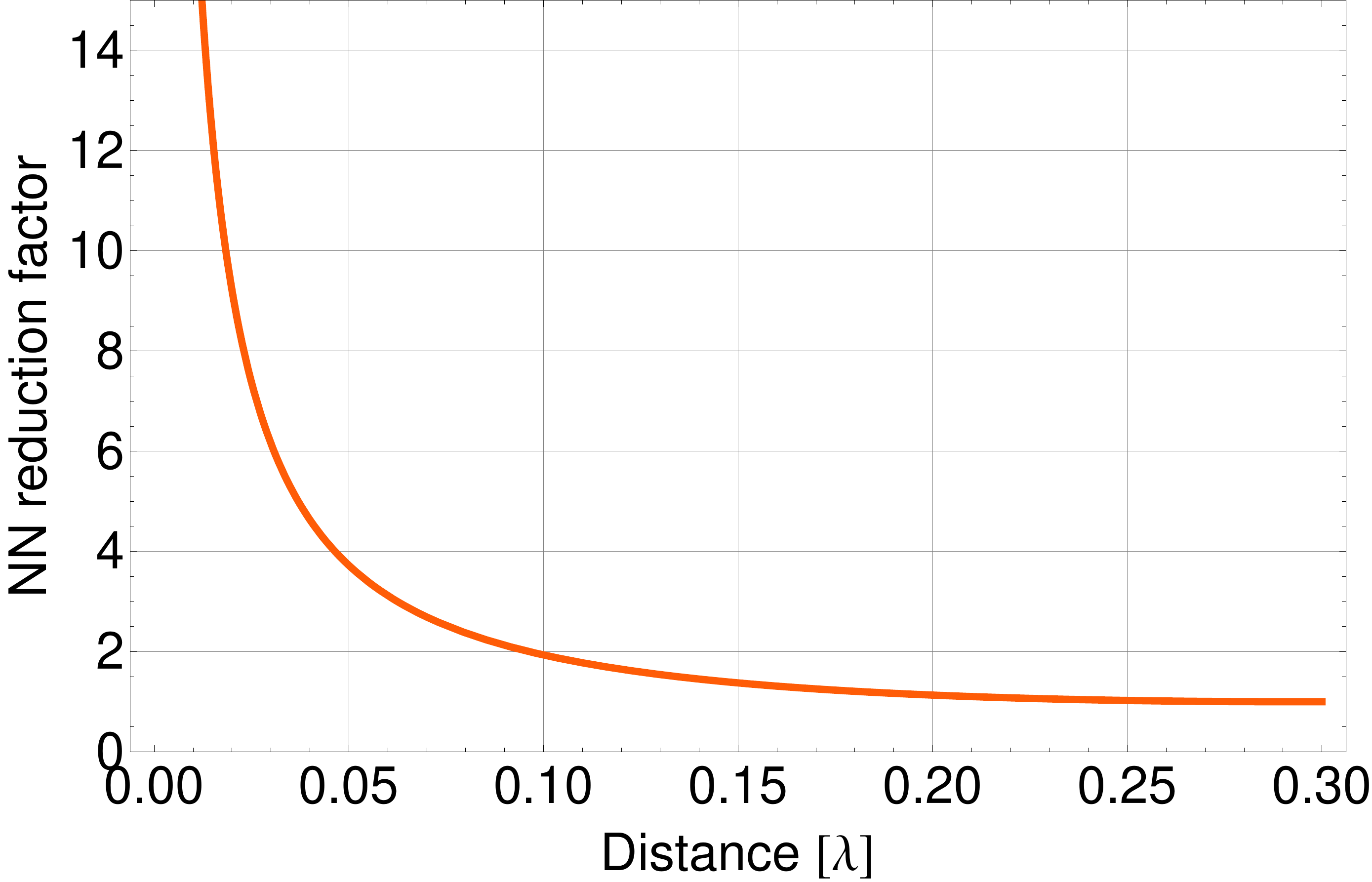}}
    \caption[Newtonian-noise cancellation using gravity gradiometers]{Left: coherence of Newtonian noise between two test masses according to Equation (\ref{eq:corrNNacc}) as a function of distance $\varrho$ in units of seismic wavelength ($\phi=0$). Right: maximal noise reduction that can be achieved with the channel $\partial_z\delta a_x$ of a gravity gradiometer as a function of maximal distance between test masses of the gravity gradiometer and the large-scale detector.}
    \label{fig:gradLIGOCancel}
    \end{figure}}
The idea is now to place one full-tensor gravity gradiometer at each test mass of the large-scale detector, and to cancel Newtonian noise of each mass. In this way, it is also impossible to cancel GW signals since GW signals of the gradiometers cancel each other. The limitations of this scheme are determined by the distance between the test mass of the large-scale detector and test-masses of the gravity gradiometer. The smaller the distance, the better the correlation and the higher the achievable noise reduction. Using Equation (\ref{eq:singlechcoh}), the maximal noise reduction can be calculated as a function of the coherence. In Figure \ref{fig:gradLIGOCancel}, right plot, the achievable noise suppression is shown as a function of distance between test masses. For example, at 10\,Hz, and assuming a Rayleigh-wave speed of 250\,m/s, the distance needs to be smaller than 1\,m for a factor 5 noise reduction. This also means that the size of the gradiometer must be of order 1\,m. 

Let us calculate what the required sensitivity of the gradiometer has to be. From Equations (\ref{eq:singlechcoh}) and (\ref{eq:corrNNacc}), we find that the maximal noise-suppression factor is given by
\beq
s \sim \frac{1}{k_\varrho r},
\eeq
where $r\ll\lambda$ is the distance between test masses of the large-scale detector and the gradiometer, which one can also interpret as maximal size of the gradiometer to achieve a suppression $s$. A numerical factor of order unity is omitted. Given a Newtonian strain noise $h_{\rm NN}$ of the large-scale detector with arm length $L$, the gradiometer observes
\beq
h_{\rm NN}^{\rm grad} = h_{\rm NN} k_\varrho L=\xi_{\rm NN} k_\varrho.
\eeq
Here, $\xi_{\rm NN}$ denotes the relative displacement noise in the large-scale detector. Now, the relative displacement noise in the gradiometer is
\beq
\xi_{\rm NN}^{\rm grad}=\frac{1}{k_\varrho s}\xi_{\rm NN} k_\varrho=\frac{\xi_{\rm NN}}{s}
\eeq
While the gravity gradiometer observes much stronger Newtonian noise in units strain, its displacement sensitivity needs to match the displacement sensitivity of the large-scale detector, and even exceed it by a factor $s$. One could raise well-justified doubts at this point if a meter-scale detector can achieve displacement sensitivity of large-scale GW detectors. Nonetheless, the analysis of this section has shown that Newtonian-noise cancellation using gravity sensors is in principle possible.

\subsection{Site selection}
\label{sec:siteselect}
An elegant way to reduce Newtonian noise is to select a detector site with weak gravity fluctuations. It should be relatively straightforward to avoid proximity to anthropogenic sources (except maybe for the sources that are necessarily part of the detector infrastructure), but it is not immediately obvious how efficient this approach is to mitigate seismic or atmospheric Newtonian noise. With the results of Sections \ref{sec:ambient} and \ref{sec:atmos}, and using numerous past observations of infrasound and seismic fields, we will be able to predict the possible gain from site selection. The aim is to provide general guidelines that can help to make a site-selection process more efficient, and help to identify suitable site candidates, which can be characterized in detail with follow-up measurements. These steps have been carried out recently in Europe as part of the design study of the Einstein Telescope \cite{BeEA2012,BBR2015}, and promising sites were indeed identified.

Already with respect to the minimization of Newtonian noise, site selection is a complicated process. One generally needs to divide into site selection for gravity measurements at low and high frequencies. The boundary between these two regimes typically lies at a few Hz. The point here is that at sufficiently low frequencies, gravity perturbations produced at or above surface are negligibly suppressed at underground sites with respect to surface sites. At higher frequencies, a detailed site-specific study is required to quantify the gain from underground construction since it strongly depends on local geology. In general, sources of gravity perturbations have different characteristics at lower and higher frequencies. Finally, to complicate the matter even further, one may also be interested to identify a site where one can expect to achieve high noise cancellation through Wiener filtering or similar methods. 

\subsubsection{Global surface seismicity}\index{seismic noise!surface}
We start with the assessment of ambient seismicity. Today this can be done systematically and easily for many surface locations since publicly available data from a global network of seismometers is continuously recorded and archived on servers. For example, Coughlin and Harms have characterized thousands of sites world-wide in this way, processing years of data from broadband seismometers \cite{CoHa2012b}. Among others, the data are provided by the US-based IRIS Data Management Service (archiving global seismic data), \url{http://www.iris.edu/ds/nodes/dmc/}, and the Japanese seismic broadband network F-Net operated by NIED \url{http://www.fnet.bosai.go.jp/}. Seismic data cannot be easily obtained from countries that have not signed the Comprehensive Nuclear-Test-Ban Treaty (which are few though). The results of their analysis were presented in the form of spectral histograms for each site, accessible through a Google Earth kmz file. An example is shown in Figure \ref{fig:seismicityUS} for a seismic station in the US.
\epubtkImage{}{%
    \begin{figure}[htbp]
    \centerline{\includegraphics[width=0.8\textwidth]{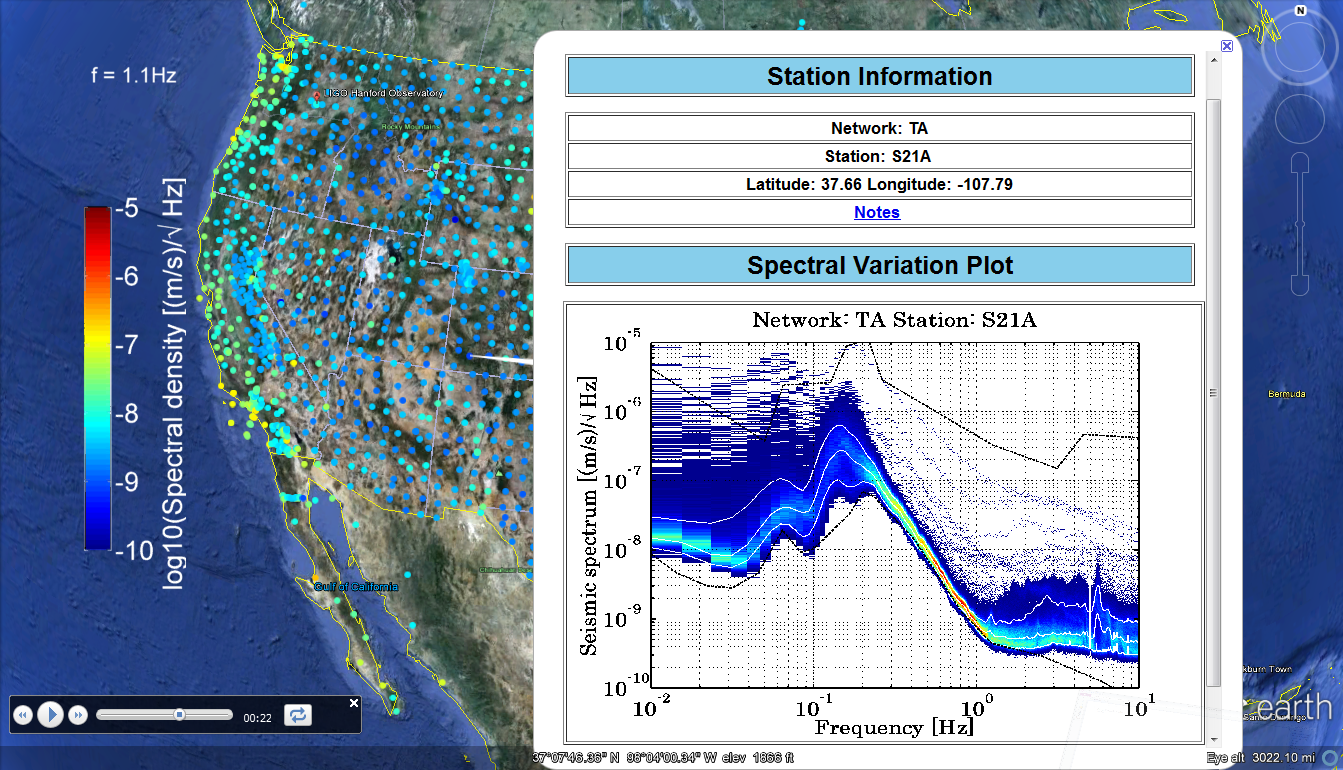}}
    \caption[US seismicity on Google Earth]{Information of world-wide ambient seismicity as a function of frequency was made available as Google Earth kmz files by Coughlin and Harms. The files can be downloaded at \url{http://www.ligo.caltech.edu/~jharms/data/GoogleEarth/}.}
    \label{fig:seismicityUS}
    \end{figure}}
The colors of the markers on the map signify the median of the spectral histograms at a specific frequency. The frequency can be changed with a video slider. Clicking on a marker pops up additional site-specific information. Studying these maps gives an idea where to find quiet places on Earth, and helps to recognize generic patterns such as the influence of mountain ranges, and the proximity to oceans. A more detailed analysis based on these data can be found in \cite{CoHa2012b}. It should be noted that especially in Japan, many seismic stations used in this study are built a few meters underground, which may lead to substantial reduction of observed ambient seismicity above a few Hz with respect to surface sites. Nonetheless, there are regions on all continents with very low surface seismicity above 1\,Hz, approaching a global minimum often referred to as global low-noise model \cite{BDE2004,CoHa2012b}. This means that one should not expect that a surface or underground site can be found on Earth that is significantly quieter than the identified quietest surface sites. Of course, underground sites may still be attractive since the risk is lower that seismicity will change in the future, while surface sites can in principle change seismicity over the course of many years, because of construction or other developments. For the same reason, it may be very challenging to find quiet surface sites in densely populated countries. As a rule of thumb, a site that is at least 50\,km away from heavy traffic and seismically active faults, and at least 100\,km away from the ocean, has a good chance to show low ambient seismicity above a few Hz. To be specific here, ambient seismicity should be understood as the quasi-stationary noise background, which excludes for example the occasional strong earthquake. Larger distances to seismically active zones may be necessary for reasons such as avoiding damage to the instrument. 

Below a few Hz, ambient seismicity is more uniform over the globe. Oceanic microseisms between 0.1\,Hz and 1\,Hz are stronger within 200\,km to the coast, and then decreasing weakly in amplitude towards larger distances. This implies that it is almost impossible to find sites with a low level of oceanic microseisms in countries such as Italy and Japan. At even lower frequencies, it seems that elevated seismic noise can mostly be explained by proximity to seismically active zones, or extreme proximity to cities or traffic. Here one needs to be careful though with the interpretation of data since quality of low-frequency data strongly depends on the quality of the seismic station. A less protected seismometer exposed to wind and other weather phenomena can have significantly increased low-frequency noise. In summary, the possibility to find low-noise surface sites should not be excluded, but underground sites are likely the only seismically quiet locations in most densely populated countries (which includes most countries in Europe).

\subsubsection{Underground seismicity}\index{seismic noise!underground}
Seismologists have been studying underground seismicity at many locations over decades, and found that high-frequency seismic spectra are all significantly quieter than at typical surface sites. This can be explained by the exponential fall off of Rayleigh-wave amplitudes according to Equation (\ref{eq:Rayfield}), combined with the fact that high-frequency seismicity is typically generated at the surface, and most surface sites are covered by a low-velocity layer of unconsolidated ground. The last means that amplitude decreases over relatively short distances to the surface. Seismic measurements have been carried out in boreholes \cite{Dou1964,SaHa1964}, and specifically in the context of site characterization for future GW detectors at former or still active underground mines \cite{HaEA2010,BeEA2012,NaEA2014,BBR2015}. There are however hardly any underground array measurements to characterize the seismic field in terms of mode composition. This is mostly due to the fact that these experiments are very costly, and seismic stations have to be maintained under unusual conditions (humidity, temperature, dust,...). Currently, a larger seismic array is being deployed for this purpose as part of the DUGL (Deep Underground Gravity Laboratory) project at the former Homestake mine, now known as the Sanford Underground Research Facility, equipped with broadband seismometers, state-of-the-art data acquisition, and auxiliary sensors such as infrasound microphones. As a consequence of the high cost, the effort could only be realized as collaboration between several groups involving seismologists and GW scientists.

The picture seems to be very simple. Underground seismicity above a few Hz is generally very small approaching the global low-noise model. Variations can however be observed, and have in some cases been identified as anthropogenic noise produced underground \cite{HaEA2010}. Therefore, it is important to evaluate how much noise is produced by the underground infrastructure that is either already in place, or is brought to the site for the underground experiment itself. Pumps and ventilation are required for the maintenance of an underground site, which may lead to excess noise. Measurements were carried out in the context of the design study of the Einstein Telescope in Europe \cite{ET2011}. Some of the collected seismic spectra were presented in \cite{BeEA2012}, which is shown again here in Figure \ref{fig:seismicityET}.
\epubtkImage{}{%
    \begin{figure}[htbp]
    \centerline{\includegraphics[width=0.7\textwidth]{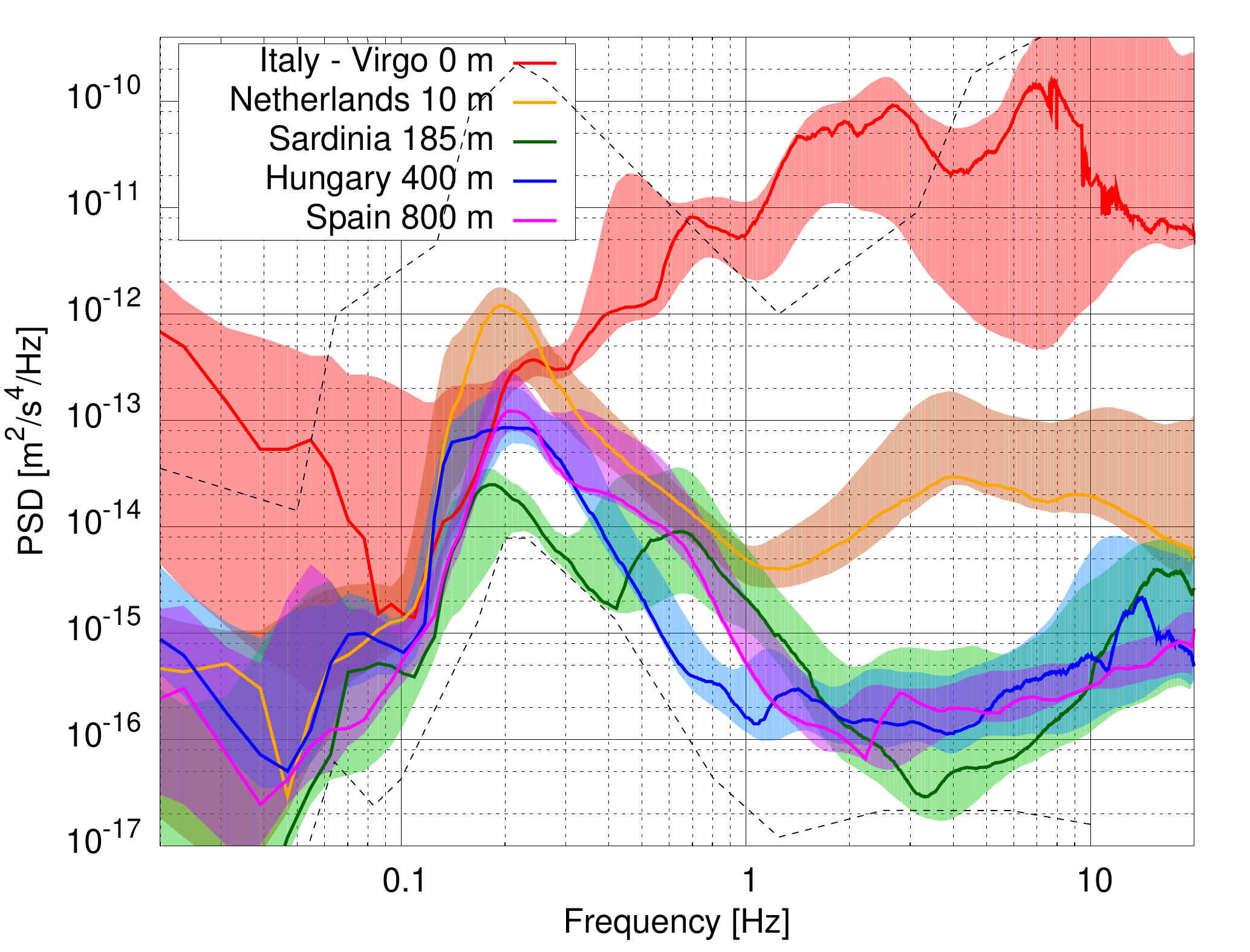}}
    \caption[Underground seismicity in Europe]{Spectral densities and typical variation of ambient seismic noise at underground sites in Europe. The depths of the seismometers are indicated in the legend. Courtesy of Beker et al \cite{BeEA2012}.}
    \label{fig:seismicityET}
    \end{figure}}
The underground sites have similar seismic spectra above about 1\,Hz, which are all lower by orders of magnitude compared to the surface spectrum measured inside one of the Virgo buildings. The Virgo spectrum however shows strong excess noise even for a surface site. This can be seen immediately since the spectrum exceeds the global high-noise model drawn as dashed curve between 1\,Hz and 3\,Hz, which means that there is likely no natural cause for the seismic energy in this range. The Virgo infrastructure may have enhanced response to ambient noise at these frequencies, or the seismic sources may be part of the infrastructure. The Netherland spectrum is closer to spectra from typical surface locations, with somewhat decreased noise level though above a few Hz since the measurement was taken 10\,m underground. Nevertheless, the reduction of seismic Newtonian noise to be expected by building a GW detector underground relative to typical surface sites is about 2 orders of magnitude, which is substantial. Whether the reduction is sufficient to meet the requirements set by the ET sensitivity goal is not clear. It depends strongly on the noise models. While results presented in \cite{Bek2012} indicate that the reduction is sufficient, results in \cite{Har2013b} show that further reduction of seismic Newtonian noise would still be necessary.
\epubtkImage{}{
    \begin{figure}[htb]
    \centerline{\includegraphics[width=0.7\textwidth]{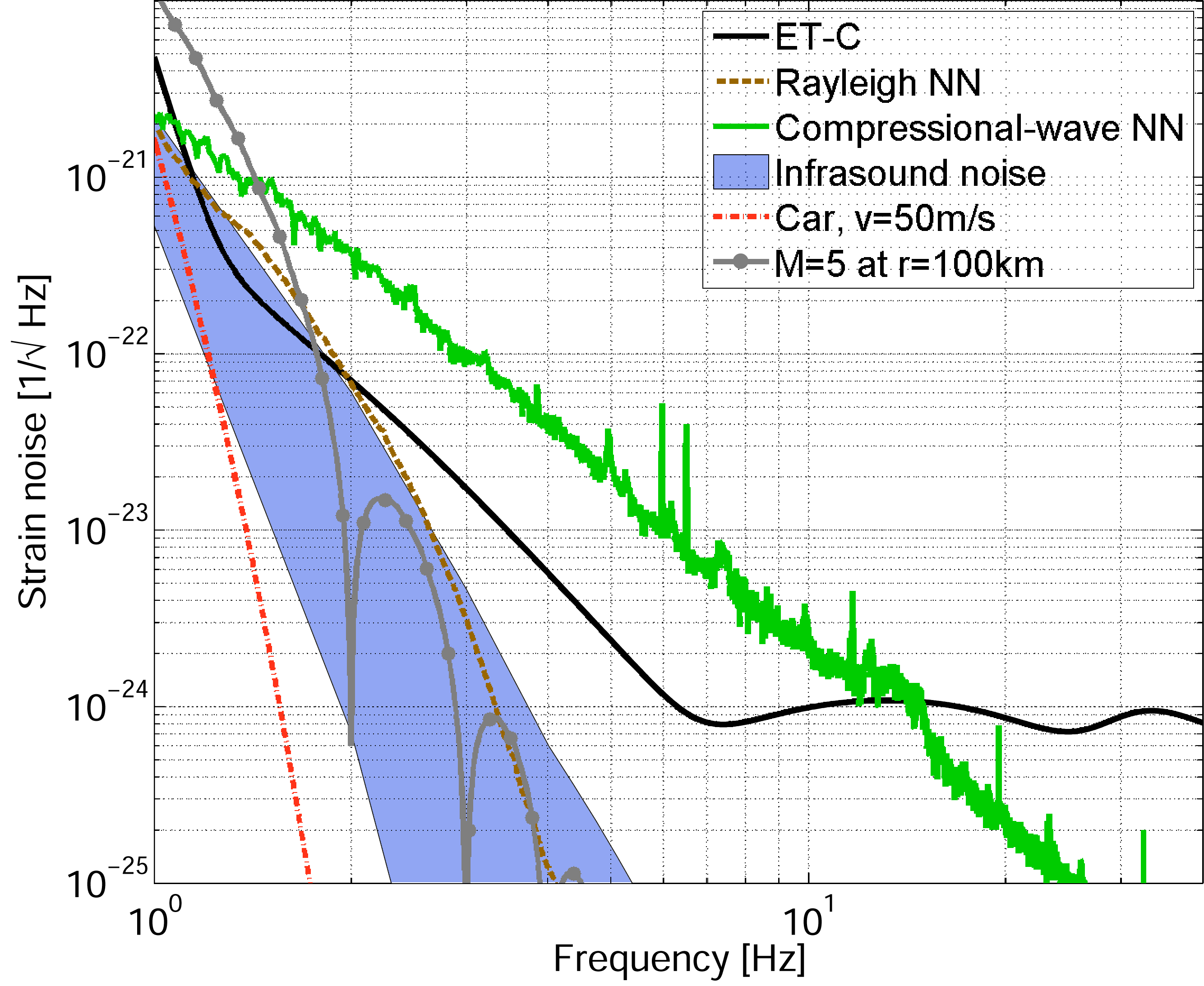}}
    \caption[Newtonian noise budget for the Einstein Telescope]{Newtonian-noise budget together with the ET-C instrumental-noise model \cite{Har2013b}. The detector is located 200\,m underground. Surface (Rayleigh NN) and underground (Compressional-wave NN) seismic spectra were measured at different sites. The surface data are from a seismometer at a quiet site in the US: TA-V34A. The underground measurement was carried out at the 4100\,ft level of the former Homestake mine in South Dakota.}
    \label{fig:NNET}
    \end{figure}}
The plot presented in \cite{Har2013b} is shown in Figure \ref{fig:NNET}. Seismic Newtonian noise from the surface, Equation (\ref{eq:RayNN}), and also infrasound Newtonian noise, Equation (\ref{eq:gravinfra}), are sufficiently suppressed according to these results. The body-wave Newtonian noise however lies above the targeted noise level (according to the ET-C model). In addition, spectra are shown for gravity perturbations from a car passing right above one test mass of the detector with 180\,km/h using Equation (\ref{eq:movsingle}). Finally, based on the first 5\,s of the simple model in Equation (\ref{eq:earlytime}), a signal spectrum of a magnitude 5 earthquake is also plotted. It may be possible to find underground sites that are seismically quieter than Homestake, but not by a large factor. According to these results, it is likely that some form of noise cancellation is still required, but only by a modest factor, which, according to Section \ref{sec:arrayNNP}, should be easier to achieve underground than at surface sites. 

\subsubsection{Site selection criteria in the context of coherent noise cancellation}
\label{sec:sitecancel}
An important aspect of the site selection that has not been considered much in the past is that a site should offer the possibility for efficient coherent cancellation of Newtonian noise. From Section \ref{sec:cohcancel} we know that the efficiency of a cancellation scheme is determined by the two-point spatial correlation of the seismic field. If it is well approximated by idealized models, then we have seen that efficient cancellation would be possible. However, if scattering is significant, or many local sources contribute to the seismic field, then correlation can be strongly reduced, and a seismic array consisting of a potentially large number of seismometers needs to be deployed. The strongest scatterer of seismic waves above a few Hz is the surface with rough topography. This problem was investigated analytically in numerous publications, see for example \cite{GiKn1960,Abu1962,Hud1967,Ogi1987}.  If the study is not based on a numerical simulation, then some form of approximation needs to be applied to describe topographic scattering. The earliest studies used the Born approximation, which means that scattering of scattered waves is neglected. In practice, it leads to accurate descriptions of seismic fields when the seismic wavelength is significantly longer than the topographic perturbation, and the slope of the topography is small in all directions. 

With this approximation, a systematic evaluation of sites in the US was carried out \cite{CoHa2012}. A topographic map of the US was divided into 10\,km $\times$ 10\,km squares. The elevation rms was calculated for each square. The rms map is shown in Figure \ref{fig:topoUS}. The hope was that flat squares can be found in low-seismicity regions, which would combine the requirements on scattering and seismicity. High elevation sites typically show weak seismic noise (above a few Hz), mostly likely because of smaller population density. 
\epubtkImage{}{%
    \begin{figure}[htb]
    \centerline{\includegraphics[width=0.63\textwidth]{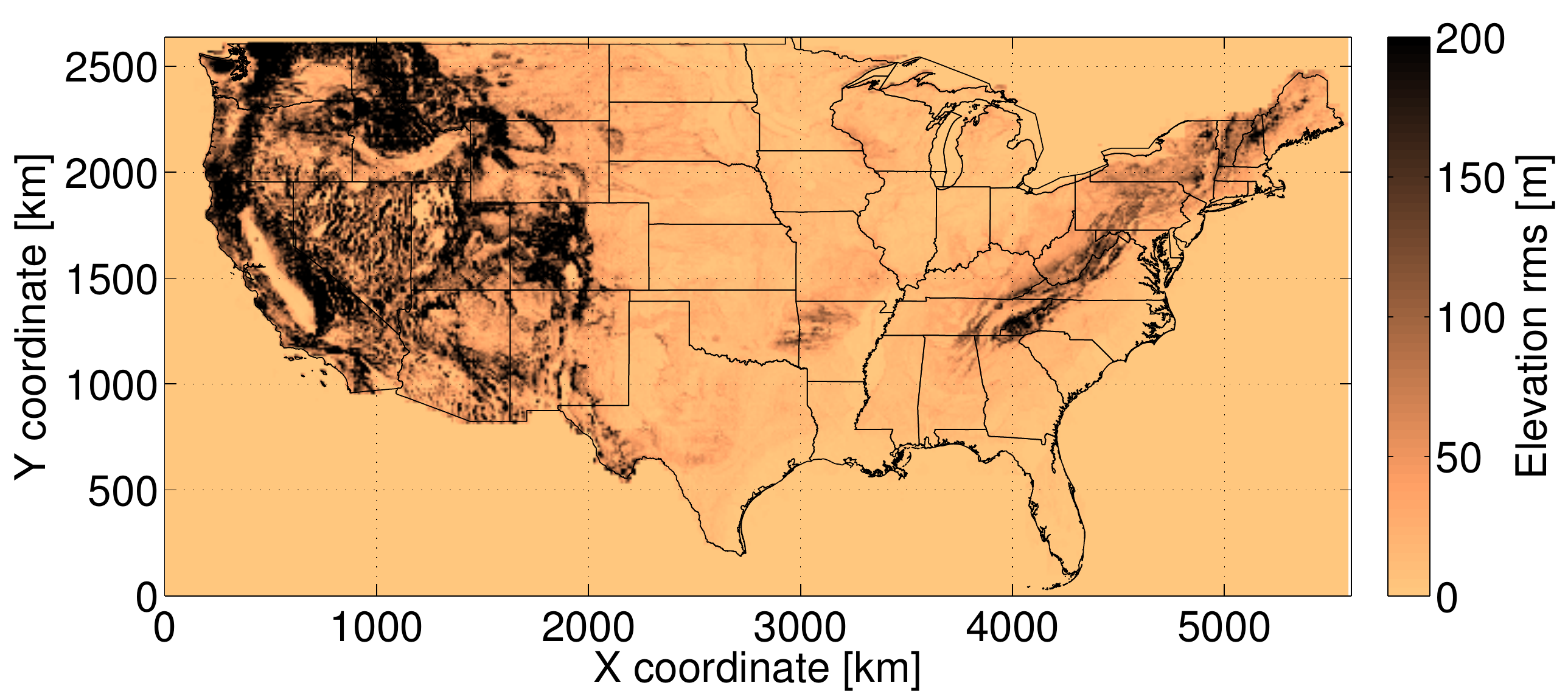}
                \includegraphics[width=0.37\textwidth]{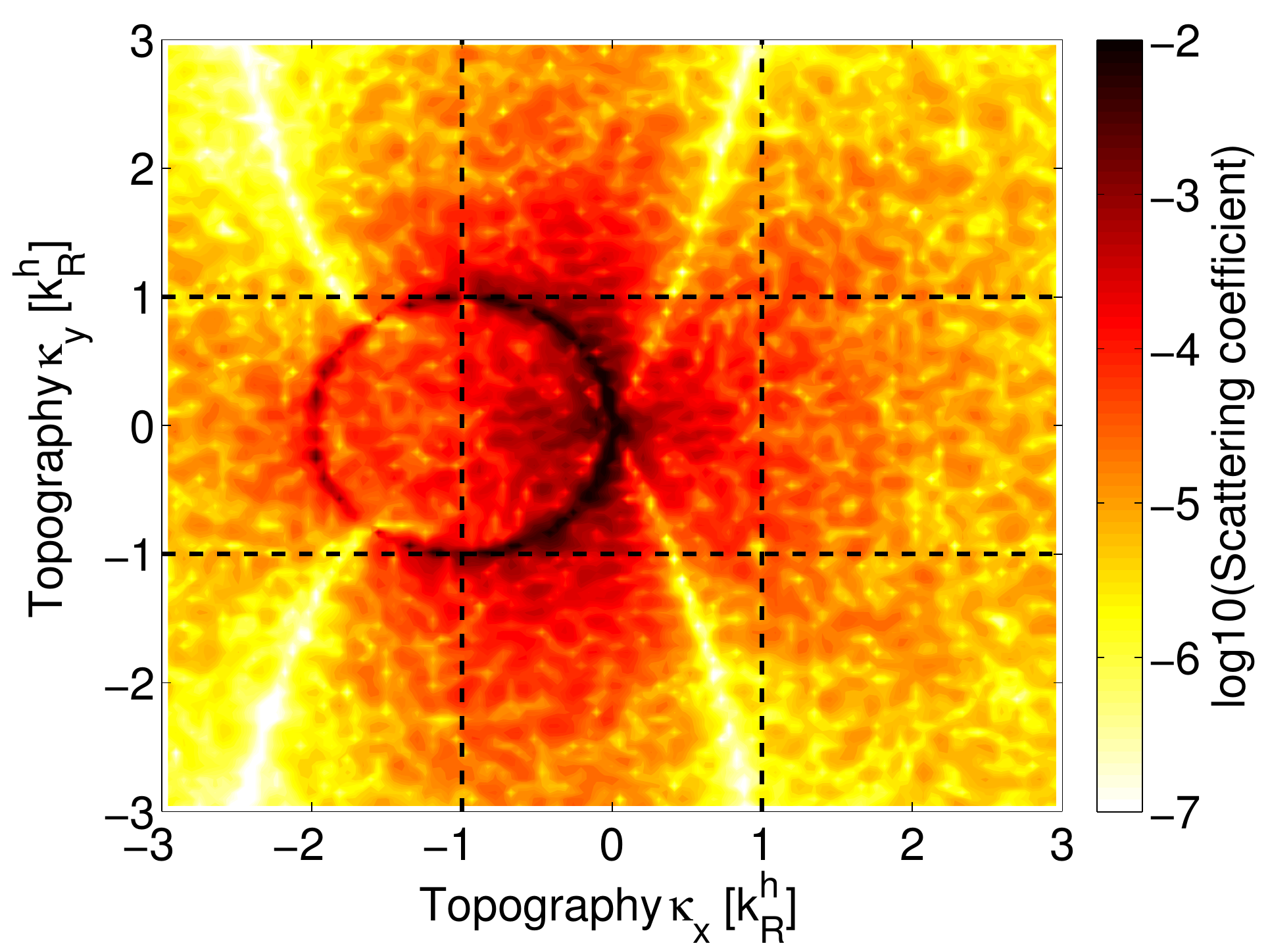}}
    \caption[Topographic site selection]{Investigation of topographic scattering for site selection. The map in the left plot shows the rms of topographies evaluated on 10\,km $\times$ 10\,km squares. Scattering coefficients of incident Rayleigh waves for a high-rms site in Montana (station F13A) are shown in the contour plot on the right.}
    \label{fig:topoUS}
    \end{figure}}
Combining the rms map with knowledge of ambient seismicity, it was in fact possible to find many sites fulfilling the two requirements. Figure \ref{fig:topoUS} shows the scattering coefficients for incident Rayleigh waves at a high-rms site in Montana. Excluding the Rayleigh-to-Rayleigh scattering channel (which, as explained in the study, does not increase the complexity of a coherent cancellation), a total integrated scatter of 0.04 was calculated. Including the fact that scattering coefficients for body waves are expected to be higher even, this value is large enough to influence the design of seismic arrays used for noise cancellation. Also, it is important to realize that the seismic field in the vicinity of the surface is poorly represented by the Born approximation (which is better suited to represent the far field produced by topographic scattering), which means that spatial correlation at the site may exhibit more complicated patterns not captured by their study. As a consequence, at a high-rms site a seismic array would likely have to be 3D and relatively dense to observe sufficiently high correlation between seismometers. Heterogeneous ground may further add to the complexity, but we do not have the theoretical framework yet to address this problem quantitatively. For this, it will be important to further develop the scattering formalism introduced in Section \ref{sec:scatterNN}. 

Underground sites that were and are being studied by GW scientists are all located in high-rms regions. This is true for the sites presented in the ET design study, for the Homestake site that is currently hosting the R\&D efforts in the US, and also for the Kamioka site in Japan, which hosts the KAGRA detector. Nonetheless, a careful investigation of spatial correlation and Wiener filtering in high-rms sites has never been carried out, and therefore our understanding of seismic scattering needs to be improved before we can draw final conclusions. 

\subsection{Noise reduction by constructing recess structures or moats}
\label{sec:shieldseismic}
Hughes and Thorne suggested that one way to reduce Newtonian noise at a surface site may be to dig moats at some distance around the test masses \cite{HuTh1998}. The purpose is to reflect incident Rayleigh waves and thereby create a region near the test masses that is seismically quieter. The reflection coefficient depends on the depth of the moat \cite{MaKn1965,FuMa1980,BDV1986}. If the moat depth is half the length of a Rayleigh wave, then the wave amplitude behind the moat is weakened by more than a factor 5. Only if the moat depth exceeds a full length of a Rayleigh wave, then substantially better reduction can be achieved. If the distance of the moat to the test mass is sufficiently large, then the reduction factor in wave amplitude should translate approximately into the same reduction of Newtonian noise from Rayleigh waves. There are two practical problems with this idea. First, the length of Rayleigh waves at 10\,Hz is about 20\,m (at the LIGO sites), which means that the moat needs to be very deep to be effective. It may also be necessary to fill moats of this depth with a light material, which can slightly degrade the isolation performance. The second problem is that the scheme requires that Rayleigh waves are predominantly produced outside the protected area. This seems unlikely for the existing detector sites, but it may be possible to design the infrastructure of a new surface site such that sources near the test masses can be avoided. For example, fans, pumps, building walls set into vibration by wind, and the chambers being connected to the arm vacuum pipes are potential sources of seismicity in the vicinity of the test masses. The advantage is that the moats do not have to be wide, and therefore the site infrastructure is not strongly affected after construction of the moats. Another potential advantage, which also holds for the recess structures discussed below, is that the moat can host seismometers, which may facilitate coherent cancellation schemes since 3D information of seismic fields is obtained. This idea certainly needs to be studied quantitatively since seismic scattering from the moats could undo this advantage.

\epubtkImage{}{%
    \begin{figure}[htbp]
    \centerline{\includegraphics[width=0.8\textwidth]{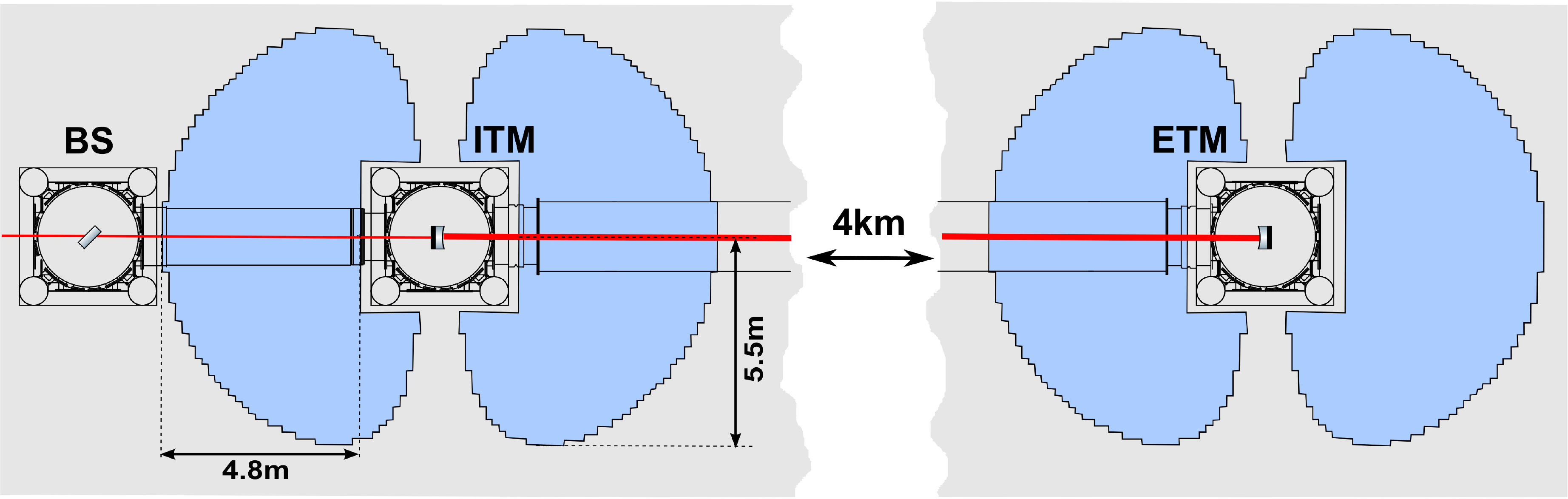}}
    \caption[Recess structure to reduce Newtonian noise]{Recess structures around test masses reduce mass, which would otherwise carry seismic waves that act as sources of gravity perturbations. }
    \label{fig:recessLIGO}
    \end{figure}}
Another approach is to dig recess structures around the test masses \cite{HaHi2014}. Here the primary goal is not to reflect Rayleigh waves, but to remove mass around the test masses that would otherwise be perturbed by seismic fields to produce Newtonian noise. A sketch of how a recess structure may look like at a detector site is shown in Figure \ref{fig:recessLIGO}. A central pillar needs to be left to support the test-mass chambers. The recess should have a depth of about 4\,m, provided that the speed of Rayleigh waves is about 250\,m/s at 10\,Hz \cite{HaOR2011}. If the speed is higher by a factor 2, then recess dimensions in all three directions need to be increased by a factor 2 to maintain the same noise reduction. This means that it is infeasible to construct effective recesses at sites with much higher Rayleigh-wave speeds (at Newtonian-noise frequencies). For a 4\,m deep recess and horizontal dimensions as shown in Figure \ref{fig:recessLIGO}, the reduction factor is plotted in the left of Figure \ref{fig:recessNN}.
\epubtkImage{}{%
    \begin{figure}[htbp]
    \centerline{\includegraphics[width=0.45\textwidth]{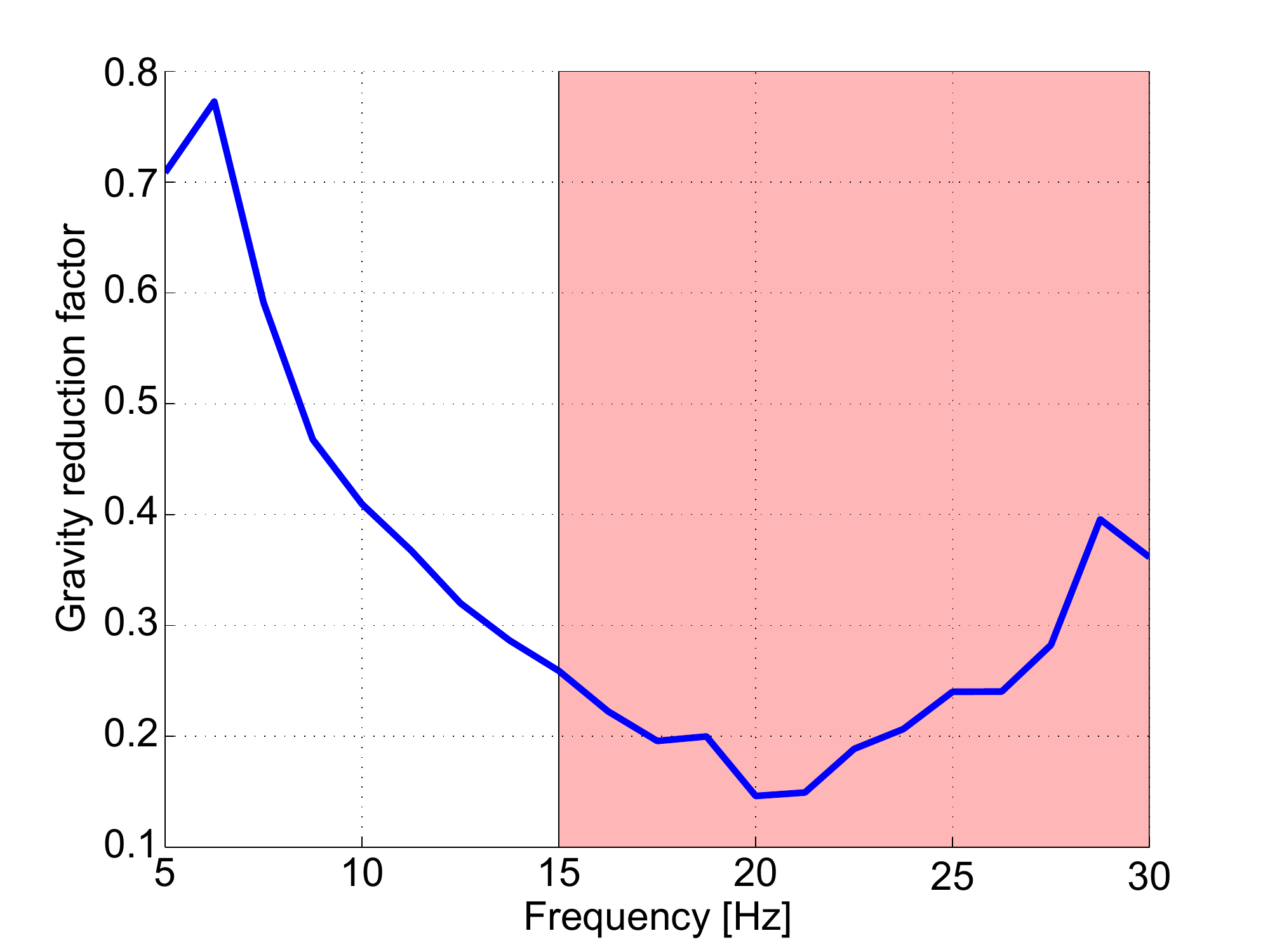}
                \includegraphics[width=0.45\textwidth]{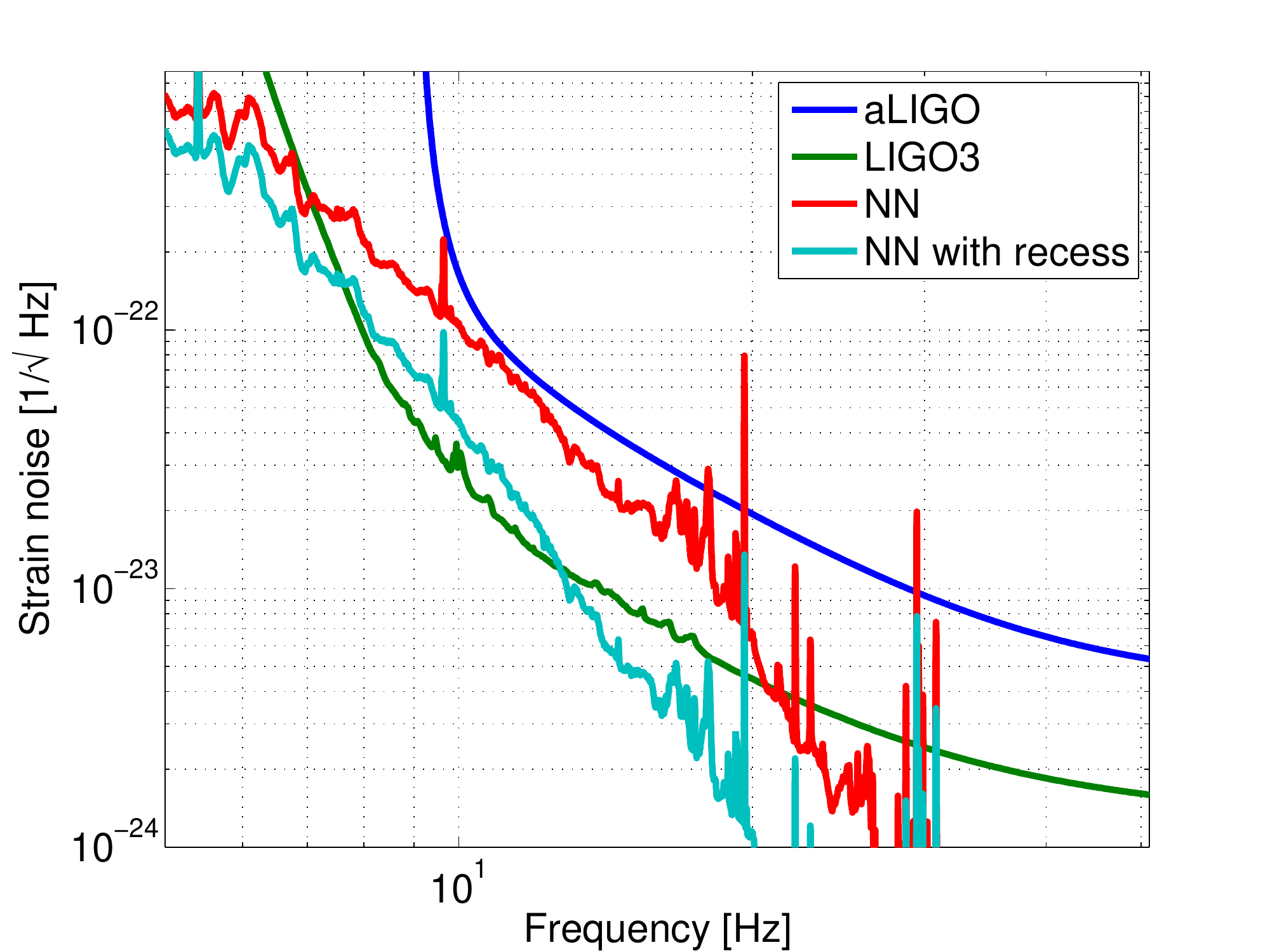}}
    \caption[Noise reduction performance of a recess structure]{The plot to the left shows the noise reduction factor from a recess. The red regime marks frequencies where significant seismic scattering from the recess may occur. The plot to the right shows the corresponding Newtonian-noise spectrum together with sensitivity models for aLIGO and a possible future version of LIGO.}
    \label{fig:recessNN}
    \end{figure}}
Even though the primary purpose of the recess is not to reflect Rayleigh waves, seismic scattering can be significant. Due to the methods chosen by the authors, scattering could not be simulated, and validity of this approximation had to be explained. Above some frequency, the wavelength is sufficiently small so that scattering from a 4\,m deep recess is significant. This regime is marked red in the plot, and the prediction of noise reduction may not be accurate. Above 20\,Hz it can be seen that reduction gets weaker. This is because the gravity perturbation starts to be dominated by density perturbations of the central pillar. It is possible that the recess already acts as a moat at these frequencies, and that the central pillar has less seismic noise than simulated in their study. A detailed simulation of scattering from the recess structure using dynamical finite-element methods is necessary to estimate the effect (see Section \ref{sec:numsim} for details). The Newtonian-noise spectrum calculated from the reduction curve is shown in the right of Figure \ref{fig:recessNN}. The green curve models the sensitivity of a possible future version of a LIGO detector. Without noise reduction, it would be strongly limited by Newtonian noise. With recess, Newtonian noise only modestly limits the sensitivity and implementation of coherent noise cancellation should provide the missing noise reduction. It is to be expected that the idea of removing mass around test masses only works at the surface. The reason is that seismic speeds are much larger underground (by a factor 10 at least compared to 250\,m/s). The idea would be to place test masses at the centers of huge caverns, but Figure \ref{fig:cavNNr} tells us that the radius of such a cavern would have to be extremely large (of the order 100\,m for a factor 2 Newtonian-noise reduction at 10\,Hz).

\subsection{Summary and open problems}
\label{sec:mitisummary}
In this section, we have described Newtonian-noise mitigation schemes including coherent noise cancellation using Wiener filters, and passive mitigation based on recess structures and site-selection. While some of the mitigation strategies are well understood (for example, coherent cancellation of Rayleigh-wave Newtonian noise, or site selection with respect to ambient seismicity), others still need to be investigated in more detail. Especially the coherent cancellation of Newtonian noise from seismic body waves depends on many factors, and in this section we could only develop the tools to address this problem systematically. The role of S-waves as coherent noise contribution among seismic sensors serving as reference channels in Wiener filtering has been described in Section \ref{sec:arrayNNP}. Since the cancellation performance presented in Figure \ref{fig:residualXiStrain} is relatively poor and possibly insufficient for future GW detectors that rely on substantial reduction of Newtonian noise, it can be said that developing an effective scheme is one of the top priorities of future investigations in this field. Possible solutions may be to combine seismometers and strainmeters in sensor arrays, and to use multi-axes sensors instead of the single-axis sensors modelled here. Nonetheless, it is remarkable that a simple approach does not lead to satisfactory results as we have seen for the cancellation of Rayleigh-wave and infrasound Newtonian noise in Figures \ref{fig:residualsRay} and \ref{fig:residualSound}. However, we have also been conservative with the body-wave modelling in the sense that we assumed isotropic fields and relatively low P-wave content. Since P-waves experience weaker damping compared to S-waves, it may well be possible that P-wave content is higher in seismic fields. We have also reviewed our current understanding of site-dependent effects on coherent noise cancellation in Section \ref{sec:sitecancel}, which adds to the complexity of the site-selection process. In this context, sites should be avoided where significant seismic scattering can be expected. This is generally the case in complex topographies typical for mountains. It should be emphasized though that a extensive and conclusive study of the impact of scattering on coherent cancellation has not been carried out so far. 

Concerning passive mitigation strategies, site selection is the preferred option and should be part of any design study of future GW detectors. The potential gain in low-frequency noise can be orders of magnitude, which cannot be guaranteed with any other mitigation strategy. This fact is of course well recognized by the community, as demonstrated by the detailed site-selection study for the Einstein Telescope and the fact that it was decided to construct the Japanese GW detector KAGRA underground. Alternative passive mitigation schemes such as the construction of recess structures around test masses are likely effective at surface sites only as explained in Section \ref{sec:shieldseismic}. The impact of these structures strongly depends on the ratio of structure size to seismic wavelength. Newtonian noise at underground sites is dominated by contributions from body waves, which can have lengths of hundreds of meters even at frequencies as high as 10\,Hz. At the surface, smaller-scale structures may turn out to be sufficient since Rayleigh-wave lengths at 10\,Hz can be a factor 10 smaller than the lengths of body waves underground. Results from finite-element simulations are indeed promising, and more detailed follow-up investigations should be carried out to identify possible problems with this approach.



\newpage


\section{Acknowledgements}
\label{sec:acknowledgements}
I was lucky to have been given the opportunity to enter the field of Newtonian noise and terrestrial gravity perturbations at a time when outstanding experimental problems had to be addressed for future GW detector concepts. I took my first steps in this field as part of the group of Prof Vuk Mandic at the University of Minnesota, Twin Cities. I have to thank Prof Mandic for his continuous support, and especially for taking the time to return comments on this manuscript. With his DUGL project currently proceeding at a steep rise, his time is very precious. During these first two years, I started to collaborate with Prof Giancarlo Cella, who by that time had already written seminal papers on Newtonian-noise modelling and mitigation. I thank Prof Cella for the many discussions on Newtonian noise, and also for pointing out important past work on Newtonian noise missing in an earlier version of the manuscript. While working on the experimental realization of an underground seismic array at the former Homestake mine as part of the DUGL project, I had the privilege to collaborate with and learn from my colleagues Dr Riccardo DeSalvo and Dr Mark Beker (at that time graduate student). Their motivation to do science in its best way without hesitation in complicated situations has inspired me since then. I thank both of them for comments and contributions to this manuscript. Starting in 2010, I was given the opportunity at Caltech to apply my experience with seismic fields and gravity modelling to investigate Newtonian noise for the LIGO detectors. I have to thank Prof Rana Adhikari for supporting me not only with my LIGO work, but also for making sure that I keep an open mind and broad view on science. I am especially thankful that I could work with one of Prof Adhikari's graduate students, Jennifer Driggers, with whom I was able to lie the foundation for future work on Newtonian noise at the LIGO sites. I thank Jenne for her dedication and for keeping me focussed on the important problems. I am currently supported by a Marie-Curie Fellowship (FP7-PEOPLE-2013-IIF) at the Universit\`a di Urbino, which gives me the freedom to contribute to the development of the field in any possible direction. Therefore, I want to thank the committee of the European Commission who evaluated my past accomplishments in a favorable way. I want to thank Prof Flavio Vetrano and my colleagues at the INFN Firenze, who involved me in exciting experimental developments in Europe on low-frequency gravity sensing, especially atom interferometry. As I could hopefully demonstrate in this paper, terrestrial gravity perturbations is a complex problem, which means that observations in the future should be expected to hold surprises for us and unexpected applications may emerge. 

Last but not least, I want to thank Marica Branchesi who made sure that I never lose motivation to write this article, and whose dedication to science and people is always an inspiration to me.
\\[1cm]
I acknowledge the use of Mathematica and Matlab for the generation of the plots in this paper, and as a help with some analytical studies.

\newpage

\appendix
\section{Mathematical formalism}
\label{sec:math} 
The purpose of this appendix is to define the mathematical quantities used in the paper, and to provide the key equations to master the more complex calculations. Only the basic properties are described here. More complex applications can be found throughout the paper. Comparing results in this article with results from other publications, one should pay attention especially to the definition of spherical scalar and vector harmonics, and multipoles. Various normalization conventions can be found in the literature, which can cause final results to look different. 

Also, to share valuable experience, here is how almost all complicated calculations can be carried out very efficiently. First, a problem needs to be calculated with pencil and paper. Half of the time, the results will be wrong. Reasons are typically a wrong sign or other mistakes in simple steps. However, with a good understanding of the structure of the calculation, it is always possible to translate the calculation efficiently into a symbolic computational software program to obtain the result (the author used \emph{Mathematica} for this purpose). This scheme has worked for all calculations in this article. Not solving the problem by hand first often leads to the situation that the symbolic software is programmed in a way that it cannot find the solution. In some cases, solutions are found by the software only for specific parameter ranges, and one needs to generalize the solution using one's understanding from the calculation by hand. More satisfactorily even, knowing the solution helps to identify the mistakes done in the first calculation by hand.

\subsection{Bessel functions}
\label{sec:cylindrical}
Bessel functions exist in two types, the Bessel function of the first kind $J_n(\cdot)$, and the Bessel function of the second kind $Y_n(\cdot)$. The latter is irregular at the origin, and will not be used in this article. A common definition of the Bessel function is\index{Bessel function!first kind}
\beq
J_n(x)=\frac{1}{2\pi}\int\limits_{-\pi}^\pi\drm\tau\,\e^{\irm(n\tau-x\sin(\tau))}
\label{eq:besselint}
\eeq
Here, the order $n$ needs to be integer. Only the $J_0(\cdot)$ is non-zero at the origin. Many important properties of $J_n(\cdot)$ can be derived from this equation. For example, negative integer orders can be re-expressed as positive orders:
\beq
J_{-n}(x)=(-1)^nJ_n(x)
\eeq
In this paper, the Bessel function will find application in cylindrical harmonics expansions. Any field that fulfills the Laplace equation $\Delta f(\vec r\,)=0$, i.~e.~a harmonic function\index{harmonic function}, can be expanded into cylindrical harmonics $C_n(k;\varrho,\phi,z)$. Cylindrical harmonics are linearly independent solutions to the Laplace equation. Throughout this article, we will only need harmonics that are regular at the origin. In this case, they can be written as:
\beq
C_n(k;\varrho,\phi,z)=J_n(k\varrho)\e^{\irm n\phi}\e^{-kz},
\eeq 
where $\varrho,\,\phi,\,z$ are the cylindrical coordinates. A arbitrary harmonic function $f(\vec r\,)$ that is regular at the origin can then be expanded according to
\beq
f(\vec r\,)=\sum\limits_{n=-\infty}^\infty\int\limits\drm k\,a_n(k)C_n(k;\varrho,\phi,z)
\eeq
Cylindrical harmonics find application in calculations of fields in a half space. The integration range of the parameter $k$ is not further specified here, since it depends on the specific physical problem. The parameter $k$ can in general be complex valued, and in some cases, i.~e.~when the field is constrained to a finite range of radii $\varrho$, it can also take on discrete values. 

Most of the important, non-trivial relations used in this article involving Bessel functions concern semi-infinite integrals. The first relation can be obtained as a limiting case of the Hankel integral \cite{Han1875,Wat1922}
\beq
\begin{split}
\int\limits_0^\infty\drm \varrho\,\varrho^p J_n(k\varrho) &= \lim\limits_{a\rightarrow 0}\int\limits_0^\infty\drm \varrho\,\e^{-a\varrho}\varrho^p J_n(k\varrho)\\
&= \frac{1}{2}\left(\frac{2}{k}\right)^{p+1} \frac{\Gamma((n+p+1)/2)}{\Gamma((n-p+1)/2)},
\end{split}
\eeq
with $-n-1<p<1/2$ and $k>0$. A related integral can be derived consistent with the last equation, even though the conditions on the parameters are not fulfilled with $p=1$:
\beq
\int\limits_0^\infty\drm \varrho\,\varrho J_n(k\varrho)=\frac{n}{k^2}
\eeq
Finally, an integral that is useful in calculations with cylindrical harmonic expansions, see Section \ref{sec:sourcehalf}, is given by \cite{ArWe2005}
\beq
\int\limits_0^\infty\drm \varrho\,\varrho J_n(k\varrho)J_n(s\varrho)=\frac{\delta(s-k)}{k},
\label{eq:closureJ}
\eeq
with $\delta(\cdot)$ being the Dirac $\delta$-distribution. This equation allows us to reduce the number of Bessel functions in more complicated integrals, and is known as \emph{closure relation}\index{Bessel function!closure relation}.

Bessel functions can also be defined for non-integer orders, which requires a modification of the definition in Equation (\ref{eq:besselint}). Using the generalized definition, one can define the spherical Bessel functions of the first kind according to \index{Bessel function!spherical}
\beq
j_n(x)=\sqrt{\frac{\pi}{2x}}J_{n+1/2}(x)
\label{eq:sphericalj}
\eeq
The spherical Bessel functions have a form very similar to the Bessel functions. Also here, $j_0(\cdot)$ is the only spherical Bessel function that does not vanish at the origin. Spherical Bessel functions of the first kind appear in correlation functions of 3D fields (see Section \ref{sec:arrayNNP}). For example, correlations between scalar fields are given in terms of $j_0(\cdot)$, correlations of vector fields include $j_0(\cdot),\,j_2(\cdot)$. They also appear in the vector plane-wave expansions, see Equations (\ref{eq:expandPWlong}) and (\ref{eq:expandPWtrans}). In many calculations involving spherical Bessel functions, the following to recurrence relations are useful
\beq
\begin{split}
j_{l+1}(x) &= \frac{2l+1}{x}j_l(x)-j_{l-1}(x)\\
\partial_xj_l(x) &= \frac{l}{x}j_l(x)-j_{l+1}(x)\\
&= \frac{l}{2l+1}j_{l-1}(x)-\frac{l+1}{2l+1}j_{l+1}(x)
\end{split}
\label{eq:sphBesselrec}
\eeq 
The first relation means that it is always possible to express a sum over spherical Bessel functions with arbitrarily many different orders as a sum over two orders only. Therefore, it is first of all a great tool to reduce complexity of a result. The second equation is often applied in calculations with integrals following integration by parts.

\subsection{Spherical harmonics}
\label{sec:spherical}
Spherical harmonics are the independent solutions to the Laplace equation in spherical coordinates. We distinguish between surface spherical harmonics and solid spherical harmonics. Two-dimensional scalar harmonic fields on spheres can be expanded into surface spherical harmonics. Three dimensional scalar harmonic fields can be expanded into solid spherical harmonics. We will also introduce the vector surface spherical harmonics used to expand vector fields on spheres. Spherical harmonics find wide application. In this article, we will use them to calculate seismic fields scattered from spherical cavities or gravity perturbations from seismic point sources (see Sections \ref{sec:scattershear} and \ref{sec:disgravity}). Furthermore, solid spherical harmonics are the constituents of the multipole expansion, which is an elegant means to describe gravity perturbations from objects with arbitrary shape (see Sections \ref{sec:vibobj} and \ref{sec:rotobj}).
 
\subsubsection{Legendre polynomials}
\index{Legendre polynomials}
Legendre polynomials are introduced since they are part of the definition of spherical harmonics. They also directly serve in expansions of harmonic fields in spherical coordinates when the fields have cylindrical symmetry. The Legendre polynomial of integer order $l$ is defined as
\beq
P_l(x)=\dfrac{1}{2^ll!}\partial_x^l(x^2-1)^l
\eeq
In order to evaluate integrals involving Legendre polynomials, it is often convenient to express powers of the argument $x$ in terms of Legendre polynomials. Table \ref{tab:legendre} summarizes the relations for the first 4 orders. 
\begin{table}[htbp]
\caption{Legendre polynomials}
\label{tab:legendre}
\renewcommand{\arraystretch}{2.5}
\centerline{
\begin{tabular}{|l|l|}
\hline
$P_0(x)=1$ & $1=P_0(x)$\\
$P_1(x)=x$ & $x=P_1(x)$\\
$P_2(x)=\frac{1}{2}(3x^2-1)$ & $x^2=\frac{1}{3}(2P_2(x)+P_0(x))$\\
$P_3(x)=\frac{1}{2}(5x^3-3x)$ & $x^3=\frac{1}{5}(2P_3(x)+3P_1(x))$\\
\hline
\end{tabular}}
\end{table}
Naturally, any polynomial of order $l$ can be expressed in terms of Legendre polynomials up to the same order. In most applications, the domain of the Legendre polynomials is the interval $[-1;1]$. In this case, the Legendre polynomials have interesting integral properties such as the orthogonality relation
\beq
\int\limits_{-1}^1\drm xP_m(x)P_n(x)=\frac{2}{2m+1}\delta_{mn}
\label{eq:legorth}
\eeq
Making use of the orthogonality relation of Equation (\ref{eq:legorth}), the inverse expansion of monomials $x^m$ into Legendre polynomials $P_l(x)$, as shown for the first few orders in Table \ref{tab:legendre}, can be obtained from the integrals
\beq
\int\limits_{-1}^1\drm x\,x^mP_l(x)=\frac{2^{2+l}(1+(m+l)/2)!m!}{((m-l)/2)!(2+m+l)!},
\eeq
for $m\geq l$ and $m+l$ even. The integral vanishes for all other pairs $m,\,l$. The Legendre polynomials obey Bonnet's recursion formula\index{Legendre polynomials!Bonnet's recursion}:
\beq
(l+1) P_{l+1}(x)=(2l+1)x P_l(x)-lP_{l-1}(x)
\eeq
Also the derivative of a Legendre polynomial can be expressed as a sum of Legendre polynomials according to
\beq
\begin{split}
\partial_xP_l(x) &= \frac{1+l}{1-x^2}(x P_l(x)-P_{l+1}(x))\\
&= \frac{1}{1-x^2}\frac{l(l+1)}{2l+1}(P_{l-1}(x)-P_{l+1}(x))
\end{split}
\label{eq:derivlegendre}
\eeq
In analytical calculations of seismic fields, it enters the equations through its role in the scalar plane-wave expansion:\index{plane-wave expansion!scalar}
\beq
\e^{\irm \vec k\cdot\vec r}=\e^{\irm kr\cos(\theta)}=\sum\limits_{l=0}^\infty \irm^l(2l+1)j_l(kr)P_l(\cos(\theta))
\label{eq:pwscalar}
\eeq
Here, $j_n(\cdot)$ is the spherical Bessel function defined in Equation (\ref{eq:sphericalj}). For example, in Section \ref{sec:scattercomp}, we will calculate the scattered seismic field from a cavity with incident longitudinal wave. This problem has cylindrical symmetry.

More important for spherical harmonics are the associated Legendre polynomials $P_l^m(\cdot)$\index{Legendre polynomials!associated}. They are parameterized by a second integer index $m=-l,\ldots,l$. Their definition is given in terms of Legendre polynomials:
\beq
\begin{split}
P_l^m(x) &= (-1)^m(1-x^2)^{m/2}\partial_x^mP_l(x)\\
&= \dfrac{(-1)^m}{2^ll!}(1-x^2)^{m/2}\partial_x^{l+m}(x^2-1)^l
\end{split}
\eeq
Definitions of the associated Legendre polynomials can vary in terms of their $l,m$-dependent normalization. For example, some authors would normalize $P_l^m(\cdot)$ such that the factor in front of the Kronecker-$\delta$ in Equation (\ref{eq:orthoasLeg}) is equal to 1. While this choice of normalization has greater aesthetic appeal, we choose the more conventional definition since we will never work explicitly with the associated Legendre polynomials. In this article, they merely serve as building block of the spherical harmonics. Defined over the domain $x\in[-1;1]$, the associated Legendre polynomials obey the orthogonality relation
\beq
\int\limits_{-1}^1\drm x\,P_k^m(x)P_l^m(x)=\frac{2}{2l+1}\frac{(l+m)!}{(l-m)!}\delta_{k,l}
\label{eq:orthoasLeg}
\eeq
Finally, positive and negative orders $m$ are linked via
\beq
P_l^{-m}(x)=(-1)^m\dfrac{(l-m)!}{(l+m)!}P_l^m(x)
\label{eq:asLegsign}
\eeq
Associated Legendre polynomials will never be used explicitly in this article, but only as part of the definition of spherical harmonics. From the theory of spherical harmonics it will become clear that cylindrically symmetric fields can always be expanded in terms of the polynomials $P_l^0(x)=P_l(x)$.

\subsubsection{Scalar surface spherical harmonics}
\index{spherical harmonics!scalar, surface}
Scalar surface spherical harmonics $Y_l^m(\theta,\phi)$ are eigenfunctions of the Laplace operator with respect to the angular coordinates
\beq
\left(\frac{1}{\sin(\theta)}\partial_\theta\sin(\theta)\partial_\theta +\frac{1}{\sin^2(\theta)}\partial^2_\phi\right)Y_l^m(\theta,\phi)=-l(l+1)Y_l^m(\theta,\phi)
\label{eq:eigenY}
\eeq
As such, they form an important part in the expansion of harmonic functions expressed in spherical coordinates (see Sections \ref{sec:solidharm} and \ref{sec:multipole}). The degree $l$ of the spherical harmonic can assume all non-negative integer values, while the order $m$ lies in the range $m=-l,\ldots,l$. Their explicit form is given by
\beq
Y_l^m(\theta,\phi)=\sqrt{\dfrac{2l+1}{4\pi}\dfrac{(l-m)!}{(l+m)!}}P_l^m(\cos(\theta))\e^{\irm m\phi}
\label{eq:surfharm}
\eeq
The first 4 degrees of the harmonics are listen in Table \ref{tab:sphereharm}.
\begin{table}[htbp]
\caption{Spherical surface harmonics}
\label{tab:sphereharm}
\renewcommand{\arraystretch}{2.5}
\centerline{
\begin{tabular}{|l|l|}
\hline
$Y_0^0$ & $\dfrac{1}{2}\sqrt{\dfrac{1}{\pi}}$\\
\hline
$Y_1^0$ & $\dfrac{1}{2}\sqrt{\dfrac{3}{\pi}}\cos(\theta)$\\
$Y_1^{\pm 1}$ & $\mp\dfrac{1}{2}\sqrt{\dfrac{3}{2\pi}}\sin(\theta)\e^{\pm\irm\phi}$\\
\hline
$Y_2^0$ & $\dfrac{1}{4}\sqrt{\dfrac{5}{\pi}}(3\cos^2(\theta)-1)$\\
$Y_2^{\pm 1}$ & $\mp\dfrac{1}{2}\sqrt{\dfrac{15}{2\pi}}\sin(\theta)\cos(\theta)\e^{\pm\irm\phi}$\\
$Y_2^{\pm 2}$ & $\dfrac{1}{4}\sqrt{\dfrac{15}{2\pi}}\sin^2(\theta)\e^{\pm 2\irm\phi}$\\
\hline
\end{tabular}
\begin{tabular}{|l|l|}
\hline
$Y_3^0$ & $\dfrac{1}{4}\sqrt{\dfrac{7}{\pi}}(5\cos^2(\theta)-3)\cos(\theta)$\\
$Y_3^{\pm 1}$ & $\mp \dfrac{1}{8}\sqrt{\dfrac{21}{\pi}}(5\cos^2(\theta)-1)\sin(\theta)\e^{\pm \irm\phi}$\\
$Y_3^{\pm 2}$ & $\dfrac{1}{4}\sqrt{\dfrac{105}{2\pi}}\cos(\theta)\sin^2(\theta)\e^{\pm 2\irm\phi}$\\
$Y_3^{\pm 3}$ & $\mp \dfrac{1}{8}\sqrt{\dfrac{35}{\pi}}\sin^3(\theta)\e^{\pm 3\irm\phi}$\\
\hline
\end{tabular}}
\end{table}
Another related role of the spherical harmonics is that, on the unit sphere, any (square-integrable) function can be expanded according to
\beq
f(\theta,\phi)=\sum\limits_{l=0}^\infty\sum\limits_{m=-l}^lf_l^mY_l^m(\theta,\phi)
\label{eq:harmexpand}
\eeq 
In expansions with cylindrical symmetry, it is convenient to define the angle $\theta$ with respect to the symmetry axis, in which case the order $m$ can be set to 0, and the associated Legendre polynomials reduce to ordinary Legendre polynomials. 
    
In this article, the normalization of spherical harmonics is chosen such that
\beq
\int\drm\Omega\, Y_l^m(Y_{l'}^{m'})^*=\delta_{ll'}\delta_{mm'}
\label{eq:normalY}
\eeq
In other words, the surface spherical harmonics form on orthonormal basis of (square-integrable) functions on the unit sphere.  The relation between positive and negative orders can be found using Equations (\ref{eq:asLegsign}) and (\ref{eq:surfharm}):
\beq
Y_l^{-m}(\theta,\phi)=(-1)^m(Y_l^m(\theta,\phi))^*
\label{eq:conjugateY}
\eeq
Finally, we conclude this section with a few obvious and not so obvious relations. The first three relations are evaluations of the spherical harmonics at specific points:
\beq
\begin{split}
&Y_l^m(\theta,0) = \sqrt{\pi(2l+1)}P_l(\cos(\theta))\delta_{m,0}\\
&Y_l^m(0,\phi) = \frac{1}{2}\sqrt{\frac{2l+1}{\pi}}\delta_{m,0}\\
&Y_l^m(\pi/2,\phi) =
\begin{cases} 
    0 & l+m\quad\mbox{odd} \\
    \frac{1}{2^l}(-1)^{(l+m)/2}\sqrt{\dfrac{2l+1}{4\pi}}\dfrac{\sqrt{(l+m)!(l-m)!}}{((l+m)/2)!((l-m)/2)!} & l+m\quad\mbox{even}
\end{cases}
\end{split}
\label{eq:pointrelY}
\eeq
All three relations can be useful if fields are to be expanded on planes. Useful integrals of the spherical harmonics are
\beq
\begin{split}
&\int\limits_0^{2\pi}\drm\phi\, Y_l^m(\theta,\phi)= 2\pi Y_l^0(\theta,0)\delta_{m,0}=\sqrt{\pi(2l+1)}P_l(\cos(\theta))\delta_{m,0}\\
&\int\limits_0^{2\pi}\drm\phi\int\limits_0^{\pi/2}\drm\theta\,\sin(\theta) Y_l^m(\theta,\phi) =\sqrt{\pi(2l+1)}
\begin{cases} 
    1 & l=0\\
    0 & l>0\quad\mbox{and}\quad l\quad\mbox{even} \\
    (-1)^{(l-1)/2}\dfrac{l!!}{l(l+1)(l-1)!!} & l\quad\mbox{odd}
\end{cases}
\end{split}
\label{eq:intrelY}
\eeq
The latter integral can be found in \cite{Bye1893}. These equations demonstrate the typical situation that integrals over angles constrain the degrees and orders of spherical harmonics in infinite expansions as in Equation (\ref{eq:harmexpand}). The second relation is quite exotic, but could be useful in some half-space problems, for example, to predict the performance of coherent cancellation of infrasound Newtonian noise (see Section \ref{sec:arrayNNatm}).

\subsubsection{Vector surface spherical harmonics}
\index{spherical harmonics!vector,surface} Vector spherical harmonics form a basis of square-integrable vector fields. One can find various definitions of vector spherical harmonics that do not only differ in normalization. The fact that so many definitions exist is because different classes of differential operators are applied to these harmonics depending on the physical problem. If the interest lies in angular momentum operators, then one defines the harmonics to be eigenfunctions of the Laplace operator as shown in \cite{Tho1980}, or, from the perspective of rotation operators invariant under rotations of a spherical coordinate system, \cite{KoJo1993}. The convention chosen here is similar to definitions typically used in seismology text books, see for example \cite{BMSi1981,AkRi2009}, and a nice introduction to these harmonics can be found in \cite{BEG1985}. Here, they are defined as
\beq
\begin{split}
\vec Y_l^m(\theta,\phi) &=Y_l^m(\theta,\phi)\vec e_r \\
\vec \Psi_l^m(\theta,\phi) &=\frac{1}{\sqrt{l(l+1)}}r\nabla Y_l^m(\theta,\phi) \\
\vec \Phi_l^m(\theta,\phi) &=\frac{1}{\sqrt{l(l+1)}}\vec r\times\nabla Y_l^m(\theta,\phi)
\end{split}
\label{eq:vectharm}
\eeq
Note that even though the radial coordinate $r$ appears explicitly in these definitions, it cancels when carrying out the gradient operations. The normalization differs from most other publications since it is chosen to make the vector spherical harmonics orthonormal:
\beq
\begin{split}
\int\drm\Omega\,\vec Y_l^m(\theta,\phi) \cdot(\vec Y_{l'}^{m'}(\theta,\phi))^* &= \delta_{ll'}\delta_{mm'}\\
\int\drm\Omega\,\vec \Psi_l^m(\theta,\phi) \cdot(\vec \Psi_{l'}^{m'}(\theta,\phi))^* &= \delta_{ll'}\delta_{mm'}\\
\int\drm\Omega\,\vec \Phi_l^m(\theta,\phi) \cdot(\vec \Phi_{l'}^{m'}(\theta,\phi))^* &= \delta_{ll'}\delta_{mm'}
\end{split}
\eeq
Integrals involving the product of two different vector spherical harmonics vanish. Using the orthogonality relations, one can also calculate the integrals
\beq
\begin{split}
\int\drm\Omega\,\vec Y_l^m(\theta,\phi) &= \sqrt{\frac{2\pi}{3}}\delta_{l,1}\left(\delta_{m,-1}(\vec e_x-\irm\vec e_y)-\delta_{m,1}(\vec e_x+\irm\vec e_y)+\sqrt{2}\delta_{m,0}\vec e_z\right)\\
\int\drm\Omega\,\vec \Psi_l^m(\theta,\phi) &= \sqrt{\frac{4\pi}{3}}\delta_{l,1}\left(\delta_{m,-1}(\vec e_x-\irm\vec e_y)-\delta_{m,1}(\vec e_x+\irm\vec e_y)+\delta_{m,0}\vec e_z\right)\\
\int\drm\Omega\,\vec \Phi_l^m(\theta,\phi) &= 0
\end{split}
\label{eq:intvecharm}
\eeq
Vector spherical harmonics are essential in calculations of scattered seismic fields. In some cases, the scattering problem can be formulated in terms of scalar quantities, but in general, as shown in Section \ref{sec:scattershear}, the calculation requires the vector harmonics. The most important properties of vector spherical harmonics are expressed by the equations that involve differential operators. For our purposes, the gradient and divergence operators are the most important ones. For example, the gradient of a scalar spherical harmonic has the following form
\beq
\phi(\vec r\,) = f(r)Y_l^m(\theta,\phi),\quad\nabla\phi(\vec r\,)=(\partial_r f(r))\vec Y_l^m(\theta,\phi)+\sqrt{l(l+1)}\dfrac{f(r)}{r}\vec\Psi_l^m(\theta,\phi),
\label{eq:gradientscalar} 
\eeq
while the divergence of the vector spherical harmonics reads
\beq
\begin{split}
{\rm div}(f(r)\vec Y_l^m(\theta,\phi)) &=\left((\partial_rf(r))+2\dfrac{f(r)}{r}\right)Y_l^m(\theta,\phi)\\
{\rm div}(f(r)\vec \Psi_l^m(\theta,\phi)) &=-\dfrac{\sqrt{l(l+1)}}{r}f(r) Y_l^m(\theta,\phi) \\
{\rm div}(f(r)\vec \Phi_l^m(\theta,\phi)) &=0
\end{split}
\label{eq:divvecharm}
\eeq
As a second example, we give expansions of simple vector fields that we will need later again. Expressed in vector harmonics as defined in this paper, the solution for a longitudinal plane wave reads:\index{plane-wave expansion!vector,longitudinal}
\beq
\begin{split}
\e^{-\irm kz}\vec e_z = \sum\limits_{l=0}^\infty\bigg[&\sqrt{\frac{4\pi}{2l+1}}(-\irm)^{l+1}\left((l+1)j_{l+1}(kr)-lj_{l-1}(kr)\right)\vec Y_l^0(\theta,\phi)\\
&-\sqrt{\frac{4\pi}{2l+1}}\sqrt{l(l+1)}(-\irm)^{l+1}\left(j_{l+1}(kr)+j_{l-1}(kr)\right)\vec \Psi_l^0(\theta,\phi)\bigg]
\end{split}
\label{eq:expandPWlong}
\eeq
As usual, expansion coefficients can be calculated by integrating products of the left-hand side of the equation with vector spherical harmonics. The exact form of the result given here can be obtained by subsequently using the recurrence relations of spherical Bessel functions as given in Equation (\ref{eq:sphBesselrec}). Transversal waves have a more complicated expansion into vector spherical harmonics:\index{plane-wave expansion!vector,transversal}
\beq
\begin{split}
\e^{-\irm kz}\vec e_x = \sum\limits_{l=1}^\infty\bigg[&\sqrt{\frac{\pi l(l+1)}{2l+1}}(-\irm)^{l+1}(j_{l+1}(kr)+j_{l-1}(kr))(\vec Y_l^1(\theta,\phi)-\vec Y_l^{-1}(\theta,\phi))\\
&+\sqrt{\frac{\pi}{2l+1}}(-\irm)^{l+1}(-lj_{l+1}(kr)+(l+1)j_{l-1}(kr))(\vec \Psi_l^1(\theta,\phi)-\vec \Psi_l^{-1}(\theta,\phi))\\
&+\sqrt{\pi(2l+1)}(-\irm)^{l+1}j_l(kr)(\vec \Phi_l^1(\theta,\phi)+\vec \Phi_l^{-1}(\theta,\phi))\bigg]\\
\e^{-\irm kz}\vec e_y = \sum\limits_{l=1}^\infty\bigg[&-\sqrt{\frac{\pi l(l+1)}{2l+1}}(-\irm)^l(j_{l+1}(kr)+j_{l-1}(kr))(\vec Y_l^1(\theta,\phi)+\vec Y_l^{-1}(\theta,\phi))\\
&-\sqrt{\frac{\pi}{2l+1}}(-\irm)^l(-lj_{l+1}(kr)+(l+1)j_{l-1}(kr))(\vec \Psi_l^1(\theta,\phi)+\vec \Psi_l^{-1}(\theta,\phi))\\
&-\sqrt{\pi(2l+1)}(-\irm)^lj_l(kr)(\vec \Phi_l^1(\theta,\phi)-\vec \Phi_l^{-1}(\theta,\phi))\bigg]
\end{split}
\label{eq:expandPWtrans}
\eeq
As complicated as these expressions may seem, they greatly simplify more complicated calculations, especially of scattering problems as shown in Section (\ref{sec:scatterNN}).

\subsubsection{Solid scalar spherical harmonics}
\label{sec:solidharm}
\index{spherical harmonics!scalar, solid}
Expanding a square-integrable field in terms of spherical harmonics, the expansion coefficients will generally be functions of the radial coordinate $r$. If the field is a solution of the Laplace equation, then it is easy to show using Equation (\ref{eq:eigenY}) that the radial dependence can only have the two forms $r^l$ and $1/r^{l+1}$. Therefore, it is convenient to define so-called solid spherical harmonics, which directly incorporate $r$ into the expansion. A nice review of solid spherical harmonics can be found in \cite{StRu1973}. To introduce the solid spherical harmonics, we start with a well-known expansion of the inverse distance:
\beq
\frac{1}{|\vec r-\vec r\,'|}=\dfrac{1}{(r^2+(r')^2-2rr'\cos(\gamma))^{1/2}}= \frac{1}{r_>}\sum\limits_{l=0}^\infty\left(\frac{r_<}{r_>}\right)^lP_l(\cos(\gamma))
\label{eq:potexpand}
\eeq
where $r_>\equiv \max(r,r')$, $r_<\equiv \min(r,r')$, and $\gamma$ is the angle between the two vectors $\vec r,\,\vec r\,'$. This equation is known as Laplace expansion\index{Laplace expansion} of the distance between two points. The expansion was later generalized to arbitrary powers of the distance, which can often serve as a short cut for calculations \cite{Sac1964a,Sac1964b,Sac1964c}.

This equation is not always directly helpful since the two position vectors $\vec r,\vec r\,'$ are often defined in a coordinate system that does not allow us to provide a simple expression of the angle $\gamma$. This can make it very difficult to calculate integrals of this expansion over angular coordinates. Another important relation, known as spherical harmonic addition theorem, can solve this problem:
\beq
P_l(\cos(\gamma))=\frac{4\pi}{2l+1}\sum\limits_{m=-l}^l\left(Y_l^m(\theta',\phi')\right)^*Y_l^m(\theta,\phi),
\label{eq:spheraddition}
\eeq
where $\gamma$ is now reexpressed in terms of the angular spherical coordinates $(\theta,\phi)$ and $(\theta',\phi')$ of the two position vectors. Together with Equation (\ref{eq:potexpand}), the Laplace expansion can be rewritten as
\beq
\frac{1}{|\vec r-\vec r\,'|}=\sum\limits_{l=0}^\infty\sum\limits_{m=-l}^l\left(I_l^m(\vec r_>)\right)^*R_l^m(\vec r_<)
\eeq
with the solid spherical harmonics defined in Racah's normalization 
\beq
R_l^m(\vec r\,)\equiv\sqrt{\dfrac{4\pi}{2l+1}}r^lY_l^m(\theta,\phi),\quad
I_l^m(\vec r\,)\equiv\sqrt{\dfrac{4\pi}{2l+1}}\dfrac{Y_l^m(\theta,\phi)}{r^{l+1}}
\label{eq:solidharm}
\eeq
The functions $R_l^m(\cdot),\,I_l^m(\cdot)$ are the regular and irregular solid spherical harmonics, respectively. The explicit expressions of the first three degrees are listed in Table \ref{tab:solidreg}.
\begin{table}[ht!]
\caption{Regular and irregular solid harmonics in Racah normalization}
\label{tab:solidreg}
\renewcommand{\arraystretch}{2.5}
\centerline{
\begin{tabular}{|l|l|}
\hline
$R_0^0$ & $1$\\
\hline
$R_1^0$ & $r\cos(\theta)$\\
$R_1^{\pm 1}$ & $\mp\dfrac{r}{\sqrt{2}}\sin(\theta)\e^{\pm\irm\phi}$\\
\hline
$R_2^0$ & $\dfrac{r^2}{2}(3\cos^2(\theta)-1)$\\
$R_2^{\pm 1}$ & $\mp r^2\sqrt{\dfrac{3}{2}}\sin(\theta)\cos(\theta)\e^{\pm\irm\phi}$\\
$R_2^{\pm 2}$ & $\dfrac{r^2}{2}\sqrt{\dfrac{3}{2}}\sin^2(\theta)\e^{\pm 2\irm\phi}$\\
\hline
\end{tabular}
\begin{tabular}{|l|l|}
\hline
$I_0^0$ & $\dfrac{1}{r}$\\
\hline
$I_1^0$ & $\dfrac{1}{r^2}\cos(\theta)$\\
$I_1^{\pm 1}$ & $\mp\dfrac{1}{r^2\sqrt{2}}\sin(\theta)\e^{\pm\irm\phi}$\\
\hline
$I_2^0$ & $\dfrac{1}{2r^3}(3\cos^2(\theta)-1)$\\
$I_2^{\pm 1}$ & $\mp \dfrac{1}{r^3}\sqrt{\dfrac{3}{2}}\sin(\theta)\cos(\theta)\e^{\pm\irm\phi}$\\
$I_2^{\pm 2}$ & $\dfrac{1}{2r^3}\sqrt{\dfrac{3}{2}}\sin^2(\theta)\e^{\pm 2\irm\phi}$\\
\hline
\end{tabular}}
\end{table}
With an appropriate definition of vector surface spherical harmonics, different from Equation (\ref{eq:vectharm}), since the surface harmonics need to be eigenfunctions of the Laplace operator, one could also define solid vector spherical harmonics. They will not be required in this review article.

\subsection{Spherical multipole expansion}
\label{sec:multipole}
The expansion of scalar and vector fields into spherical harmonics is an example of a so-called multipole expansion\index{multipole expansion}. We will see interesting applications in Section \ref{sec:objects}, but a simple example is discussed in this section already to illustrate the method. In the context of calculating gravity perturbations between two objects, the goal is to provide the multipole expansion of their mass distributions. These expansions come in two forms. If the two objects are much smaller than the distance between them, then it is possible to solve the problem in terms of the so-called exterior multipole moments\index{multipole moments!exterior} 
\beq
X_l^m\equiv \int\drm V\,\rho(\vec r\,)R_l^m(\vec r\,),
\label{eq:multiext}
\eeq
which require the regular solid harmonics. The moment $X_0^0$ is always equal to the total mass of the object. As outlined in Section \ref{sec:solidharm}, the coordinate vector $\vec r$ needs to be ``shorter'', in this case shorter than the distance between the two objects. However, since the length of the vector depends on the location of the origin of the coordinate system, and since only one of two distant objects can be close to the origin, a more complicated expansion scheme is required to make use of the exterior multipole moments of both objects. This problem is discussed in Section \ref{sec:momentinter}. Another possible scenario is that one mass is located inside another hollow mass. In this case, it is impossible to calculate their gravitational attraction using only exterior mass multipole moments. At least one mass distribution needs to be described in terms of its interior multipole moments\index{multipole moments!interior}
\beq
N_l^m\equiv \int\drm V\,\rho(\vec r\,)I_l^m(\vec r\,),
\label{eq:multiint}
\eeq
In the remainder of this section, the calculation of an example will highlight the effect of symmetry of mass distributions on their multipole moments. For this purpose, we consider $N$ point masses regularly distributed on a circle as shown in Figure \ref{fig:pointring}. The results could for example be used to approximate the mass multipole moments of a rotor. The mass density of a point mass $m_i$ at $\vec r_i=(r_i,\theta_i,\phi_i)$ can be written in spherical coordinates as
\beq
\rho(\vec r\,)=\frac{M_i}{r^2\sin(\theta)}\delta(r-r_i)\delta(\theta-\theta_i)\delta(\phi-\phi_i)
\eeq
We want to use this example to explore the effect of simple symmetries in multipole expansion. The mass is considered to lie on a circle with radius $R$, so that we can choose $r_i=R$ and $\theta_i=\pi/2$. Together with Equation (\ref{eq:pointrelY}), we find
\beq
\begin{split}
R_l^m(r=R,\theta=\pi/2,\phi) &= R^lK_l^m\e^{\irm m\phi}\\
K_l^m &=
\begin{cases} 
    0 & l+m\quad\mbox{odd} \\
    \frac{1}{2^l}(-1)^{(l+m)/2}\dfrac{\sqrt{(l+m)!(l-m)!}}{((l+m)/2)!((l-m)/2)!} & l+m\quad\mbox{even}
\end{cases}
\end{split}
\eeq
Therefore the exterior multipoles of a point mass $M_i$ at $\vec r_i=(R,\pi/2,\phi_i)$ can be written
\beq
X_l^m(\vec r_i)=M_iR^lK_l^m\e^{\irm m\phi_i}
\eeq
This result means that all multipole moments of a point mass with odd $l+m$ vanish, whereas moments with even $l+m$ are nonzero independent of $\phi_i$. Now we consider two point masses at antipodal locations $\phi_1=0,\,\phi_2=\pi$ at the same distance $R$ to the origin and the same mass $M$. The multipoles are given by
\beq
X_l^m=MR^lK_l^m\left(1+(-1)^m\right)
\eeq
Therefore, $m$ needs to be even for non-vanishing multipole moments, which also means that $l$ needs to be even. 
\epubtkImage{}{%
    \begin{figure}[htbp]
    \centerline{\includegraphics[width=0.3\textwidth]{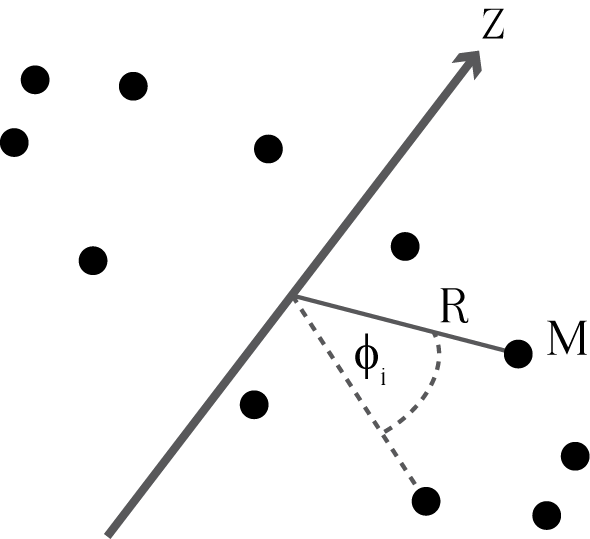}}
    \caption[Point masses in a plane]{Symmetric configuration of point masses in a plane.}
    \label{fig:pointring}
    \end{figure}}
As a last example, we add two more point masses, so that the configuration now consists of four equal masses at $\phi_1=0,\,\phi_2=\pi/2,\,\phi_3=\pi,\,\phi_4=3\pi/2$. The multipoles moments are
\beq
X_l^m=MR^lK_l^m\left(1+(-1)^m\right)\left(1+\irm^m\right)
\eeq
Now $m$ needs to be a multiple of 4 to generate a non-vanishing moment, and $l$ needs to be even as in the previous case. For $N$ point masses, we have
\beq
X_l^m=MR^lK_l^m\dfrac{1-\e^{2\pi\irm m}}{1-\e^{2\pi\irm m/N}}
\eeq
The fraction is equal to $N$ for $m$ being a multiple of $N$ (including $m=0$), and 0 otherwise. As before $l+m$ needs to be even for non-vanishing $K_l^m$. The limit $N\rightarrow\infty$ turns the collection of point masses into a continuous ring, which can be obtained as finite limit by expressing the individual point mass in terms of the total mass of the ring as $M=M_{\rm ring}/N$. In this case only the $m=0$ moments do not vanish. This is a property of multipole moments of all axially symmetric mass distributions provided that the angle $\theta$ is measured with respect to the symmetry axis. The only non-vanishing moment of spherically symmetric mass distributions with total mass $M$ is $X_0^0=M$. 

\subsection{Clebsch-Gordan coefficients}
\label{sec:clebsch}\index{Clebsch-Gordan coefficients}
Clebsch-Gordan coefficients $\langle l_1,m_1;l_2,m_2|L,M\rangle$ are required for the bipolar expansion discussed in Section \ref{sec:momentinter}. In general, they can be calculated recursively according to
\beq
\begin{split}
& C_\pm(L,M)\langle l_1,m_1;l_2,m_2|L,M\pm 1\rangle = \\
&\qquad C_\pm(l_1,m_1\mp 1)\langle l_1,m_1\mp 1;l_2,m_2|L,M\rangle+C_\pm(l_2,m_2\mp 1)\langle l_1,m_1;l_2,m_2\mp 1|L,M\rangle
\end{split}
\label{eq:recclebsch}
\eeq
where the integer parameters can assume the values $l_1\geq 0$, $l_2\geq 0$, $m_1=-l_1,\ldots l_1$, $m_2=-l_2,\ldots l_2$, $0\leq L\leq l_1+l_2$, $M=m_1+m_2$ and
\beq
C_\pm(l,m)\equiv\sqrt{l(l+1)-m(m\pm 1)},
\eeq
in Condon-Shortley phase convention. The Clebsch-Gordan coefficients obey the orthogonality relation:
\beq
\sum\limits_{m_1+m_2=M}\langle L,M|l_1,m_1;l_2,m_2\rangle\langle l_1,m_1;l_2,m_2|L,M\rangle=1
\label{eq:normclebsch}
\eeq
A practical method to calculate the coefficients using the recursion relation is based on a graphical scheme, which we are going to outline with the help of Figure \ref{fig:clebsch} for the case of $l_1=l_2=1$.
\epubtkImage{}{%
    \begin{figure}[htbp]
    \centerline{\includegraphics[width=0.6\textwidth]{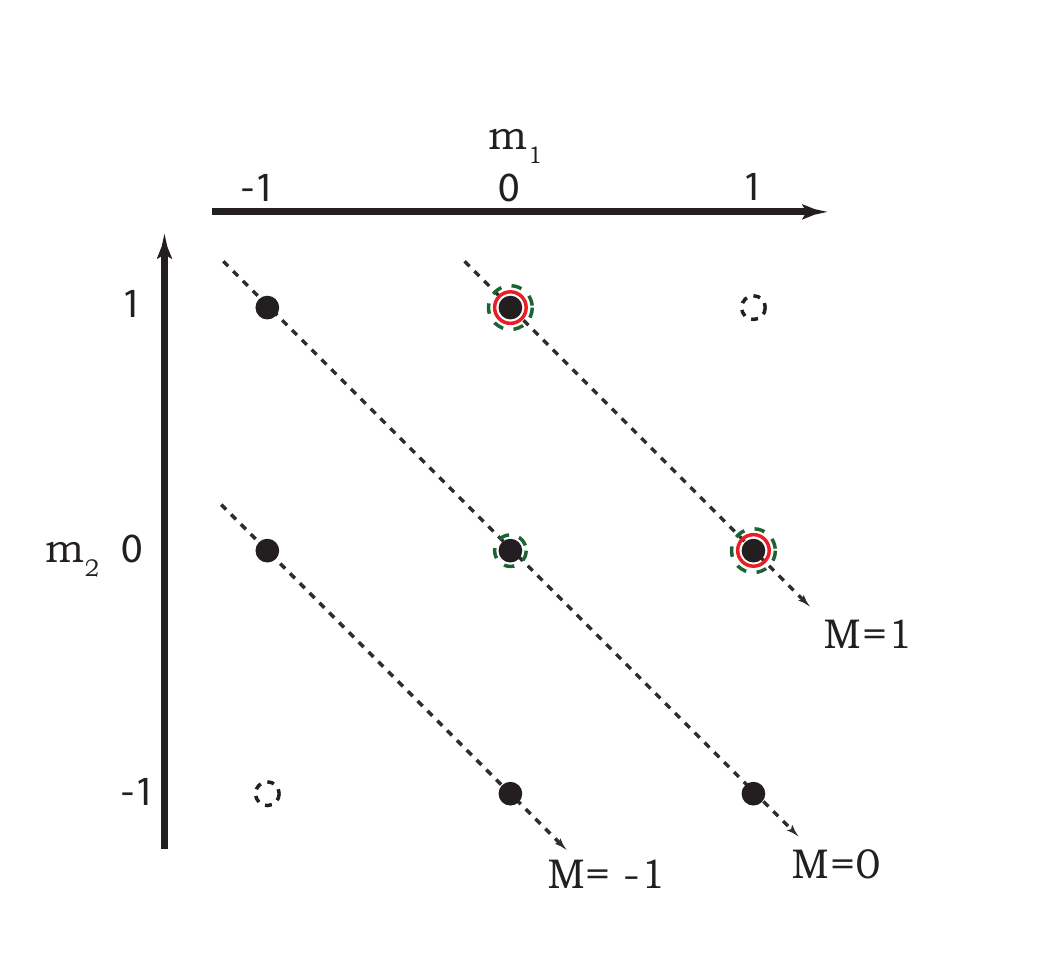}}
    \caption[Clebsch-Gordan coefficients]{Illustration of recursion relation for Clebsch-Gordan coefficients: $l_1,\,l_2=1$.}
    \label{fig:clebsch}
    \end{figure}}
The diagram shows a table with row index $m_2=-1,0,1$, and column index $m_1=-1,0,1$. The points in the diagram represent Clebsch-Gordan coefficients. Clebsch-Gordan coefficients are zero unless $M=m_1+m_2$. We know that the upper right corner must belong to $L=2$ since $M=2$. The two points marked with solid, red rings either belong to $L=2$ or $L=1$. Let us pick the value $L=1$ as example. Only the filled points represent possible coefficients in this case with $M=-1,0,1$. Now, inserting $M=1$ into Equation (\ref{eq:recclebsch}) and choosing the lower sign, the recursion relation links three points as for example the three marked with dashed, green rings. If the values of two points of a triangle are known, then the value of the third can be calculated. If we choose the point $m_1=1,\,m_2=0$ as upper corner of such a triangle, then the recursion relation only involves two coefficients. The lower-right corner of the triangle is off the diagram and therefore zero.  Starting from there, one can fill in the values of all other points using the recursion relation. The orientation of the triangle formed by the green marked points can be flipped across the diagonal by using the other sign in Equation (\ref{eq:recclebsch}). We can set the value of one coefficient equal to 1, and later use Equation (\ref{eq:normclebsch}) to give all coefficients the correct normalization. Equation (\ref{eq:normclebsch}) says that the sum of squares of coefficients along a $M=\rm const$ diagonal is equal to 1. Note that all coefficients need to be recalculated for a different value of $L$. Nonetheless, the procedure is straight-forward, and one only needs to set up a new diagram for each combination of values of $l_1$, $l_2$.
    
\subsection{Noise characterization in frequency domain}
\label{sec:noisefreq}
In this section, we give a brief introduction into frequency-domain functions used to characterize random processes. We will assume throughout this section that the random processes are Gaussian and stationary. Gaussianity implies that variances, correlations, and their spectral variants, i.~e.~power spectral densities and cross spectral densities, give a complete characterization of the noise. The role of stationarity is explained below. This does not mean that the presented equations are not useful in practice, when noise is non-stationary, and non-Gaussian, but then one needs to be more careful than we want to be in this article. For stationary random processes the auto-correlation between measurements at two different times $t_1,\,t_2$ is only a function of the difference $\tau=t_2-t_1$. In this case, the power spectral density can be defined as the Fourier transform of the auto-correlation with respect to $\tau$:\index{spectral density}
\beq
S(x;\omega)=2\int\drm\tau \langle x(t)x(t-\tau)\rangle \e^{-\irm\omega\tau}
\label{eq:defspecdens}
\eeq
This equation assumes stationary noise $x(t)$. If noise is non-stationary, then the spectrum $S(x;\omega)$ explicitly depends on the time $t$. Another property of stationary noise is that Fourier amplitudes of the random process at different frequencies are uncorrelated:
\beq
\langle x(\omega)x^*(\omega')\rangle=2\pi \frac{1}{2}S(x;\omega)\delta(\omega-\omega')
\label{eq:gaussian}
\eeq
The left-hand side is an ensemble average over many noise realizations. Since a stationary random process has a constant expectation value of its noise power for all times $-\infty<t<\infty$, its Fourier transform does not exist strictly speaking. This is the reason why the right-hand side involves a $\delta$-distribution. A more suitable form is obtained by integrating the last equation over frequency $\omega'$. Considering the product of Fourier amplitudes of two different random processes, one obtains 
\beq
2\int\limits_0^\infty\frac{\drm\omega'}{2\pi}\langle x(\omega)y^*(\omega')\rangle= S(x,y;\omega)
\eeq
The cross spectral density\index{spectral density!cross} $S(x,y;\omega)$ is equal to the Fourier transform of the cross-correlation $\langle x(t)y(t-\tau)\rangle$. In this article, the cross spectral density will often be denoted as $\langle x(\omega),y(\omega)\rangle$ and referred to as correlation function. 

A typical case is that the two quantities $x(t),\,y(t)$ represent measurements of a field at two potentially different locations. In this case, the correlation function can be cast into the form
\beq
\langle x(\vec r_1,\omega),x(\vec r_2,\omega)\rangle= S(x;\omega)\mathpzc{r}(\vec r_1,\vec r_2\,)
\label{eq:twopoint}
\eeq
with $\mathpzc{r}(\vec r,\vec r\,)=1$. In practice, correlation functions are calculated based on plane-wave (or normal-mode) solutions. A field can then be represented as a superposition of plane waves, and field correlations are obtained by averaging the plane-wave correlations over wave parameters such as propagation directions $\vec e_k$ and polarizations $\vec p$. If the random field is isotropic, stationary, and unpolarized, then different modes $\vec k,\,\vec p$ are uncorrelated \cite{AlRo1999}. Consequently, we can focus on correlations between waves that are described by the same parameters:
\beq
\langle x_{\vec k,\vec p}(\vec r_1,\omega),y_{\vec k,\vec p}(\vec r_2,\omega)\rangle = S(x,y;\omega)\mathpzc{s}(\vec k,\vec p)\e^{\irm \vec k\cdot(\vec r_2-\vec r_1)}
\eeq
Note that this expression is evaluated for fixed wave parameters, and the only random quantities in this equation are the complex (scalar) amplitudes of the two waves. As a next step, we consider the field as a superposition of waves with random polarization and propagation direction. Averaging over directions and polarizations, we find
\beq
\langle x_k(\vec r_1,\omega),y_k(\vec r_2,\omega)\rangle = S(x,y;\omega)\frac{1}{4\pi P}\int\drm\vec p\int\drm\Omega_k\,\mathpzc{s}(\vec k,\vec p)\e^{\irm \vec k\cdot(\vec r_2-\vec r_1)}
\label{eq:twoproccorr}
\eeq
Here, $P\equiv\int\drm\vec p$ is the measure of the integral over all polarization parameters, and since the number of polarizations is finite, the integral can also be rewritten as a sum over polarizations. The last equation is formally identical to the definition of the so-called overlap reduction function, which describes correlations between measurements of a stochastic GW background at two different locations \cite{Chr1992,Fla1993}. If the two random processes represent measurements of the same (scalar) field, then together with Equation (\ref{eq:twopoint}), we have
\beq
\mathpzc{r}(\vec r_1,\vec r_2\,)=\frac{1}{4\pi P}\int\drm\vec p\int\drm\Omega_k\,\mathpzc{s}(\vec k,\vec p)\e^{\irm \vec k\cdot(\vec r_2-\vec r_1)}
\eeq
Two-point correlation functions can have rich structure if the two random processes in Equation (\ref{eq:twoproccorr}) represent projections of a vector or tensor field at two different locations. Examples of this case can be found in Section \ref{sec:cohcancel}.


\bibliography{c:/MyStuff/references}

\newpage
\printindex

\end{document}